%% file: fernando-phd-thesis.tex
\documentclass[UKenglish]{uiophd}
\usepackage[subpreambles=true,mode=buildnew]{standalone}
\usepackage[utf8]{inputenc}
\usepackage{amsmath,amssymb,mathtools}
\usepackage{tikz}
\usepackage{subfiles}
\usepackage{xr-hyper}
\usepackage{hyperref}
\usepackage{import}
\usepackage{dsfont}
\usepackage{graphicx}
\usepackage{caption}
\usepackage{subcaption}
\usepackage{wrapfig}
\usepackage{etoolbox}
\usepackage{tabularx}
\usepackage[color,all]{xy}
\usepackage{listings}
\usepackage{multirow}
\usepackage{algorithm}
\usepackage{algpseudocode}
\usepackage{fancybox}
\usepackage{afterpage} 
\usepackage[characters]{styles/ltl}

\subimport{styles/}{styles.tex}
\subimport{styles/}{formalisation-commands.tex}
\subimport{styles/}{commutative-diagrams-styles.tex}

\subimport{styles/}{ltl-styles.tex}

\title{MULTILEVEL MODELLING AND DOMAIN-SPECIFIC LANGUAGES}

\author{Doctoral Dissertation by\\Fernando Macías}

\begin{document}

\newcommand{\mainfile}{} 

\frontmatter

\pagenumbering{gobble} 

\maketitle

\clearpage
\vspace*{\fill}
\noindent
Fernando Macías: \textit{Multilevel Modelling and Domain-Specific Languages}

\noindent
\textcopyright\ February 2019

\clearpage
\begin{center}
    \thispagestyle{empty}
    \vspace*{\fill}
    \hfill\textit{To Marina}
    \vspace*{\fill}
\end{center}
\clearpage

\afterpage{\null\newpage}

\subfile{thesis/00-preface}

\tableofcontents

\mainmatter

\pagenumbering{arabic} 

\subfile{thesis/01-introduction}

\subfile{thesis/02-mlm}

\subfile{thesis/03-mcmt}

\subfile{thesis/04-tooling}

\subfile{thesis/05-validation}

\subfile{thesis/06-discussion}

\bibliographystyle{abbrv}
\bibliography{bib/fmac,bib/bibliography}

\end{document}

%% file: styles/styles.tex
\usepackage{amsthm}

\newcommand{\algstyle}[1]{\ensuremath{\mathit{#1}}}

\newcommand{\elementname}[1]{\texttt{#1}}
\newcommand{\potencyseparator}{@}
\newcommand{\typename}[2]{\texttt{#1\potencyseparator#2}}
\newcommand{\element}[3][]{\ifstrempty{#1}{\texttt{#2:#3}}{\texttt{#2:#3\potencyseparator#1}}}

\theoremstyle{plain} 
\newtheorem{definition}{Definition} 

\theoremstyle{definition} 

\newtheorem{lemma}{Lemma} 
\newtheorem{corollary}{Corollary}
\newtheorem{condition}{Condition}

\newtheorem{proof}{Proof}

\theoremstyle{remark} 
\newtheorem*{remark}{Remark}

%% file: styles/formalisation-commands.tex
\usepackage{xargs}
\usepackage{etoolbox} 

\newcommand{\graphname}[3][]{\ensuremath{#2_{#3}^{#1}}} 
\newcommand{\chain}[3][]{(\overline{#2}_{#1},{#3},\tau^{#2}_{#1})} 
\newcommandx{\cat}[2][2]{\mbox{\boldmath\ensuremath{\mathsf{#1}}\unboldmath}#2}
\newcommand{\chainname}[2][]{\ensuremath{\mathcal{#2}_{#1}}}
\newcommand{\source}[1][]{\mathit{sc}^{#1}}
\newcommand{\target}[1][]{\mathit{tg}^{#1}}
\newcommand{\typeof}[2][]{\ensuremath{\mathit{ty}^{#1}(#2)}}

\newcommand{\typemorph}[3][]{\ifstrempty{#2}{\ensuremath{\tau^{#1}}}{\ifstrempty{#3}{\tau_{#2}^{#1}}{\ensuremath{\tau_{#2,#3}^{#1}}}}}
\newcommand{\domain}[1]{D(#1)}
\newcommand{\tc}[1]{\overline{#1}}
\newcommand{\partialmap}{{\mbox{\(\hspace{0.7em}\circ\hspace{-1.3em}\longrightarrow\hspace{0.1em}\)}}}
\newcommand{\difference}[2][]{\ensuremath{\mathit{df}^{#1}\!(#2)}}
\newcommand{\step}[1]{\ensuremath{s_{#1}}}

\newcommand{\chainmorph}[2][]{\ensuremath{\sigma_{#1}^{#2}}}

\newcommand{\examplenodeone}{x}
\newcommand{\examplenodetwo}{y}
\newcommand{\examplearrowone}{f}
\newcommand{\indexone}{\ensuremath{i}}
\newcommand{\indextwo}{\ensuremath{j}}
\newcommand{\indexthree}{\ensuremath{k}}
\newcommand{\exampleindexone}{\ensuremath{a}}
\newcommand{\exampleindextwo}{\ensuremath{b}}
\newcommand{\chaindepthone}{\ensuremath{n}}
\newcommand{\chaindepthtwo}{\ensuremath{m}}
\newcommand{\chaindepththree}{\ensuremath{l}}
\newcommand{\variabledepthone}{\ensuremath{x}}
\newcommand{\examplegraphone}{\ensuremath{G}}
\newcommand{\examplegraphtwo}{\ensuremath{H}}
\newcommand{\examplegraphthree}{\ensuremath{K}}
\newcommand{\examplegraphfour}{\ensuremath{P}}
\newcommand{\variablegraphone}{\ensuremath{X}}
\newcommand{\graphnodes}{\ensuremath{N}}
\newcommand{\grapharrows}{\ensuremath{A}}
\newcommand{\morphone}{\ensuremath{\phi}}
\newcommand{\morphtwo}{\ensuremath{\psi}}
\newcommand{\morphthree}{\ensuremath{\kappa}}
\newcommand{\maplevelone}{\ensuremath{f}}
\newcommand{\mapleveltwo}{\ensuremath{g}}
\newcommand{\maplevelthree}{\ensuremath{h}}
\newcommand{\rulegraphleft}{\ensuremath{L}}
\newcommand{\rulegraphmiddle}{\ensuremath{I}}
\newcommand{\rulegraphright}{\ensuremath{R}}
\newcommand{\hierarchygraphleft}{\ensuremath{S}}
\newcommand{\hierarchygraphmiddle}{\ensuremath{D}}
\newcommand{\hierarchygraphright}{\ensuremath{T}}
\newcommand{\rulegraph}{\ensuremath{MM}}
\newcommand{\hierarchygraph}{\ensuremath{TG}}
\newcommand{\rulechain}{\ensuremath{\chain{\rulegraph}{\chaindepthone}}}
\newcommand{\hierarchychain}{\ensuremath{\chain{\hierarchygraph}{\chaindepthtwo}}}

\newcommand{\maprulelefttomiddle}{\ensuremath{\lambda}}
\newcommand{\maprulerighttomiddle}{\ensuremath{\rho}}
\newcommand{\maphierarchylefttomiddle}{\ensuremath{\varsigma}}
\newcommand{\maphierarchyrighttomiddle}{\ensuremath{\theta}}
\newcommand{\mapruletohierarchybinding}{\ensuremath{\beta}}
\newcommand{\mapruletohierarchyleft}{\ensuremath{\mu}}
\newcommand{\mapruletohierarchymiddle}{\ensuremath{\delta}}
\newcommand{\mapruletohierarchyright}{\ensuremath{\nu}}
\newcommand{\po}{\ensuremath{PO}}
\newcommand{\pb}{\ensuremath{PB}}

\newcommand{\fpbc}{\ensuremath{FPBC}}

%% file: styles/commutative-diagrams-styles.tex
\usetikzlibrary{arrows,positioning,decorations.markings,backgrounds}

\newcommand{\subscripttt}[1]{\scalebox{.7}{#1}}
\newcommand{\subsubscripttt}[1]{\scalebox{.6}{#1}}
\newcommand{\subsubsubscripttt}[1]{\scalebox{.4}{#1}}

\definecolor{cdblue}{RGB}{0,0,255}
\definecolor{cdgreen}{RGB}{0,163,0}
\definecolor{cdred}{RGB}{255,0,0}

\tikzset{math/.style={execute at begin node=$, execute at end node=$}}
\tikzset{element/.style={inner sep=2pt,minimum height=1.3em}}
\tikzset{el-math/.style={element,math}}
\tikzset{label/.style={auto,midway,inner sep=2pt}}
\tikzset{la-math/.style={label,math}}
\tikzset{map-base/.style={->,>=stealth',semithick}}
\tikzset{map/.style={map-base}}
\tikzset{map2/.style={map-base,color=black!20}}
\tikzset{mapdots/.style={map-base,densely dotted}}
\tikzset{incmapl/.style={map-base,right hook->}}
\tikzset{incmapr/.style={map-base,left hook->}}
\tikzset{typemap/.style={map-base,dashed,blue}}
\tikzset{partmap/.style={map-base,decoration={markings,mark=at position 0.5 with {\draw circle [radius=.4ex];}},postaction={decorate}}}
\tikzset{partmap2/.style={partmap,color=black!20}}
\tikzset{partmapdots/.style={mapdots,decoration={markings,mark=at position 0.5 with {\node[draw,solid,name=s,shape=circle,inner sep=0pt,minimum size=.8ex] {};}},postaction={decorate}}}
\tikzset{mapto/.style={map-base,|->}}
\tikzset{doublemap/.style={->,double,>=stealth'}}

%% file: styles/ltl-styles.tex
\newcommand{\www}[1]{\href{http://#1}{\nolinkurl{#1}}}

\newcommand{\LTLprophecy}[1]{\triangleright_{[#1]}}

\let\phi\varphi

\newcommand{\circled}[1]{
  \setbox0=\hbox{#1}%
  \dimen0\wd0%
  \divide\dimen0 by 2%
  \tikz[baseline=(a.base)]{%
      \useasboundingbox (-\the\dimen0,0pt) rectangle (\the\dimen0,1pt);%
      \node[circle,draw,outer sep=0pt,inner sep=0.1ex] (a) {#1};%
  }\,}

\newcommand{\circledblue}[1]{
  \setbox0=\hbox{#1}%
  \dimen0\wd0%
  \divide\dimen0 by 2%
  \tikz[baseline=(a.base)]{%
      \useasboundingbox (-\the\dimen0,0pt) rectangle (\the\dimen0,1pt);%
      \node[circle,blue,draw,outer sep=0pt,inner sep=0.1ex] (a) {#1};%
  }\,}

\newcommand{\bluecircled}[1]{
  \setbox0=\hbox{#1}%
  \dimen0\wd0%
  \divide\dimen0 by 2%
  \tikz[baseline=(a.base)]{%
      \useasboundingbox (-\the\dimen0,0pt) rectangle (\the\dimen0,1pt);%
      \node[circle,draw,outer sep=0pt,inner sep=0.1ex,white,fill=blue] (a) {#1};%
  }\,}

%% file: thesis/00-preface.tex
\chapter*{Preface}

Hey there, dear reader!
Glad to see that you are browsing these pages.
It took me quite some time to get this thesis to the final state that you can see, and I hope you find it interesting and maybe even useful.
In any case, once you are done with it, I hope you can appreciate the amount of work and dedication that it took me to complete it, and that you can find a small spot for it in your bookshelf.
Thank you, and enjoy the reading!

\vspace{1ex}

\hfill
\textit{Bergen, 08.02.2019}

\chapter*{Scientific environment}

The research presented in this dissertation has been conducted at the Department of Computing, Mathematics and Physics at Western Norway University of Applied Sciences, in cooperation with the Department of Informatics at University of Oslo.

The work in this thesis was partially supported by the bilateral projects ``Modern Refactoring'' UTF-2016-CAPES-SIU/10032 and ``Methods and Tool Support for Refinement, Model Transformation and Verification of Network Systems'' UTF-2016-short-term/10049, and the ERASMUS Staff Mobility program.

\chapter*{Acknowledgements}

My first and biggest thanks belong to Adrian Rutle and Volker Stolz, my supervisors during the last four years.
They have always helped me in every step of the way, and I can never thank them enough for their wisdom, patience, hard work and kindness.
Not only were they there whenever I needed some wise words, motivation or simply a chance to talk.
But both of them, and their families (thanks to them too!), have welcomed me at their homes when I needed to push some extra work hours, or just to have a chat and relax over delicious home-made food.
And last but not least, thanks to Adrian for teaching me to ski and to Volker for taking me with him around the world.

Not only I had the best official supervisors I could wish for, but I am also indebted to those who helped me so much during my time as PhD candidate, that I consider them my ``unofficial co-supervisors''.
First, I must thank Uwe Wolter for his perseverance when trying to push Category Theory into my underprepared engineer mind.
The amount of time he selflessly used to guide me inside a forest of pushouts and functors is unrivalled.
Without him, the formal aspects of this thesis would not be the same.
Next, I want to express my gratitude to Roberto Rodriguez Echeverría, who supervised my BSc and MSc theses, and never stopped being a mentor since then.
It is thanks to him that I dared to take a PhD in Norway in the first place.
And to Francisco Durán, I am grateful for his help on figuring out key aspects of the work presented here, for helping me put them into words, and for welcoming me in Málaga.

Thanks to my opponents Thomas Kühne and Reiko Heckel, for all their dedication and accurate comments aimed at making this thesis better, and to Birger Møller-Pedersen for leading the committee.
Also to all those reviewers, both known and anonymous, who reviewed the earlier versions of this manuscript and all the papers that helped shape the work it presents.

To my two closest colleagues during my first years, then friends and finally also flatmates, Ajith and Rui, I am forever thankful for their everyday company and their support.
They have been the best travel companions one could wish for, and thanks to them I had a home in Bergen, not just a house.
Thanks to Rabbi, who always helped me with his experience regarding both paperwork and science.
Also thank you to all the other PhD candidates and post-docs at HVL, past and present, for the great working atmosphere and the very-much-needed lunch breaks that helped me relax and focus when I needed it the most.
Special mentions here to my (e)Spanish friends Álex and Ángela for everything they did for me, including offering me a place to stay in the last weeks of my PhD; to Patrick, who was a great (and far more experienced) skiing companion and a great friend; to Simon and Lucas, who were always willing to have a chat over some beers; and to Svetlana, for a lot of helpful advice and great conversations.
Thank you also to Iril, Alejandro and all the nice people from BSI Dans for all the fun activities and for letting me enjoy their company.

I have been involved with several institutions, inside and outside Norway, and in all of them I have met excellent people who made the experience rewarding.

Having such a great workplace would not be possible without the other people at the ICT department, too many to mention them here but I am still grateful to every single one of them.
This includes everyone from the Department of Informatics at the University of Oslo, with special mention to Violet Ka I Pun, and also everyone at the miso group from the Autonomous University of Madrid, who were my closest colleagues and great friends for half a year.
I would like to also thank all the people from University of Cáceres for their warm welcomes every time I visited them and cheering me up towards the end of my PhD.
They were my first colleagues at the beginning of my career in research, and I wish to stay in contact for many more years.
Thank you also to Alessandro Rossini for his help an advice at several crucial points during these four years, and for being a great reference to follow.
And finally, thank you to Antonio Vallecillo, for showing interest and sympathy about my progress every time we met and for providing me with guidance during the first steps of my PhD.

I am really lucky to have a lot of people who supported me from Spain all this time.
Without them, getting through a PhD would have been much more difficult.

First of all, thanks to Marina, to whom I dedicate this thesis.
She has patiently endured years of long-distance relationship just so that I could have this opportunity, and never stopped pushing me to work harder and better, while providing me with unconditional affection, support and happiness.
I cannot thank you enough.

Thank you to my parents, for being my best role models and an unlimited source of comfort and advice.
I have tried to put to work the values they have taught me all my life, and any success I achieved I owe it to them.
And to my sister and brother, for the fun and positive mindset that I got from them every time I visited home.
Thank you also to my grandmother, always concerned about my happiness and well-being, for her wisdom and love.
Thanks to Marina's parents and brother, for almost adopting me and caring for me like just another member of their family.
And thanks to all my uncles, aunts and cousins who always ask about my experiences and wonder how it is to live among Vikings.

And last but not least, thank you to all my friends from Badajoz: Alberto, Dani, Juanmi y Prieto.
Once again, I went far from home for a long time, and once again you made sure to come here and prove that I can count on you for anything and everything.
Thanks to Sánchez, for his encouragement and for showing me that he was proud of me taking a PhD.
And thanks to all my friends from university, for staying in touch and willing to meet even if we are now spread all across Europe.
Thanks for Mario and Melania for visiting me before anyone else and insist that I visited them in London too.
To Carlos, for giving me a mug that fed me coffee all through these four years and always making me laugh when talking about music or bad movies.
And to Jose and Judit, for their visit, but most of all for their friendship.

I hope I do not forget anyone in these lines.
If I do, those who know me will now that the only reason is (again) my absent-mindedness and nothing else.
Forgotten or not, to all of you, a million thanks.

\chapter*{Abstract}

Modern software engineering must deal with growingly demanding problems that force the software solutions to increase in size and complexity.
The area of Model-Driven Software Engineering (MDSE) tackles this complexity by using models throughout the whole software development process and raises the level of abstraction at which software can be developed and maintained.
Although using models addresses some of the communication problems among the different stakeholders involved in the process, the use of Domain-Specific Modelling Languages (DSML) is a promising means for empowering domain experts, usually with non-technical profiles, to participate in the development.

Concepts with different levels of abstraction are required in the development of DSMLs, but traditional modelling approaches do not allow enough flexibility to organise such concepts adequately in multiple levels.
Conversely, Multilevel Modelling (MLM) approaches provide an unbounded number of levels of abstraction, together with other features which are a perfect fit for DSML development.
Moreover, the manipulation of domain-specific models can benefit from Model Transformations (MT), especially when such transformations encode the behavioural semantics of DSMLs.
MTs can be exploited in a multilevel setting, becoming a precise and reusable definition of the behaviour of modelling languages.

This thesis presents a Multilevel Modelling and Multilevel Model Transformation approach which tackles open issues in this area of research, and compares it with the state of the art, both through literature review and empirical experiments.
We provide the underlying formalisation of our approach using Category Theory and its implementation using MultEcore---a set of Eclipse-based tools.
Several case studies are also presented to exemplify our explanations and validate the approach.

\chapter*{List of abbreviations}

The following table lists the acronyms and abbreviations used in this thesis.

\vspace{1ex}

\begin{table}[h]
\begin{tabular}{ll}
	\hline
	\bfseries Abbreviation	& \bfseries Meaning							\\
	\hline
	MDSE					& Model-Driven Software Engineering			\\
	DSML					& Domain-Specific Modelling Language		\\
	EMF						& Eclipse Modelling Framework				\\
	UML						& Unified Modelling Language				\\
	MT						& Model Transformation						\\
	MLM						& Multilevel Modelling						\\
	MCMT					& Multilevel Coupled Model Transformation	\\
	LTL						& Linear Temporal Logic						\\
	PLS						& Product Line System						\\
	PO						& Pushout									\\
	PB						& Pullback									\\
	FPBC					& Final Pullback Complement					\\
	OCA						& Orthogonal Classification Architecture	\\
	
	\hline
\end{tabular}
\end{table}

%% file: thesis/01-introduction.tex
\chapter{Introduction}
\label{chap:introduction}

The discussion of abstract ideas and of the physical reality that surrounds us can be traced back, at least, to the early days of Western Philosophy~\cite{russell2013philosophy}.
Even before the Common Era, philosophers like Plato~\cite{gulley2013plato} pondered about the nature of the reality that surrounds us, and how that reality related to the ideas in our head and to our representations of those ideas.
In these early days, the concepts that now we call abstractions, classifications or ontologies~\cite{harvey1973archelogia}, started to emerge.
Furthermore, the need to structure these concepts appeared, and one natural way to do it is to somehow assign them to layers and create a kind of hierarchy.
Thanks to such hierarchical orderings, the discussion about these concepts and their relation to reality could start to advance.

Let us fast-forward a couple of millennia, and we get to the appearance of computing.
At first, the electromechanical machines that eventually evolved into today's computers were merely big calculators~\cite{williams1997history}, so that the discussion about programming them in a language with concepts that represented reality as we see and understand it was out of the question.
We can consider this \emph{step zero} in a bird's-eye view of the history of computing, where computers consisted of configurable hardware through physical rewiring of its components.
When this initial approach was outgrown by the increasingly demanding tasks assigned to computers, low-level programming arose.
This \emph{step one} for software consisted in the creation of assembly languages, whose code permitted the direct manipulation of the hardware's memory registries.

It is at this point where we can start talking about the discipline of software engineering.
Assembly languages evolved and grew more complex, eventually leading into \emph{step two}: high-level programming languages.
As the code grew in complexity, different paradigms to manage the architecture of software were proposed.
Arguably the most prevailing paradigm in this stage was object orientation, beginning with the Simula language~\cite{nygaard1978simula}.
This method attempted to imitate our understanding of reality by providing two well-defined levels of abstraction: the level of concepts (classes) that a language can manipulate, and the concrete realisation of such concepts.
One could now design, for example, the concept of Robot as a class, and include its details as attributes, like an integer with the number of wheels it has, or as references to other classes, like Sensor.
These classes could then be instantiated into objects representing specific entities, like the object Robot\_1 which represents a specific robot, which has a number of wheels equal to three and a reference to an object Proximity, of type Sensor, with the details of a sensor that this particular robot has.

Object orientation has been the main paradigm for software development in the last decades of software engineering.
Object-oriented programming languages are rich enough to represent abstract, real-world concepts in the software and use them to configure the behaviour of the hardware.
However, they are not a silver bullet.
Modern software engineering faces a growing complexity of the requirements and the solutions they demand, especially in terms of maintenance, which is in line with the trends identified in the first software crisis~\cite{randell1996nato} and later documented by other authors, like Boehm~\cite{boehm1976se}.

In a nutshell, the cost in money and time of software development and, especially, maintenance can be reduced drastically if we change the way we build software.
Moreover, modern approaches to software development can also improve the overall quality of the final software product.
One of the approaches towards this goal is Model-Driven Software Engineering (MDSE~\cite{brambilla2012mdse}).
In MDSE, models are the first-class citizens of the development process, instead of mere references or blueprints for coding.
A model can be any kind of high-level description of the system under development, both textual or (most commonly) graphical, and are normally created with general-purpose modelling languages.
Examples of these are Class Diagrams, Activity Diagrams and Sequence Diagrams, all part of the Unified Modelling Language (UML~\cite{odell1998advanced}), which have been in use for decades, which we can consider \emph{step three} in this overview.
The two main goals of models at the end of the MDSE process are either \emph{interpretation} or \emph{code generation}.
In the first case, the models can be directly understood by the modelling framework and therefore replace code altogether.
In the second case, a kind of compiler is used to (semi)automatically generate code from the models, avoiding many bugs and suboptimal code solutions which are often introduced in handwritten code.

\begin{figure}[ht]
	\centering
	\includegraphics{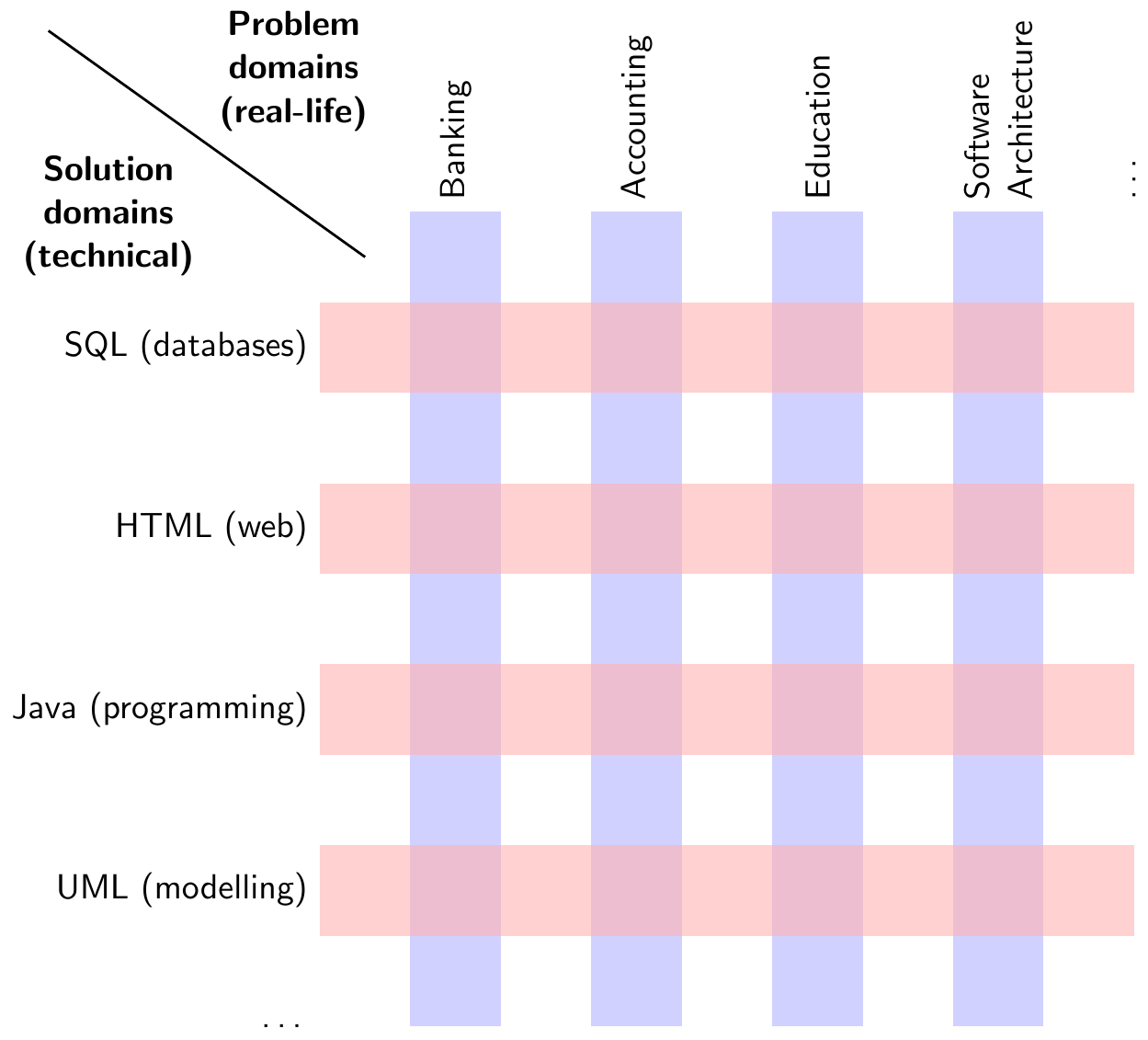}
	\caption{Problem and solution domains}
	\label{fig:problem-and-solution-domains}
\end{figure}

The MDSE approach is, however, not exempt of problems.
As with traditional Software Engineering, the use of general-purpose languages forces the engineer to specify many details about the software solution, to the point where the complexity of models is comparable to that of code, rendering the approach redundant.
The need to specify so many details comes from the fact that the technologies used to implement the software solutions use concepts from their own \emph{solution domain}.
For example, the Structured Query Language (SQL,~\cite{beaulieu2009sql}) allows to create and manipulate databases and the tables inside them, and therefore all the concepts in SQL are related to tables, rows, columns, etc.
But software clients use concepts from real-life domains in the requirements for the software solution they require, and all these technological domains are completely alien to them.
These real-life concepts belong to the \emph{problem domains}, and one example is banking, where the clients need to specify the requirements for the software they need in terms of accounts, balances, transactions and so on.
As Figure~\ref{fig:problem-and-solution-domains} shows, problem and solution domains are two different aspects of software development which can overlap in different ways: a software application for education may require both web and database technologies, and modelling languages can be used in both accounting and software architecture domains.

\begin{figure}[ht]
	\centering
	\includegraphics{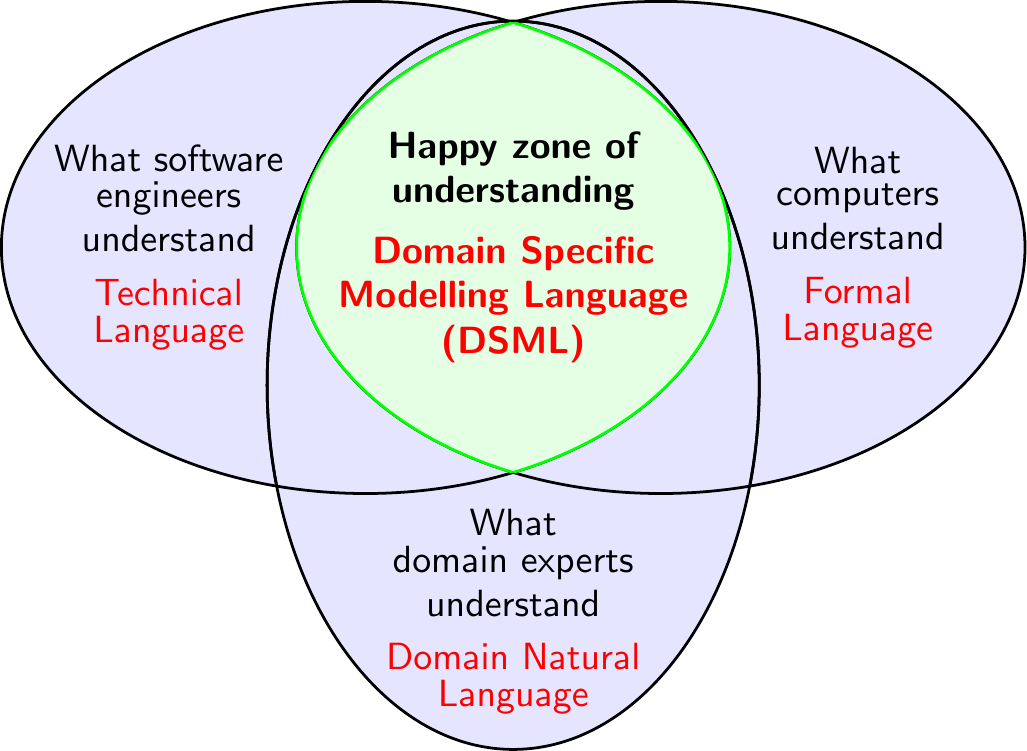}
	\caption{Motivation for Domain-Specific Modelling Languages}
	\label{fig:dsml-motivation-venn-diagram}
\end{figure}

Moreover, the difference of domains of expertise between the clients (experts in their problem domain) and the software engineers (experts in one or more solution domains) causes many misunderstandings and misspecified requirements which ultimately result in project overruns, both in time and budget.
We can also include the actual machines that run the software, and the task of ``translating'' requirements into code that software engineers have.
So we get, roughly speaking, three different parties (or stakeholders) involved in the process, with possible communication problems among them.
As Figure~\ref{fig:dsml-motivation-venn-diagram} hints, Domain-Specific Modelling Languages (DSMLs) aim at bridging these gaps.

DSMLs propose a new line of research within MDSE.
As shown in Figure~\ref{fig:dsml-motivation-venn-diagram}, the main goal of these languages is creating a common area of understanding for the three parties aforementioned.
The general idea is creating languages which are tailored to each problem domain, and can therefore hide the technicalities of the solution domain from the clients by designing models which use concepts from this problem domain.
First, the clients are involved in the process of creating the language by providing the concepts that should appear in it and their semantics.
Hence, the language is completely natural for them.
Second, the modelling engineers who develop the DSML are familiar with it and with the technicalities behind it, and can use it to communicate both among themselves and with the client.
And third, the language can still be interpreted or compiled into technologies used in the solution domain, so its information can be processed by, for example, the machines that run the software.
So we can talk about DSMLs as a \emph{step four} in our overview of software history, where the paradigm is shifted from software engineering to language engineering: the main task of the engineers is not using technical languages to create solutions for the clients, but developing domain-specific languages that the users can understand and use, and create solutions with these in the problem domain.
It is worth mentioning at this point that the broader concept of Domain-Specific Language (DSL) is also used in the literature~\cite{fowler2011dsl}.
Since we use modelling techniques to develop DSLs in this thesis, and those languages are model-based, we only employ the term DSML in the following.

In this thesis, we focus on the development of DSMLs in a way that covers both the syntactic and semantic aspects.
In most of the examples we show, the languages are used to define some kind of scenario that evolves through time, to show that our approach covers both the structural and behavioural specifications of the language.
For example, the workflow that an autonomous robot follows can be specified by building a model of its initial state (structure) and a series of rules which create new models, which represent states, based on the previous ones and the input from the robot's sensors (behaviour).
Therefore, we sometimes use the expression \emph{behavioural} DSML, although our approach is not restricted to only that kind of language.
A sensible approach to representing this duality of behavioural DSMLs, and the more widespread, is using models for the structural part and model transformations (MTs) for the behavioural one.

In the first paragraph of this introduction, we mentioned how layering concepts in different levels of abstraction is a useful way to represent knowledge.
Nowadays, we could use modelling techniques to do so, in a way that such knowledge can be understood by computers and turned into actual software.
In such a way, one can imagine families of DSMLs with different levels of abstraction, organised in a hierarchical manner, and with behaviour defined for them as model transformations, also exploiting the existence of multiple levels of abstraction.
However, the evolution in software engineering from step two (object-oriented, high-level programming languages) to step three (MDSE) caused the latter to be influenced by the former.
The world-view of software engineers was biased towards object orientation, and the organisation of all concepts in just two levels: classes and objects.
So this two-level paradigm is actually a limitation for DSML development and only caused by historical reasons, not practical ones.
The area of Multilevel Modelling (MLM) has appeared in recent years to challenge the notion of two-level modelling and address some of its limitations.

MLM exploits the idea of layering in modelling environments, by allowing a higher, even potentially unbounded, number of levels of abstraction (i.e.\ layers).
That is, the abstraction relation among objects and their classes becomes in MLM a pivotal concept to create multilevel hierarchies of models.
This technique opens new possibilities for modelling, especially applied to DSML specification, and enhanced expressive capabilities.
Moreover, the advantages of ``traditional'' MDSE are maintained: models are still formal, structured representations which can be understood by computers, verified, compiled, etc.

Summarising, this thesis focuses in the areas of Multilevel Modelling, Model Transformations and Domain-Specific Modelling Languages.
We understand a DSML as a user-defined domain-specific language, where the user is a software engineer, ideally in close cooperation with a domain expert.
Hence, DSMLs are used in this work as a fitting target to realise our approach to MLM and MTs, and we purposely leave the door open to apply these techniques in activities unrelated to DSML development.
With these notions in mind, we used the following research questions to guide our research:

\begin{itemize}
	\item \textbf{RQ1:}
	Is there a common framework which generalises different state-of-the-art techniques and approaches to MLM in a consistent manner?
	\item \textbf{RQ2:}
	Is there a formalisation of MLM which accounts for different state-of-the-art techniques and approaches to MLM?
	\item \textbf{RQ3:}
	Is there an MLM transformation approach which integrates different state-of-the-art techniques and approaches to MLM?
\end{itemize}

These questions lead us to the creation of a formal framework for MLM and MT.
The contributions which be believe that our framework provides are several:

\begin{itemize}
	\item An approach to Multilevel Modelling and Model Transformations in a multilevel setting, called Multilevel Coupled Model Transformations (MCMTs), which accounts for different state-of-the-art approaches to MLM and with a strong focus on flexibility and reusability.
	\item A formalisation of Multilevel Modelling and Multilevel Coupled Model Transformations using Graph Theory and Category Theory.
	\item The implementation of an EMF-based framework for MLM and MCMTs to evaluate our approach, which removes the need for custom-made environments and tools.
	\item The elaboration of several case studies (DSMLs) to analyse the strengths and weaknesses of our framework.
	\item The completion of several experiments which validate MLM in general, and our approach to it in particular.
\end{itemize}

These contributions and research questions are discussed in the rest of this thesis.
In the remaining part of this chapter, we introduce some basic concepts and terminology.
First, we give a short introduction to traditional MDSE.
Second, we outline the techniques that MLM uses to enhance traditional modelling.
And third, we briefly present the area of model transformations and some of its related concepts.

The rest of the thesis is structured as follows:
Chapters~\ref{chap:mlm} and~\ref{chap:mcmt} present our formal approach to Multilevel Modelling and Multilevel Model Transformations, respectively.
In Chapter~\ref{chap:tooling} we introduce our tooling environment MultEcore, which reifies our proposed approach.
Then, in Chapter~\ref{chap:validation} we discuss the case studies, experiments and formal constructions that we used to validate our work.
Finally, Chapter~\ref{chap:discussion} compares our approach with the state of the art, outlines future lines of research, discusses our contributions, and concludes this thesis.

\section{Modelling Basics}
\label{sec:modelling-basics}

The word ``model'' has different meanings depending on the field where it is used.
Even if we stay within the area of mathematics and computing, a model can be a 3D blueprint of a real-life object for a mechanical engineer or a probability distribution for a mathematician working in statistical analysis.
The second entry in the Oxford English Dictionary at the moment of writing these lines defines the concept as ``Something which accurately resembles or represents something else, especially on a small scale; a person or thing that is the likeness of another.''%
\footnote{\footnotesize\url{http://www.oed.com/view/Entry/120577}}%
In this work, we adopt this definition of model to the way that it is used in MDSE.
Therefore, we define model in this thesis as: ``A structured, formal representation of both abstract and real-life concepts, in a textual or graphical way, such that they can be interpreted and manipulated by modelling frameworks and model transformations.''
Moreover, models in MDSE can be prescriptive or descriptive.
A prescriptive model is the one which describes something to be build, and aligns with the early-days use of models as blueprints for software, e.g.\ a UML class diagram~\cite{odell1998advanced}.
A descriptive model aims to be an abstract representation of something which exists already, so that this representation can be manipulated and used for reasoning instead of the real thing, e.g.\ a simulation for robot behaviour before the actual robot is put to work.
Models created with DSMLs are often both descriptive, in the sense they represent reality, and prescriptive, since they guide the creation of software or act as software artefacts themselves.

Even if the area of MDSE is relatively modern, the things we call for models now have existed in Software Engineering long before they were grouped under the same word.
For example, state machines have been around since the late-forties~\cite{shannon1948mathematical} and Petri Nets~\cite{reisig1985petri} may have appeared even a decade earlier.
Those two examples, as well as many of the models depicted in this thesis are graphical, meaning that they have a sort of pictorial representation.
Moreover, in the next chapters we relate models to graphs, which are also depicted in a visual manner in most cases.
However, the reader must note that a model does not need to have a graphical syntax.
We can talk in this regard about the \emph{abstract syntax} of models, their underlying representation, and \emph{concrete syntax}, which is a specific way to represent a model in a way that it is easily created, read and manipulated by humans.
A model has one single abstract syntax, but can have several concrete ones, which can be textual or graphical.
The following list contains the main concepts related to models that are used in this thesis, where the influence of object-oriented techniques can still be perceived.

\begin{itemize}
	\item \textbf{Model}
	The central piece of MDSE.
	A model is a container for certain information, generally structured in a way that allows it to be parsed and/or understood by humans.
	A common way of structuring models is using graph-like elements like nodes and arrows, where the nodes represent pieces of information and the arrows relate nodes to each other.
	
	\item \textbf{Metamodel}
	A model, usually created using a general-purpose metamodelling language like UML~\cite{odell1998advanced} or Ecore~\cite{steinberg2008emf}, which defines concepts that can be used in order to define other models.
		
	\item \textbf{Class/Entity}
	Similar to the object-oriented paradigm, classes are defined within a metamodel to represent some abstract concept which can be instantiated.
	Classes are normally depicted as boxes with a name, and can contain attributes and be related to other classes by references.
	In some cases, the name entity can also be used, like in the Entity-Relation model used in conceptual modelling~\cite{chen1976entity}.
	
	\item \textbf{Node}
	Our models are internally represented using graphs, and classes are rendered as graph nodes, so we use this graph-related terminology extensively in this thesis.
	
	\item \textbf{Relation/Reference} 
	A connection between two classes which represents some sort of relationship from one of them, called source, to the other one, called target.
	They are depicted using arrows, since relations are not symmetric (i.e.\ commutative), and the source and target can be the same class (i.e.\ loops are allowed).
	In this thesis, we use the former term more frequently.
	
	\item \textbf{Edge/Arrow}
	Due to the graph-based nature of our models, relations are represented as graph edges, but still visualised as arrows.
	Therefore, although we employ both terms interchangeably, the latter appears more frequently in this thesis.
	
	\item \textbf{Element}
	A class or a relation contained within a model.
	Or, conversely, a node or edge contained within a graph.
	We use this term to refer to all contents within a model without needing to specify to which of the two categories they belong.
	
	\item \textbf{Attribute}
	A property associated to a class used to contain useful information about such class.
	They are usually represented inside the class, and are identified by their name.
	In many cases, included this thesis, attributes must also have a data type to be correctly specified.
	
	\item \textbf{Data type}
	The kind of information contained in an attribute.
	Usually, data types can be chosen from a fixed, predefined collection of ``basic'' ones, inspired by the primitive types of programming languages.
	In our approach, we work with the data types allowed by our tool, which suffice for all the examples presented in this thesis.
	These data types are \emph{Integer} (number without decimals), \emph{Real} (number with decimals), \emph{Boolean} (true or false) and \emph{String} (any sequence of characters).
	
	\item \textbf{Inheritance and Generalisation/Specialisation}
	A special kind of relation between two classes, where the source class becomes a ``child'' of the target class, which becomes the ``parent''.
	When this relation is established, the children inherits all attributes and references from its parent.
	It is usually permitted that several classes have the same parent, so that they become ``siblings'', and we also allow for a class to have several parents (multiple inheritance), as long as there are no name conflicts with the parents' attributes and references and certain constraints are respected.
	Transitive inheritance is also possible, so that the child of a certain class becomes parent of another class, as long as the relations do not become cyclic.
	This constraint also excludes the case of a class being its own parent.
	
	Inheritance is mainly employed as an useful mechanism to avoid repetitions of the same attributes and relations in different classes and to handle sibling classes uniformly.
	
	We use the name inheritance in general, but when we talk about this relation from the point of view of the child we can also use it (the child inherits from the parents), as well as specialisation (the child specialises its parents).
	We use generalisation for the parent's point of view (the parent generalises its children).
	
	\item \textbf{Containment}
	Another special kind of relation between two classes, when the source class becomes the container of the target one.
	While this relation does not have any real effect, it is supported to ensure compatibility with the EMF framework~\cite{steinberg2008emf}, and it is useful in some implementation-related scenarios.
	
	A class can contain other classes and, conversely, a class can be contained by more than one class.
	As with inheritance, transitive containment is allowed, but this relation also allows cyclic containment and self containment.
	
	\item \textbf{Attribute value}
	The actual value given to an attribute, which must be consistent with its data type.
	For example, an integer attribute cannot be given the values ``2.5'' or ``foo''.
	
	\item \textbf{Instance/Object and Classification/Instantiation}
	An object is, as in object-oriented programming, a specific configuration of a class, where its attributes are given values and related to other objects.
	When an object is created from a class, we can say that the object instantiates (i.e.\ is an instance of) that class or, conversely, that the object is classified by its class.
	The meaning of this relation, in plain language, is that the class defines the structure of a certain kind of elements, and all instances of that class conform to such structure, which also provides them with a certain kind of semantics.
	
	Since objects are contained in a model and every object is said to be an instance of a class, the ``instance-of'' relations from all objects in a model to all (or some) of the classes in the model's metamodel can be generalised, so we can say that the model is an instance of its metamodel.
	
	A class can have several instances (in the same model or different ones), and an object can be an instance of more than one class.
	Likewise, a model has at least one metamodel, and several models can share the same metamodel.
	
	\item \textbf{Typing and Abstraction}
	Typing is the main relation that we use in this work to connect models, and the elements within them, to each other.
	This relation expresses that the type of an element, defined in the metamodel of the element's model, defines an abstract concept that is materialised into that element.
	That is, a (concrete) element is ``typed by'' its (abstract) type, 
	
	Our modelling framework is intended to support both classification and generalisation, but in the examples in this thesis we do not enforce a particular school of thought regarding when and how to use them.
	That is, typing relations have the semantics of abstraction, but we do not claim whether those semantics align with classification, generalisation, both or neither.
	We still reuse, for the sake of clarity, the terms ``class'', ``object'' and ``instance'', so we can say that the typing relation indicates the connection between objects and their classes, and in general between models and their metamodels.
	Hence, we can use the expression ``a class is the type of an object'' and reuse, conversely, terms like ``an object instantiates a class'' or ``an object is an instance of a class''.
	
	We also apply these notions to relations, so that a reference between two objects can be typed by a reference between the classes of each object.
	Regarding attributes, we also consider typing the process of assigning a value to an attribute inside an object which has been defined in the object's type.
	
	\item \textbf{Reflexive and Self-defining metamodel}
	We consider a metamodel to be reflexive when the concepts it contains are so generic that every concept defines itself, and therefore the whole metamodel is an instance of itself.
	That is, every class in the metamodel is an instance of itself, and the same applies to references.
	
	We can call a metamodel self-defining when it is an instance of itself, but all the concepts it contains can be instances of other concepts.
	In this second case, a class can be an instance of a different class, and a reference can be typed by another element.
			
	\item \textbf{Snapshot}
	Since some of our model instances represent a particular state (or point in time) of a behavioural DSML, such instances are called snapshots in some parts of this thesis in order to highlight their purpose, i.e.\ being a state on a model-based simulation.
	
	\item \textbf{Abstract class}
	Influenced by object-oriented design, MDSE also includes the concept of abstract classes in many cases.
	A class declared as abstract cannot be instantiated, but can be used in an interface-like manner for children classes that may inherit from it, as well as contain attributes common to all those children.
	For the sake of EMF compatibility and due to the usefulness of this concept, our approach also supports abstract classes.
	However, abstract classes should not be confused with the more general concept of abstraction as a mechanism to define typing relations between concepts (which are more concrete) and their types (which are more abstract).
	
	\item \textbf{Abstract and Concrete syntax}
	The structure of a model must follow a certain set of rules, and can therefore be represented using a certain syntax.
	The abstract syntax of a model is that which only depicts the language's concepts and how they relate to each other.
	A common abstract syntax in modelling is class diagrams (or similar box-and-arrow notations), although tree structures and grammar notations can also be employed.
	In contrast, the concrete syntax of a language is related to how the language is supposed to be used and read.
	For example, a concrete textual syntax can approximate natural language to ease the process of reading and writing, like the statements of SQL which employ verbs, adverbs and nouns to approximate common speech.
	And a concrete graphical syntax may use symbols which are familiar in the language's domain, like a drag-and-drop graphical editor for electrical circuits where resistances, switches, etc. are represented with their standard symbols.
	That is, the concrete syntax of a language facilitates manipulating the concepts of the language's abstract syntax in a more comfortable and familiar way.
	
\end{itemize}

We conclude this short introduction to MDSE by briefly presenting the Ecore metamodel, the central piece around which the EMF framework~\cite{steinberg2008emf} is built.
Ecore is a self-defining, general-purpose, class-based metamodel.
First, the concepts it defines are generic enough for Ecore to be self-defining (but not reflexive).
Second, these concepts are not tied to any of the specific domains where EMF-based modelling can be applied.
And third, the Ecore metamodel defines concepts that resemble object-oriented languages  to some extent, like using classes and references.

Ecore can be used to build metamodels, which in turn allow to create models.
Hence, we could actually call Ecore a meta-metamodel.
Due to the fact that Ecore is based in EMOF (Essential MOF), which is in turn a simplification of the MOF standard (Meta-Object Facility~\cite{omg2016mof}), all elements defined in Ecore are preceded by ``E''.
Hence, the concepts of class, reference and attribute are called in Ecore ``EClass'', ``EReference'' and ``EAttribute'', respectively.
Ecore also defines other concepts like enumerations (``EEnum''), operations (''EOperation'') and annotations for its elements (``EAnnotation'').
While only EClass and EReference are relevant for the formal part of our approach, most of the elements in Ecore are used in the tooling part in order to ensure compatibility with the EMF framework, and hence benefit from its maturity and widespread usage.

\section{Multilevel Modelling}
\label{sec:mlm}

Multilevel modelling is fundamentally based on the idea of expanding the type-instance relation between the level of metamodels and the level of their instances to more than these two levels.
In this manner, a model (or an element within a model) can be, at the same time, the instance of another model (or element) in a higher level of abstraction and act as type for another model (or element) in a lower abstraction level.
If we relate this duality to the concepts of traditional MDSE aforementioned, we can say that an element within a model can be, at the same time, an object of some class and the class of another object.
Hence, the term \emph{clabject} is widely used in MLM as a portmanteau of \emph{class} and \emph{object} to define such entities.
This term is included in the following list, together with other concepts from MLM that we use in this thesis.
However, the reader should note that the terminology related to MLM may vary depending on the authors~\cite{gerbig2016comparison}.

\begin{itemize}
	\item \textbf{Clabject}
	An element in a model that is, at the same time, an instance of a class (which is also a clabject) and a class that can be instantiated to create objects (which are usually also clabjects).
	We use the word ``node'' defined previously instead of ``clabject'' in most of this thesis.
	
	\item \textbf{Metamodel, Model and Instance}
	Given that MLM blurs the line between model and metamodel, we employ extensively the term ``model'' since a model can play both roles in a multilevel hierarchy.
	We reserve the terms ``metamodel'' and ``instance'' to be used in a contextual manner, so that we talk in this thesis about the ``metamodel of a particular model'' or the ``instance of a particular model''.
	
	\item \textbf{Level}
	Since MLM opens the door to chains of models related to each other by typing relations, the traditional two levels give way to a potentially unbounded number of them.
	Therefore, MLM can use the concept of levels, which are usually numbered sequentially and identified by such number.
	Models, and the elements inside them, belong to the level where they are located.
	In our approach, we use these levels as an organisational tool whose semantics are not strictly related to classification relations.
	
	\item \textbf{Multilevel hierarchy}
	Since a model can have more than an instance, and these instances can in turn have more instances, we use this term extensively in this thesis to refer to a whole family of models which connect to each other via typing relations.
	
	\item \textbf{Potency}
	A typing mechanism related to MLM where the amount of times a node can be transitively instantiated, and where, is somehow constrained.
	For example, a node can be limited to only create instances from it, but not instances of those instances.
	Also, potency can control how many levels below a node can be instantiated (by default, clabjects are usually instantiated in the level immediately below them).
	In this work, we also apply this notion to relations and attributes.
	
	\item \textbf{Depth and Deep instantiation}
	These concepts are closely related to potency, and are sometimes used interchangeably in the literature.
	In this thesis, we use ``depth'' to refer to the specific use of potency where the transitive instantiations of an element are limited and ``deep instantiation'' when talking about elements being instantiated in a level different from the one directly below.
	
	\item \textbf{Linguistic metamodel, linguistic and ontological typing and linguistic extension}
	An important mechanism closely related to MLM is the existence of a linguistic metamodel which exists outside of the multilevel hierarchy and actually acts as a metamodel for the complete structure.
	Most MLM approaches require this metamodel and use it to define all the concepts that are used to create multilevel hierarchies, like clabject, typing, potency and attribute.
	Hence, the complete hierarchy is somehow ``flattened'' and becomes a single instance of the linguistic metamodel.
	All elements in the hierarchy are therefore linguistically typed by this metamodel, whereas their typing relations among them become a sort of ``synthetic'' relation called ontological typing, which is in turn used to name every model and typing inside the hierarchy as ontological.
	
	The concept of linguistic extension implies creating elements without an ontological type and only the linguistic one.
	This technique is useful for creating elements in the hierarchy without a pre-existing ontological type, which increases the flexibility and expressive power of MLM.
	
	The linguistic approach can also be used to simulate MLM by just using two levels of abstraction: the linguistic metamodel and the whole MLM hierarchy as an instance of it.
	However, that dependence in a linguistic metamodel ties the whole modelling apparatus to a single metamodel, which is too rigid and not ideal.
	The existence of a linguistic metamodel is one of the MLM notions that is challenged in our approach.
	
\end{itemize}

The idea of multiple levels of abstraction applied to MDSE started to gain momentum in the 2000s ~\cite{atkinson2001processes,atkinson2001essence,atkinson2002rearchitecting,atkinson2008reducing,delara2010automating,rossini2014formalisation} and consolidated in the next decade in the MULTI workshop series~\cite{multievents}.
However, using several abstraction layers was already a well-established technique for the specification of information systems in the previous decades~\cite{mylopoulos1980relationships,borgida1984generalization,mylopoulos1990telos,odell1994power,atkinson1997distributed,bezivin1997ontology,odell1998advanced}.

Still, MLM has several challenges which hamper its wide-range adoption, such as lack of a clear consensus on fundamental concepts of the paradigm, which has in turn led to lack of common focus in current multilevel tools.
One of the aims of this thesis is to create a foundation for MLM, both practical and formal, which has its roots in well-established approaches and tries to move the state of the art towards the convergence of Multilevel Modelling approaches.

Moreover, MLM is inherently suited for DSML development, since it allows to design increasingly specific languages, and this can be done until the level of specificness that suits the target application or domain is reached, without technological limitations.
Therefore, all examples and case studies in this thesis are related to the creation of different DSMLs, as a way to validate the approach while showcasing its strengths and benefits.

\section{Model Transformation}
\label{sec:mt}

One common way to manipulate models in MDSE are model transformations.
Such transformations are expressed by means of transformation rules, which are usually declarative.
That is, the rule has at least two parts: one part which specifies a pattern to be matched against a model, and another one which uses the information from the match to generate the desired result.
We use model transformations in this thesis as a way of manipulating the instances of our DSMLs in order to provide them with behaviour (recall the example with the robot's workflow from the beginning of this chapter).
At the same time, we exploit the mechanisms of MLM in the definition of such transformations and explore its advantages regarding flexibility, conciseness and reusability.

We have stated that the formalisation of models is commonly done using Graph Theory, and that also applies to model transformations, which can be formalised by means of graph transformations.
Therefore, the terminology that we use for model transformations, listed below, is also applicable to the graph transformations we employ for their formalisation in this thesis.

\begin{itemize}
	\item \textbf{Rule}
	A model transformation can be composed of a set of rules, each of which specifies a certain transformation to be performed.
	For example, a transformation that renames all elements in a model may have one rule for renaming classes, another for references, another for attributes and so on.
	Usually, each rule has a name which briefly describes the kind of manipulation which the rule performs.
	\item \textbf{Input/Source and Output/Target}
	In a very broad sense, model transformations take a model and modify it according to a set of rules.
	So one can see model transformations as a black box where an input model is provided as a source and an output model is generated as a target, and we use this terminology at some points of this thesis.
	\item \textbf{Model-to-model and Model-to-text}
	When we refer to model transformations in general, we generally assume that the input and output of the transformations are both models.
	In this case, we can refer to them as model-to-model transformations.
	Another possibility is that the output of the transformation is textual, like a source code file in a given programming language.
	This process can be understood as a sort of high-level compilation and it is commonly called ``code generation''.
	
	The idea of using model transformations to generate runnable software is related to the concepts of Platform-Independent Model (PIM) and Platform-Specific Model (PSM), where the same information is exchanged between formats that can be parsed by existing interpreters using model transformations.
	\item \textbf{Endogenous and exogenous}
	In two-level modelling, these two terms are used, respectively, if the transformation manipulates models in the same language (i.e.\ the input an output models have the same metamodel) or in different languages (i.e.\ the metamodels for the input and the output are different).
	In a multilevel setting, these terms can be reused to refer to whether the transformations manipulate a single model, which is usually in the bottommost level of a hierarchy, in which case we could talk of endogenous multilevel MTs.
	Since these transformations open the door to managing models in more than one level at once, we can also reuse the term exogenous to refer to those multilevel MTs which manipulate several levels of a multilevel hierarchy at once.
	\item \textbf{In-place and Out-place}
	In some scenarios, it is desirable that the input and output models are actually the same model, so that the transformation does not generate a new model but rather modifies the existing one.
	In this case, we can refer to in-place transformations.
	In contrast, out-place transformations are those where the output model is freshly created and the input one remains unchanged.
	However, this difference is more related to implementation, since each kind can be manipulated to behave like the other: an in-place transformation can preserve a copy of the input model before it is modified or, conversely, an out-place transformation can replace the input model by the output one.
	\item \textbf{Match}
	In order to be applied, a transformation rule must find a certain set of elements which conform to the provided patterns and constraints.
	When that happens, we can say that the rule matches the input model, or that a match is found.
	\item \textbf{Left-Hand Side and Right-Hand Side}
	A common way to define model transformations is using a declarative notation.
	Our approach to model transformations conforms to this paradigm.
	In such a way, our transformations also borrow the concept of two sides.
	First, the side which specifies the patterns and constraints that the elements in the source model must fulfil in order for the rule to match.
	And second, the side which specifies which elements must be created, deleted or modified in the matched input pattern in order to generate the target model.
	Given that transformation rules are commonly depicted horizontally, the convention is to show the pattern for the source in the left and the pattern for the target in the right, hence the terms left- and right-hand side. 
	\item \textbf{Negative Application Condition}
	In addition to the source and target patterns, some approaches also allow for the definition of an additional pattern that disables the rule application in case we find a match of it.
	That is, negative application conditions (NAC) can prevent the application of a rule even if a match of the left-hand side is found in the source model.
	Although our approach does not explicitly allow for the definition of NACs, we present this concept since it is briefly mentioned later in this thesis.

\end{itemize}

%% file: thesis/02-mlm.tex
\chapter{DSML Structure with Multilevel Modelling}
\label{chap:mlm}

We present here the foundational concepts of our approach for Multilevel Modelling.
These are illustrated with examples from our case studies, and rely in formal definitions using Graph Theory~\cite{diestel2018graph} which have been revised from the definitions we presented originally in~\cite{macias2018mlm}.
This chapter contains three sections.
We justify the usage of MLM by highlighting the features which make it fitting for the development of Domain-Specific Modelling Languages in Section~\ref{sec:why-mlm}.
In Section~\ref{sec:hierarchies} we present the concepts and theory behind the constructions of a single multilevel hierarchy.
Then, in Section~\ref{sec:three-dimensions} we show how to combine hierarchies in three different so-called \emph{dimensions}, in order to increase expressibility and modularity.

\section{Why Using MLM?}
\label{sec:why-mlm}

Multilevel Modelling techniques are a great fit for the creation of DSMLs, especially when we focus on behavioural languages.
When it comes to creating the structure of a domain-specific language as a model, we usually need to represent concepts with different degrees of abstraction into these models.
The technique we use to create DSMLs should ideally allow us to reflect these layers of abstraction into the language implementation.

\begin{figure}[ht]
	\centering
	\includegraphics{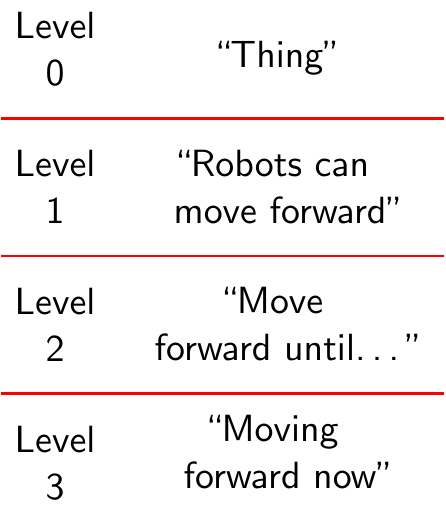}
	\caption{Four levels of abstraction of a simple behavioural DSML}
	\label{fig:behavioural-dsml-four-levels}
\end{figure}

If we focus on a simple language to specify behaviour, we can easily identify four levels of abstraction.
Recall the language for specifying the workflows of autonomous robots that we used as an example in Chapter~\ref{chap:introduction}.
In order to create such a language, we first need to use a metalanguage generic enough to define any kind of concept.
For example, this metalanguage can contain the concept ``Thing''.
Then we can instantiate this metalanguage to develop our own behavioural language with concepts like ``move forward'' or ``scan the environment with infrared''.
Then, a specific workflow can be created, with instructions such as ``stop moving forward if an obstacle is detected ahead'', which are instances of the concepts in the level above.
Moreover, the state of the robot at a specific point in time can also be represented using the concepts of the language, so that we can represent that ``at this point, the robot is moving forward''. 
In a nutshell, when developing and using a behavioural, domain-specific language, even if it is a simple one, we can identify at least four different levels of abstraction, as Figure~\ref{fig:behavioural-dsml-four-levels} shows: general-purpose metalanguage (level 0), behavioural language (level 1), specific model defined using the language (level 2) and the state of such model at a particular point (level 3).
If we are less conservative, we may identify even more levels if the domain is complex enough and we need to separate our concepts in more layers.
For example, we could create a model in a level between 0 and 1 with more abstract behavioural concepts like ``task'' or ``sensor input'' which are reusable for different robots.

A fixed-level framework only allows for two or three user-accessible levels, which is not enough for even the simpler scenarios.
When several layers of abstraction are fit into one single level (i.e.\ the same model), modelling anti-patterns emerge, like the type-object pattern described, among others, in~\cite{delara2014whenandhow}.
This anti-pattern, for example, appears when a concept and its metaconcept are defined in the same model, and the relation between them is encoded in a convoluted manner.
A comprehensive list of such anti-patterns can be found at~\cite{delara2014whenandhow}, and multilevel alternatives to these anti-patterns are proposed in~\cite{delara2018refactoring}.

We propose using MLM as a way of creating DSMLs so that their structure resembles the layered fashion in which we organise concepts with different levels of abstraction within a domain.
It could be argued that using specialisation and generalization~\cite{borgida1984generalization,mylopoulos1980relationships} could be used instead of multilevel typing to reduce the amount of levels to those available in fixed-level approaches.
As some authors have already analysed, classification (i.e.\ instantiation relations) and specialisation (i.e.\ inheritance relations) should not be confused or considered as equivalent~\cite{kuhne2006matters, kuhne2009contrast}.
Hence, the approach we present here allows for both types of relations without enforcing a particular modelling style, so that modellers can combine them at their own will.
Moreover, we do not adopt a particular or strict semantics for our typing relations among elements and models.

In general, we argue that MLM has the following advantages when compared to fixed-level approaches, which motivate its use in our approach:

\begin{itemize}
	\item \textbf{Separation of concerns}
	Fixed-level approaches force the user to group together concepts that do not really belong in the same model.
	Apart from avoiding the anti-patterns we previously mentioned, MLM also allows us to separate elements that, while related to the same abstract concepts, do not belong to the same exact domain.
	Consequently, we can prevent our models from becoming polluted with concepts that do not really relate to the task at hand.
	Take for example the autonomous robots we mentioned earlier, and imagine that we have some robots that only fly and some that only have wheels.
	A non-multilevel DSML would force us to use an inheritance relation (instead of typing) among the abstract ``task'' concept and both the ``fly up'' task the first group robots can perform and the ``roll forward'' task that the second group can do.
	So we would end up with a language that has unnecessary concepts in both scenarios, or with two different versions of the same model with slight variations, which complicates maintainability, version control and reusability.
	\item \textbf{Reusability}
	Another positive consequence of modularising a DSML using typing relations (instead of overusing inheritance) is the possibility to define a concept in a high level of abstraction, and then reuse it or override it in the lower levels.
	In our example, having ``task'' as an abstract concept allows us to refine it in different models, e.g.\ as `fly up'' for flying robots and ``roll forward'' for robots with wheels.
	Therefore, we avoid pollution while keeping the relation among abstract and specific concepts and we are able to reuse abstract semantics.
	\item \textbf{Extendibility}
	MLM also allows us to easily extend a language by creating new instances of its abstract concepts without altering those instantiations which already exist and the possible artefacts that are derived from them.
	For example, creating a new language for water-borne robots with instances of task like ``swim deeper'' would affect neither of the languages for rolling or flying robots.
	We can call this kind of extension \emph{horizontal}.
	But \emph{vertical} extendibility, where we can create growingly specific instances of our languages as desired, is also facilitated by MLM.
	This is due to the fact that we provide an unbounded number of abstraction levels.
	\item \textbf{Modularization}
	In most MLM approaches, families of models are organised in tree-shaped hierarchies with the more abstract concepts on top.
	This has the positive consequence that each metamodel in the hierarchy can be used as a module for itself without paying much considerations to neighbouring branches.
	That is, even if a bird's-eye view of the whole multilevel structure has all advantages aforementioned, we can use any model in an almost standalone fashion within its own branch, ignoring other instantiations of the abstract languages for different specific domains.
\end{itemize}

As we see in Chapter~\ref{chap:mlm}, the definition of DSML behaviour can also benefit from a multilevel framework.

As a final remark, notice that even if we focus in behavioural DSMLs we want to leave the door open for multilevel model-driven engineering in general.
Therefore, the definitions in this chapter might seem to be more generic than required.

\section{Multilevel Metamodelling Hierarchies}
\label{sec:hierarchies}

Our multilevel metamodelling approach is based on a flexible typing mechanism based on graphs.
In this section, we introduce the kind of graphs used to represent our models, as well as the way to organize them in hierarchies, by means of relations among graphs and the elements that they contain.
We use one of the case studies created in the context of this thesis as an illustrative, incremental example from the domain of workflow-like languages for simple autonomous robots.
Note that some of the names in the example are shortened for display purposes when compared with the full version, which is shown in Section~\ref{subsec:application-dimension}.

\subsection{Models as directed multigraphs}
\label{subsec:directed-multigraphs}

Let us suppose that we are using models to run a simulation of the workflow that a simple robot follows.
One of the most useful models would be that one representing the workflow that the robot must follow.
Our models are represented by means of graphs, since graphs are widely used to represent software models due to the fact that graphs are a very natural way of explaining complex situations on an intuitive level, e.g.\ for data and control flow diagrams, for entity-relationship diagrams and for UML diagrams~\cite{ehrig2006fundamentals}.
So all models in our approach, including the one representing our robot's workflow, are represented with a name and considered abstractly as directed multigraphs, meaning that all edges have an explicit direction and that two nodes can be connected by several edges.
For our formal definitions in this thesis, an arbitrary graph is usually called \(\graphname{\examplegraphone}{}\), but the one representing the state of our robot is called \elementname{robot\_1}.

\begin{definition}[Graph]
\label{def:graph}
A graph \(\graphname{\examplegraphone}{} = (\graphname[\graphnodes]{\examplegraphone}{} , \graphname[\grapharrows]{\examplegraphone}{} , \source[\examplegraphone] , \target[\examplegraphone])\) consists of a set of nodes \(\graphname[\graphnodes]{\examplegraphone}{}\), a set of arrows \(\graphname[\grapharrows]{\examplegraphone}{}\) and two morphisms \(\source[\examplegraphone] : \graphname[\grapharrows]{\examplegraphone}{} \to \graphname[\graphnodes]{\examplegraphone}{}\) and \(\target[\examplegraphone] : \graphname[\grapharrows]{\examplegraphone}{} \to \graphname[\graphnodes]{\examplegraphone}{}\) that assign to each arrow its source and target node, respectively.
We use the notations \(\elementname{\examplenodeone} \xrightarrow{\elementname{\examplearrowone}} \elementname{\examplenodetwo}\) or \(\elementname{\examplearrowone} : \elementname{\examplenodeone} \to \elementname{\examplenodetwo}\) to indicate that \(\source[\examplegraphone](\elementname{\examplearrowone}) = \elementname{\examplenodeone}\) and \(\target[\examplegraphone](\elementname{\examplearrowone}) = \elementname{\examplenodetwo}\).
These two morphisms must be total for the graph to be considered valid.
\end{definition}

That is, graphs consist of nodes and arrows.
A node represents a class, and an arrow represents a relation between two classes.
Hence, an arrow always connects two nodes in the same graph, and any two nodes can be connected by an arbitrary number of arrows.
Arrows with source and target in the same node (loops) are also allowed, and a node can likewise have any number of loops.
We will use the word \emph{element} to refer to both nodes and arrows, and assume that all elements are named and identified by such name, for which we use typewriter font.
Since the name is their unique identifier, the names of any two nodes in the same graph must be different.
For the arrows, we allow for equal names as long as the source, the target or both are different, in order to be able to differentiate them.
Due to the use of graphs as the underlying representation of models, both terms \emph{graph} and \emph{model} could be used interchangeably in this thesis.
We will however differentiate by using the former for the formal definitions, and the latter when discussing specific examples or talking from a more practical perspective (like tooling aspects).

\begin{figure}[ht]
	\centering
	\includegraphics[page=1]{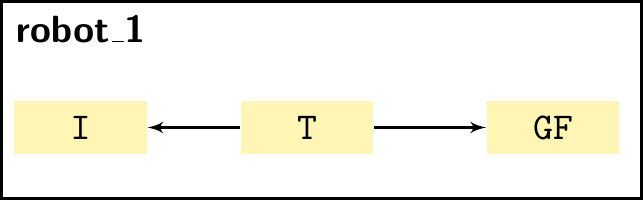}
	\caption{Directed multigraph with named elements}
	\label{fig:hierarchy-introductory-example-1}
\end{figure}

Using this kind of graphs, we now have a clear structure to represent our robot's workflow.
Figure~\ref{fig:hierarchy-introductory-example-1} shows the small graph \elementname{robot\_1} which is used as a model of the workflow.
This graph contains three nodes, depicted as yellow squares, which represent a transition \elementname{T} between the initial state of the execution \elementname{I} and the first task \elementname{GF} that the robot must perform, which consists of going forward.
The arrows connecting these nodes represent the task's input and output tasks, hence the names \elementname{in} and \elementname{out}.

The \elementname{robot\_1} graph does not convey much information by itself, since it only consists of some nodes an arrows with names on them.
In our approach, an useful model is always part of a hierarchy of them, and the relations among these models, and among the elements inside them, is what gives us the full picture of the scenario we are modelling.
Therefore, we need a way to define relations like typing and matching between the graphs which we use to represent our models.
The most fitting way to define a relation between any two graphs is by using \emph{graph homomorphisms}.

\begin{definition}[Graph Homomorphism]
\label{def:graph-homomorphism}
A homomorphism \(\morphone : \graphname{\examplegraphone}{} \to \graphname{\examplegraphtwo}{}\) between graphs is given by the two maps \({\morphone^\graphnodes : \graphname[\graphnodes]{\examplegraphone}{} \to \graphname[\graphnodes]{\examplegraphtwo}{}}\) and \({\morphone^\grapharrows : \graphname[\grapharrows]{\examplegraphone}{} \to \graphname[\grapharrows]{\examplegraphtwo}{}}\) such that \(\source[\examplegraphone] ; \morphone^\graphnodes = \morphone^\grapharrows ; \source[\examplegraphtwo]\) and \(\target[\examplegraphone] ; \morphone^\graphnodes = \morphone^\grapharrows ; \target[\examplegraphtwo]\).
Note, that we use the symbol \(\_\, ; \_\) to denote composition in diagrammatic order.
\end{definition}

Additionally, further along this chapter we need to identify and manage only fragments of our graphs, for which we need to define a new concept.

\begin{definition}[Subgraph]
\label{def:subgraph}
A graph \(\graphname{\examplegraphone}{} = (\graphname[\graphnodes]{\examplegraphone}{} , \graphname[\grapharrows]{\examplegraphone}{} , \source[\examplegraphone] , \target[\examplegraphone])\) is a subgraph of a graph \(\graphname{\examplegraphtwo}{} = (\graphname[\graphnodes]{\examplegraphtwo}{} , \graphname[\grapharrows]{\examplegraphtwo}{} , \source[\examplegraphtwo] , \target[\examplegraphtwo])\), denoted as \(\graphname{\examplegraphone}{} \sqsubseteq \graphname{\examplegraphtwo}{}\), iff \(\graphname[\graphnodes]{\examplegraphone}{} \subseteq \graphname[\graphnodes]{\examplegraphtwo}{}\), \(\graphname[\grapharrows]{\examplegraphone}{} \subseteq \graphname[\grapharrows]{\examplegraphtwo}{}\) and, for all \(\elementname{\examplearrowone} \in \graphname[\grapharrows]{\examplegraphone}{}\) we have \(\source[\examplegraphone](\elementname{\examplearrowone}) = \source[\examplegraphtwo](\elementname{\examplearrowone})\) and \(\target[\examplegraphone](\elementname{\examplearrowone}) = \target[\examplegraphtwo](\elementname{\examplearrowone})\).
\end{definition}

And since a subgraph is still a graph, it can be related to the full graph of which it is a fragment by a special type of relation, which again can be defined using graph homomorphisms.

\begin{definition}[Partial Graph Homomorphism]
For any two graphs \(\graphname{\examplegraphone}{}\) and \(\graphname{\examplegraphtwo}{}\) with \(\graphname{\examplegraphone}{} \sqsubseteq \graphname{\examplegraphtwo}{}\), if the inclusions \(\graphname[\graphnodes]{\examplegraphone}{} \subseteq \graphname[\graphnodes]{\examplegraphtwo}{}\) and \(\graphname[\grapharrows]{\examplegraphone}{} \subseteq \graphname[\grapharrows]{\examplegraphtwo}{}\) define a graph homomorphism, we call this a partial or \emph{inclusion} morphism, represented as \(\graphname{\examplegraphone}{} \hookrightarrow \graphname{\examplegraphtwo}{}\).
\end{definition}

\subsection{Tree-shaped hierarchies, levels and typing chains}
\label{subsec:tree-shape-levels-and-tc}

Continuing with our robot example, suppose that we now need to explicitly define the language that we use to design the robot's workflow.
We also have robots from different platforms, so that the abstract workflow language needs to be specialised in more specific versions according to the capabilities of those platforms.
These versions of the language with different levels of abstraction need to be organised together with the workflows created using them.
In Chapter~\ref{chap:introduction} we defined loosely the concept of multilevel hierarchy as a family of graphs (models) connected to each other via typing relations.
In our approach, every hierarchy starts with a graph of choice that must be reflexive or, at least, self-defining.
This initial graph, called root, can have several instances, which in turn can have several instances, and so on.
Hence, our hierarchies are tree-shaped with a single root graph.
Later on, we show that our way of characterising typing necessarily yields this shape.
Also, implicit in that assumption is the fact that each graph, except the one at the root, has exactly one ``parent'' graph in the hierarchy.

A hierarchy has \(l + 1\) \emph{abstraction levels} where \(l\) is the maximal length of paths in the hierarchy tree.
Each level in the hierarchy represents a different degree of abstraction.
Levels are indexed with increasing integers starting from the uppermost one, with index \(0\), so that the levels do not need to be re-numbered as the hierarchy grows down.
Each graph in the hierarchy is placed at some level \(\indexone\), where \(\indexone\) is the length of the path from the root to the graph.
We will use the notation \(\graphname{\examplegraphone}{\indexone}\) to indicate that a graph is placed at level \(\indexone\).

\begin{definition}[Typing Chain]
\label{def:typing-chain}
For any graph \(\graphname{\examplegraphone}{\indexone}\) we call the unique path \(\tc{\graphname{\examplegraphone}{\indexone}} = [\graphname{\examplegraphone}{\indexone}, \graphname{\examplegraphone}{\indexone-1}, \dots, \graphname{\examplegraphone}{1}, \graphname{\examplegraphone}{0}]\) from this graph to the root graph of the hierarchy the typing chain of \(\graphname{\examplegraphone}{\indexone}\).
\end{definition}

Level \(0\) contains, in any hierarchy, just the root \(\graphname{\examplegraphone}{0}\) of the tree.
For implementation reasons, we use Ecore~\cite{steinberg2008emf} as root graph at level \(0\) in all example hierarchies, since Ecore is based on the concept of graph which makes it powerful enough to represent the structure of software models.
We will not fully display this model in our figures for the sake of clarity.
Another possible option for our root graph worth mentioning is the node-and-arrow graph described, for example, in~\cite{rutle2012formal}.

Although it might be more appropriate, conceptually, to use Ecore as a linguistic metamodel~\cite{atkinson2001essence}, we leave Ecore on top of our hierarchy since (i) Ecore does not provide any concepts for defining levels and typing between levels and (ii) our implementation strives to minimise the threshold for migration from fixed-level EMF-based modelling to MLM.
Also, the modelling hierarchies presented in this thesis may have a mixture of linguistic and ontological nature, or not be ontological at all since they can act as prescriptive models for software implementation~\cite{assmann2006ontology}.

\begin{figure}[ht!]
	\centering
	\includegraphics[page=2,width=\linewidth]{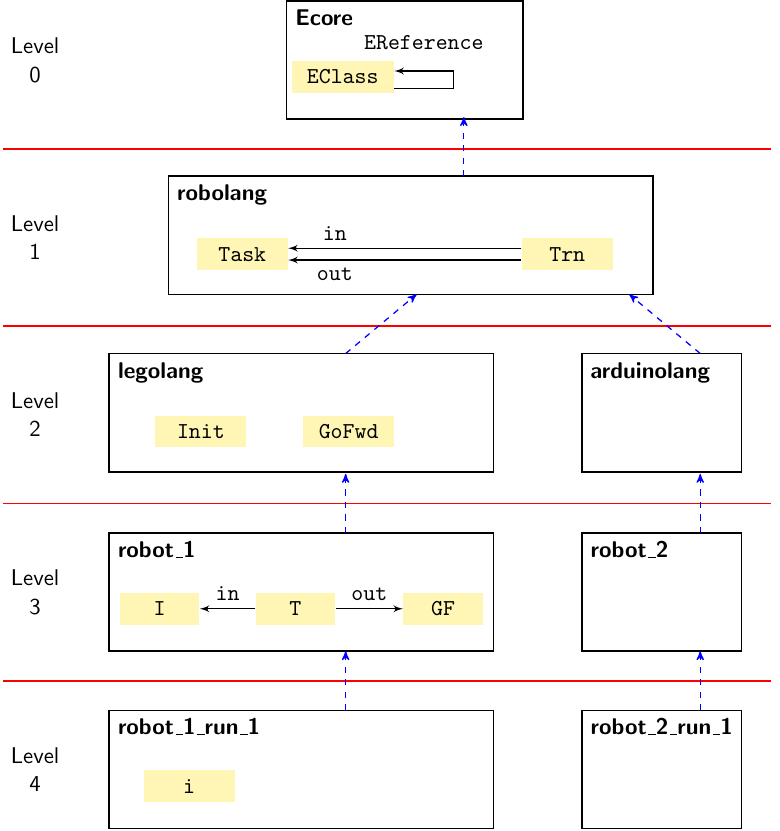}
	\caption{Tree-shaped hierarchy of graphs with levels and typing chains}
	\label{fig:hierarchy-introductory-example-2}
\end{figure}

In Figure~\ref{fig:hierarchy-introductory-example-2} we show several graphs that constitute a tree-shaped hierarchy, including the one from Figure~\ref{fig:hierarchy-introductory-example-1}.
We locate the graph \elementname{robot\_1}, representing a particular configuration for a robot, in level \(3\), together with another graph \elementname{robot\_2}, defining a different configuration for another robot.
These two graphs, although conceptually similar, are independent from each other in the sense that they neither belong to the same branch of the tree nor share their parent graph (that is, their metamodel).
They serve the role of parent graphs, respectively, for \elementname{robot\_1\_run\_1} and \elementname{robot\_2\_run\_1}, located at level \(4\).
Since we in this thesis focus on behavioural modelling languages, the models we define evolve through time, representing the execution of the modelled system.
Thus, the purpose of \elementname{robot\_1\_run\_1} is to store, at any particular point in time, the state of the execution of the specific process defined in \elementname{robot\_1}.
Likewise for \elementname{robot\_2\_run\_1} and \elementname{robot\_2}.

Back to our hierarchy, the parent graphs of \elementname{robot\_1} and \elementname{robot\_2} are located at level \(2\), where most of the types of their elements are defined.
These two parent graphs, \elementname{legolang} and \elementname{arduinolang}, define the type of elements that we can use to define a specific process for a robot designed in the Lego EV3~\cite{garber2015lego} or Arduino~\cite{banzi2008arduino} platforms, respectively.
Both models share the common parent graph \elementname{robolang}, located at level \(1\).
This language contains basic concepts for process modelling, independently from the robot hardware or platform.
Finally, as we previously mentioned, we locate Ecore, the topmost graph and root of the tree, at level \(0\).
This graph is the parent graph of \elementname{robolang}.
Note that we use red horizontal lines to indicate the separation between two adjacent levels, and blue dashed arrows for sequences of graphs that constitute typing chains and provide the required tree shape.
Since the purpose of Figure~\ref{fig:hierarchy-introductory-example-2} is to illustrate the tree shape of multilevel hierarchies, the contents of the branch on the right are not displayed.
The full diagram for the left branch is shown in Figure~\ref{fig:robolang-multilevel-hierarchy}, where we also explain in more detail the concepts defined in the graphs.
More examples of hierarchies with branches can also be found in Section~\ref{sec:case-studies}.

\subsection{Individual typing}
\label{subsec:individual-typing}

So far we have organised the family of graphs which represent our languages and workflows for robots in a tree-shaped hierarchy.
Following, we need to define how these graphs relate to each other via instantiation, i.e.\ with typing relations.
First, we do so element-wise for all elements within a graph.

Any element \elementname{e} in any graph \(\graphname{\examplegraphone}{\indexone}\) has a unique type denoted by \(\typeof{\elementname{e}}\).
In that case, we can say that \elementname{e} is \emph{typed by} \(\typeof{\elementname{e}}\) or, equivalently, that \elementname{e} is an \emph{instance} of \(\typeof{\elementname{e}}\).
The \(\typeof{\elementname{e}}\) relation has the semantic meaning of the \emph{type} of \elementname{e}, e.g.\ \(\typeof{\elementname{Task}} = \elementname{EClass}\) and is defined as follows.

\begin{definition}[Individual Type]
For any \elementname{e} in a graph \(\graphname{\examplegraphone}{\indexone}\), with \(\indexone \geq 1\), its \emph{individual} (or \emph{direct}) \emph{type} \(\typeof{\elementname{e}}\) is found in a graph \(\graphname{\hierarchygraph}{}(\elementname{e})\), which is one of the graphs \(\graphname{\examplegraphone}{\indexone-1}, \dots, \graphname{\examplegraphone}{1}, \graphname{\examplegraphone}{0}\) in its typing chain.
Similarly, we can say that \elementname{e} is a \emph{direct instance} of \typeof{\elementname{e}}.
\end{definition}

This definition of the typing relation allows it to jump over levels, which is required later on for the definition of potency, a key concept in MLM (see Section~\ref{subsec:potency}).
As a consequence, the graphs in which we locate the types of different elements in \(\graphname{\examplegraphone}{\indexone}\) may also be different.
Also, the difference between the level of an element and the level of its type is not the same for every element, which yields the concept of level difference.

\begin{definition}[Level Difference]
For any \elementname{e} in a graph \(\graphname{\examplegraphone}{\indexone}\), we denote the difference between \(\indexone\) and the level where \(\graphname{\hierarchygraph}{}(\elementname{e})\) is located by \(\difference{\elementname{e}}\).
\end{definition}

In most cases, this difference is \(1\), meaning that the type of \elementname{e} is located at the level directly above it.
In short, we can say that, for any element \elementname{e} in a given graph \(\graphname{\examplegraphone}{\indexone}\), with \(\indexone \geq 1\), its type \(\typeof{\elementname{e}}\) is an element in the graph \(\graphname{\examplegraphone}{\indexone - \difference{\elementname{e}}}\), where \(1 \leq \difference{\elementname{e}} \leq \indexone\).

\begin{figure}[ht!]
	\centering
	\includegraphics[page=3,width=\linewidth]{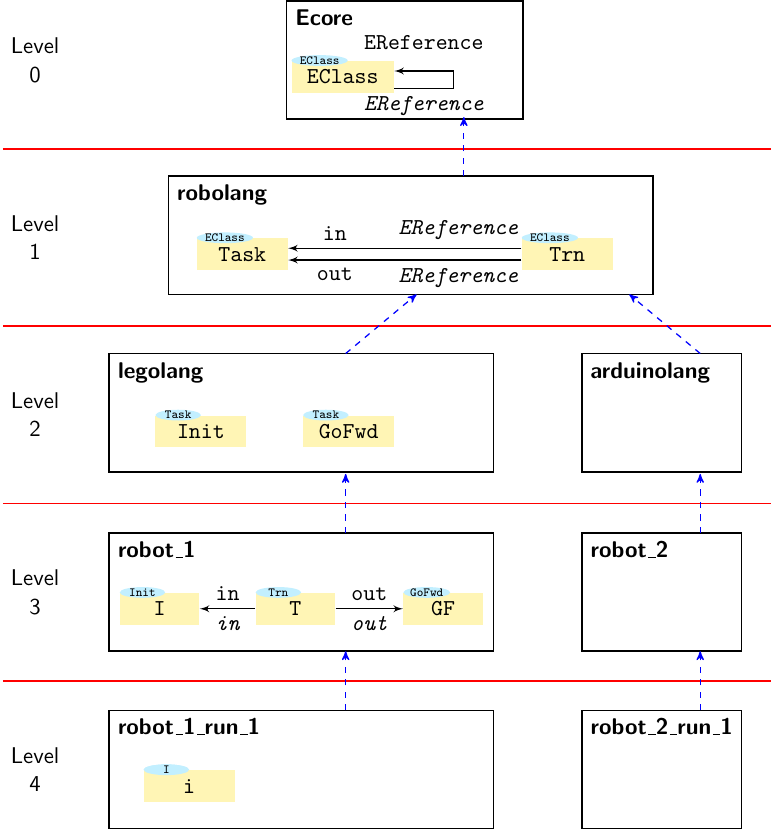}
	\caption{Graph hierarchy with typed elements}
	\label{fig:hierarchy-introductory-example-3}
\end{figure}

Figure~\ref{fig:hierarchy-introductory-example-3} displays our example hierarchy, including the type of each element.
To avoid polluting the diagram with too many arrows connecting instances with their types, we use alternative representations for them.
For every node, its type is identified by name and depicted in a blue ellipse attached to the node.
For example, the type of \elementname{I} is \elementname{Initial}, which could also be represented as an arrow between these two nodes.
For the arrows, the type is represented as another label with the name of the type.
This label is distinguished from the one with the element's own name by using italics font.
For example, the type of the \elementname{out} arrow in \elementname{robolang} is \elementname{EReference}, located in the \elementname{Ecore} graph.

Recall that the root graph \(\graphname{\examplegraphone}{0}\) in our approach is self-defining.
Hence, instead of requiring that \(\typeof{\elementname{e}} = \elementname{e}\) for all \(\elementname{e} \in \graphname{\examplegraphone}{0}\), we just need to ensure that if \(\elementname{e} \in \graphname{\examplegraphone}{0}\) then \(\typeof{\elementname{e}} \in \graphname{\examplegraphone}{0}\).

From a more general point of view, we obtain for any \elementname{e} in \(\graphname{\examplegraphone}{\indexone}\) a sequence
\begin{center}
\begin{tikzpicture}[on grid,node distance=25mm]

\def\vd{7mm}

\node[el-math]	(e)						{\elementname{e}};
\node[el-math]	(tye)	[right of=e]	{\typeof{\elementname{e}}};
\node[el-math]	(ty2e)	[right of=tye]	{\typeof[2]{\elementname{e}}};
\node[el-math]	(dots)	[right of=ty2e]	{\dots};
\node[el-math]	(tyse)	[right of=dots]	{\typeof[\step{\subsubscripttt{\elementname{e}}}]{\elementname{e}}};

\node[el-math]	(in1)	[below=\vd of e]	{\rotatebox[origin=c]{-90}{\(\in\)}};
\node[el-math]	(in2)	[below=\vd of tye]	{\rotatebox[origin=c]{-90}{\(\in\)}};
\node[el-math]	(in3)	[below=\vd of ty2e]	{\rotatebox[origin=c]{-90}{\(\in\)}};
\node[el-math]	(in4)	[below=\vd of tyse]	{\rotatebox[origin=c]{-90}{\(\in\)}};

\node[el-math]	(gi)	[below=\vd of in1]	{\graphname{\examplegraphone}{\indexone}};
\node[el-math]	(gid)	[below=\vd of in2]	{\graphname{\examplegraphone}{\indexone-\difference{\subscripttt{\elementname{e}}}}};
\node[el-math]	(gid2)	[below=\vd of in3]	{\graphname{\examplegraphone}{\indexone-{\difference[2]{\subscripttt{\elementname{e}}}}}};
\node[el-math]	(gids)	[below=\vd of in4]	{\graphname{\examplegraphone}{\indexone-{\difference[\step{\subsubsubscripttt{\elementname{e}}}]{\subscripttt{\elementname{e}}}}}};
\node[el-math]	(eq)	[right=12mm of gids]	{=};
\node[el-math]	(go)	[right=7mm of eq]		{\graphname{\examplegraphone}{0}};

\draw[mapto]	(e)		to	node [la-math]	(m1)	{}	(tye);
\draw[mapto]	(tye)	to	node [la-math]	(m2)	{}	(ty2e);
\draw[mapto]	(ty2e)	to	node [la-math]	(m3)	{}	(dots);
\draw[mapto]	(dots)	to	node [la-math]	(m4)	{}	(tyse);

\end{tikzpicture}
\end{center}
of typing assignments of length \(1 \le \step{\elementname{e}} \le \indexone\) with \((\indexone-{\difference[\step{\elementname{e}}]{\elementname{e}}}) = 0\).
The number \(\step{\elementname{e}}\) of steps depends individually on the item \elementname{e}.
For convenience, we use the following abbreviations:

\begin{equation*}
\begin{aligned}[t]
\typeof[2]{\elementname{e}}&=\typeof{\typeof{\elementname{e}}}\\
\typeof[3]{\elementname{e}}&=\typeof{\typeof[2]{\elementname{e}}}\\
\dots
\end{aligned}
\hspace{1cm}
\begin{aligned}[t]
\difference[2]{\elementname{e}}&=\difference{\elementname{e}}+\difference{\typeof{\elementname{e}}}\\
\difference[3]{\elementname{e}}&=\difference[2]{\elementname{e}}+\difference{\typeof[2]{\elementname{e}}}\\
\dots
\end{aligned}
\end{equation*}

\begin{definition}[Transitive Type]
For any \elementname{e}, we call any of the elements \(\typeof[2]{\elementname{e}}\), \(\typeof[3]{\elementname{e}}\), \ldots , \(\typeof[\step{\subsubscripttt{\elementname{e}}}]{\elementname{e}}\) a \emph{transitive type} of \elementname{e}.
Conversely, we can refer to \elementname{e} as a \emph{transitive} (or \emph{indirect}) \emph{instance} of \(\typeof[\step{\subsubscripttt{\elementname{e}}}]{\elementname{e}}\).
\end{definition}

Let us consider an arbitrary arrow \(\elementname{\examplenodeone} \xrightarrow{\elementname{\examplearrowone}} \elementname{\examplenodetwo}\), together with its source and target nodes in a graph \(\graphname{\examplegraphone}{\indexone}\).
The types of the three elements may be located in three different graphs, e.g.\ the edge \elementname{in} in the graph at level 3 in Figure~\ref{fig:hierarchy-introductory-example-3} has its type at level 1, while the type of its source is also at level 1 and the type of its target is at level 2.
The typing of arrows should, however, be compatible with the typing of sources and targets, that is, the source and the target of an arrow \(\typeof{\elementname{\maplevelone}} \in \graphname{\examplegraphone}{\indexone - \difference{\elementname{\maplevelone}}}\) must be provided by the types of \(\elementname{\examplenodeone}\) and \(\elementname{\examplenodetwo}\), respectively.

\begin{center}
	\includegraphics[width=\linewidth]{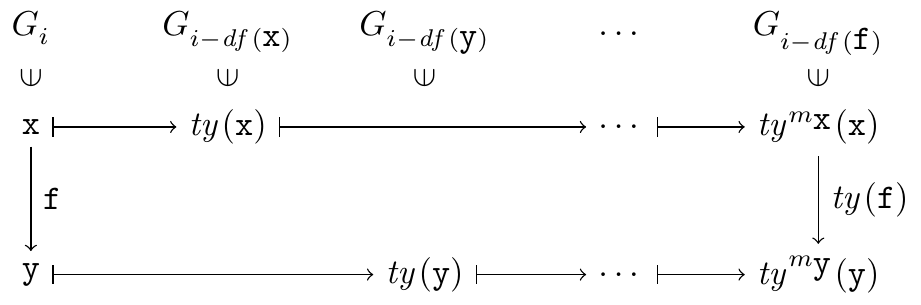}
\end{center}

Specifically, we require that the following \emph{non-dangling typing} condition is satisfied:

\begin{condition}[Non-dangling Typing]
\label{cond:non-dangling}
There exist \(1 \le m_\elementname{\examplenodeone} \le \step{\elementname{\examplenodeone}}\) and \(1 \le m_\elementname{\examplenodetwo} \le \step{\elementname{\examplenodetwo}}\) such that

\begin{itemize}
	\item \(\difference[m_\elementname{\examplenodeone}]{\elementname{\examplenodeone}} = \difference[m_\elementname{\examplenodetwo}]{\elementname{\examplenodetwo}} = \difference[]{\elementname{\examplearrowone}}\) and
	\item \(\typeof[m_\elementname{\examplenodeone}]{\elementname{\examplenodeone}}\) is the source of \(\typeof{\elementname{\examplearrowone}}\) and
	\item \(\typeof[m_\elementname{\examplenodetwo}]{\elementname{\examplenodetwo}}\) is the target of \(\typeof{\elementname{\examplearrowone}}\).
\end{itemize}
\end{condition}

For example, for the arrow \(\elementname{T} \xrightarrow{\elementname{in}} \elementname{I}\) in the graph \elementname{robot\_1} in our example hierarchy in Figure~\ref{fig:hierarchy-introductory-example-3} we do have \(\typeof{\elementname{in}} = \elementname{in}\) with \(\difference{\elementname{in}} = 2\), \(m_\elementname{T} = 2\) with \(\typeof{\elementname{T}} = \elementname{Trn}\) and  \(m_\elementname{I} = 2\) with \(\typeof[2]{\elementname{I}} = \elementname{Task}\).

\subsection{Typing morphisms and domains of definition}
\label{subsec:typing-morphisms-and-domain-definition}

Thanks to the previous characterisation of individual typing, we can now define how our graphs in a hierarchy relate to each other, by generalising the relations from one graph to another based on the individual typing relations of the elements they contain.
That is, we can define typing relations between any two graphs in a typing chain as an abstraction of the individual typing that we defined for their elements in Section~\ref{subsec:individual-typing}.
The vocabulary defined for individual typing can be reused here, so that a graph can be an \emph{instance} of another graph, or be \emph{typed by} it.
These relations among graphs are defined by means of graph homomorphisms.
Since the individual typing maps are allowed to jump over levels, two different elements in the same graph may have their types located in different graphs along the typing chain.
Hence, the typing morphisms established between graphs become partial graph homomorphisms~\cite{rossini2014formalisation}.
We first define the concept of domain of definition, which helps in the construction of partial graph homomorphisms.

\begin{definition}[Domain of Definition]
\label{def:domain-definition}
For any levels \(\indexone\), \(\indextwo\) such that \(0 \le \indexone < \indextwo \le \chaindepthone\), there is a partial typing morphism \(\typemorph{\indextwo}{\indexone} : \graphname{\examplegraphone}{\indextwo} \partialmap \graphname{\examplegraphone}{\indexone}\) given by a subgraph \(\domain{\typemorph{\indextwo}{\indexone}} \sqsubseteq \graphname{\examplegraphone}{\indextwo}\), called the \emph{domain of definition} of \(\typemorph{\indextwo}{\indexone}\), and a \emph{total typing homomorphism} \(\typemorph{\indextwo}{\indexone} : \domain{\typemorph{\indextwo}{\indexone}} \to \graphname{\examplegraphone}{\indexone}\).
\end{definition}

\begin{figure}[ht]
	\centering
	\includegraphics{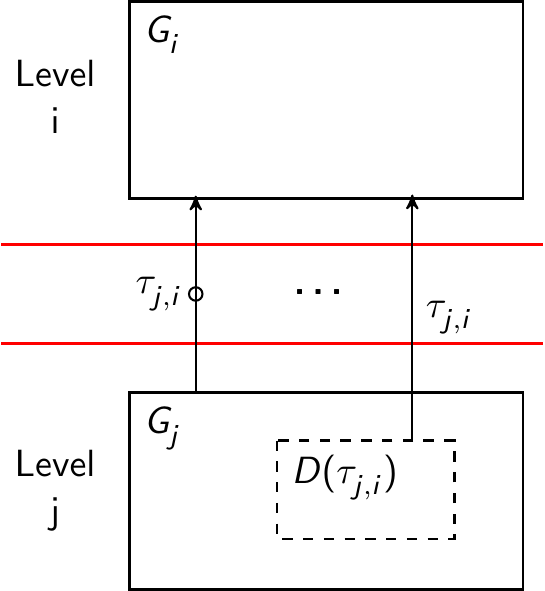}
	\caption{Typing morphisms and domain of definition}
	\label{fig:tau-and-domain-of-definition}
\end{figure}

The domain of definition may be empty (and, consequently, \(\typemorph{\indextwo}{\indexone}\) is just the inclusion of the empty graph in \(\graphname{\examplegraphone}{\indexone}\)) in case no \(\elementname{e} \in \graphname{\examplegraphone}{\indextwo}\) has a transitive type \(\typeof[m]{\elementname{e}} \in \graphname{\examplegraphone}{\indexone}\) for some \(m \geq 1\).
Also, in abuse of notation, we use the same name for both morphisms, since they represent the same typing information.
Using the same syntax as the examples, Figure~\ref{fig:tau-and-domain-of-definition} depicts these concepts in a generic yet visual manner.

Since a graph is given by its set of nodes and its set of arrows, we can define the domain of definition component-wise.
For any \(0 \le \indexone < \indextwo \le \chaindepthone\) we can define the set \(\domain{\typemorph[\graphnodes]{\indextwo}{\indexone}} \subseteq \graphname[\graphnodes]{\examplegraphone}{\indextwo}\) of all nodes in \(\graphname{\examplegraphone}{\indextwo}\) that are transitively typed by nodes in \(\graphname{\examplegraphone}{\indexone}\).
That is, \(\elementname{e} \in \domain{\typemorph[\graphnodes]{\indextwo}{\indexone}}\) iff there exists \(1 \le m \le \step{\elementname{e}}\) such that \(\indextwo - \indexone = \difference[m]{\elementname{e}}\), and thus \(\typeof[m]{\elementname{e}} \in \graphname[\graphnodes]{\examplegraphone}{\indexone}\).
We set \(\typemorph[\graphnodes]{\indextwo}{\indexone}(\elementname{e}) := \typeof[m]{\elementname{e}}\) for all nodes in \(\domain{\typemorph[\graphnodes]{\indextwo}{\indexone}}\) and obtain, in such a way, a map \(\typemorph[\graphnodes]{\indextwo}{\indexone} : \domain{\typemorph[\graphnodes]{\indextwo}{\indexone}} \to \graphname[\graphnodes]{\examplegraphone}{\indexone}\).
This total map defines a partial map \(\typemorph[\graphnodes]{\indextwo}{\indexone} : \graphname[\graphnodes]{\examplegraphone}{\indextwo} \partialmap \graphname[\graphnodes]{\examplegraphone}{\indexone}\) with the domain of definition \(\domain{\typemorph[\graphnodes]{\indextwo}{\indexone}}\).
This component-wise construction is required to ensure the compatibility between the way that domains of definition are calculated and the individual typing.

We can follow an analogous procedure for the set of arrows \(\graphname[\grapharrows]{\examplegraphone}{}\) to obtain a partial map \(\typemorph[\grapharrows]{\indextwo}{\indexone} : \graphname[\grapharrows]{\examplegraphone}{\indextwo} \partialmap \graphname[\grapharrows]{\examplegraphone}{\indexone}\) with domain of definition \(\domain{\typemorph[\grapharrows]{\indextwo}{\indexone}}\).

The non-dangling typing condition is now equivalent to the requirement that the pair \((\domain{\typemorph[\graphnodes]{\indextwo}{\indexone}}, \domain{\typemorph[\grapharrows]{\indextwo}{\indexone}})\), constitutes a subgraph \(\domain{\typemorph{\indextwo}{\indexone}}\) of \(\graphname{\examplegraphone}{\indextwo}\).
That is, for any arrow \(\elementname{e} \in \domain{\typemorph[\grapharrows]{\indextwo}{\indexone}}\) we have \(\source[\graphname{\examplegraphone}{\indextwo}](\elementname{e}), \target[\graphname{\examplegraphone}{\indextwo}](\elementname{e}) \in \domain{\typemorph[\graphnodes]{\indextwo}{\indexone}}\).
Consequently, the pair of morphisms \((\typemorph[\graphnodes]{\indextwo}{\indexone}, \typemorph[\grapharrows]{\indextwo}{\indexone})\) provides a total graph homomorphism \({\typemorph{\indextwo}{\indexone} : \domain{\typemorph{\indextwo}{\indexone}} \to \graphname{\examplegraphone}{\indexone}}\) and thus a partial typing morphism \(\typemorph{\indextwo}{\indexone} : \graphname{\examplegraphone}{\indextwo} \partialmap \graphname{\examplegraphone}{\indexone}\).

All the typing morphisms \(\typemorph{\indextwo}{0} : \graphname{\examplegraphone}{\indextwo} \to \graphname{\examplegraphone}{0}\) with \(1 \leq \indextwo \leq \chaindepthone\) are total, since every element has a type.
The uniqueness of typing is reflected on the abstraction level of type morphisms by the \emph{uniqueness condition}:

\begin{condition}[Typing Uniqueness]
\label{cond:typing-uniqueness}
For all \(0 \le \indexone < \indextwo < \indexthree \leq \chaindepthone\), we have that \(\typemorph{\indexthree}{\indextwo} ; \typemorph{\indextwo}{\indexone} \preceq \typemorph{\indexthree}{\indexone}\), i.e., \(\domain{\typemorph{\indexthree}{\indextwo} ; \typemorph{\indextwo}{\indexone}} \sqsubseteq \domain{\typemorph{\indexthree}{\indexone}}\) and, moreover, \(\typemorph{\indexthree}{\indextwo} ; \typemorph{\indextwo}{\indexone}\) and \(\typemorph{\indexthree}{\indexone}\) coincide on \(\domain{\typemorph{\indexthree}{\indextwo};\typemorph{\indextwo}{\indexone}}\).
\end{condition}

\begin{figure}[ht]
	\centering
	\includegraphics{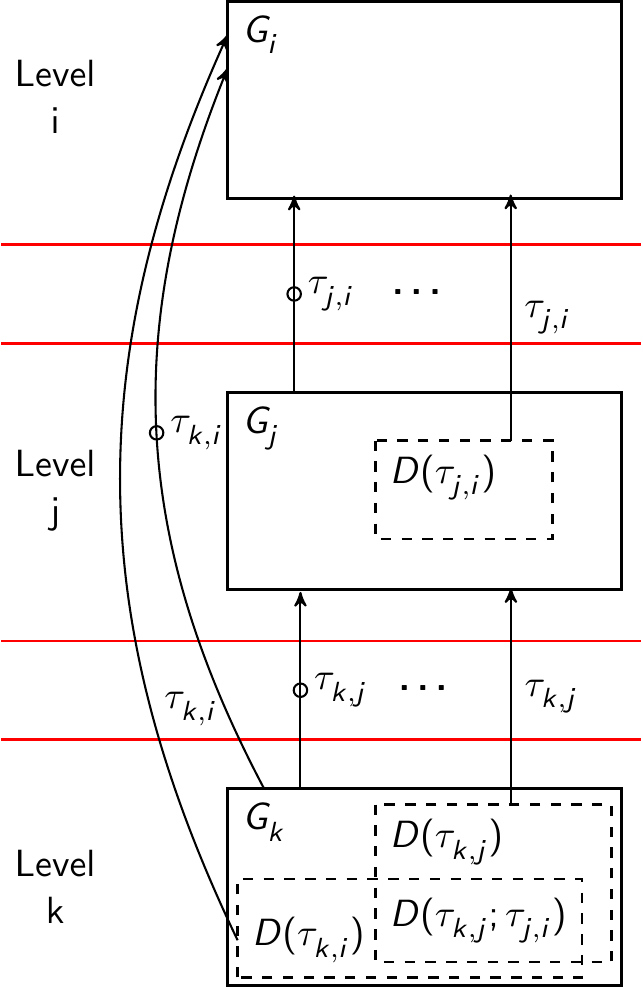}
	\caption{Typing uniqueness and domains of definition}
	\label{fig:tau-and-domain-of-definition-transitivity-and-uniqueness}
\end{figure}

Condition~\ref{cond:typing-uniqueness} is illustrated in Figure~\ref{fig:tau-and-domain-of-definition-transitivity-and-uniqueness} with three arbitrary levels and the domains of definition for each level.
Note, that the domain of definition of the composition \(\domain{\typemorph{\indexthree}{\indextwo} ; \typemorph{\indextwo}{\indexone}}\) is obtained by a pullback (inverse image) (see Definition~\ref{def:graph-chain} in Section~\ref{subsec:multilevel-typing}).
Informally, the uniqueness condition means that whenever an element in \(\graphname{\examplegraphone}{\indexthree}\) is transitively typed by an element in \(\graphname{\examplegraphone}{\indextwo}\) such that this element in \(\graphname{\examplegraphone}{\indextwo}\) is, in turn, transitively typed by an element in \(\graphname{\examplegraphone}{\indexone}\) then the element in \(\graphname{\examplegraphone}{\indexthree}\) is also transitively typed by the same type in \(\graphname{\examplegraphone}{\indexone}\).

The other way around, we can reconstruct individual typings from a family of partial typing morphisms between graphs that satisfy the totality and uniqueness conditions.
For any item \elementname{e} in a graph \(\graphname{\examplegraphone}{\indexthree}\) there exists a maximal (least abstract) level \(0 \le \indexone_\elementname{e} < \indexthree\) such that \elementname{e} is in \(\domain{\typemorph{\indexthree}{\indexone_\elementname{e}}}\) but not in \(\domain{\typemorph{\indexthree}{\indextwo}}\) for all \({\indexone_\elementname{e}} < \indextwo < \indexthree\), since \(\typemorph{\indextwo}{0}\) is total and \(\indexthree\) a finite number.
Hence, the individual type of \elementname{e} is given by \(\typeof{\elementname{e}} := \typemorph{\indexthree}{\indexone_\elementname{e}}(\elementname{e})\) and \(\difference{\elementname{e}} := \indexthree - \indexone_\elementname{e}\).

\begin{figure}[ht!]
	\centering
	\includegraphics[page=4,width=\linewidth]{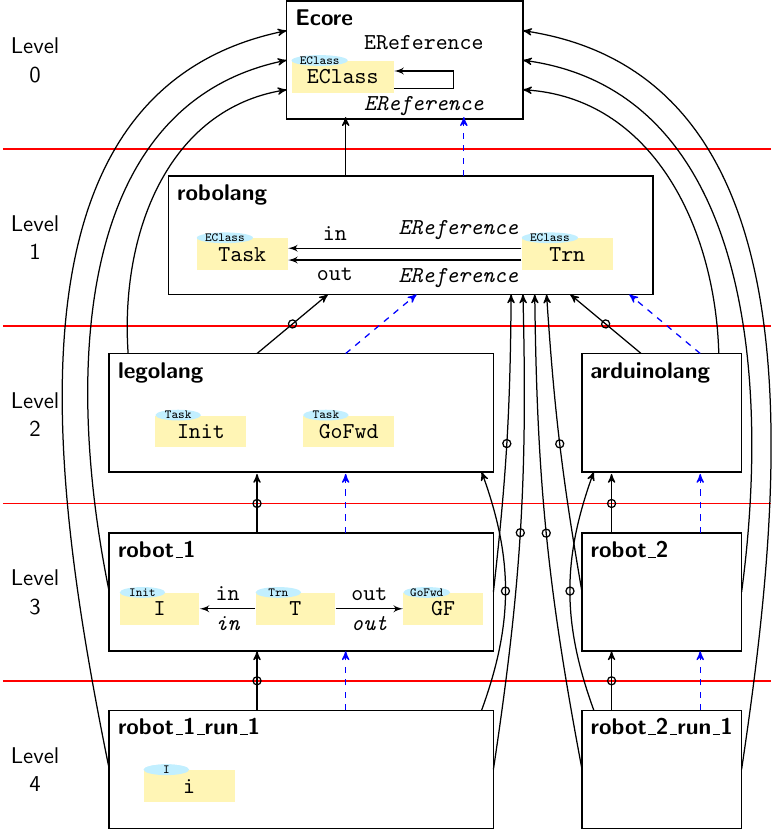}
	\caption{Graph hierarchy with typing morphisms}
	\label{fig:hierarchy-introductory-example-4}
\end{figure}

Figure~\ref{fig:hierarchy-introductory-example-4} depicts the typing morphisms between all the graphs in the hierarchy.
Note that the morphism from \elementname{robolang} to \elementname{Ecore} is total, since all the types used in the former can only be defined in the latter, and all elements must have a type.
As shown in the figure, it is also possible to define a total morphism from any graph in the hierarchy to the root graph via composition of partial typing morphisms with respect to the individual typing morphisms that they represent.

In summary, we define multilevel typing, first, by individual typing of graph elements, and then by abstracting it away by means of partial typing morphisms.

\subsection{Potency}
\label{subsec:potency}

The way we define typing relations and typing morphisms up to this point is too loose to ensure that we obtain coherent tree-shaped hierarchies, since typing relations can jump an arbitrary number of levels, which makes them redundant.
Hence, levels become less useful in practice if the individual typing relations are unbounded.
However, typing jumps over several levels allows for convenient solutions like defining \elementname{Trn} only once (recall Figure~\ref{fig:hierarchy-introductory-example-4}) and then connect transitive instances of \elementname{Task} with direct instances of \elementname{Trn}, like we do in the \elementname{robot\_1} graph.
So we need a mechanism to restrict the jumps of typing morphisms across levels in some cases, and allow it in some others.
For this purpose, we introduce in the following our modification and formalisation of the concept of \emph{potency of an element} originally defined in \cite{atkinson2002rearchitecting}.

Potency is used on elements as a means of restricting the length of the jumps of typing morphisms for their instances.
Other authors represent the potency of an element with just a number, whereas we employ an interval that allows for a higher degree of expressiveness, using the notation \elementname{min--max--depth}.
These values may appear after the declaration of an element, using \elementname{\potencyseparator} as a separator.
Furthermore, some existing realizations of potency have different effects on classes than on attributes.
However, we do not require to differentiate between them since attributes and their data types are also represented as nodes and arrows (see Section~\ref{subsec:data-type-dimension}).

To explain the meaning of the three values which we use to specify potency, we first focus on the first two, which act as a range (hence the names \elementname{min} and \elementname{max}).
The intuition behind these values is specifying how many levels below a direct instance of an element can be created, bounding the valid range within the two specified values.
Formally, we can define the constraint imposed by the first two values of potency as follows:

\begin{condition}[Potency Values (a)]
\label{cond:potency-values-1}
For any element \elementname{\examplenodeone} in \(\graphname{\examplegraphone}{\indexone}\) we require \(\elementname{min} \le \difference{\elementname{\examplenodeone}} \le \elementname{max}\) where \(\typename{\typeof{\elementname{\examplenodeone}}}{min--max}\) is the declared potency of the type \(\typeof{\elementname{\examplenodeone}}\) in \(\graphname{\examplegraphone}{\indexone-\difference{\elementname{\examplenodeone}}}\).
\end{condition}

This condition can be reformulated using partial typing morphisms:

\begin{condition}[Potency Values (b)]
\label{cond:potency-values-2}
For any element \elementname{\examplenodetwo} in \(\graphname{\examplegraphone}{\indextwo}\) with a potency declaration \typename{\examplenodetwo}{min--max} we require that \(\typemorph[-1]{\indextwo}{\indexone}(\elementname{\examplenodetwo})\) is empty for all \(\indextwo\) with \(\indextwo - \indexone < \elementname{min}\) or \(\indextwo - \indexone > \elementname{max}\).
\end{condition}

In all other cases, that is, \(\elementname{min} \le \indextwo - \indexone \le \elementname{max}\), there is no strict requirement.
\(\typemorph[-1]{\indextwo}{\indexone}(\elementname{\examplenodetwo})\) may be empty or not.

As for the third value, which resembles the concept of potency understood as \emph{depth}~\cite{atkinson2001essence}, it allows to control the number of times that an instance of an instance (and so on) can be created.
In a formal way, we define this third value as an addition to the previous condition, as follows:

\begin{condition}[Potency Values (c)]
\label{cond:potency-values-3}
For any element \elementname{\examplenodeone} in \(\graphname{\examplegraphone}{\indexone}\) we require \(\elementname{min} \le \difference{\elementname{\examplenodeone}} \le \elementname{max}\) and \(\elementname{depth} \ge 1\) where \(\typename{\typeof{\elementname{\examplenodeone}}}{min--max--depth}\) is the declared potency of the type \(\typeof{\elementname{\examplenodeone}}\) in \(\graphname{\examplegraphone}{\indexone-\difference{\elementname{\examplenodeone}}}\).
\end{condition}

This new version of the potency condition can finally be reformulated using partial typing morphisms and typing relations:

\begin{condition}[Potency Values (d)]
\label{cond:potency-values-4}
For any element \elementname{\examplenodetwo} in \(\graphname{\examplegraphone}{\indextwo}\) with a potency declaration \typename{\examplenodetwo}{min--max-depth} we require that \(\typemorph[-1]{\indextwo}{\indexone}(\elementname{\examplenodetwo})\) is empty for all \(\indextwo\) with \(\indextwo - \indexone < \elementname{min}\) or \(\indextwo - \indexone > \elementname{max}\) or also for all \(\typeof[-\step{\subsubscripttt{\elementname{\examplenodetwo}}}]{\elementname{\examplenodetwo}}\) with \(\step{\subsubscripttt{\elementname{\examplenodetwo}}} > \elementname{depth}\).
\end{condition}

Condition~\ref{cond:potency-values-4} provides a definition of three-valued potency which can accommodate the aforementioned realisations of the concept given in other prominent MLM approaches.
Additionally, for the value of depth to be consistent, the depth of a given element \elementname{e} is always strictly less than the depth of \(\typeof{\elementname{e}}\).
In most cases, if the depth of \(\typeof{\elementname{e}}\) equals a number \(d\) such that \(d > 0\), then the depth of \elementname{e} equals \(d - 1\).
If the value for depth is unbounded (represented as \elementname{*}), the depth of its instances is allowed to be unbounded, a positive integer, or zero.

\begin{figure}[ht!]
	\centering
	\includegraphics[page=5,width=\linewidth]{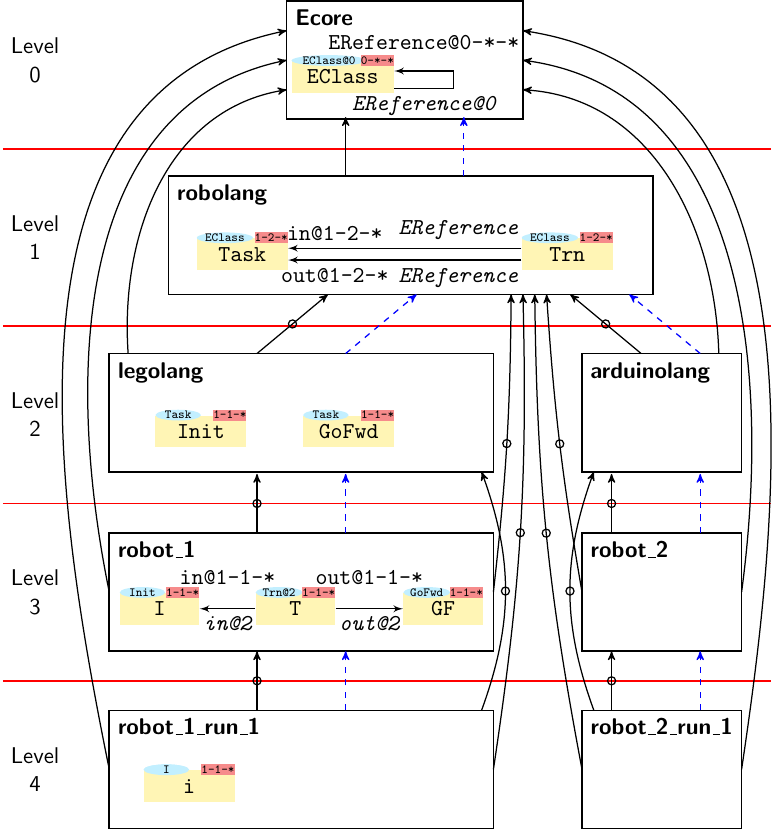}
	\caption{Full graph hierarchy with potency}
	\label{fig:hierarchy-introductory-example-5}
\end{figure}

For the typing relations used in our example to be correct, and given that we assume the default potency \elementname{1--1--*} unless otherwise specified, we require that at least the node \elementname{Trn} and the arrows \elementname{in} and \elementname{out} in \elementname{robolang} to have increased depth (see Figure~\ref{fig:hierarchy-introductory-example-5}). 
That way, it becomes possible to use these three elements respectively as types for \elementname{T}, \elementname{in} and \elementname{out}, which are located in the \elementname{robot\_1} graph two levels below.
Besides, and to ease the interpretation of a diagram, the type of an element which is specified in a level different from the one immediately above reuses the \elementname{\potencyseparator} notation to indicate it.
For nodes, the annotation is located in the blue ellipse.
For arrows, the \(\typename{ty(e)}{\difference{\elementname{e}}}\) annotation is displayed in the declaration of the type (in italics).
In both cases, we also define a non-graphical representation of the element and its type in the form \(\element[\difference{\elementname{e}}]{e}{ty(e)}\).
For the default value \elementname{1}, used in all the other cases, the notation \element{e}{ty(e)} is used as a shorthand for \element[1]{e}{ty(e)}.
Figure~\ref{fig:hierarchy-introductory-example-5} shows the final version of the example hierarchy in which potencies have been added to all elements.
Note that the potencies of the Ecore elements \elementname{EClass} and \elementname{EReference} are not the default \elementname{1--1--*}, defined for ``user-accessible'' levels but, \elementname{0--*--*}.
The minimum potency of \elementname{0} is required to allow for self-definition and the maximum unbounded value is required to be able to create direct instances of Ecore elements at any level below, without forcing the creation of an intermediate type (which mimics the functionality of linguistic extensions).
Moreover, the unbounded depth is necessary so that further instantiations are allowed as we keep adding levels to the hierarchy.

\subsection{Inheritance and multiplicity}
\label{subsec:inheritance-and-cardinalities}

These two techniques allow to further control the processes of specification and instantiation of the elements defined in a model.
Although they are not used in the Robolang example hierarchy, we include them here for the sake of completeness.
Examples of inheritance between nodes and multiplicities can be found in Sections~\ref{sec:case-studies} and~\ref{subsec:supplementary-dimension}.

The inheritance (i.e.\ specialisation) relation is a special type of arrow among any two nodes within the same level, which imposes on the child node the same typing and potency as the parent node.
Moreover, the inheritance relation gives the child node access to the incoming and outgoing arrows of the parent node, while still allowing the child node to define additional attributes or arrows.
However, in order to avoid conflicts, the child node cannot redefine an arrow of its parent node by reusing its name.
Using the concepts defined in the previous subsections, we can reformulate this requirement as follows:

\begin{condition}[Inheritance of Typing]
\label{cond:inheritance-typing}
For any two nodes \(\elementname{\examplenodeone} , \elementname{\examplenodetwo} \in \graphname[\graphnodes]{\examplegraphone}{\indextwo}\) and \(0 \le \indexone < \indextwo\), if \elementname{\examplenodeone} inherits from \elementname{\examplenodetwo}, then it is required that \(\typemorph[\graphnodes]{\indextwo}{\indexone}(\elementname{\examplenodeone}) = \typemorph[\graphnodes]{\indextwo}{\indexone}(\elementname{\examplenodetwo})\).
\end{condition}

Furthermore, cyclic inheritance and inheritance between arrows are forbidden, as it is usual in most realisations of the inheritance relation, both in modelling tools and object-oriented programming languages.

In the case of multiplicity (i.e.\ cardinality) of arrows, we consider them as an additional constraint that can be added as an annotation on top of the arrow to further control the instantiation process.
That is, the formalisation presented above does not constrain the amount of direct instances of an arrow that can be created in a model in a given level.
For this purpose, we introduce the concept of multiplicity of arrows, which also affects the multiplicity of attributes (see Section~\ref{subsec:data-type-dimension}).
So, the same way as potency allows the modeller to control \emph{where} instantiation can happen, multiplicities serve the purpose of specifying \emph{how many times} a direct instance of an arrow can, or should, be created in a model in a given level.
Multiplicities are expressed in this work in the traditional notation of \elementname{min..max}, where the \elementname{min} (and \elementname{max}) value restricts the minimum (and maximum) number of direct instances of the arrow in the context of its source node.
The default value for multiplicity is \elementname{0..*}, where, again, \elementname{*} represents \emph{unbounded}.

\section{A Three-Dimensional Framework}
\label{sec:three-dimensions}

All the concepts defined in Section~\ref{sec:hierarchies} are used to specify a single hierarchy that contains a family of related models with different levels of abstraction but a common domain.
Such is the case of the hierarchy used in the examples so far, which we call simply \emph{Robolang} in the following.

Nevertheless, complex scenarios from behavioural modelling may comprise more than one domain.
For example, it could be useful to introduce verification aspects into the language, so that we can specify correctness properties, or enhance it with logging capabilities that gather information about the behaviour of a particular instance.
In addition, data types such as integers, strings or boolean values are elements that may appear in any model and should be reusable in any hierarchy we create.
These three ``aspects'' of modelling (the base language, its possible additional aspects and the data types) can be represented as different modelling hierarchies in a consistent manner using the definitions from Section~\ref{sec:hierarchies}.

In this section, we introduce three dimensions where hierarchies can be located, depending on their nature and purpose, as well as the relations that can be defined between elements in different dimensions.
First, we consider the hierarchies that define behavioural models and instances to be the ``main'' ones.
We define these hierarchies to be located in the \emph{application dimension} and, in consequence, we name them \emph{application hierarchies}.
These are presented in more detail in Section~\ref{subsec:application-dimension}.
Second, the hierarchies representing additional aspects of an application hierarchy are located in a secondary dimension.
We call this the \emph{supplementary dimension}, which is therefore inhabited by \emph{supplementary hierarchies}.
Section~\ref{subsec:supplementary-dimension} delves into this concept and introduces an example of supplementary hierarchy.
And third, we represent data types, commonly used for attributes in other modelling approaches, as yet another dimension.
Data types are defined in a fixed hierarchy with three levels, which are presented in Section~\ref{subsec:data-type-dimension}.
The \emph{data type hierarchy} is the only one located in the \emph{data type dimension}.

The relations between elements in different dimensions are expressed using \emph{multiple typing}.
This concept is defined as an extension of the individual typing presented in Section~\ref{subsec:individual-typing}.
If there are one or more supplementary hierarchies for a given application hierarchy, any element \elementname{e} defined in the application hierarchy may have several types, combined as follows:

\begin{itemize}
	\item Exactly one typing map to another element in the same application hierarchy.
	In the uncommon case that this type is not specified, it will be assigned a default one.
	In our implementation using Ecore, the type would be \elementname{EClass} or \elementname{EReference} for nodes and arrows, respectively.
	That is, the requirement that every element has at least one type must hold.
	\item An arbitrary number of supplementary typing maps, with at most one type per supplementary hierarchy.
	Each hierarchy represents a different \emph{aspect}, so the element \elementname{e} can have one supplementary type in each supplementary hierarchy related to the application hierarchy where \elementname{e} is located.
	Section~\ref{subsec:supplementary-dimension} shows an example of such case and explains how to use supplementary typing, comparing our definition of multiple typing with the concepts of linguistic metamodels and extensions.
	\item At most one typing map to the data type hierarchy.
	This is the case when declaring attributes.
	In Section~\ref{subsec:data-type-dimension} we illustrate how to apply single or double typing to represent attributes and their data types.
\end{itemize}

It is worth pointing out the fact that the typing map \(\elementname{e}\to\typeof{\elementname{e}}\), with \elementname{e} and \(\typeof{\elementname{e}}\) belonging to the same application \emph{hierarchy}, is the only one allowed in the application \emph{dimension}.
That is, the type of \elementname{e}, or any of its types, cannot be in a different application hierarchy than \elementname{e}.
This means that any two application hierarchies are totally unrelated to each other, in the sense that it is not possible to create typing maps between their elements.
In other words, there cannot be two application hierarchies in the same system.
It is possible, however, that two application hierarchies in different systems have the same supplementary hierarchy (one ``copy'' each), and even that the same hierarchy acts as application hierarchy in one system and as supplementary in a different one.
Summing up, we can group these constraints under the requirement that there cannot be typing relations among different hierarchies located in the same dimension.

Each one of the typing maps of \elementname{e} has its own typing chain associated.
Hence, the concept of typing chains is also extended to allow the graphs that form it to belong to different hierarchies in different dimensions.
For the sake of simplicity, the examples in this thesis just illustrate cases with \emph{double typing}, but there are no conceptual or technical obstacles to adding more than two types to an element in an application hierarchy.

\subsection{Application dimension}
\label{subsec:application-dimension}

In our examples, the hierarchies used to define behavioural languages are located in this dimension.
One of the defining characteristics of application hierarchies is their independence from the supplementary hierarchies.
That is, while some scenarios cannot be modelled without using a combination of both kinds of hierarchies, the basic concepts of an application hierarchy should be independent of any supplementary ones.
For example, the temporal properties introduced in Section~\ref{subsec:supplementary-dimension} require to double type application elements.
However, those application elements can constitute a valid model without the double types.

Besides, an element with multiple typing where one of the types belongs to an application hierarchy must always be inside an application hierarchy.
That means that the supplementary and data type dimensions always act as secondary or complementary aspects, and never as the main language of a specification.
In a more general way, we can reformulate this as follows:

\begin{condition}[Priority of Application Typing]
\label{cond:application-typing}
Any graph \(\graphname{\examplegraphone}{}\), containing an element \elementname{e} (\(\elementname{e} \in \graphname[\graphnodes]{\examplegraphone}{}\) or \(\elementname{e} \in \graphname[\grapharrows]{\examplegraphone}{}\)) with double type, will necessarily be part of an application hierarchy.
\end{condition}

As pointed out before, we focus on the definition of structure of behavioural models and the specification of behaviour as model transformations.
The fact that the models in an application dimension are independent from those in the supplementary dimension is reflected also in those transformations, in the sense that they are independent from the supplementary models and model transformations.

\begin{figure}[ht]
	\centering
	\includegraphics[width=\textwidth]{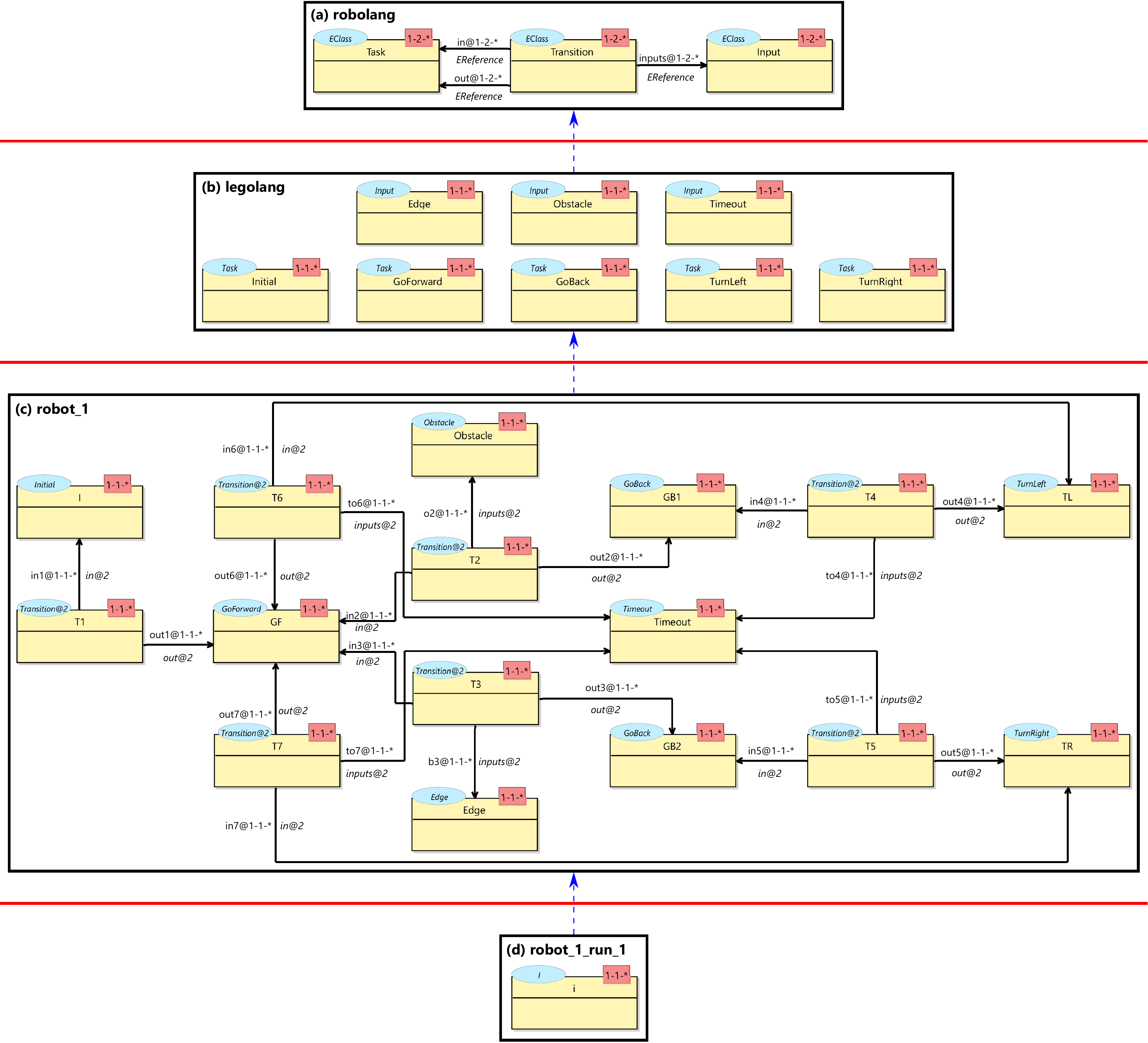}
	\caption{Full hierarchy for Robolang case study}
	\label{fig:robolang-multilevel-hierarchy}
\end{figure}

In Figure~\ref{fig:robolang-multilevel-hierarchy}, we display a hierarchy of models used for the definition of behaviour on simple robots, in order to demonstrate the properties and purpose of application hierarchies.
This figure is the full version of the left branch of the hierarchy used as an example in Section~\ref{sec:hierarchies}, but we do not show Ecore on top of the hierarchy since it is not relevant for the explanation.
The figure has been generated from the real implementation of the Robolang hierarchy created with the tool MultEcore, that we present in Section~\ref{sec:modelling-tooling}.
The default value for the multiplicity of an arrow is \elementname{0..n}, which is not displayed for the sake of clarity.

In Figure~\ref{fig:robolang-multilevel-hierarchy}(a) we define a DSML for process modelling in the domain of robots, hence the name \elementname{robolang}.
The main concept in this language is \elementname{Task}, which represents an action that a robot can perform.
\elementname{Transitions} are used to connect any number of \elementname{in} tasks to any number of \elementname{out} ones.
These transitions are triggered by \elementname{Inputs}, related to transitions by the \elementname{inputs} relation.
Note that there may still exist instances of \elementname{Transition} without an associated instance of \elementname{Input}, in which case it gets fired instantaneously.

The concepts defined in the \elementname{robolang} language are adapted for simple Lego EV3 robots in \elementname{legolang}.
This kind of robots are capable of moving in a flat surface, and detect both physical obstacles in their way and the edges of that surface.
The \elementname{legolang} language, depicted in Figure~\ref{fig:robolang-multilevel-hierarchy}(b), defines specific tasks for the movement possibilities of such robots: \elementname{GoForward}, \elementname{GoBack}, \elementname{TurnLeft} and \elementname{TurnRight}.
It also contains \elementname{Initial}, which represents the starting point in the process of performing those tasks.
Besides, the language defines the required elements for the detection of an \elementname{Edge} in the surface, an \elementname{Obstacle} in front of the robot, or the expiration of the time assigned to a task (\elementname{Timeout}).
One of the key points in this model is the lack of arrows, since the specification of \elementname{robolang} and the potencies of its elements allow us to refine only some of their concepts (\elementname{Task} and \elementname{Input}) while leaving the rest unaffected.

In Figure~\ref{fig:robolang-multilevel-hierarchy}(c), we define a specific behaviour for a robot, named \elementname{robot\_1}.
The robot starts by executing its initial task \elementname{I}, which fires automatically the transition \elementname{T1} to \elementname{GF}, indicating the robot to go forward.
Then, the robot may take two different courses of action.
In the first one, it will fire \elementname{T2} and start going back (\elementname{GB1}) if an obstacle (\elementname{Obstacle}) is detected.
After it goes back for a while, the timeout (\elementname{Timeout}) fires the transition (\elementname{T4}) and it starts turning left (\elementname{TL}).
After another timeout, \elementname{T6} is fired, and the robot resumes to going forward again until another input is detected.
The second course of action is similar to the previous one, but it starts by detecting an (\elementname{Edge}), which fires \elementname{T3} and commands the robot to go back (\elementname{GB2}).
After timeout, it will fire \elementname{T5} and turn right (\elementname{TR}) until the next timeout, which will fire \elementname{T7} and return to \elementname{GF}, completing the loop.
Note that all the arrows are typed by \elementname{in}, \elementname{out} and \elementname{input}, defined in \elementname{robolang}.
As explained in Section~\ref{sec:hierarchies}, these arrows can be defined without requiring an intermediate redefinition in \elementname{legolang}, thanks to the use of potency.
Thus, the separation in levels of abstraction is satisfied, since it is not desirable that concepts from the level of abstraction of \elementname{robolang} have to be unnecessarily repeated in \elementname{legolang}.

The bottommost level on this hierarchy, depicted in Figure~\ref{fig:robolang-multilevel-hierarchy}(d), represents the state of a particular execution of the \elementname{robot\_1} model, hence the name \elementname{robot\_1\_run\_1}.
The represented state is actually the initial state, where the robot begins its tasks by running an instance of the initial task of the model (\element{i}{I}).
This is the level that will be modified during the execution of the behavioural semantics defined as model transformations (see Section~\ref{subsec:mcmts-robolang}).

\begin{figure}[ht]
	\centering
	\includegraphics{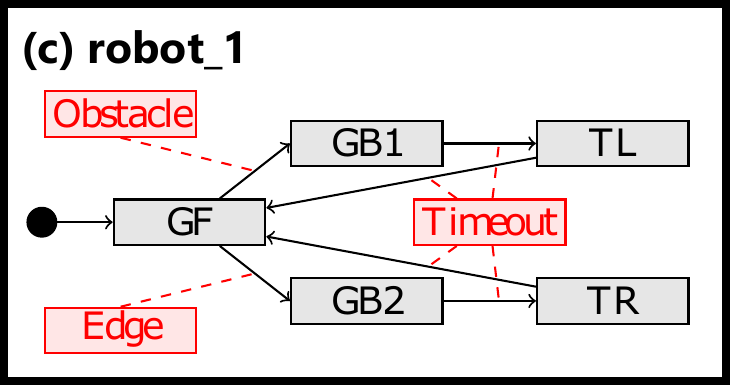}
	\caption{Model \elementname{robot\_1} with concrete syntax}
	\label{fig:robolang-level-3-robot-1-concrete-syntax}
\end{figure}

We conclude the presentation of the Robolang hierarchy by illustrating the usage of concrete syntax to facilitate the understanding of DSMLs.
The workflow in Figure~\ref{fig:robolang-multilevel-hierarchy}(c) can be difficult to read and manage using the basic syntax of MultEcore, which we can consider a sort of abstract syntax.
In contrast, we can define new visualisations for the elements based in their transitive types.
For example, in Figure~\ref{fig:robolang-level-3-robot-1-concrete-syntax} we display the model \elementname{robot\_1}.
In this alternative, we choose to represent tasks with grey rectangles that contain the node's name and use a black dot to distinguish the initial one, we use arrows (which connect the input and output tasks) for transitions and inputs appear as labelled red boxes connected with dashed lines to the transitions they trigger.

\subsection{Supplementary dimension}
\label{subsec:supplementary-dimension}

The purpose of hierarchies in this dimension is defining additional aspects that can be introduced in behavioural languages defined in the application dimension.

Compared with other MLM approaches (see Section~\ref{subsec:mlm}), the purpose of supplementary hierarchies can be understood as an evolution of the concept of \emph{linguistic metamodels}, including \emph{linguistic typing} and \emph{linguistic extension}.
First, the inspiration for separating dimensions for different hierarchies of languages, and having one of them acting as the ``main'' one, while the other adds some new concepts, stems from the idea of linguistic metamodels~\cite{atkinson2001essence,rossini2014formalisation}.
In these models, some concepts are defined so that they can be used in the main \emph{ontological} hierarchy.
These linguistic types must be applied on ontological elements, providing elements with double ontological-linguistic typing.
This means that the ontological hierarchy is dependent on the linguistic metamodel in the sense that every ontological element requires a linguistic type, which is a consequence of employing the \emph{clabject} paradigm~\cite{atkinson1997distributed}.
Alternatively, elements defined in a linguistic metamodel can be used as the only type of elements in the ontological hierarchy, hence creating elements with only one linguistic type and no ontological one.
This technique is known as linguistic extension.
Both concepts can be emulated in our conceptual framework by means of using an additional hierarchy, which serves the role of linguistic metamodel.
In addition, our framework does not constrain other possibilities like an element having just ontological type (on its own hierarchy) or more than two types (one additional type per supplementary hierarchy), while keeping the constraint of exactly one type per hierarchy.
This also allows for addition of initially unforeseen types, as other modelling approaches do.
See, for example,~\cite{delara2015aposteriori}.

An important consideration in our framework is related to the combination of supplementary typing with the value of depth, which is the third component of our three-valued potency.
When an element is provided with a new supplementary type, its depth must be the minimum of all the depths of its types (application and supplementary) minus one.
That is, the depth is calculated in the more restrictive manner.
If that element already have instances, such double-typing would pose a contradiction which must be fixed in advance.

We illustrate one possible use of supplementary hierarchies in the following.
In this new example, we define a property specification language in order to describe temporal correctness properties for behavioural languages, as we presented originally in~\cite{macias2016integration}.
As stated in Section~\ref{sec:hierarchies}, Ecore is the topmost level of the hierarchy from a technical point of view, although usually not user-accessible or modifiable.

The LTL (Linear-time Temporal Logic) specification language defines a propositional logic~\cite{manna1995tlspec}. 
This implies that one of the elements in the language are atomic propositions which can be evaluated at any particular point in time to \emph{true} or \emph{false}.
Since we apply this language to behavioural models, the result of the evaluation is based on the state of the model, that is, the existence of specific elements on the running instance of the model.
The LTL language contains the usual boolean connectives and additional operators, which define the meaning of a formula over a trace of observations for the propositions.

\begin{figure}[ht!]
  \centering
  \includegraphics[width=\linewidth]{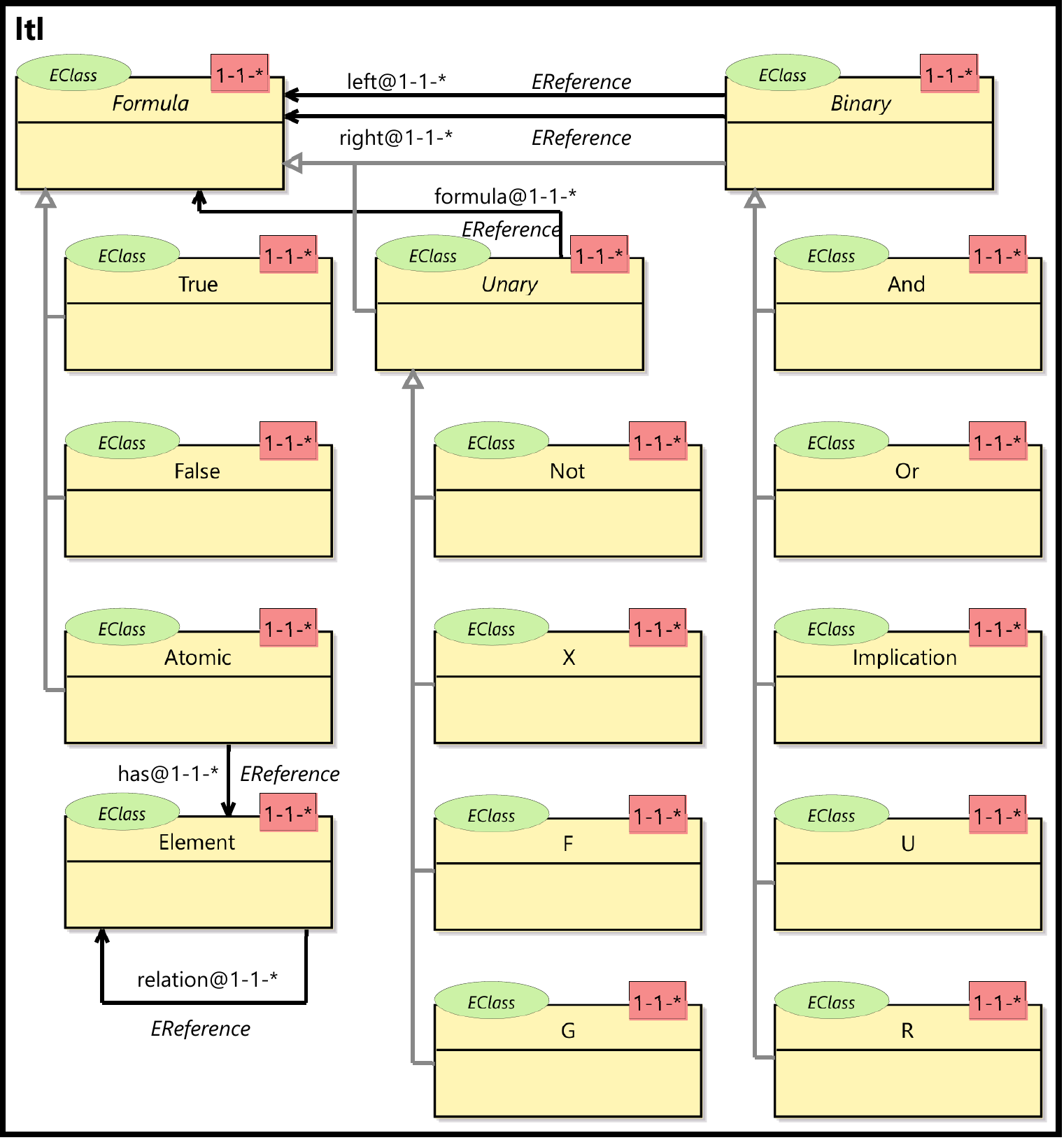}
  \caption{Supplementary LTL model}
  \label{fig:ltl-supplementary-model}
\end{figure}

In Figure~\ref{fig:ltl-supplementary-model} we represent the model that defines all the LTL concepts, which will allow us to specify LTL properties as double-typed models (see~\cite{macias2016integration}).
One of the advantages of this technique is that the supplementary models' properties evolve together with the actual model they refer to, while this application model can be also used for simulation, code generation and evaluation of such properties.
We used the same abstract syntax in Figure~\ref{fig:ltl-supplementary-model} as in the rest of the examples, to show that our framework is capable of also representing the grammar and semantics of a second-order logic.
In practice, it would be more convenient to specify a concrete (probably textual) syntax, synchronised with these abstract concepts.

\begin{figure}[ht]
  \centering
  \includegraphics[width=\linewidth]{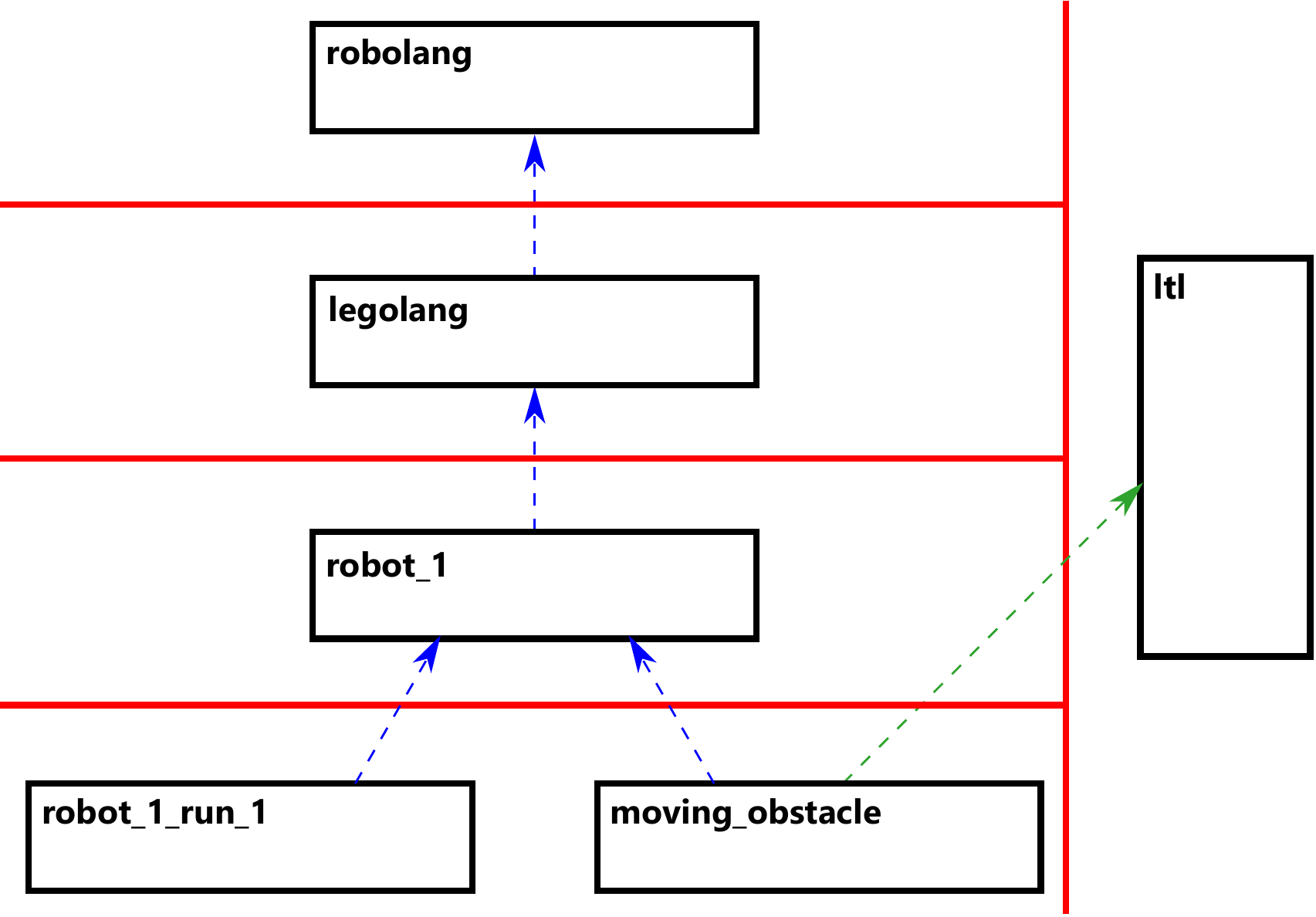}
  \caption{Overview of the Robolang and LTL hierarchies}
  \label{fig:ltl-multilevel-hierarchy-outline}
\end{figure}

The reasons to depict the single \elementname{ltl} model as a means to present this hierarchy are two.
First, we omit the Ecore metamodel on top, as we do in all the other hierarchies presented in this section.
And second, the properties are specified as new models typed by the elements in \elementname{ltl} but, as explained below, some of its elements (actually all, in this example) will have a double typing relation: one application type and one supplementary type.
In this case of double-typed elements, and as explained at the beginning of this section, the model will be located in the application hierarchy whose types are being used and, as a consequence, the models representing specific LTL properties will not be part of the LTL supplementary hierarchy.
The outline showing how these two hierarchies connect to each other is shown in Figure~\ref{fig:ltl-multilevel-hierarchy-outline}.

In the \elementname{ltl} model, all elements are instances of Ecore types, but this time we display that type in a green ellipse, instead of blue, to clarify that they belong to a supplementary hierarchy.
The main element is \elementname{Formula}, a node which can represent the whole LTL property, as well as any of its subexpressions, connected by operators.
These operators can be \elementname{Unary} or \elementname{Binary}, in which case they are connected to one or two subformulas, respectively.
These connections are represented with the relation \elementname{formula} in the first case, and with \elementname{left} and \elementname{right} in the second.
A unary operator can be a boolean negation (\elementname{Not}), or the temporal operators \elementname{X}, \elementname{F} and \elementname{G}.
The element \elementname{X} represents the \emph{next} LTL operator, which requires its subformula to hold in the next state.
\elementname{F} and \elementname{G} represent the temporal operators \emph{eventually} (the subformula should hold at some point in the future) and \emph{globally} (the subformula must always hold, from the current state on).

The binary elements can be boolean operators, like \elementname{Or} and \elementname{Implication}, or the temporal operators \elementname{U} and \elementname{R}.
\elementname{U} represents the binary operator \emph{until}, which captures that the left subformula must hold up to a point of transition where the right subformula holds.
And the operator \elementname{R} represents the temporal operator \emph{release}, similar to \emph{until}, but with the additional requirement that the left and right subformulas should hold simultaneously at the moment of the transition.

All the operators are used to connect \emph{atomic propositions}.
These are the most basic type of formulas, and eventually evaluate to true or false in a particular model.
An atomic proposition which is not evaluated is represented by the \elementname{Atomic} element, and once it is evaluated, the result is represented as either \elementname{True} or \elementname{False}.

Note that \elementname{Formula}, \elementname{Unary} and \elementname{Binary} have potency \typename{}{1-1-*} since that is the potency required by the operators which inherit from them.
However, we define them as abstract indicated by the name in italics and meaning that we cannot instantiate them.
In other cases, we can achieve a similar effect by using zero potency (\typename{}{0-0-0}).
Alternatively, we could have refined \elementname{Formula}, \elementname{Unary} and \elementname{Binary} by means of typing, separating them from the most specific elements (\elementname{Not}, \elementname{And}, \elementname{X}, etc.) by using two models in two different levels, in a similar fashion as the \elementname{robolang} and \elementname{legolang} models in Section~\ref{subsec:application-dimension}.
However, we wanted to illustrate that, thanks to the reuse of Ecore as the root of our hierarchies, we can define inheritance relations between nodes.
In our approach, we do not enforce a particular way of layering elements and allow for both inheritance/specialisation and metalevels/classification (see~\cite{kuhne2009contrast,borgida1984generalization,mylopoulos1980relationships}).
Instead, we choose to provide the necessary flexibility to allow both and leave the final choice to the model designer.
In this regard, it is worth pointing out that strict modelling guidelines and semantic-related sanity checks can use our framework as a starting point and merely add constraints ``on top'' of it without compromising its soundness.
In our LTL example, the way of modelling that we use makes it easier to read \elementname{ltl} as the model version of the EBNF grammar of the LTL language.
This way, all operators define non-terminal symbols, \elementname{True}, \elementname{False} and \elementname{Atomic} define terminals, and the relations define the structure of the language.

\begin{figure}[ht]
  \centering
  \includegraphics[width=.7\linewidth]{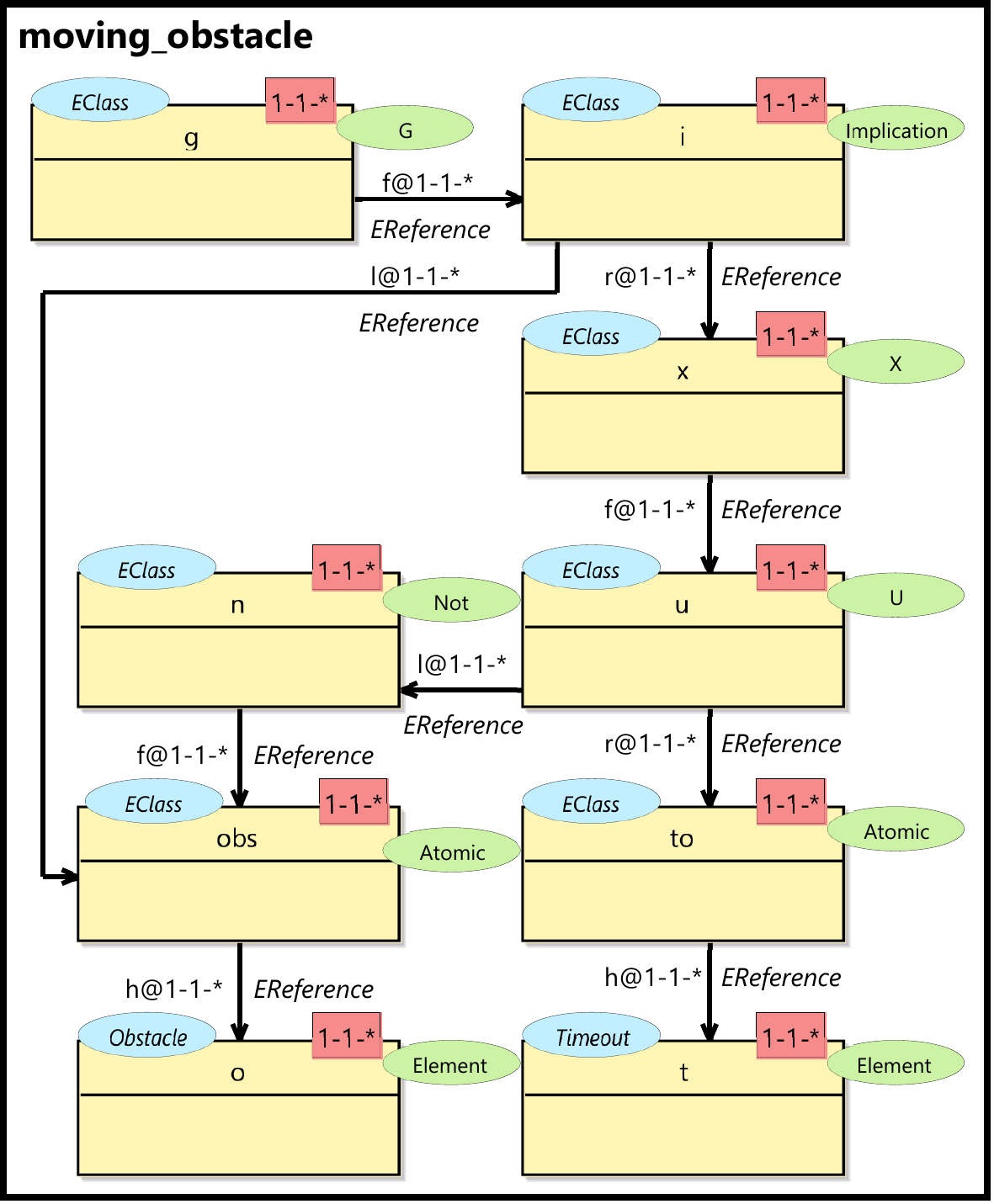}
  \caption{A temporal property with supplementary single-typed and double-typed elements}
  \label{fig:moving-obstacle-model}
\end{figure}

In Figure~\ref{fig:moving-obstacle-model}, we display a model which represents a specific temporal property, called \elementname{moving\_obstacle}, created as an instance of \elementname{robolang} in the application dimension, but also using types from \elementname{ltl} in the supplementary dimension.
The model encodes a safety condition on the behaviour of the robot:
when approaching an obstacle, backing away from it should clear the sensor input again.
If that is not the case, the robot's assumption about the environment, i.e.\ that obstacles should not move, is incorrect.
If that happened, the property would be violated.
This property, visualized in LTL syntax, is written as \(G(\mathit{obs} \rightarrow X(\neg \mathit{obs} \, U \, \mathit{to}))\).

The model \elementname{moving\_obstacle} also shows examples of double typing, especially in the \elementname{o} and \elementname{t} elements, which illustrate the way we use our framework to connect temporal properties to behavioural models.
First, the property as a model is part of an application hierarchy, and is modelled using types from the supplementary dimension.
And secondly, the atomic propositions get evaluated by matching them to the models representing executions of the behaviour, making use of the behavioural semantics for LTL, defined as model transformations in Section~\ref{subsec:mcmts-ltl}.
We omit the double-typing for arrows to simplify the visualization.

\begin{figure}[ht]
  \centering
  \includegraphics[width=\linewidth]{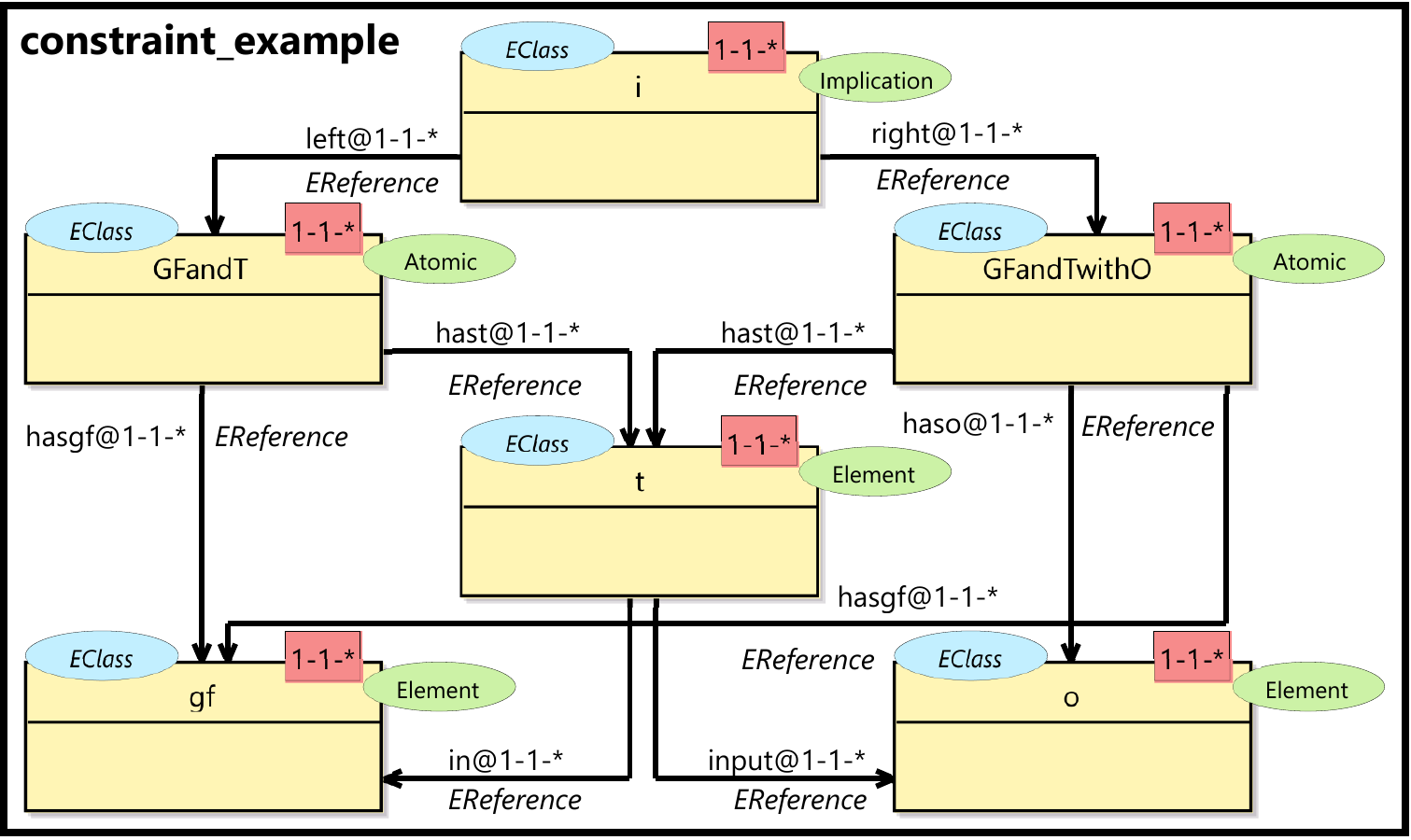}
  \caption{Example constraint using multilevel typing}
  \label{fig:constraint-example}
\end{figure}

We could also use LTL, or some other language, to define constraints in any of the levels of the hierarchy, exploiting the multilevel capabilities of the framework.
The actual implementation of the MultEcore tool can be instrumented to apply the language's semantics to evaluate such constraints against the models, without requiring any additional modelling mechanisms.
For instance, just by using the non-temporal operators of LTL, it is possible to define propositional constraints like the one shown in Figure~\ref{fig:constraint-example}.
In fact, the boolean logic language can be seen as a meta-language for temporal logics like LTL, and could be separated into different levels of a multilevel hierarchy, which we do not show here for the sake of brevity.

Let us assume that we want to define a constraint that forces instances of \elementname{GoForward} to only be connected to transitions triggered by an input of type \elementname{Obstacle}.
This constraint, specified in the \elementname{constrain\_example} model in Figure~\ref{fig:constraint-example}, is expressed by means of an implication, where the left-hand side contains an atomic proposition \elementname{GFandT} with the instances of \elementname{GoForward} and \elementname{Transition}, and the right-hand side reuses the same pattern (actually, the same elements) and adds the required instance of \elementname{Obstacle} inside the atomic proposition \elementname{GFandTwithO}.
The textual version of this constraint is written as \(\mathit{GFandT} \rightarrow \mathit{GFandTwithO}\), where \(\mathit{GFandT}\) and \(\mathit{GFandTwithO}\) are the names of the model fragments representing both atomic propositions.

\subsection{Data type dimension}
\label{subsec:data-type-dimension}

\begin{figure}[ht]
	\centering
	\includegraphics[width=\textwidth]{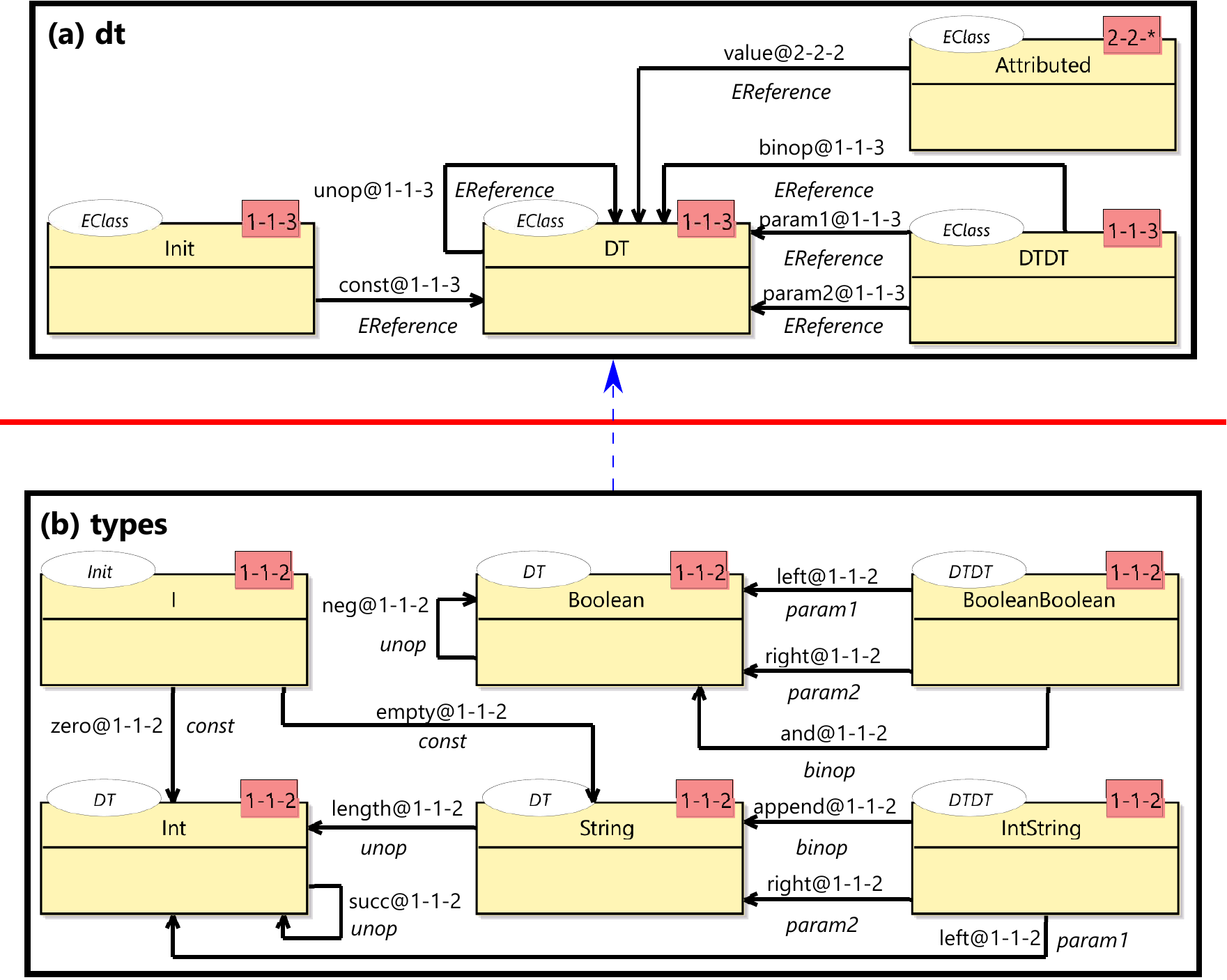}
	\caption{Fragment of the data type hierarchy}
	\label{fig:data-types-multilevel-hierarchy}
\end{figure}

In this work, we do not use a different formalisation for the attributes of a node, their data types and their instantiations, but represent them as nodes, in a similar fashion to~\cite{mylopoulos1990telos}.
Therefore, we can just use multiple typing and a third orthogonal dimension for handling attributes in a way which is consistent with the rest of our approach to MLM.
This third dimension contains a single and fixed hierarchy of three levels, and defines the basic and most common data types used in most modelling tools, such as integers, strings and booleans.
As with the two dimensions aforementioned, we choose Ecore for the root of the hierarchy for implementation reasons but do not display it.
By being orthogonal to the two aforementioned dimensions, data types are available to use in any other hierarchy.
Figure~\ref{fig:data-types-multilevel-hierarchy} shows an excerpt of the two bottom levels of this hierarchy.

The top level displayed on Figure~\ref{fig:data-types-multilevel-hierarchy}(a) defines the concept of data type (\elementname{DT}), combined data type (\elementname{DT,DT}) and initial element (\elementname{Init}).
The arrows represent operations, or definitions of them, between the different data types.
For example, \elementname{Init} is used to define constants for some data types, hence the arrow \elementname{const}.
The arrow \elementname{unop} indicate that an unary operation can relate two specific instances of \elementname{DT}.
Similarly, \elementname{binop} represents a binary operation from a combination of two data types (\elementname{DTDT}) to a single one.
The two arrows \elementname{param1} and \elementname{param2} indicate how a combined data type is constructed as a combination of two single ones.
The element \elementname{Attribute} can be used to double-type a node in any other hierarchy, indicating that it is an attribute.
The \elementname{value} relation indicates the connection of such a node with its value.
Note that these last two elements have a potency of \typename{}{2-2-*} and \typename{}{2-2-2}, respectively, which indicates that they must be instantiated exactly two levels below, and only twice for \elementname{value}, whereas the rest of the elements have a potency \typename{}{1-1-3} which lets us define (1) the level below of specific data types, then (2) use these data types to double-type application nodes as attributes two levels below, and finally (3, hence the value of depth) instantiate those attributes.

In the second level, three data types are defined (\elementname{Boolean}, \elementname{Int} and \elementname{String}) as well as two of their combined types (\elementname{BooleanBoolean} and \elementname{IntString}).
These combined data types can be used to define, for example, the binary operation \elementname{append} that takes a string and generates a new one by attaching a given integer to its end, and the \elementname{and} operation for boolean values.
Note that all operations are defined as arrows between two nodes, but their semantics are specified using model transformations (see Section~\ref{subsec:mcmts-data-types}).
The arrows from \elementname{I} are used to define the constants \elementname{zero} for integers and \elementname{empty} for strings.
Finally, the unary operation \elementname{succ} is defined to get the successor of an integer.
All types in this model have depth 2 (one less than their type), indicating that they can type a node as attribute and then be instantiated one last time to give them a value.

\begin{figure}[ht]
	\centering
	\includegraphics{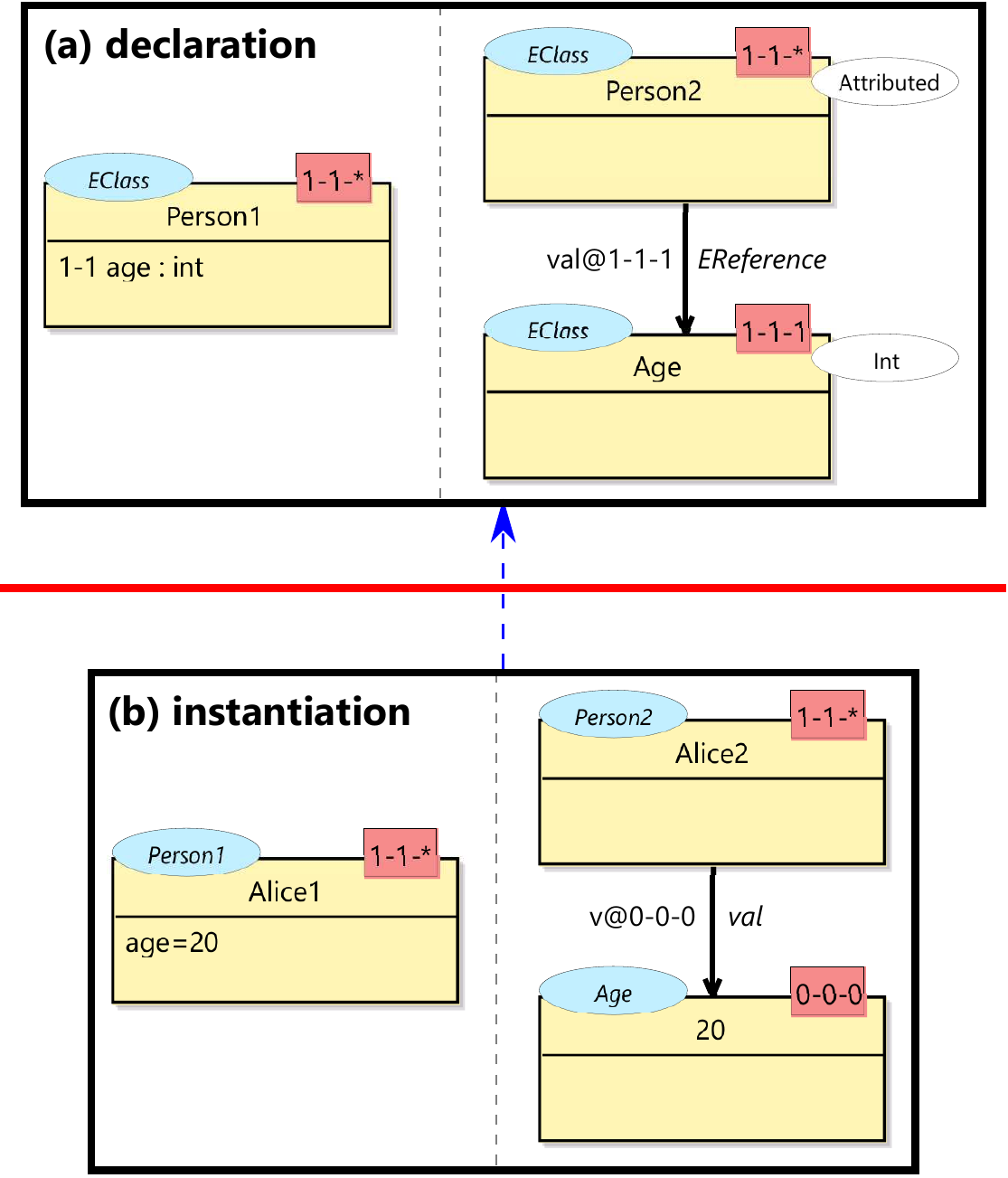}
	\caption{Attribute instantiation, with two alternative syntaxes}
	\label{fig:attributes-multilevel-hierarchy}
\end{figure}

The models on Figure~\ref{fig:data-types-multilevel-hierarchy} are mostly theoretical, and our proposal uses this hierarchy to establish a conceptual interface to the actual implementation of data types, in order to keep the formalisation separated from implementation particularities.
Hence, this representation is consistent with the rest of the framework, including the definition of the semantics of the operations defined as morphisms, like \elementname{succ} for integers, as we show in Section~\ref{subsec:mcmts-data-types}.
Figure~\ref{fig:attributes-multilevel-hierarchy} contains an application hierarchy where we illustrate how attributes can be declared and instantiated in our approach, with the syntax we actually use (left of the dashed lines) and with the explicit one which uses double typing (application + data type) to declare a node as instance of a data type.
In Figure~\ref{fig:attributes-multilevel-hierarchy}(a) we declare a simple node \elementname{Person1} which contains an attribute of type integer called \elementname{age}.
On the right-hand side, \elementname{Person2} shows the formal way in which this is done in our theoretical framework, by double-typing it with \elementname{Attributed} from the data type hierarchy, which allows to connect it with the node \elementname{Age}, also double-typed with \elementname{Int}, by creating an arrow \elementname{val} of application type \elementname{EReference} (displayed) and double-typed with \elementname{value} (not displayed).
Both \elementname{val} and \elementname{Age} have depth 1, enforced by their types from the data type hierarchy, as desired.
Still, the level(s) where the attribute can finally given a value can be controlled with the potency range \elementname{min--max}.

To complete the example, we illustrate how to instantiate attributes in our approach, with both syntaxes: the one we use in practice and the one which illustrates formally the double types and potency depths.
In Figure~\ref{fig:attributes-multilevel-hierarchy}(b) we instantiate the \elementname{age} attribute from \elementname{Person1} in an instance of it, named \elementname{Alice1}, and do the same in the right-hand side with \elementname{Age}, \elementname{Person2} and \elementname{Alice2}.
Note that this notation shows clearly how the potency depth reaches value 0, since it does not make sense to create an instance of an integer value such as \elementname{20}.

As with the LTL syntax in Section~\ref{subsec:supplementary-dimension}, this way of representing attributes would be cumbersome in practice.
Therefore, an alternate representation is used, such that attributes are represented, in a more familiar way, as a list inside the node that contains the attributes, excluding potency depth in the declaration (since it is always 1) and using the \elementname{name = value} notation for instantiation.
Also, for the actual implementation (see Section~\ref{sec:modelling-tooling}), we ensure compatibility with the Ecore data types by reusing them, avoiding redefinition.

%% file: thesis/03-mcmt.tex
\chapter{DSML Behaviour with Multilevel Coupled Model Transformations}
\label{chap:mcmt}

In this chapter we present our approach to model transformations in the context of multilevel modelling hierarchies.
This chapter can be understood as a direct continuation from Chapter~\ref{chap:mlm}, and follows a similar structure: we illustrate our explanations with excerpts from a case study to showcase the advantages of our proposal and to serve as examples.
Also, we provide a formal foundation to our approach by based on our previous graph-based constructions.
For this purpose, we use Category Theory, a mathematical formalism that can be understood as an universal language based on defining objects inside a category as ``black boxes'' and analysing them through their relations to other objects inside the same category~\cite{barr1990category}.
We presented earlier versions of the contents of this chapter in~\cite{macias2019mcmt,macias2017chains}.

This chapter is divided in three sections.
Section~\ref{sec:why-mcmt} justifies the need for an approach which can properly exploit the advantages of MLM in the context of model transformations.
Then, Section~\ref{sec:chain-category} presents the formalisation of MCMT rule definition and application.
Finally, Section~\ref{sec:mcmts-three-dimensions} includes example MCMT rules for the example hierarchies presented in the previous chapter.

\section{Why Using MCMTs?}
\label{sec:why-mcmt}

To enable the integration of MLM in MDSE projects, model-to-model transformations are key.
In the literature, we find notions such as multilevel to two-level model transformations~\cite{atkinson2012towards}, deep transformations (defined at a specific level but with references to upper levels~\cite{delara2015dsmml}), or coupled transformations, which operate on a model and its metalevels simultaneously~\cite{herrmannsdorfer2014coupled,delara2015dsmml,rutle2012behavioural}.
We propose in this thesis a generic way of specifying multilevel transformations, operating on models that belong to a multilevel hierarchy.
Specifically, we use them to define the behaviour of DSMLs.
Our MLM approach also opens a door for the reusability of model's behaviour.
Since most behavioural models have some commonalities both in concepts and their semantics, reusing these model transformations across behavioural models would mean a significant gain.
By utilising MLM in a metamodelling process for the definition of modelling languages, we can exploit commonalities among these languages through abstraction, genericness and reusability of behaviour.
We will build on our running example \emph{Robolang} (see Section~\ref{subsec:application-dimension}) to explain our approach to reusable model transformations, namely, \emph{Multilevel Coupled Model Transformations} (MCMTs).

In order to offer a gentle example to justify the need for MCMTs, let us recall the \elementname{robolang} model from Figure~\ref{fig:robolang-multilevel-hierarchy}(a), and abstract away from the fact that a particular workflow for a robot could be defined several levels below this model.
We already have the language syntax, provided as a model (or hierarchy of them), and we wish now to define its semantics, starting by the behaviour of firing transitions.
In a model specified using a \elementname{robolang} language, an instance of a transition connects two indirect%
\footnote{Hereafter we drop the word \emph{indirect}.}
instances of task together with at least one instance of input.
This can be seen in Figure~\ref{fig:robolang-multilevel-hierarchy}(c).
In a running instance of a robot, such as Figure~\ref{fig:robolang-multilevel-hierarchy}(d), whenever we detect an instance of an input which is connected to a transition which in turn is connected to an instance of a source task, we can fire the transition and hence create an instance of its target task.
By source and target instances of task of a transition we mean instances of task which are connected by the \elementname{in} and \elementname{out} relations to the transition, respectively.

To express the semantics of this behaviour using model transformation rules, we can consider three options: two-level rules, multilevel rules and our MCMT rules.
For the display of the example rules in the following, we use a declarative way of specifying transformation rules, and we do this using the same graphical syntax for nodes, arrows and types as Chapter~\ref{chap:mlm}.
Note that some elements may have a type without the actual type (node or arrow) being depicted, since the existence of the type can be inferred from the declaration of the instance.
We apply this simplification since it makes the graphical notation more concise and straightforward.

\subsection{Two-level model transformation rules}
\label{subsec:two-level-mts}

\begin{figure}
    \begin{minipage}{0.45\textwidth}
		\centering
		\includegraphics[width=\linewidth]{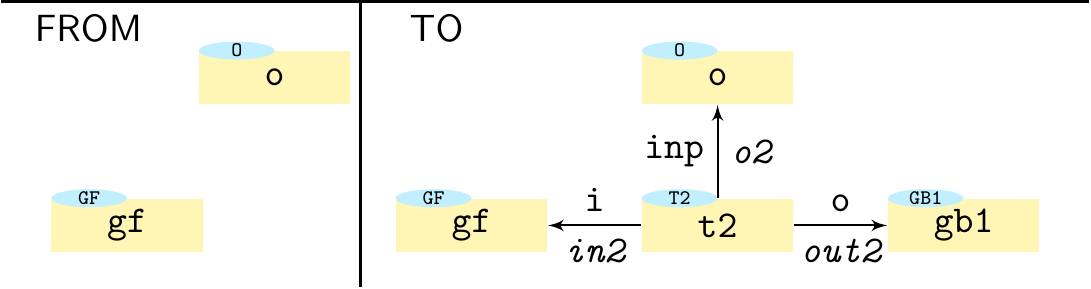}
		\caption{Behaviour for the \elementname{T2} transition}
		\label{fig:robolang-rule-fire-transition-2-level-1}
    \end{minipage}\hfill
    \begin{minipage}{0.45\textwidth}
		\centering
		\includegraphics[width=\linewidth]{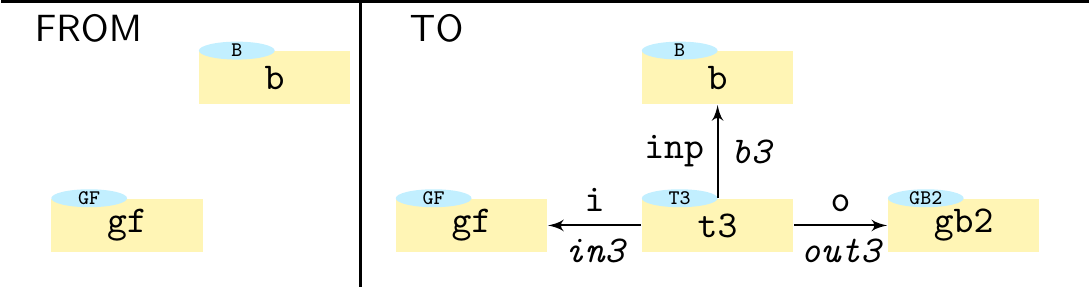}
		\caption{Behaviour for the \elementname{T3} transition}
		\label{fig:robolang-rule-fire-transition-2-level-2}
	\end{minipage}
\end{figure}

Without the usage of multilevel rules, that is, using two-level transformation rules, we would need to define one rule per instance of transition.
Following their declarative nature, these two-level transformation rules consist of a FROM block and a TO block.
The FROM block contains the pattern that the rule must match in order to be executed.
If a match is found, then the matched subgraph is modified in the way defined in the TO block, which may comprise the creation, deletion or modification of nodes and arrows.
For example, Figures~\ref{fig:robolang-rule-fire-transition-2-level-1} and~\ref{fig:robolang-rule-fire-transition-2-level-2} show the rules for the instances of transition \elementname{T2} and \elementname{T3}.
Only for the model \elementname{robot\_1} in Figure~\ref{fig:robolang-multilevel-hierarchy}(c) we would already need seven transformation rules, one for each node of type \elementname{Transition}.
And this number would keep increasing every time we added new elements to our model, when we created another robot model or when we extended the Robolang metamodel or any of its instances, e.g.\ a the language for Arduino robots hinted in Section~\ref{sec:hierarchies}.

In our example, it is easy to see that problems arise easily regarding reusability with a two-level approach.
Two-level rules would be too specific, tied to the types \elementname{T2} and \elementname{T3} (and their connected elements), and the firing of other transitions in this or other models would require additional specific rules.
As a consequence, a significant number of very similar rules are required, leading to a \emph{proliferation problem}: each transition instance would need a rule and each branch in the hierarchy would need its set of almost identical rules.

\begin{figure}[ht]
	\centering
	\includegraphics{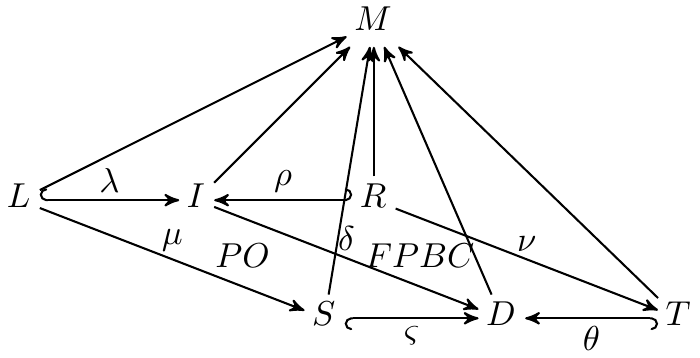}
	\caption{Two-level model transformation rule}
	\label{fig:two-level-rule-formal}
\end{figure}

The basic, formal structure of these rules is outlined in Figure~\ref{fig:two-level-rule-formal}.
In its most general terms, a graph transformation rule is defined as a left \(\rulegraphleft\) and a right \(\rulegraphright\) pattern.
These patterns are graphs which are mapped to each other via graph morphisms \(\maprulelefttomiddle\) and \(\maprulerighttomiddle\) from or into a third graph \(\rulegraphmiddle\), such that \(\rulegraphleft, \rulegraphmiddle, \rulegraphright\) constitute either a span or a co-span, respectively~\cite{ehrig2006fundamentals,ehrig2009alternative,mantz2015coevolving}.
We use here the co-span version, which facilitates the manipulation of model elements without the two phases of \emph{delete-then-add}.
This same construction will be reused for the next two alternatives.
In order to apply a rule to a source graph \(\hierarchygraphleft\), first a match of the left pattern must be found in \(\hierarchygraphleft\), that is, a graph homomorphism \(\mapruletohierarchyleft : \rulegraphleft \to \hierarchygraphleft\), then using a pushout construction followed by a final pullback complement construction~\cite{corradini2006sesqui} will create a target graph \(\hierarchygraphright\).
In case of several rules being applicable at the same time, there are different strategies to get rid of non-determinism in the literature of graph transformations and model transformations, which we do not tackle in this thesis.
Two examples are layering (prioritization of rules) and negative application conditions.
Details of application conditions and theoretical results on graph transformations can be found in~\cite{ehrig2006fundamentals}.

In typed transformation rules, the calculation of matches needs to fulfil a typing condition: in the construction in Figure~\ref{fig:two-level-rule-formal}, all the triangles must be commutative.
In other words, since both the rules and the models are typed by the same type graph, the rules are defined at the type-graph level and are applied to typed graphs.

\subsection{Multilevel rules}
\label{subsec:multilevel-mts}

\begin{figure}
	\centering
	\includegraphics{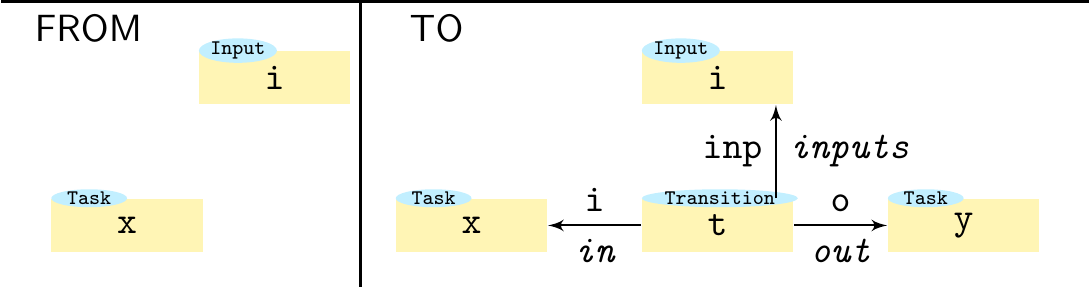}
	\caption{Abstract behaviour for firing transitions}
	\label{fig:robolang-rule-fire-transition-multilevel}
\end{figure}

Multilevel rules, also known as deep rules (see for example~\cite{delara2015dsmml}), enable us to specify a common behaviour at upper levels of abstraction.
In our example, multilevel rules enable us to define a single rule, by using the more abstract types \elementname{Task}, \elementname{Transition} and \elementname{Input}, as shown in Figure~\ref{fig:robolang-rule-fire-transition-multilevel}).
Thus, this rule can be applied to all the instances of transition (\elementname{T2}, \elementname{T3}, etc.) in Figure~\ref{fig:robolang-multilevel-hierarchy}(c), other robot models typed by \elementname{legolang} in Figure~\ref{fig:robolang-multilevel-hierarchy}(b), and, in general, any other model which is transitively typed by \elementname{robolang} in Figure~\ref{fig:robolang-multilevel-hierarchy}(a).

While multilevel model transformations solve the proliferation problem, they introduce another problem which is related to case distinction.
That is, the approach works fine in cases where the model on which the rules are applied contains the structure that is required by the rule and where all types required by the rule are existing in the same metalevel.
If that is not the case, the rules will only be able to express behaviour in a generic way, but they will not be precise enough.
In other words, the rules become too generic and imprecise.
In this example, we get the side effect that all transitions can fire at any time, even if they are redefined in subsequent levels and given new behaviour, with the same conditions (detecting an input) and, even worse, the rule does not take into account the type of tasks which the transition connects: finding an instance of any input and an instance of any (source) task would lead to creating an arbitrary instance of transition connected to an arbitrary (target) instance of a task.
For example, in Figure~\ref{fig:robolang-multilevel-hierarchy}(c), an instance of \elementname{Obstacle} together with an instance of \elementname{TL} (which actually are not directly related) might trigger firing the transition \elementname{T5} and create an instance of \elementname{TR}, which is not the correct or desired behaviour for firing transitions.

\begin{figure}[ht]
	\centering
	\includegraphics{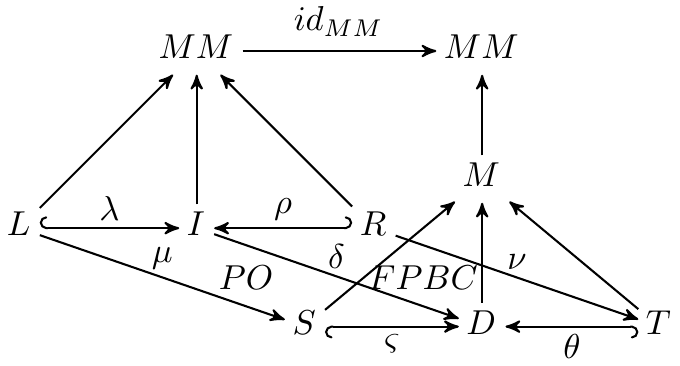}
	\caption{Multilevel model transformation rule}
	\label{fig:multilevel-rule-formal}
\end{figure}

The general structure of multilevel transformation rules is shown in Figure~\ref{fig:multilevel-rule-formal}.
This could be considered as a method to relax the strictness of two-level model transformations through multilevel model transformations (see for example~\cite{atkinson2012towards,atkinson2015enhancing}).
This approach works only on multilevel metamodelling hierarchies, which are not restricted to a fixed number of levels. 
To achieve flexibility, the rules can be defined over a type graph somewhere at a higher level in the hierarchy and applied to running instances at the bottom of the hierarchy.
Types are resolved by composing typing graph homomorphisms from the model on which the rule is applied and upwards to the level on which the type graph is defined.

\subsection{Multilevel coupled model transformation rules}
\label{subsec:mcmts}

\begin{figure}[hb]
	\centering
	\includegraphics{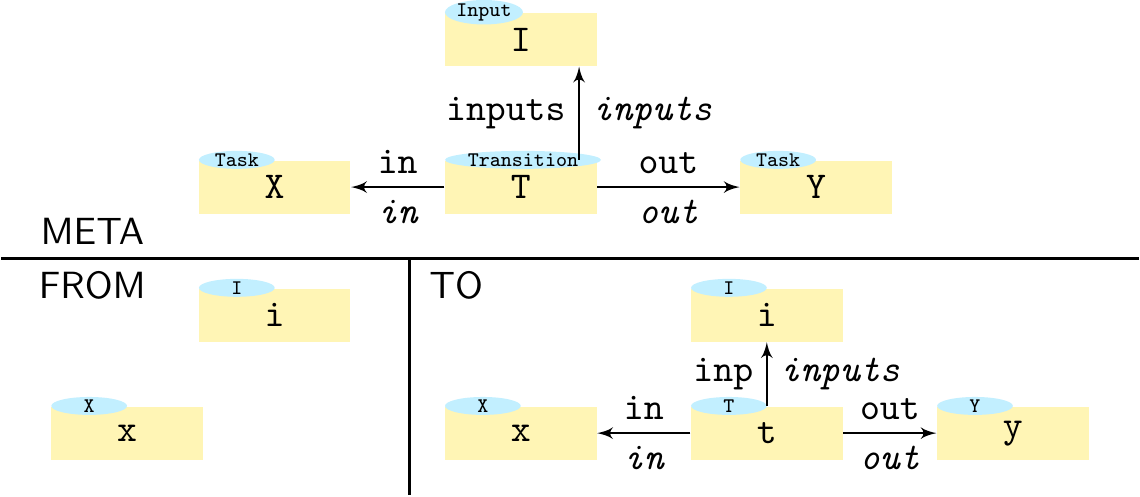}
	\caption{MCMT rule \emph{Fire Transition}: a transition gets fired by the associated input}
	\label{fig:robolang-rule-fire-transition}
\end{figure}

We propose MCMT rules, as a means to overcome the issues of the two approaches aforementioned.
Using MCMTs, the desired rule for firing transitions would be as shown in Figure~\ref{fig:robolang-rule-fire-transition}.
We introduce a new component into the rules, namely the META block.
This new part of the rule allows us to locate types in any level of the hierarchy that can then be used in the FROM and TO blocks.
Moreover, the real expressive power from this new block comes from the fact that we can define a whole pattern that these types must satisfy in order for the rule to be applied.
This feature allows us to create case distinctions in the rules easily, allowing for a finer tuning of the rules that prevents the aforementioned side effects.

This rule can be applied to fire any transition (except for the one which has an initial task as its source) in any robot model defined by the \elementname{legolang} language. 
Similar to the multilevel rules, the variables \elementname{T}, \elementname{X}, \elementname{Y} and \elementname{I} can match different instances of task, transition and input.
However, the difference in this case is that each instance is matched together with its type, hence the name \emph{coupled}.
That is, when a variable is bound to a type, it will keep its type through the transformation. 
For example, consider applying the rule to the robot model in Figure~\ref{fig:robolang-multilevel-hierarchy}(c): if \elementname{I} is bound to \elementname{Obstacle}, then the only choice for binding \elementname{X} would be \elementname{GF}.
This would lead to the binding of \elementname{T} to \elementname{T2} and \elementname{Y} to \elementname{GB1}.
In this way, the right transition (\element{t}{T2}) will be fired and the right target instance of task (\element{y}{GB1}) will be created and connected to the right source instance of task (\element{x}{GF}).

The usage of the new META block allows for a finer tuning, when compared to other techniques like inheritance.
For example, the possibility of referring to the transitions connecting instances of \elementname{Task} and \elementname{Input} when more abstraction is required, while still being able to fire them only in some cases (like \elementname{GF} and \elementname{Obstacle}) is a level of granularity which cannot be achieved by turning all instances of \elementname{Task} and \elementname{Input} into subclasses.
This alternate approach using inheritance would render impossible to retain the abstraction ``inputs fire transitions'' while specifying that ``the transition with \elementname{GF} as source (\elementname{T2}) can only be fired by \elementname{Obstacle}''.

\begin{figure}[ht!]
	\centering
	\includegraphics{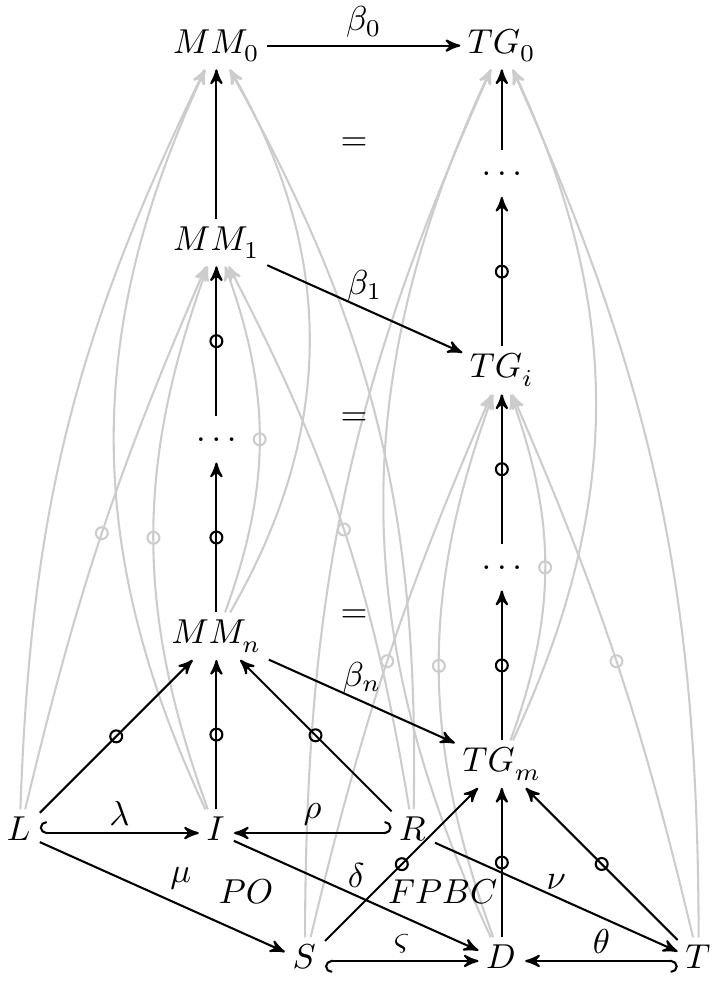}
	\caption{Multilevel Coupled Model Transformation Rule}
	\label{fig:multilevel-coupled-rule-formal}
\end{figure}

The META section may include several metamodel levels. 
The general, formal structure of an MCMT rule is displayed in Figure~\ref{fig:multilevel-coupled-rule-formal}.
The figure can be visualized as two flat trees, each of them defined by typing chains and connected to each other by matching morphisms.

The tree on the left contains the pattern that the user defines in the rule.
It consists of the left and right parts of the rule (FROM and TO, respectively), represented as \(\graphname{\rulegraphleft}{}\) and \(\graphname{\rulegraphright}{}\) in the diagram, and the interface \(\graphname{\rulegraphmiddle}{}\) that contains the union of both \(\graphname{\rulegraphleft}{}\) and \(\graphname{\rulegraphright}{}\), hence the inclusion morphisms.
It is important to remark that, in all examples in this thesis, the interface \(\graphname{\rulegraphmiddle}{}\) is the union of \(\graphname{\rulegraphleft}{}\) and \(\graphname{\rulegraphright}{}\), i.e., the pushout of the span \(\graphname{\rulegraphleft}{} \hookleftarrow \graphname{\rulegraphleft}{} \cap \graphname{\rulegraphright}{} \hookrightarrow \graphname{\rulegraphright}{}\), thus the morphisms \(\maprulelefttomiddle\) and \(\maprulerighttomiddle\) are inclusion graph homomorphisms.
But in the general case, \(\graphname{\rulegraphmiddle}{}\) is obtained by the pushout of a span \(\graphname{\rulegraphleft}{} \leftarrow \graphname{C}{} \rightarrow \graphname{\rulegraphright}{}\) of injective  graph homomorphisms, thus the graph homomorphisms \(\maprulelefttomiddle\) and \(\maprulerighttomiddle\) become injective and jointly surjective.
The three graphs \(\graphname{\rulegraphleft}{}\), \(\graphname{\rulegraphmiddle}{}\), \(\graphname{\rulegraphright}{}\) are typed by elements in the same typing chain, which is represented as a sequence of metamodels \(\graphname{\rulegraph}{x}\) that ends with the root of the hierarchy tree \(\graphname{\rulegraph}{0}\) (Ecore in our case).
The multilevel typing arrows from \(\graphname{\rulegraphleft}{}\), \(\graphname{\rulegraphmiddle}{}\), \(\graphname{\rulegraphright}{}\) to the typing chain \([\graphname{\rulegraph}{\chaindepthone}, \graphname{\rulegraph}{\chaindepthone-1}, \dots, \graphname{\rulegraph}{1}, \graphname{\rulegraph}{0}]\) (see Section~\ref{subsec:multilevel-typing}) should be compatible with typing (see Section~\ref{subsec:type-compatibility-as-chain-morphism}).
That is, both triangles at the bottom must be commutative.

The tree on the right side represents the actual path in the type hierarchy on which the rule is applied.
Before applying the rule, we only have the graph \(\graphname{\hierarchygraphleft}{}\) and its multilevel type to the typing chain \([\graphname{\hierarchygraph}{\chaindepthtwo} , \graphname{\hierarchygraph}{\chaindepthtwo-1}, \dots, \graphname{\hierarchygraph}{1} , \graphname{\hierarchygraph}{0}]\).
In order for the rule to be applied, it is required to find matches of all metamodel graphs \(\graphname{\rulegraph}{x}\) into the actual typing chain graphs \(\graphname{\hierarchygraph}{y}\).
These matches do not require to be \emph{parallel} in the sense that the difference of levels between the two sources of any two matching morphisms is not required to be equal to the difference of levels between the targets of those two morphisms, although we do not allow two matching morphisms to have the same target.
This is due to the flexibility in the specification of the number of levels that separate two metamodel graphs in the pattern.
In terms of our example, if we add more intermediate levels to the Robolang hierarchy, and consequently the depth becomes bigger, the defined rules would still be applicable.
Moreover, when the rules are defined with a flexible depth, they would fit or match different branches of the hierarchy regardless their depth.
That is, the graph representing the pattern could be matched in several different ways to the same hierarchy, hence providing the flexibility that we require in our approach.

\(\graphname{\rulegraph}{0}\) has to be matched with \(\graphname{\hierarchygraph}{0}\).
The match \(\mapruletohierarchybinding_{0}\) is trivially the identity in the implementation since both \(\graphname{\rulegraph}{0}\) and \(\graphname{\hierarchygraph}{0}\) are Ecore.
If all the metamodel graphs can be matched into the type graphs, such that every resulting square commutes, we proceed to find a match (a graph homomorphism \(\mapruletohierarchyleft\)) of the pattern graph \(\graphname{\rulegraphleft}{}\) into the instance \(\graphname{\hierarchygraphleft}{}\).
This match must be type compatible, i.e., the types of elements in \(\graphname{\rulegraphleft}{}\) which are matched to elements in \(\graphname{\hierarchygraphleft}{}\) must be matched by one of the bindings \(\mapruletohierarchybinding_{x}\) (see Section~\ref{subsec:type-compatibility-as-chain-morphism}).
If this match \(\mapruletohierarchyleft\) is successful, i.e., a type compatible graph homomorphism from \(\graphname{\rulegraphleft}{}\) to \(\graphname{\hierarchygraphleft}{}\) is found, we construct the intermediate instance \(\graphname{\hierarchygraphmiddle}{}\) by pushout, and then proceed to generate the target instance \(\graphname{\hierarchygraphright}{}\) by final pullback complement.

To summarise this section, while existing approaches which employ reusable model transformations for the definition of behavioural models focus on traditional two-level modelling hierarchies and their affiliated two-level model transformations (see~\cite{kusel2015reuse,kusel2013reality} for a survey), multilevel model transformations~\cite{atkinson2015enhancing,atkinson2012towards} are relatively new and are not yet proven suitable for reuse and definition of model's behaviour.
Our approach to multilevel model transformations builds on top of these approaches to combine reusability with flexibility in several modelling levels.
In other words, the depth of the modelling hierarchy on which the rules are applied is not important when defining the rules, since no matter how deep the hierarchy gets, the rules can still be matched and applied.
Moreover, our approach leaves the door open to transforming several levels at once (i.e.\ co-transformations), although we do not explore that possibility in this thesis.

It is also important to notice that, so far, we have not identified a scenario in which MCMTs have more expressive power than two-level MTs.
That is, one MCMT can conveys the same information as a set of two-level MTs with commonalities, and the size of this set depends on the amount of types it uses and the possibility of abstracting such types.
The advantage of using MCMT rules is that such information is conveyed in a much more concise, reusable and maintainable way, as illustrated in our previous example.

\section{The Category of Graph Chains and Graph Chain Morphisms}
\label{sec:chain-category}

Since every branch in our MLM hierarchies is represented by a graph chain, the application of our model transformation rules will rely on pushout and final pullback complement constructions in the category \cat{Chain}.
By constructing this category for multilevel hierarchies and MCMTs, we are able build upon the already existing co-span approach to graph transformations~\cite{ehrig2009alternative}.
We introduce appropriate definitions for the category \cat{Chain} in this section.

This formalisation is required since the semantics of model transformations needs to be adapted to our multilevel setting, keeping flexibility of rules in mind.
The formalism can be later used as a reference for the expected behaviour of MCMTs, regardless of the mechanism used to implement them.

\subsection{Multilevel typing}
\label{subsec:multilevel-typing}

MCMTs are defined as graphs \(\graphname{\rulegraphleft}{}\), \(\graphname{\rulegraphmiddle}{}\), \(\graphname{\rulegraphright}{}\) which are located at the bottom of a graph chain representing the metamodels in the rule hierarchy.
Moreover, the instance graph \(\graphname{\hierarchygraphleft}{}\) on which we apply the MCMTs is also located at the bottom of the graph chain.
We call the relation between these graphs and their typing chains \emph{multilevel typing}.
In fact, the relation between each graph \(\graphname{\examplegraphone}{\indexone}\) in a typing chain \(\tc{\examplegraphone}\) and the rest of the chain \([\graphname{\examplegraphone}{\indexone-1} ... \graphname{\examplegraphone}{0}]\) above the graph is a multilevel typing.
Further, applying an MCMT will produce the intermediate graph \(\graphname{\hierarchygraphmiddle}{}\) and the target graph \(\graphname{\hierarchygraphright}{}\) which must have a multilevel typing relation to the target typing chain.
To be able to use the well-established constructions of pushout and final pullback complement in the lowest level graphs \(\graphname{\rulegraphleft}{}\), \(\graphname{\rulegraphmiddle}{}\), \(\graphname{\rulegraphright}{}\), \(\graphname{\hierarchygraphleft}{}\), \(\graphname{\hierarchygraphmiddle}{}\) and \(\graphname{\hierarchygraphright}{}\), and indeed get the multilevel typing of the constructed graphs, we will define some restrictions on typing chains.
First of all, we will generalise the concept of typing chain and formally define the concept ``graph chain''.
Then, we will define graph chain morphisms and multilevel typing of a graph over a graph chain.

\begin{definition}[Graph chain]
\label{def:graph-chain}
A graph chain \(\chainname{\examplegraphone} := \chain{\examplegraphone}{\chaindepthone}\) is given by a natural number \(\chaindepthone\), a sequence \(\tc{\examplegraphone} = [\graphname{\examplegraphone}{\chaindepthone}, \graphname{\examplegraphone}{\chaindepthone-1}, \dots, \graphname{\examplegraphone}{1}, \graphname{\examplegraphone}{0}]\) of graphs of length \(\chaindepthone+1\)
together with a family \(\typemorph[\examplegraphone]{}{} = (\typemorph[\examplegraphone]{\indextwo}{\indexone} : \graphname{\examplegraphone}{\indextwo} \partialmap \graphname{\examplegraphone}{\indexone} \mid 0 \le \indexone < \indextwo \le \chaindepthone)\) of partial graph homomorphisms where
\begin{itemize}
	\item each partial graph homomorphism \(\typemorph[\examplegraphone]{\indextwo}{\indexone} : \graphname{\examplegraphone}{\indextwo} \partialmap \graphname{\examplegraphone}{\indexone}\) is given by a subgraph \(\domain{\typemorph[\examplegraphone]{\indextwo}{\indexone}} \sqsubseteq \graphname{\examplegraphone}{\indextwo}\), called the domain of definition of \(\typemorph[\examplegraphone]{\indextwo}{\indexone}\), and a total graph homomorphism with the same name (in an abuse of notation) \(\typemorph[\examplegraphone]{\indextwo}{\indexone} : \domain{\typemorph[\examplegraphone]{\indextwo}{\indexone}} \xrightarrow{} \graphname{\examplegraphone}{\indexone} \),
	\item all the morphisms \(\typemorph[\examplegraphone]{\indextwo}{0} : \graphname{\examplegraphone}{\indextwo} \to \graphname{\examplegraphone}{0}\) with \(1 \le \indextwo \le \chaindepthone\) to the top level graph are total, i.e., \(\domain{\typemorph[\examplegraphone]{\indextwo}{0}} = \graphname{\examplegraphone}{\indextwo}\), and
	\item for all \(0 \le \indexone < \indextwo < \indexthree \leq \chaindepthone\) the uniqueness condition \(\typemorph[\examplegraphone]{\indexthree}{\indextwo};\typemorph[\examplegraphone]{\indextwo}{\indexone} \preceq \typemorph[\examplegraphone]{\indexthree}{\indexone}\) is satisfied, i.e., \(\domain{\typemorph[\examplegraphone]{\indexthree}{\indextwo};\typemorph[\examplegraphone]{\indextwo}{\indexone}} \sqsubseteq \domain{\typemorph[\examplegraphone]{\indexthree}{\indexone}}\) and, moreover, \(\typemorph[\examplegraphone]{\indexthree}{\indextwo};\typemorph[\examplegraphone]{\indextwo}{\indexone}\) and \(\typemorph[\examplegraphone]{\indexthree}{\indexone}\) coincide on \(\domain{\typemorph[\examplegraphone]{\indexthree}{\indextwo};\typemorph[\examplegraphone]{\indextwo}{\indexone}}\).
	The composition \(\typemorph[\examplegraphone]{\indexthree}{\indextwo};\typemorph[\examplegraphone]{\indextwo}{\indexone}\) is defined by pullback (inverse image) of total graph homomorphisms as follows:
	
	\begin{center}
	\begin{tikzpicture}[on grid,node distance=25mm]
	
	\def\vd{10mm}
	\def\hd{25mm}
	
	\node[el-math] (dtki)										{\domain{\typemorph[\examplegraphone]{\indexthree}{\indextwo};\typemorph[\examplegraphone]{\indextwo}{\indexone}}};
	\node[el-math] (dtkj)	[below left=\vd and \hd of dtki]	{\domain{\typemorph[\examplegraphone]{\indexthree}{\indextwo}}};
	\node[el-math] (dtji)	[below right=\vd and \hd of dtki]	{\domain{\typemorph[\examplegraphone]{\indextwo}{\indexone}}};
	\node[el-math] (gk)		[below left=\vd and \hd of dtkj]	{\graphname{\examplegraphone}{\indexthree}};
	\node[el-math] (gj)		[below right=\vd and \hd of dtkj]	{\graphname{\examplegraphone}{\indextwo}};
	\node[el-math] (gi)		[below right=\vd and \hd of dtji]	{\graphname{\examplegraphone}{\indexone}};
	
	\draw[incmapr]			(dtki) to node[la-math,above left]	(in1)	{\sqsubseteq}																								(dtkj);
	\draw[map]				(dtki) to node[la-math,above left]	(map1)	{}																											(dtji);
	\draw[incmapr]			(dtkj) to node[la-math,above left]	(in2)	{\sqsubseteq}																								(gk);
	\draw[map]				(dtkj) to node[la-math,above]		(tkj)	{\typemorph[\examplegraphone]{\indexthree}{\indextwo}}														(gj);
	\draw[incmapr]			(dtji) to node[la-math,above left]	(in3)	{\sqsubseteq}																								(gj);
	\draw[map]				(dtji) to node[la-math,above]		(tji)	{\typemorph[\examplegraphone]{\indextwo}{\indexone}}														(gi);
	\draw[map,bend left]	(dtki) to node[la-math,above right]	(tki)	{\typemorph[\examplegraphone]{\indexthree}{\indextwo};\typemorph[\examplegraphone]{\indextwo}{\indexone}}	(gi);
	
	\node[el-math] (pb)		[below=\vd and of dtki]	{\pb};
	
	\end{tikzpicture}
	\end{center}
	
	That is, \(\domain{\typemorph[\examplegraphone]{\indexthree}{\indextwo};\typemorph[\examplegraphone]{\indextwo}{\indexone}} = (\typemorph[\examplegraphone]{\indexthree}{\indextwo})^{-1} (\domain{\typemorph[\examplegraphone]{\indextwo}{\indexone}})\) thus we have \(\domain{\typemorph[\examplegraphone]{\indexthree}{\indextwo};\typemorph[\examplegraphone]{\indextwo}{\indexone}} = \domain{\typemorph[\examplegraphone]{\indexthree}{\indextwo}}\) if \(\typemorph[\examplegraphone]{\indextwo}{\indexone}\) is total, i.e., \(\graphname{\examplegraphone}{\indextwo} = \domain{\typemorph[\examplegraphone]{\indextwo}{\indexone}}\).
\end{itemize}

\end{definition}

\begin{definition}[Multilevel typing]
\label{def:multilevel-typing}
A multilevel typing \(\chainmorph{\examplegraphtwo} : \graphname{\examplegraphtwo}{} \Rightarrow \chainname{\examplegraphone}\) of a graph \(\graphname{\examplegraphtwo}{}\) over a graph chain \(\chainname{\examplegraphone} := \chain{\examplegraphone}{\chaindepthone}\) is given by a family \(\chainmorph{\examplegraphtwo} = (\chainmorph[\indexone]{\examplegraphtwo} : \graphname{\examplegraphtwo}{} \partialmap \graphname{\examplegraphone}{\indexone} \mid \indexone \in [\chaindepthone])\), with \([\chaindepthone] \in \{0, 1, 2, \ldots, \chaindepthone\} \), of partial graph homomorphisms such that
\begin{enumerate}
	\item \(\chainmorph[0]{\examplegraphtwo} : \graphname{\examplegraphtwo}{} \to \graphname{\examplegraphone}{0}\) is total and 
	\item  for all \(0 \leq \indexone < \indextwo \leq \chaindepthone\) we have \(\domain{\chainmorph[\indexone]{\examplegraphtwo}} \cap \domain{\chainmorph[\indextwo]{\examplegraphtwo}} \sqsubseteq {\chainmorph[\indextwo]{\examplegraphtwo}}^{-1}(\domain{\typemorph[\examplegraphone]{\indextwo}{\indexone}})\), i.e., there exists a graph homomorphism \(\chainmorph[\indextwo \mid \indexone]{\examplegraphtwo} : \domain{\chainmorph[\indexone]{\examplegraphtwo}} \cap \domain{\chainmorph[\indextwo]{\examplegraphtwo}} \to \domain{\typemorph[\examplegraphone]{\indextwo}{\indexone}}\) such that the following diagram of total graph homomorphisms commutes
	\begin{center}
		\includegraphics{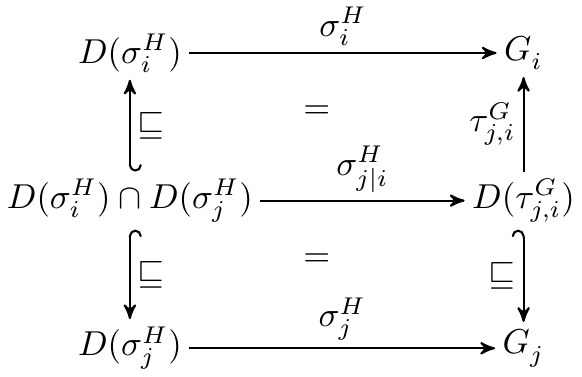}
	\end{center}
\end{enumerate}
\end{definition}

A simple but crucial observation is that all the domains of definition \(\domain{\chainmorph[\indexone]{\examplegraphtwo}}\) with \(\indexone \in [\chaindepthone]\) are subgraphs of \(\graphname{\examplegraphtwo}{}\) where \(\domain{\chainmorph[0]{\examplegraphtwo}} = \graphname{\examplegraphtwo}{}\) since \(\chainmorph[0]{\examplegraphtwo}\) is total.
Any family of subgraphs gives rise to a graph chain.

We assume now multilevel typings \(\chainmorph{\rulegraphleft} : \graphname{\rulegraphleft}{} \Rightarrow \chainname{\rulegraph}\), \(\chainmorph{\rulegraphmiddle} : \graphname{\rulegraphmiddle}{} \Rightarrow \chainname{\rulegraph}\), \(\chainmorph{\rulegraphright} : \graphname{\rulegraphright}{} \Rightarrow \chainname{\rulegraph}\) and \(\chainmorph{\hierarchygraphleft} : \graphname{\hierarchygraphleft}{} \Rightarrow \chainname{\hierarchygraph}\).
For our rule \begin{tikzpicture}[baseline=-1mm,node distance=15mm]\node[el-math](l){\graphname{\rulegraphleft}{}};\node[el-math](i)[right of=l]{\graphname{\rulegraphmiddle}{}};\node[el-math](r)[right of=i]{\graphname{\rulegraphright}{}};\draw[incmapl](l)to node[la-math,above](lm){\maprulelefttomiddle}(i);\draw[incmapr](r)to	node[la-math,above](rm){\maprulerighttomiddle}(i);\end{tikzpicture} we require that both \maprulelefttomiddle{} and \maprulerighttomiddle{} are \emph{type compatible}:
\begin{equation}\label{eq:type-compatibility-rule-maps-1}
	\chainmorph[\indexone]{\rulegraphleft} = \maprulelefttomiddle;\chainmorph[\indexone]{\rulegraphmiddle} \mbox{ for all } \indexone \in [\chaindepthone]
\end{equation}
\begin{equation}\label{eq:type-compatibility-rule-maps-2}
	\chainmorph[\indexone]{\rulegraphright} = \maprulerighttomiddle;\chainmorph[\indexone]{\rulegraphmiddle} \mbox{ for all } \indexone \in [\chaindepthone]
\end{equation}

\begin{figure}[ht]
\begin{center}
	\begin{minipage}{.3\linewidth}
		\centering
		\includegraphics{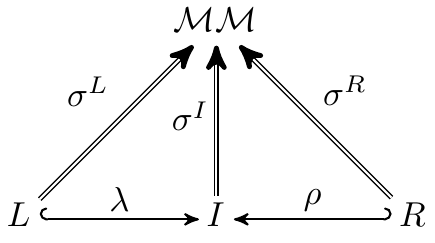}
	\end{minipage}
	\quad
	\begin{minipage}{.3\linewidth}
		\centering
		\includegraphics{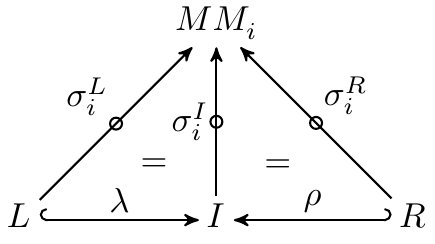}
	\end{minipage}
	\caption{Type compatibility of rule maps}
	\label{fig:type-compatibility-rule-maps}
\end{center}
\end{figure}

Applying an MCMT rule would now require finding a match of the rule, which has three characteristics: (i) a graph homomorphism \(\mapruletohierarchyleft : \graphname{\rulegraphleft}{} \to \graphname{\hierarchygraphleft}{}\), (ii) a flexible match of the graph chain \(\chainname{\rulegraph}\) into the graph chain \(\chainname{\hierarchygraph}\), and, (iii) both these matches need to be \emph{compatible} with respect to multilevel typing, i.e.\ satisfy a corresponding commutativity condition.
To formalise and ensure the intended flexibility of matching MCMT rules, and in order to define matches between graph chains, we define \emph{graph chain morphisms}.

\begin{figure}[ht!]
	\centering
	\includegraphics{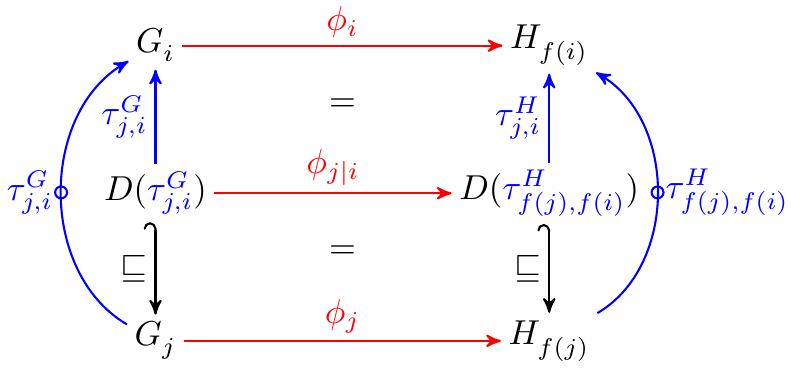}
	\caption{Establishing a morphism between two different objects in \(\cat{Chain}\), level-wise}
	\label{fig:mlm-binding-diagram-formal}
\end{figure}

\begin{definition}[Graph chain morphism]
\label{def:graph-chain-morphism}
A morphism \((\morphone,\maplevelone) : \chainname{\examplegraphone} \to \chainname{\examplegraphtwo}\) between two graph chains \(\chainname{\examplegraphone} = \chain{\examplegraphone}{\chaindepthone}\) and \(\chainname{\examplegraphtwo} = \chain{\examplegraphtwo}{\chaindepthtwo}\) with \(\chaindepthone \leq \chaindepthtwo\) is given by
\begin{itemize}
	\item a function \(\maplevelone : [\chaindepthone] \to [\chaindepthtwo] \), such that \(\maplevelone(0) = 0\) and \(\indexone < \indextwo\) implies \(\maplevelone(\indexone) < \maplevelone(\indextwo)\) for all \(\indexone, \indextwo \in [\chaindepthone]\), and
	\item a family of total graph homomorphisms \(\morphone = (\morphone_\indexone : \graphname{\examplegraphone}{\indexone} \to \graphname{\examplegraphtwo}{\maplevelone(\indexone)} \mid \indexone \in [\chaindepthone] )\) such that \(\typemorph[\examplegraphone]{\indextwo}{\indexone} ; \morphone_\indexone \preceq \morphone_\indextwo ; \typemorph[\examplegraphtwo]{\maplevelone(\indextwo)}{\maplevelone(\indexone)}\) for all \(0 \leq \indexone < \indextwo \leq \chaindepthone\), i.e., due to the definition of composition of partial morphisms (cf. Definition~\ref{def:graph-chain}), we assume for any \(0 \leq \indexone < \indextwo \leq \chaindepthone\) the commutative diagram of total graph homomorphisms displayed in Figure~\ref{fig:mlm-binding-diagram-formal}.
\end{itemize}
\end{definition}

A flexible match of the rule graph chain \(\chainname{\rulegraph} := \rulechain\) to the target graph chain \(\chainname{\hierarchygraph} := \hierarchychain\) is now given by a (graph) chain morphism \((\mapruletohierarchybinding,\maplevelone) : \chainname{\rulegraph} \to \chainname{\hierarchygraph}\) with \(\mapruletohierarchybinding=(\mapruletohierarchybinding_{\indexone} : \graphname{\rulegraph}{\indexone} \to \graphname{\hierarchygraph}{\maplevelone(\indexone)} \mid \indexone \in [\chaindepthone])\), and we require type compatibility:

\begin{equation}\label{eq:type-compatibility-rule-match}
	\chainmorph[\indexone]{\rulegraphleft};\mapruletohierarchybinding_{\indexone} = \mapruletohierarchyleft;\chainmorph[\maplevelone(\indexone)]{\hierarchygraphleft} \mbox{ for all } \indexone \in [\chaindepthone]
\end{equation}

\begin{figure}[ht!]
\begin{center}
	\begin{minipage}{.3\linewidth}
		\centering
		\includegraphics{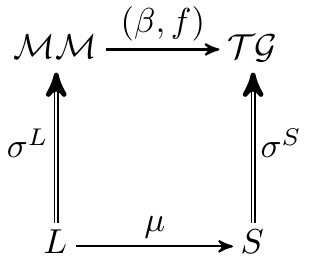}
	\end{minipage}
	\quad
	\begin{minipage}{.3\linewidth}
		\centering
		\includegraphics{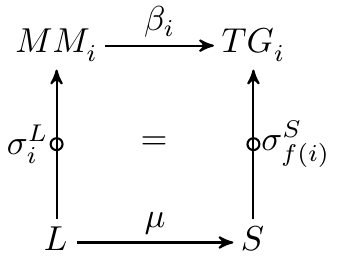}
	\end{minipage}
	\caption{Type compatibility of the match of a rule}
	\label{fig:type-compatibility-rule-match}
\end{center}
\end{figure}

This type compatibility preserves our idea of flexible matching, where a direct typing relation in the rule can be matched against a transitive typing relation in the target hierarchy, therefore allowing for high reusability.
As it is the case when building multilevel hierarchies, the flexibility of MCMT rule matching can be constrained further in order to implement sanity checks if required by providing additional conditions or requirements.

Our objective when applying a rule is to compute the pushout of graph morphisms
\begin{center}
	\includegraphics{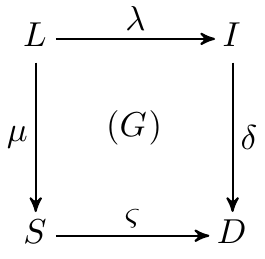}
\end{center}
and everything else should be obtained ``automatically'':
\begin{enumerate}
	\item Multilevel typing \(\chainmorph{\hierarchygraphmiddle} : \graphname{\hierarchygraphmiddle}{} \Rightarrow \chainname{\hierarchygraph}\) and
	\item that the graph homomorphisms \(\mapruletohierarchymiddle : \graphname{\rulegraphmiddle}{} \to \graphname{\hierarchygraphmiddle}{}\) and \(\maphierarchylefttomiddle : \graphname{\hierarchygraphleft}{} \hookrightarrow \graphname{\hierarchygraphmiddle}{}\) are type compatible.
\end{enumerate}

\begin{figure}[ht!]
\begin{center}
	\begin{minipage}[t]{.3\linewidth}
		\centering
		\includegraphics{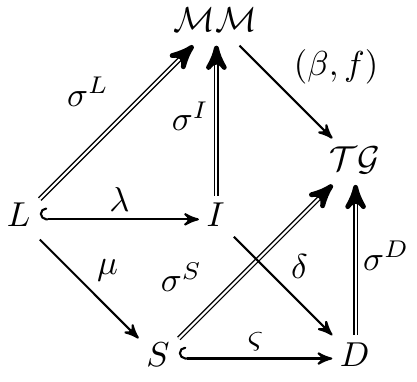}
	\end{minipage}
	\quad
	\begin{minipage}[t]{.3\linewidth}
		\centering
		\includegraphics{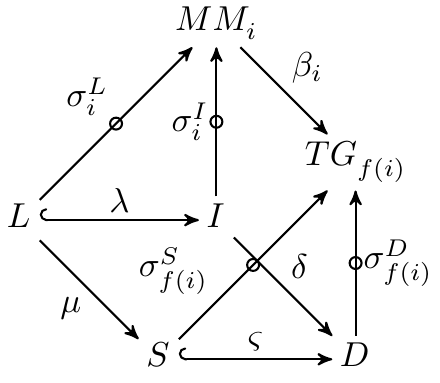}
	\end{minipage}%
	\caption{Rule application objectives}
	\label{fig:rule-application-objectives}
\end{center}
\end{figure}

That is, we want to have, simultaneously
\begin{equation}\label{eq:rule-application-objectives-1}
	\maphierarchylefttomiddle;\chainmorph[\indextwo]{\hierarchygraphmiddle} = \chainmorph[\indextwo]{\hierarchygraphleft} \mbox{ for all } \indextwo \in[\chaindepthtwo]
\end{equation}
and
\begin{equation}\label{eq:rule-application-objectives-2}
	\chainmorph[\indexone]{\rulegraphmiddle};\mapruletohierarchybinding_\indexone = \mapruletohierarchymiddle;\chainmorph[\maplevelone(\indexone)]{\hierarchygraphmiddle} \mbox{ for all } \indexone \in [\chaindepthone].
\end{equation}

If these constructions could yield several possible results (i.e.\ the rule is applicable in more than one way), and some of the resulting multilevel hierarchies violated potency or multiplicity constraints, such solutions would not be considered a valid match and rule application.
If all matches were to yield constraint-violating results, the rule itself would be considered not applicable in the target hierarchy.

On the left-hand diagrams in Figures~\ref{fig:type-compatibility-rule-maps}, \ref{fig:type-compatibility-rule-match} and \ref{fig:rule-application-objectives} we have three different kinds of arrows: graph homomorphisms, multilevel typing and graph chain morphisms.
They cannot be composed in a uniform way, thus we cannot state that they have to be commutative.
We want to describe the multilevel typings as well as the type-commutative graph homomorphisms by means of graph chain morphisms and to show that the pushout \((\graphname{\examplegraphone}{})\) of graph homomorphisms induces a pushout in the category \cat{Chain} that gives us exactly the desired result.

The condition \(\typemorph[\rulegraph]{\indextwo}{\indexone} ; \mapruletohierarchybinding_{\indexone} \preceq \mapruletohierarchybinding_{\indextwo} ; \typemorph[\hierarchygraph]{\maplevelone(\indextwo)}{\maplevelone(\indexone)}\), for all \(0 \leq \indexone < \indextwo \leq \chaindepthone\), means that we are requiring that typing is preserved.
That is, we have:
\begin{equation}\label{eq:preserve-reflect}
	\domain{\typemorph[\rulegraph]{\indextwo}{\indexone}} \sqsubseteq \mapruletohierarchybinding_{\indextwo}^{-1}(\domain{\typemorph[\hierarchygraph]{\maplevelone(\indextwo)}{\maplevelone(\indexone)}})
\end{equation}

Since composition of partial graph homomorphisms is monotonic w.r.t.\ the relation \(\preceq\), the following definition of composition of graph chain morphisms is sound.

\begin{definition}[Composition of graph chain morphisms]
\label{def:composition-graph-chain-morphisms}
The composition, represented as \((\morphone,\maplevelone);(\morphtwo,\mapleveltwo) : \chainname{\examplegraphone} \to \chainname{\examplegraphthree}\), of two graph morphisms \((\morphone,\maplevelone) : \chainname{\examplegraphone} \to \chainname{\examplegraphtwo}\), \((\morphtwo,\mapleveltwo) : \chainname{\examplegraphtwo} \to \chainname{\examplegraphthree}\) between graph chains \(\chainname{\examplegraphone} = \chain{\examplegraphone}{\chaindepthone}\), \(\chainname{\examplegraphtwo} = \chain{\examplegraphtwo}{\chaindepthtwo}\), \(\chainname{\examplegraphthree} = \chain{\examplegraphthree}{\chaindepththree}\) with \(\chaindepthone \leq \chaindepthtwo \leq \chaindepththree\) is defined by
\[(\morphone,\maplevelone);(\morphtwo,\mapleveltwo) = (\morphone;\morphtwo_{\downarrow \maplevelone},\maplevelone;\mapleveltwo)\]
where \(\morphtwo_{\downarrow \maplevelone} := (\morphtwo_{\maplevelone(\indexone)} : \graphname{\examplegraphtwo}{\maplevelone(\indexone)} \to \graphname{\examplegraphthree}{\mapleveltwo(\maplevelone(\indexone))} \mid \indexone \in [\chaindepthone])\) and thus
\[\morphone;\morphtwo_{\downarrow \maplevelone} := (\morphone_\indexone;\morphtwo_{\maplevelone(\indexone)} : \graphname{\examplegraphone}{\indexone} \to \graphname{\examplegraphthree}{\mapleveltwo(\maplevelone(\indexone))} \mid \indexone \in [\chaindepthone])\]
\end{definition}

Obviously, identity graph homomorphisms provide identity morphisms for graph chains.

\begin{definition}
\label{def:identity-graph-chain-morphisms}
For any graph chain \(\chainname{\examplegraphone} = \chain{\examplegraphone}{\chaindepthone}\) we obtain an identity graph chain morphism \((\overline{id}^{\chainname{\examplegraphone}},id_{[\chaindepthone]}):\chainname{\examplegraphone} \to \chainname{\examplegraphone}\) with the indexed family of identity graph homomorphisms
\[\overline{id}^{\chainname{\examplegraphone}} := (id_{\graphname{\examplegraphone}{\indexone}} : \graphname{\examplegraphone}{\indexone} \to \graphname{\examplegraphone}{\indexone} \mid \indexone \in [\chaindepthone])\]
\end{definition}

\begin{remark}
It is straightforward to show that the composition of graph chain morphisms is associative and that identity graph chain morphisms are neutral w.r.t.\ composition.
In such a way, we obtain a category \cat{Chain} of graph chains and graph chain morphisms.
\end{remark}

\begin{figure}[htb]
	\begin{minipage}[t]{.44\linewidth}
		\centering
		\includegraphics{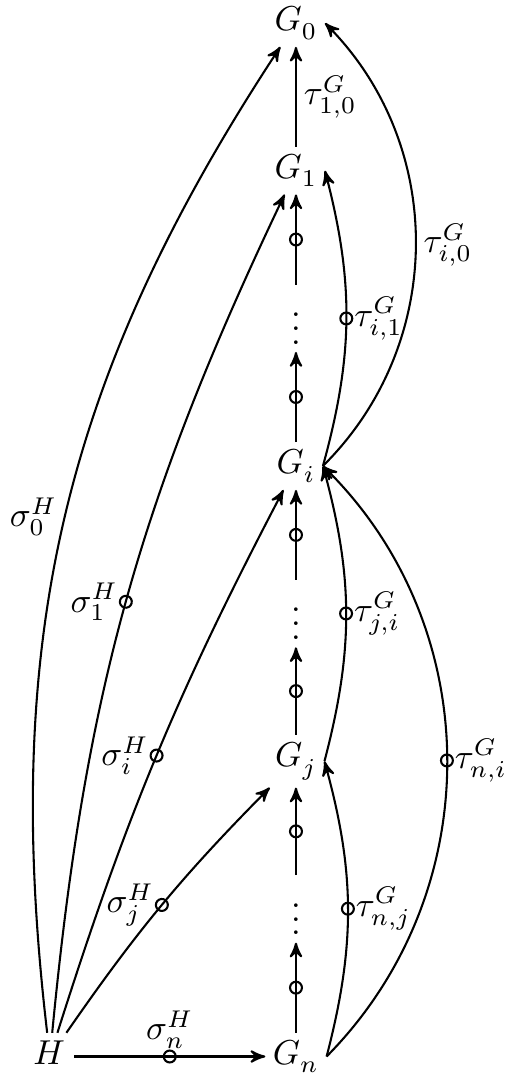}
		\subcaption{Multilevel typing of the graph \(\graphname{\examplegraphtwo}{}\) over \(\chainname{\examplegraphone}\)}
		\label{fig:multilevel-typing}
	\end{minipage}
	\quad
	\begin{minipage}[t]{.44\linewidth}
		\centering
		\includegraphics{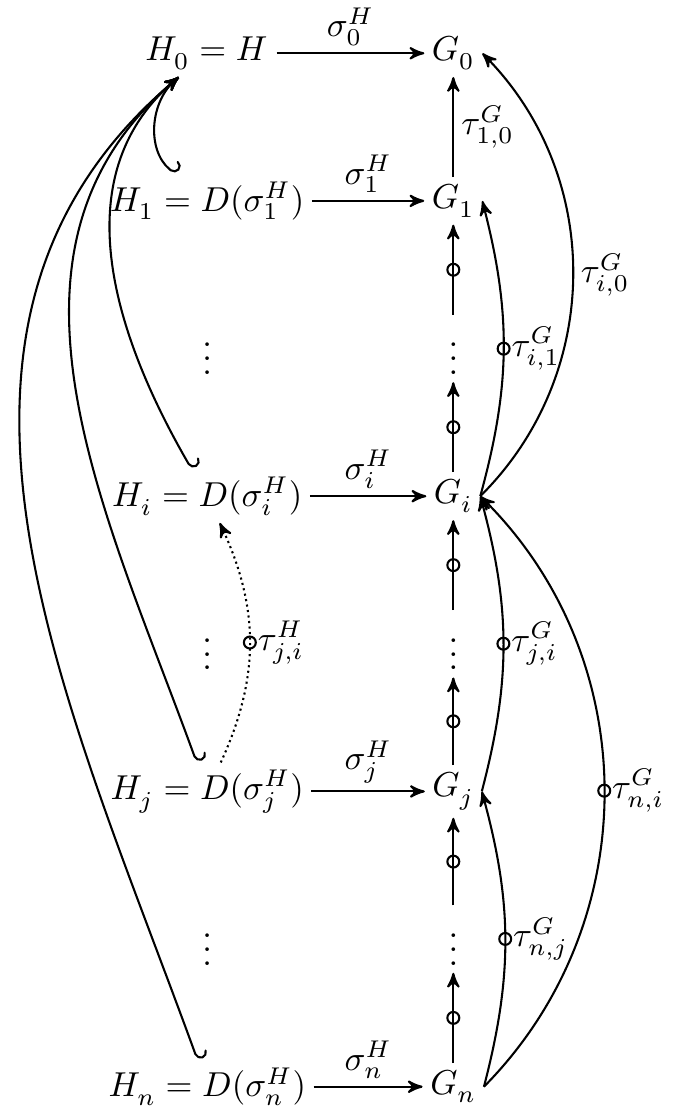}
		\subcaption{A graph and its multilevel typing represented as an inclusion chain and a chain graph morphism}
		\label{fig:multilevel-typing-as-graph-chain}
	\end{minipage}%
	\caption{Refactoring multilevel typing to chain graph and chain graph morphism}
	\label{fig:refactoring-mlm-into-chains}
\end{figure}

As mentioned, all the graphs \(\graphname{\rulegraphleft}{}\), \(\graphname{\rulegraphmiddle}{}\), \(\graphname{\rulegraphright}{}\), \(\graphname{\hierarchygraphleft}{}\), \(\graphname{\hierarchygraphmiddle}{}\), \(\graphname{\hierarchygraphright}{}\) in Figure~\ref{fig:multilevel-coupled-rule-formal} are multilevel-typed over either the graph chain \(\chainname{\rulegraph}\) or \(\chainname{\hierarchygraph}\).
To explain the multilevel typing relation we consider an arbitrary graph \(\graphname{\examplegraphtwo}{}\) typed over a graph chain \(\chainname{\examplegraphone} = \chain{\examplegraphone}{\chaindepthone}\), as depicted in Figure~\ref{fig:refactoring-mlm-into-chains}.

\begin{figure}[ht!]
	\centering
	\includegraphics{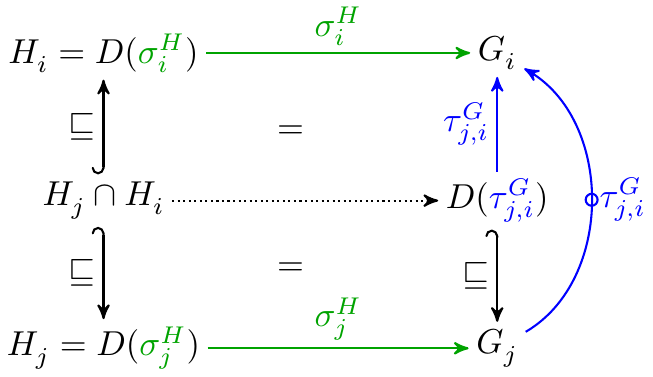}
	\caption{Refactoring multilevel typing into an object in \(\cat{Chain}\), level-wise}
	\label{fig:mlm-refactoring-diagram-formal}
\end{figure}

That is, we require the diagram in Figure~\ref{fig:mlm-refactoring-diagram-formal} to commute for all \(0 \leq \indexone \leq \indextwo \leq \chaindepthone\).
Colour coding has been added for later reference.

\begin{lemma}[Inclusion chain]\label{lem:inclusion-chain}
For any graph \(\graphname{\examplegraphtwo}{}\) we can extend any sequence \(\tc{\examplegraphtwo} = [\graphname{\examplegraphtwo}{\chaindepthone}, \graphname{\examplegraphtwo}{\chaindepthone-1}, \dots, \graphname{\examplegraphtwo}{1}, \graphname{\examplegraphtwo}{0}]\) of subgraphs of \(\graphname{\examplegraphtwo}{}\), with \(\graphname{\examplegraphtwo}{0} = \examplegraphtwo\), to a graph chain \(\chainname{\examplegraphtwo} = \chain{\examplegraphtwo}{\chaindepthone}\) where for all \(0 \leq \indexone \leq \indextwo \leq \chaindepthone\), \(\typemorph[\examplegraphtwo]{\indextwo}{\indexone} : \graphname{\examplegraphtwo}{\indextwo} \partialmap \graphname{\examplegraphtwo}{\indexone}\) (also called a partial inclusion morphism) is given by \(\domain{\typemorph[\examplegraphtwo]{\indextwo}{\indexone}} := \graphname{\examplegraphtwo}{\indextwo} \cap \graphname{\examplegraphtwo}{\indexone}\) and the span of inclusions \({\graphname{\examplegraphtwo}{\indextwo} \hookleftarrow \domain{\typemorph[\examplegraphtwo]{\indextwo}{\indexone}} := \graphname{\examplegraphtwo}{\indextwo} \cap \graphname{\examplegraphtwo}{\indexone} \hookrightarrow \graphname{\examplegraphtwo}{\indexone}}\).
We call the graph chain \(\chainname{\examplegraphtwo} = \chain{\examplegraphtwo}{\chaindepthone}\) also an \emph{inclusion chain}.
\end{lemma}

\begin{proof}

The \(\typemorph[\examplegraphtwo]{\indextwo}{0}\) are just total inclusions since \(\graphname{\examplegraphtwo}{\indextwo} \sqsubseteq \graphname{\examplegraphtwo}{0}\) for all \(0 < \indextwo \leq \chaindepthone\).
By construction we have
\begin{alignat*}{3}
\domain{\typemorph[\examplegraphtwo]{\indexthree}{\indextwo};\typemorph[\examplegraphtwo]{\indextwo}{\indexone}} \quad & =
		& \quad \domain{\typemorph[\examplegraphtwo]{\indexthree}{\indextwo}} \cap \domain{\typemorph[\examplegraphtwo]{\indextwo}{\indexone}} \\
\quad & =
		& \quad (\graphname{\examplegraphtwo}{\indexthree} \cap \graphname{\examplegraphtwo}{\indextwo}) \cap (\graphname{\examplegraphtwo}{\indextwo} \cap \graphname{\examplegraphtwo}{\indexone}) \\
\quad & =
		& \quad (\graphname{\examplegraphtwo}{\indexthree} \cap \graphname{\examplegraphtwo}{\indextwo} \cap \graphname{\examplegraphtwo}{\indexone}) \\
\quad & \sqsubseteq
		& \quad \graphname{\examplegraphtwo}{\indexthree} \cap \graphname{\examplegraphtwo}{\indexone} \\
\quad & =
		& \quad \domain{\typemorph[\examplegraphtwo]{\indexthree}{\indexone}}
\end{alignat*}
Thus the uniqueness condition is ensured since all the \(\typemorph[\examplegraphtwo]{\indextwo}{\indexone}\) are partial inclusions.
\qed
\end{proof}

A comparison of Definitions~\ref{eq:type-compatibility-rule-maps-1} and~\ref{eq:type-compatibility-rule-maps-2} makes obvious that multilevel typing can be equivalently described by graph chain morphisms.

\begin{corollary}\label{cor:multilevel-typing-as-chain-morphisms}
Let be given a graph \(\graphname{\examplegraphtwo}{}\), a graph chain \(\chainname{\examplegraphone}\) and a family \(\chainmorph{\examplegraphtwo} = (\chainmorph[\indexone]{\examplegraphtwo} : \graphname{\examplegraphtwo}{} \partialmap \graphname{\examplegraphone}{\indexone} \mid \indexone \in [\chaindepthone])\) of partial graph homomorphisms.
Moreover, let \(\chainname{\examplegraphtwo} = \chain{\examplegraphtwo}{\chaindepthone}\) be the inclusion chain with \(\graphname{\examplegraphtwo}{\chaindepthone} := \domain{\chainmorph[\indexone]{\examplegraphtwo}}\) for all \(\indexone\in[\chaindepthone]\), according to Lemma~\ref{lem:inclusion-chain}.
The following two conditions are equivalent:
\begin{enumerate}
	\item \(\chainmorph{\examplegraphtwo} = (\chainmorph[\indexone]{\examplegraphtwo} : \graphname{\examplegraphtwo}{} \partialmap \graphname{\examplegraphone}{\indexone} \mid \indexone \in [\chaindepthone])\) establishes a multilevel typing \(\chainmorph{\examplegraphtwo} : \graphname{\examplegraphtwo}{} \Rightarrow \chainname{\examplegraphone}\) according to Definition~\ref{def:multilevel-typing}.
	\item The family \((\chainmorph[\indexone]{\examplegraphtwo} : \domain{\chainmorph[\indexone]{\examplegraphtwo}} \to \graphname{\examplegraphone}{\indexone} \mid \indexone \in [\chaindepthone])\) of the corresponding total graph homomorphisms establishes a graph chain morphism \(\chainmorph{\examplegraphtwo} : \chainname{\examplegraphtwo} \to \chainname{\examplegraphone}\).
\end{enumerate}
\end{corollary}

To summarise, for any of the pairs \((\graphname{\rulegraphleft}{}, \chainmorph{\rulegraphleft})\), \((\graphname{\rulegraphmiddle}{}, \chainmorph{\rulegraphmiddle})\), \((\graphname{\rulegraphright}{}, \chainmorph{\rulegraphright})\) and \((\graphname{\hierarchygraphleft}{}, \chainmorph{\hierarchygraphleft})\), of a graph and a multilevel typing, Corollary~\ref{cor:multilevel-typing-as-chain-morphisms} ensures that we get four corresponding inclusion chains
\(\chainname{\rulegraphleft} = \chain{\rulegraphleft}{\chaindepthone}\),
\(\chainname{\rulegraphmiddle} = \chain{\rulegraphmiddle}{\chaindepthone}\),
\(\chainname{\rulegraphright} = \chain{\rulegraphright}{\chaindepthone}\) and
\(\chainname{\hierarchygraphleft} = \chain{\hierarchygraphleft}{\chaindepthtwo}\), respectively,
together with chain morphisms
\((\chainmorph{\rulegraphleft},id_{[\chaindepthone]}) : \chainname{\rulegraphleft} \to \chainname{\rulegraph}\),
\((\chainmorph{\rulegraphmiddle},id_{[\chaindepthone]}) : \chainname{\rulegraphmiddle} \to \chainname{\rulegraph}\),
\((\chainmorph{\rulegraphright},id_{[\chaindepthone]}) : \chainname{\rulegraphright} \to \chainname{\rulegraph}\), and
\((\chainmorph{\hierarchygraphleft},id_{[\chaindepthtwo]}) : \chainname{\hierarchygraphleft} \to \chainname{\hierarchygraph}\).
Note that the \(\chainmorph{}\)'s in the chain morphisms are the total components of the partial typing morphisms in the original pairs.

\begin{lemma}\label{lem:inclusion-chains-establish-chain-morphism}
Let be given two inclusion chains \(\chainname{\examplegraphone} = \chain{\examplegraphone}{\chaindepthone}\), \(\chainname{\examplegraphtwo} = \chain{\examplegraphtwo}{\chaindepthtwo}\), as described in Lemma~\ref{lem:inclusion-chain}, with \(\chaindepthone \leq \chaindepthtwo\).
Moreover, let be given a function \(\maplevelone : [\chaindepthone] \to [\chaindepthtwo]\) such that \(\maplevelone(0) = 0\) and \(\indexone < \indextwo\) implies \(\maplevelone(\indexone) < \maplevelone(\indextwo)\), and a family \(\morphone = (\morphone_\indexone : \graphname{\examplegraphone}{\indexone} \to \graphname{\examplegraphtwo}{\maplevelone(\indexone)} \mid \indexone \in [\chaindepthone])\) of graph homomorphisms.

If for all \(\indexone \in [\chaindepthone]\) the left-hand square below is a pullback, then the pair \((\morphone,\maplevelone)\) constitutes a graph chain morphism \((\morphone,\maplevelone) : \chainname{\examplegraphone} \to \chainname{\examplegraphtwo}\) where for all \(0 \leq \indexone < \indextwo \leq \chaindepthone\) the right-hand diagram below consist of two pullbacks

\begin{figure}[ht!]
\begin{center}
	\begin{minipage}{.4\linewidth}
	\centering
	\begin{tikzpicture}[on grid,node distance=25mm]
	
	\node[el-math] (g0)								{\graphname{\examplegraphone}{0}};
	\node[el-math] (h0)	[right of=g0]				{\graphname{\examplegraphtwo}{0}};
	\node[el-math] (gi)	[below of=g0]				{\graphname{\examplegraphone}{\indexone}};
	\node[el-math] (hi)	[below of=h0]				{\graphname{\examplegraphtwo}{\indexone}};
	\node[el-math] (pb)	[below right=17mm of g0]	{\pb};
	
	\draw[map]		(gi) to node[la-math]	(pi)	{\morphone_\indexone}	(hi);
	\draw[map]		(g0) to node[la-math]	(p0)	{\morphone_0}			(h0);
	\draw[incmapl]	(gi) to node[la-math]	(ing)	{}						(g0);
	\draw[incmapl]	(hi) to node[la-math]	(inh)	{}						(h0);
	
	\end{tikzpicture}
	\end{minipage}
	\quad
	\begin{minipage}{.5\linewidth}
	\centering
	\begin{tikzpicture}[on grid,node distance=30mm]
	
	\node[el-math] (gi)						{\graphname{\examplegraphone}{\indexone}};
	\node[el-math] (hi)		[right of=gi]	{\graphname{\examplegraphtwo}{\indexone}};
	\node[el-math] (gji)	[below of=gi]	{\graphname{\examplegraphone}{\indextwo} \cap \graphname{\examplegraphone}{\indexone}};
	\node[el-math] (hji)	[right of=gji]	{\graphname{\examplegraphtwo}{\indextwo} \cap \graphname{\examplegraphtwo}{\indexone}};
	\node[el-math] (gj)		[below of=gji]	{\graphname{\examplegraphone}{\indextwo}};
	\node[el-math] (hj)		[right of=gj]	{\graphname{\examplegraphtwo}{\indextwo}};
	
	\draw[incmapl]	(gi) to node[la-math]	(mgihi)		{\morphone_\indexone}							(hi);
	\draw[incmapl]	(gj) to node[la-math]	(mgjhj)		{\morphone_\indextwo}							(hj);
	\draw[incmapl]	(gji) to node[la-math]	(mgjihji)	{\morphone_{\indextwo \mid \indexone}}	(hji);
	
	\draw[incmapl]	(gji) to node[la-math]			(ingjigi)	{\sqsubseteq}	(gi);
	\draw[incmapr]	(hji) to node[la-math,right]	(inhjihi)	{\sqsubseteq}	(hi);
	\draw[incmapr]	(gji) to node[la-math,left]		(ingjigj)	{\sqsubseteq}	(gj);
	\draw[incmapl]	(hji) to node[la-math,right]	(inhjihj)	{\sqsubseteq}	(hj);
	
	\draw[partmap,bend left]	(gj) to node[la-math]		(tgji)	{\typemorph[\examplegraphone]{\indextwo}{\indexone}}	(gi);
	\draw[partmap,bend right]	(hj) to node[la-math,right]	(thji)	{\typemorph[\examplegraphtwo]{\indextwo}{\indexone}}	(hi);
	
	\node[el-math] (pb1)	[below=16mm of mgihi]	{\pb};
	\node[el-math] (pb2)	[below of=pb1]			{\pb};
	
	\end{tikzpicture}
	\end{minipage}
\end{center}
\end{figure}

\end{lemma}

\begin{definition}[Reduct]\label{def:reduct}
Given a graph chain morphism \((\morphone,\maplevelone) : \chainname{\examplegraphone} \to \chainname{\examplegraphtwo}\) as described in Lemma~\ref{lem:inclusion-chains-establish-chain-morphism}, we call \(\chainname{\examplegraphone}\) the \emph{reduct} of \(\chainname{\examplegraphtwo}\) along \(\maplevelone\) and \(\morphone_0 : \graphname{\examplegraphone}{0} \to \graphname{\examplegraphtwo}{0}\) .
In case \(\graphname{\examplegraphone}{\indexone} \sqsubseteq \graphname{\examplegraphtwo}{\maplevelone(\indexone)}\) for all \(\indexone \in [\chaindepthone]\), where \(\morphone_\indexone : \graphname{\examplegraphone}{\indexone} \to \graphname{\examplegraphtwo}{\indexone}\) is the corresponding inclusion graph homomorphism, we call \(\chainname{\examplegraphone}\) a closed subchain of \(\chainname{\examplegraphtwo}\) via \(\maplevelone\).
And, if \(\chaindepthone = \chaindepthtwo\) and thus \(\maplevelone = id_{[\chaindepthone]}\), \(\chainname{\examplegraphone}\) is simply called a closed subchain of \(\chainname{\examplegraphtwo}\).
\end{definition}

\subsection{Type compatibility as chain morphism}
\label{subsec:type-compatibility-as-chain-morphism}

In the same lines of typed graph morphisms where type compatibility must be ensured (see, e.g.,~\cite{ehrig2006fundamentals}) we have to make sure that graph chain morphisms, when needed, also respect typing.
For our MCMT rules, this means that \(\maprulelefttomiddle : \graphname{\rulegraphleft}{} \to \graphname{\rulegraphmiddle}{}\) and \(\mapruletohierarchyleft : \graphname{\rulegraphleft}{} \to \graphname{\hierarchygraphleft}{}\) are compatible with respect to typing.
First we will discuss the typing compatibility of \(\maprulelefttomiddle\), then we use an analogous explanation for \(\mapruletohierarchyleft\).

In case of \(\maprulelefttomiddle : \graphname{\rulegraphleft}{} \to \graphname{\rulegraphmiddle}{}\), we do have actually a morphism \(\maprulelefttomiddle : (\graphname{\rulegraphleft}{}, \chainmorph{\rulegraphleft}) \to (\graphname{\rulegraphmiddle}{}, \chainmorph{\rulegraphmiddle})\) where we require that, for all \(0 < \indexone \leq \chaindepthone\), the following diagram is commutative
\begin{center}
\begin{tikzpicture}[on grid,node distance=25mm]
	
\node[el-math] (mm)							{\graphname{\rulegraph}{\indexone}};
\node[el-math] (l)	[below left of=mm]		{\graphname{\rulegraphleft}{}};
\node[el-math] (i)	[below right of=mm]		{\graphname{\rulegraphmiddle}{}};
\node[el-math] (e1)	[below=10mm of mm]		{=};

\draw[partmap]	(l) to node[la-math]				(tl)	{\chainmorph[\indexone]{\rulegraphleft}}	(mm);
\draw[partmap]	(i) to node[la-math,above right]	(ti)	{\chainmorph[\indexone]{\rulegraphmiddle}}	(mm);

\draw[incmapl]	(l) to node[la-math,below]			(lid)	{\maprulelefttomiddle}				(i);
	
\end{tikzpicture}
\end{center}
Since \(\maprulelefttomiddle\) is total this gives for all \(0 < \indexone \leq \chaindepthone\) a pullback square
\begin{center}
\begin{tikzpicture}[on grid,node distance=25mm]

\node[el-math] (mm)						{\graphname{\rulegraph}{\indexone}};
\node[el-math] (l0) [below left of=mm]	{\graphname{\rulegraphleft}{0} = \graphname{\rulegraphleft}{}};
\node[el-math] (i0) [below right of=mm]	{\graphname{\rulegraphmiddle}{0} = \graphname{\rulegraphmiddle}{}};
\node[el-math] (li)	[below of=l0]		{\graphname{\rulegraphleft}{\indexone} = \domain{\chainmorph[\indexone]{\rulegraphleft}}};
\node[el-math] (ii)	[below of=i0]		{\graphname{\rulegraphmiddle}{\indexone} = \domain{\chainmorph[\indexone]{\rulegraphmiddle}}};
\node[el-math] (e1)	[below=8mm of mm]	{=};
\node[el-math] (pb)	[below=22mm of e1]	{\pb};

\draw[incmapl]	(li) to node[la-math]				(ili)	{\maprulelefttomiddle_{\indexone}}					(ii);
\draw[incmapl]	(li) to node[la-math,right]			(tl)	{\typemorph[\rulegraphleft]{\indexone}{0}}			(l0);
\draw[incmapl]	(ii) to node[la-math,right]			(ti)	{\typemorph[\rulegraphmiddle]{\indexone}{0}}		(i0);
\draw[incmapl]	(l0) to node[la-math]				(il0)	{\maprulelefttomiddle_{0} = \maprulelefttomiddle}	(i0);
\draw[map]		(l0) to node[la-math]				(sl0)	{\chainmorph[\indexone]{\rulegraphleft}}			(mm);
\draw[map]		(i0) to node[la-math,above right]	(si0)	{\chainmorph[\indexone]{\rulegraphmiddle}}			(mm);

\draw[map,bend left=60]		(li.west) to node[la-math]				(sli)	{\chainmorph[\indexone]{\rulegraphleft}}	(mm.west);
\draw[map,bend right=60]	(ii.east) to node[la-math,above right]	(sii)	{\chainmorph[\indexone]{\rulegraphmiddle}}	(mm.east);

\end{tikzpicture}
\end{center}
That is, we have for all \(0 < \indexone \leq \chaindepthone\)
\begin{equation}\label{eq:li-equals-l0-ii}
\graphname{\rulegraphleft}{\indexone} = \graphname{\rulegraphleft}{0} \cap \graphname{\rulegraphmiddle}{\indexone}
\end{equation}
This condition is equivalent to condition
\begin{equation}\label{eq:li-l0-equivalence}
\graphname{\rulegraphmiddle}{\indexone} \setminus \graphname{\rulegraphleft}{\indexone} = \graphname{\rulegraphmiddle}{\indexone} \setminus \graphname{\rulegraphleft}{0}
\end{equation}
										
It is now straight forward to show that the family \(\maprulelefttomiddle=(\maprulelefttomiddle_\indexone : \graphname{\rulegraphleft}{\indexone} \to \graphname{\rulegraphmiddle}{\indexone} \mid \indexone \in [\chaindepthone])\) of inclusion homomorphisms establishes a chain morphism \((\maprulelefttomiddle, id_{[\chaindepthone]}) : \chainname{\rulegraphleft} \to \chainname{\rulegraphmiddle}\) (cf. Section~\ref{subsubsec:pushout-chain-equal-depth}).
For all \(0\leq \indexone < \indextwo \leq \chaindepthone\) we have
\begin{center}
\begin{tikzpicture}[on grid,node distance=35mm]

\node[el-math] (l0)											{\graphname{\rulegraphleft}{0} = \graphname{\rulegraphleft}{}};
\node[el-math] (i0)		[right of=l0]						{\graphname{\rulegraphmiddle}{0} = \graphname{\rulegraphmiddle}{}};
\node[el-math] (li)		[below of=l0]						{\graphname{\rulegraphleft}{\indexone}};
\node[el-math] (ii)		[right of=li]						{\graphname{\rulegraphmiddle}{\indexone}};
\node[el-math] (lj)		[below right=15mm and 15mm of l0]	{\graphname{\rulegraphleft}{\indextwo}};
\node[el-math] (ij)		[right of=lj]						{\graphname{\rulegraphmiddle}{\indextwo}};
\node[el-math] (lij)	[below of=lj]						{\graphname{\rulegraphleft}{\indexone} \cap \graphname{\rulegraphleft}{\indextwo}};
\node[el-math] (iij)	[right of=lij]						{\graphname{\rulegraphmiddle}{\indexone} \cap \graphname{\rulegraphmiddle}{\indextwo}};

\draw[incmapl]	(lj) to node[la-math,pos=.4]	(dj)		{\maprulelefttomiddle_\indextwo}					(ij);
\draw[incmapl]	(l0) to node[la-math]			(d0)		{\maprulelefttomiddle_0=\maprulelefttomiddle}		(i0);
\draw[incmapr]	(lj) to node[la-math]			(inljl0)	{}													(l0);
\draw[incmapr]	(ij) to node[la-math]			(iniji0)	{}													(i0);

\draw[mapdots]	(lij) to node[la-math]			(dji)		{\maprulelefttomiddle_{\indextwo \mid \indexone}}	(iij);
\draw[incmapl]	(li) to node[la-math,pos=.6]	(di)		{\maprulelefttomiddle_\indexone}					(ii);
\draw[incmapr]	(lij) to node[la-math]			(inlijli)	{}													(li);
\draw[incmapr]	(iij) to node[la-math]			(iniijii)	{}													(ii);

\draw[incmapr]	(lij) to node[la-math]			(inlijlj)	{}													(lj);
\draw[incmapr]	(iij) to node[la-math]			(iniijij)	{}													(ij);
\draw[incmapr]	(li) to node[la-math]			(inlil0)	{}													(l0);
\draw[incmapr]	(ii) to node[la-math]			(iniii0)	{}													(i0);

\end{tikzpicture}
\end{center}
where the top and back squares are pullbacks by equation~\ref{eq:li-equals-l0-ii}.
The left and right squares are also pullbacks since they arise by intersection.
Due to the universal property of the right square we get a unique inclusion homomorphism \(\maprulelefttomiddle_{\indextwo \mid \indexone} : \graphname{\rulegraphleft}{\indexone} \cap \graphname{\rulegraphleft}{\indextwo} \hookrightarrow \graphname{\rulegraphmiddle}{\indexone} \cap \graphname{\rulegraphmiddle}{\indextwo}\) making the cube commute.

Due to the composition of the left and back pullbacks, the diagonal square from the front-left homomorphism to the back-right morphism becomes a pullback too.
Decomposing this diagonal pullback relative to the right pullback gives us also a pullback on the front square.

Analogously, we can show that the bottom square is a pullback as well.
Finally, we get a diagram with two pullback squares for all \(0 \leq \indexone < \indextwo \leq \chaindepthone\) and thus a graph chain morphism \(\maprulelefttomiddle = (\maprulelefttomiddle_\indexone : \graphname{\rulegraphleft}{\indexone} \to \graphname{\rulegraphmiddle}{\indexone} \mid \indexone \in [\chaindepthone])\) according to Definition~\ref{def:graph-chain-morphism} and Lemma~\ref{lem:inclusion-chains-establish-chain-morphism}.

\begin{center}
\begin{tikzpicture}[on grid,node distance=30mm]

\node[el-math] (li)						{\graphname{\rulegraphleft}{\indexone}};
\node[el-math] (ii)		[right of=li]	{\graphname{\rulegraphmiddle}{\indexone}};
\node[el-math] (lij)	[below of=li]	{\graphname{\rulegraphleft}{\indextwo} \cap \graphname{\rulegraphleft}{\indexone}};
\node[el-math] (iij)	[right of=lij]	{\graphname{\rulegraphmiddle}{\indextwo} \cap \graphname{\rulegraphmiddle}{\indexone}};
\node[el-math] (lj)		[below of=lij]	{\graphname{\rulegraphleft}{\indextwo}};
\node[el-math] (ij)		[right of=lj]	{\graphname{\rulegraphmiddle}{\indextwo}};

\draw[incmapl]	(li) to node[la-math]	(di)	{\maprulelefttomiddle_\indexone}					(ii);
\draw[incmapl]	(lj) to node[la-math]	(dj)	{\maprulelefttomiddle_\indextwo}					(ij);
\draw[incmapl]	(lij) to node[la-math]	(dji)	{\maprulelefttomiddle_{\indextwo \mid \indexone}}	(iij);

\draw[incmapl]	(lij) to node[la-math]			(inlijli)	{\sqsubseteq}	(li);
\draw[incmapr]	(iij) to node[la-math,right]	(iniijii)	{\sqsubseteq}	(ii);
\draw[incmapr]	(lij) to node[la-math,left]		(inlijlj)	{\sqsubseteq}	(lj);
\draw[incmapl]	(iij) to node[la-math,right]	(iniijij)	{\sqsubseteq}	(ij);

\draw[partmap,bend left]	(lj) to node[la-math]		(tlji)	{\typemorph[\rulegraphleft]{\indextwo}{\indexone}}		(li);
\draw[partmap,bend right]	(ij) to node[la-math,right]	(tiji)	{\typemorph[\rulegraphmiddle]{\indextwo}{\indexone}}	(ii);

\node[el-math] (pb1)	[below=16mm of di]	{\pb};
\node[el-math] (pb2)	[below of=pb1]		{\pb};

\end{tikzpicture}
\end{center}

The type compatibility for \(\maprulelefttomiddle : \graphname{\rulegraphleft}{} \to \graphname{\rulegraphmiddle}{}\) can be described now by a commutative triangle of chain morphisms.

\begin{center}
\begin{tikzpicture}[on grid,node distance=25mm]

\node[el-math] (mm)							{\chainname{\rulegraph}};
\node[el-math] (l)	[below left of=mm]		{\chainname{\rulegraphleft}};
\node[el-math] (i)	[below right of=mm]		{\chainname{\rulegraphmiddle}};
\node[el-math] (l0)							{};
\node[el-math] (e1)	[below=10mm of mm]		{=};

\draw[map]		(l) to node[la-math]				(tl)	{(\chainmorph{\rulegraphleft},id_{[\chaindepthone]})}	(mm);
\draw[map]		(i) to node[la-math,above right]	(ti)	{(\chainmorph{\rulegraphmiddle},id_{[\chaindepthone]})}	(mm);

\draw[incmapl]	(l) to node[la-math,below]			(lid)	{(\maprulelefttomiddle,id_{[\chaindepthone]})}			(i);
	
\end{tikzpicture}
\end{center}

Analogous to \(\maprulelefttomiddle\), the type compatibility of \(\mapruletohierarchyleft\) can be established by showing that the family \(\mapruletohierarchyleft = (\mapruletohierarchyleft_\indexone : \graphname{\rulegraphleft}{\indexone} \to \graphname{\hierarchygraphleft}{\maplevelone(\indexone)} \mid \indexone \in [\chaindepthone])\) for all \(0 < \indexone \leq \chaindepthone\) establishes the chain morphism \((\mapruletohierarchyleft, \maplevelone) : \rulegraphleft \to \hierarchygraphleft\), thus making the following diagram commutative

\begin{center}
\begin{tikzpicture}[on grid,node distance=25mm]

\node[el-math] (mm)								{\graphname{\rulegraph}{\indexone}};
\node[el-math] (tg)	[right of=mm]				{\graphname{\hierarchygraph}{\maplevelone(\indexone)}};
\node[el-math] (l)	[below of=mm]				{\graphname{\rulegraphleft}{}};
\node[el-math] (s)	[below of=tg]				{\graphname{\hierarchygraphleft}{}};
\node[el-math] (eq)	[below right=18mm of mm]	{=};

\draw[map]		(mm) to node[la-math]		(bf)	{\mapruletohierarchybinding_\indexone}						(tg);
\draw[map]		(l) to node[la-math]		(mf)	{\mapruletohierarchyleft}									(s);
\draw[partmap]	(l) to node[la-math]		(tl)	{\chainmorph[\indexone]{\rulegraphleft}}					(mm);
\draw[partmap]	(s) to node[la-math,right]	(ts)	{\chainmorph[\maplevelone(\indexone)]{\hierarchygraphleft}}	(tg);
	
\end{tikzpicture}
\end{center}
Since \(\mapruletohierarchybinding_{\chaindepthone}\) and \(\mapruletohierarchyleft\) are total we get the following diagram
\begin{center}
\begin{tikzpicture}[on grid,node distance=25mm]

\node[el-math] (mm)						{\graphname{\rulegraph}{\indexone}};
\node[el-math] (tg)	[right=40mm of mm]	{\graphname{\hierarchygraph}{\maplevelone(\indexone)}};
\node[el-math] (l0) [below of=mm]		{\graphname{\rulegraphleft}{0} = \graphname{\rulegraphleft}{}};
\node[el-math] (s0) [below of=tg]		{\graphname{\hierarchygraphleft}{0} = \graphname{\hierarchygraphleft}{}};
\node[el-math] (li)	[below of=l0]		{\graphname{\rulegraphleft}{\indexone} = \domain{\chainmorph[\indexone]{\rulegraphleft}}};
\node[el-math] (si)	[below of=s0]		{\graphname{\hierarchygraphleft}{\maplevelone(\indexone)} = \domain{\chainmorph[\maplevelone(\indexone)]{\hierarchygraphleft}}};

\draw[map]		(mm) to node[la-math]		(bi)	{\mapruletohierarchybinding_{\indexone}}									(tg);
\draw[map]		(l0) to node[la-math]		(m0)	{\mapruletohierarchyleft_{0} = \mapruletohierarchyleft}						(s0);
\draw[map]		(li) to node[la-math]		(mi)	{\mapruletohierarchyleft_{\indexone}}										(si);
\draw[incmapl]	(li) to node[la-math,left]	(tl)	{\typemorph[\rulegraphleft]{\indexone}{0}}									(l0);
\draw[incmapl]	(si) to node[la-math]		(ts)	{\typemorph[\hierarchygraphleft]{\maplevelone(\indexone)}{\maplevelone(0)}}	(s0);
\draw[map]		(l0) to node[la-math]		(sl1)	{\chainmorph[\indexone]{\rulegraphleft}}									(mm);
\draw[map]		(s0) to node[la-math]		(ss1)	{\chainmorph[\maplevelone(\indexone)]{\hierarchygraphleft}}					(tg);

\draw[map,bend left=40]		(li) to node[la-math]		(sl2)	{\chainmorph[\indexone]{\rulegraphleft}}					(mm);
\draw[map,bend right=40]	(si) to node[la-math,right]	(ss2)	{\chainmorph[\maplevelone(\indexone)]{\hierarchygraphleft}}	(tg);

\node[el-math] (eq)	[below=15mm of bi]	{=};
\node[el-math] (pb) [below of=eq]		{\pb};

\end{tikzpicture}
\end{center}
It is now straightforward to show that the \(\mapruletohierarchyleft_{\chaindepthone}\) establish a chain morphism \((\mapruletohierarchyleft, \maplevelone) : \chainname{\rulegraphleft} \to \chainname{\hierarchygraphleft}\), hence the compatibility of typing for \(\mapruletohierarchyleft\) can be described by the following diagram

\begin{center}
\begin{tikzpicture}[on grid,node distance=30mm]

\node[el-math] (li)						{\graphname{\rulegraphleft}{\indexone}};
\node[el-math] (si)		[right of=li]	{\graphname{\hierarchygraphleft}{\maplevelone(\indexone)}};
\node[el-math] (lij)	[below of=li]	{\graphname{\rulegraphleft}{\indextwo} \cap \graphname{\rulegraphleft}{\indexone}};
\node[el-math] (sij)	[right of=lij]	{\graphname{\hierarchygraphleft}{\maplevelone(\indextwo)} \cap \graphname{\hierarchygraphleft}{\maplevelone(\indexone)}};
\node[el-math] (lj)		[below of=lij]	{\graphname{\rulegraphleft}{\indextwo}};
\node[el-math] (sj)		[right of=lj]	{\graphname{\hierarchygraphleft}{\maplevelone(\indextwo)}};

\draw[incmapl]	(li) to node[la-math]	(di)	{\mapruletohierarchyleft_\indexone}						(si);
\draw[incmapl]	(lj) to node[la-math]	(dj)	{\mapruletohierarchyleft_\indextwo}						(sj);
\draw[mapdots]	(lij) to node[la-math]	(dji)	{\mapruletohierarchyleft_{\indextwo \mid \indexone}}	(sij);

\draw[incmapl]	(lij) to node[la-math]			(inlijli)	{\sqsubseteq}	(li);
\draw[incmapr]	(iij) to node[la-math,right]	(iniijii)	{\sqsubseteq}	(si);
\draw[incmapr]	(lij) to node[la-math,left]		(inlijlj)	{\sqsubseteq}	(lj);
\draw[incmapl]	(iij) to node[la-math,right]	(iniijij)	{\sqsubseteq}	(sj);

\draw[partmap,bend left=40]		(lj) to node[la-math]		(tlji)	{\typemorph[\rulegraphleft]{\indextwo}{\indexone}}									(li);
\draw[partmap,bend right=40]	(sj) to node[la-math,right]	(tiji)	{\typemorph[\hierarchygraphleft]{\maplevelone(\indextwo)}{\maplevelone(\indexone)}}	(si);

\node[el-math] (pb1)	[below=16mm of di]	{\pb};
\node[el-math] (pb2)	[below of=pb1]		{\pb};

\end{tikzpicture}
\end{center}

\subsection{Pushouts in the category \cat{Chain}}
\label{subsec:pushouts-category-chain}

As indicated in the previous section we use model transformations to define the behaviour of our MLM hierarchies. Since every branch in our MLM hierarchies is represented by a graph chain, the application of our model transformation rules will rely on pushout and final pullback complement constructions in the category \cat{Chain}. Due to similarities between these constructions, the main focus of this section will be on the construction of the pushout of the following span
\begin{center}
\begin{tikzpicture}[on grid,node distance=35mm]

\node[el-math] (s)					{\chainname{\hierarchygraphleft}};
\node[el-math] (l)	[right of=s]	{\chainname{\rulegraphleft}};
\node[el-math] (i)	[right of=l]	{\chainname{\rulegraphmiddle}};

\draw[map]		(l) to node[la-math,above]	(map)	{(\mapruletohierarchyleft,\maplevelone)}		(s);
\draw[incmapl]	(l) to node[la-math,above]	(in)	{(\maprulelefttomiddle,id_{[\chaindepthone]})}	(i);

\end{tikzpicture}
\end{center}
of chain morphisms between inclusion chains in the category \(\cat{Chain}\), where our assumption should ensure that the pushout becomes an inclusion chain as well. Moreover, the pushout in \(\cat{Chain}\) should be fully determined by the construction of the pushout \(\graphname{\hierarchygraphmiddle}{}\) of the span
\begin{center}
\begin{tikzpicture}[on grid,node distance=35mm]

\node[el-math] (s)					{\graphname{\hierarchygraphleft}{0} = \graphname{\hierarchygraphleft}{}};
\node[el-math] (l)	[right of=s]	{\graphname{\rulegraphleft}{0} = \graphname{\rulegraphleft}{}};
\node[el-math] (i)	[right of=l]	{\graphname{\rulegraphmiddle}{0} = \graphname{\rulegraphmiddle}{}};

\draw[map]		(l) to node[la-math,above]	(map)	{\mapruletohierarchyleft_{0}}	(s);
\draw[incmapl]	(l) to node[la-math,above]	(in)	{\maprulelefttomiddle_{0}}		(i);

\end{tikzpicture}
\end{center}
in the category \cat{Graph} of graphs and total graph homomorphisms.
Especially, the multilevel typing of \(\hierarchygraphmiddle\), i.e.\ the level-wise pushouts of all the spans
\begin{center}
\begin{tikzpicture}[on grid,node distance=35mm]

\node[el-math] (s)						{\graphname{\hierarchygraphleft}{\maplevelone(\indexone)}};
\node[el-math] (l)	[right of=s]		{\graphname{\rulegraphleft}{\indexone}};
\node[el-math] (i)	[right of=l]		{\graphname{\rulegraphmiddle}{\indexone}};
\node[el-math] (ex)	[right=20mm of i]	{1 \leq \indexone \leq \chaindepthone};

\draw[map]		(l) to node[la-math,above]	(map)	{\mapruletohierarchyleft_{\indexone}}	(s);
\draw[incmapl]	(l) to node[la-math,above]	(in)	{\maprulelefttomiddle_{\indexone}}		(i);

\end{tikzpicture}
\end{center}
should be just parts of the pushout construction for the base \(\chaindepthone = 0\).

The result of the level-wise pushouts should be an inclusion chain of length \(\chaindepthone+1\).
The rule provides, however, only information about the typing at the levels \(\maplevelone([\chaindepthone]) \subseteq [\chaindepthtwo]\).
For the levels in \([\chaindepthtwo] \setminus \maplevelone([\chaindepthone])\) we have to borrow the typing from the corresponding untouched levels in \(\chainname{\hierarchygraphleft}\).
In terms of graph chain morphisms, this means that we factorize \((\mapruletohierarchyleft,\maplevelone)\) into two graph chain morphisms and that we construct the resulting inclusion chain in two pushout steps (see Figure~\ref{fig:two-po-construct-d}) where \(\chainname[\downarrow\maplevelone]{\hierarchygraphleft} = \chain[\downarrow\maplevelone]{\hierarchygraphleft}{\chaindepthone}\) with
\(\overline{\hierarchygraphleft}_{\downarrow \maplevelone} = [\graphname{\hierarchygraphleft}{\maplevelone(\chaindepthone)},  \graphname{\hierarchygraphleft}{\maplevelone(\chaindepthone-1)}, \ldots, \graphname{\hierarchygraphleft}{\maplevelone(1)}, \graphname{\hierarchygraphleft}{\maplevelone(0)}]\) and
\(\typemorph[\hierarchygraphleft]{\downarrow \maplevelone}{} = (\typemorph[\hierarchygraphleft]{\maplevelone(\indextwo)}{\maplevelone(\chaindepthone)} : \graphname{\hierarchygraphleft}{\maplevelone(\indextwo)} \partialmap \graphname{\hierarchygraphleft}{\maplevelone(\chaindepthone)} \mid 0 \le \indexone < \indextwo \le \chaindepthone)\).
Note that \(\overline{\hierarchygraphleft}_{\downarrow \maplevelone}=[\graphname{\hierarchygraphleft}{\maplevelone(\chaindepthone)}, \graphname{\hierarchygraphleft}{\maplevelone(\chaindepthone-1)},\ldots,\graphname{\hierarchygraphleft}{\maplevelone(1)},\graphname{\hierarchygraphleft}{\maplevelone(0)}]\) is just a shorthand for the statement: \((\overline{\hierarchygraphleft}_{\downarrow \maplevelone})_\indexone := \graphname{\hierarchygraphleft}{\maplevelone(\chaindepthone)}\) for all \(0 \leq \indexone \leq \chaindepthone\). 

\begin{figure}[ht!]
\begin{center}
\begin{tikzpicture}[on grid,node distance=25mm]

\node[el-math] (l)										{\chainname{\rulegraphleft}};
\node[el-math] (sf)	[right=40mm of l]					{\chainname[\downarrow\maplevelone]{\hierarchygraphleft}};
\node[el-math] (s)	[right=40mm of sf]					{\chainname{\hierarchygraphleft}};
\node[el-math] (i)	[below of=l]						{\chainname{\rulegraphmiddle}};
\node[el-math] (df)	[below of=sf]						{\chainname[\downarrow\maplevelone]{\hierarchygraphmiddle}};
\node[el-math] (si)	[below of=s]						{\chainname{\hierarchygraphmiddle}};
\node[el-math] (eq)	[below right=12mm and 20mm of l]	{(1)};
\node[el-math] (eq)	[below right=12mm and 25mm of sf]	{(2)};

\draw[map]		(l) to node[la-math]			(mu1)	{(\mapruletohierarchyleft,id_{[\chaindepthone]})}											(sf);
\draw[map]		(sf) to node[la-math]			(id1)	{(\overline{id}^{\chainname{\hierarchygraphleft}}_{\downarrow\maplevelone},\maplevelone)}	(s);
\draw[incmapl]	(l) to node[la-math,left]		(in1)	{(\maprulelefttomiddle,id_{[\chaindepthone]})}												(i);
\draw[incmapl]	(sf) to node[la-math]			(in2)	{(\maphierarchylefttomiddle_{\downarrow\maplevelone},id_{[\chaindepthone]})}				(df);
\draw[incmapl]	(s) to node[la-math]			(in3)	{(\maphierarchylefttomiddle,id_{[\chaindepthtwo]})}											(si);
\draw[map]		(i) to node[la-math,below]		(mu2)	{(\mapruletohierarchymiddle,id_{[\chaindepthone]})}											(df);
\draw[incmapl]	(df) to node[la-math,below]		(id2)	{(\overline{id}^{\chainname{\hierarchygraphmiddle}}_{\downarrow\maplevelone},\maplevelone)}	(si);
	
\end{tikzpicture}
\end{center}
\caption{Two pushout steps to construct the inclusion chain \(\chainname{\hierarchygraphmiddle}\)}
\label{fig:two-po-construct-d}
\end{figure}

The graph chain morphism \((\overline{id}^{\chainname{\hierarchygraphleft}}_{\downarrow\maplevelone},\maplevelone) : \chainname[\downarrow\maplevelone]{\hierarchygraphleft} \to \chainname{\hierarchygraphleft}\) is a level-wise identity and just embeds a chain of length \(\chaindepthone\) into a chain of length \(\chaindepthtwo\), i.e., \(\overline{id}^{\chainname{\hierarchygraphleft}}_{\downarrow\maplevelone} = (id_{\maplevelone(\chaindepthone)}: \graphname{\hierarchygraphleft}{\maplevelone(\chaindepthone)} \to \graphname{\hierarchygraphleft}{\maplevelone(\chaindepthone)} \mid \indexone \in [\chaindepthone] )\).

In the first step (1) we get the pushout of inclusion chains of equal depth \(\chaindepthone\) and obtain a chain \(\chainname[\downarrow\maplevelone]{\hierarchygraphmiddle} = (\overline{\hierarchygraphmiddle}_{\downarrow\maplevelone},\chaindepthone,\typemorph[\hierarchygraphmiddle]{\downarrow\maplevelone}{})\) with
\(\overline{\hierarchygraphmiddle}_{\downarrow\maplevelone} = [\graphname{\hierarchygraphmiddle}{\maplevelone(\chaindepthone)}, \graphname{\hierarchygraphmiddle}{\maplevelone(\chaindepthone-1)},\ldots,\graphname{\hierarchygraphmiddle}{\maplevelone(1)},\graphname{\hierarchygraphmiddle}{\maplevelone(0)}]\) and
\(\typemorph[\hierarchygraphmiddle]{\downarrow\maplevelone}{} = (\typemorph[\hierarchygraphmiddle]{\maplevelone(\indextwo)}{\maplevelone(\indexone)} : \graphname{\hierarchygraphmiddle}{\maplevelone(\indextwo)} \partialmap \graphname{\hierarchygraphmiddle}{\maplevelone(\indexone)} \mid 0 \le \indexone < \indextwo \le \chaindepthone)\).
In pushout step (2) we fill the gaps in \(\chainname[\downarrow\maplevelone]{\hierarchygraphmiddle}\) with the corresponding untouched graphs from the original chain \(\chainname{\hierarchygraphleft}\).

\subsubsection{Pushout for inclusion graph homomorphisms}
\label{subsubsec:pushout-inclusion-graph-homomorphisms}

To have the proper basis for the rest of this section we have to recapitulate the special pushout construction in category \cat{Graph} for a span with one inclusion graph homomorphism and, based on this, we present a necessary result about the structure of special mediating morphisms.

For an inclusion graph homomorphism \(\morphone : \graphname{\examplegraphone}{} \hookrightarrow \graphname{\examplegraphtwo}{}\) and an arbitrary graph homomorphism \(\morphtwo : \graphname{\examplegraphone}{} \to \graphname{\examplegraphthree}{}\) we can construct the pushout as follows, where we assume that \(\graphname{\examplegraphtwo}{}\) and \(\graphname{\examplegraphthree}{}\) are disjoint.
\begin{center}
\begin{tikzpicture}[on grid,node distance=25mm]

\node[el-math] (gi)								{\graphname{\examplegraphone}{}};
\node[el-math] (hi)	[right of=gi]				{\graphname{\examplegraphtwo}{}};
\node[el-math] (ki)	[below of=gi]				{\graphname{\examplegraphthree}{}};
\node[el-math] (pi)	[below of=hi]				{\graphname{\examplegraphfour}{}};
\node[el-math] (po) [below right=18mm of gi]	{\po};

\draw[incmapl] (gi) to	node[la-math]	(ph)	{\morphone}		(hi);
\draw[incmapl] (ki) to	node[la-math]	(phs)	{\morphone^*}	(pi);
\draw[map] (gi) to	node[la-math,left]	(ps)	{\morphtwo}		(ki);
\draw[map] (hi) to	node[la-math]		(pss)	{\morphtwo^*}	(pi);

\end{tikzpicture}
\end{center}

The pushout graph \(\graphname{\examplegraphfour}{}\) is given by \(\graphname[\graphnodes]{\examplegraphfour}{} := \graphname[\graphnodes]{\examplegraphthree}{} \cup \graphname[\graphnodes]{\examplegraphtwo}{} \setminus \graphname[\graphnodes]{\examplegraphone}{}\), \(\graphname[\grapharrows]{\examplegraphfour}{} := \graphname[\grapharrows]{\examplegraphthree}{} \cup \graphname[\grapharrows]{\examplegraphtwo}{} \setminus \graphname[\grapharrows]{\examplegraphone}{}\) and
\begin{equation*}
	\source[\examplegraphfour](e) := \begin{cases}
		\source[\examplegraphthree](e) &\text{if \(e \in \graphname[\grapharrows]{\examplegraphthree}{}\)}\\
		\morphtwo^\grapharrows(\source[\examplegraphtwo](e)) &\text{if \(e \in \graphname[\grapharrows]{\examplegraphtwo}{} \setminus \graphname[\grapharrows]{\examplegraphone}{}\)}
	\end{cases}
\end{equation*}
Moreover, \(\target[\examplegraphfour]\) is defined analogously.
\(\morphone^*: \graphname{\examplegraphthree}{} \hookrightarrow \graphname{\examplegraphfour}{}\) is an inclusion graph homomorphism by construction and \(\morphtwo^*: \graphname{\examplegraphtwo}{} \to \graphname{\examplegraphfour}{}\) is defined by
\begin{equation*}
	\morphtwo^{*\graphnodes}(v) := \begin{cases}
		\morphtwo^\graphnodes(v) &\text{if \(v \in \graphname[\graphnodes]{\examplegraphone}{}\)}\\
		v &\text{if \(v \in \graphname[\graphnodes]{\examplegraphtwo}{} \setminus \graphname[\graphnodes]{\examplegraphone}{}\)}
	\end{cases}
\end{equation*}
and
\begin{equation*}
	\morphtwo^{*\grapharrows}(e) := \begin{cases}
		\morphtwo^\grapharrows(e) &\text{if \(e \in \graphname[\grapharrows]{\examplegraphone}{}\)}\\
		e &\text{if \(e \in \graphname[\grapharrows]{\examplegraphtwo}{} \setminus \graphname[\grapharrows]{\examplegraphone}{}\)}
	\end{cases}
\end{equation*}

Note that we can have \(\source[\examplegraphtwo](e) \in \graphname[\graphnodes]{\examplegraphone}{}\) or \(\target[\examplegraphtwo](e) \in \graphname[\graphnodes]{\examplegraphone}{}\) even if \(e \in \graphname[\grapharrows]{\examplegraphtwo}{} \setminus \graphname[\grapharrows]{\examplegraphone}{}\).
The pair \(\graphname{\examplegraphone}{} \setminus \graphname{\examplegraphtwo}{} := (\graphname[\graphnodes]{\examplegraphtwo}{} \setminus \graphname[\graphnodes]{\examplegraphone}{}, \graphname[\grapharrows]{\examplegraphtwo}{} \setminus \graphname[\grapharrows]{\examplegraphone}{})\) of subsets of nodes and arrows of \(\graphname{\examplegraphtwo}{}\) is, in general, not establishing a subgraph of \(\graphname{\examplegraphtwo}{}\).
We will nevertheless use the notation \(\graphname{\examplegraphfour}{} = \graphname{\examplegraphthree}{} + \graphname{\examplegraphtwo}{} \setminus \graphname{\examplegraphone}{}\) to indicate that \(\graphname{\examplegraphfour}{}\) is constructed as described above.

We consider now two spans and graph homomorphisms such that the left and the back square in the following cube are pullbacks where the bottom and the top squares are pushouts.

\begin{center}
\begin{tikzpicture}[on grid,node distance=30mm]

\node[el-math] (g1)										{\graphname{\examplegraphone}{1}};
\node[el-math] (h1)	[right of=g1]						{\graphname{\examplegraphtwo}{1}};
\node[el-math] (k1)	[below right=15mm and 10mm of g1]	{\graphname{\examplegraphthree}{1}};
\node[el-math] (p1)	[right of=k1]						{\graphname{\examplegraphfour}{1}};
\node[el-math] (g2)	[below=35mm of g1]					{\graphname{\examplegraphone}{2}};
\node[el-math] (h2)	[below=35mm of h1]					{\graphname{\examplegraphtwo}{2}};
\node[el-math] (k2)	[below=35mm of k1]					{\graphname{\examplegraphthree}{2}};
\node[el-math] (p2)	[below=35mm of p1]					{\graphname{\examplegraphfour}{2}};
\node[el-math] (po1) [below right=8mm and 20mm of g1]	{\po};
\node[el-math] (po2) [below=35mm of po1]				{\po};

\draw[incmapl] (g1) to	node[la-math]		(ph1)	{\morphone_1}	(h1);
\draw[incmapl] (k1) to	node[la-math,below]	(ph1s)	{\morphone^*_1}	(p1);
\draw[map] (g1) to	node[la-math]			(ps1)	{\morphtwo_1}	(k1);
\draw[map] (h1) to	node[la-math]			(ps1s)	{\morphtwo^*_1}	(p1);

\draw[map]		(g2) to	node[la-math,left]				(tg)	{\typemorph[\examplegraphone]{}{}}		(g1);
\draw[map]		(h2) to	node[la-math,right,near start]	(th)	{\typemorph[\examplegraphtwo]{}{}}		(h1);
\draw[map]		(k2) to	node[la-math,left,near end]		(tk)	{\typemorph[\examplegraphthree]{}{}}	(k1);
\draw[map,dotted] (p2) to node[la-math,right]			(tp)	{\typemorph[\examplegraphfour]{}{}}		(p1);

\draw[incmapl] (g2) to	node[la-math]		(ph2)	{\morphone_2}	(h2);
\draw[incmapl] (k2) to	node[la-math,below]	(ph2s)	{\morphone^*_2}	(p2);
\draw[map] (g2) to	node[la-math,left]		(ps2)	{\morphtwo_2}	(k2);
\draw[map] (h2) to	node[la-math,left]		(ps2s)	{\morphtwo^*_2}	(p2);

\end{tikzpicture}
\end{center}

Since the bottom square is a pushout, we obtain a unique graph homomorphism \(\typemorph[\examplegraphfour]{}{} : \graphname{\examplegraphfour}{2} \to \graphname{\examplegraphfour}{1}\) making the square commutative (since it is the mediating morphism of he pushout).
And, since the top square has the van Kampen property~\cite{ehrig2006fundamentals,wolter2015topoi} the front and right squares become pullbacks as well.

If the following condition is satisfied
\begin{equation}\label{eq:type-h-minus-g-condition}
	\typemorph[\examplegraphtwo]{}{}(\graphname{\examplegraphtwo}{2} \setminus \graphname{\examplegraphone}{2}) \sqsubseteq \graphname{\examplegraphtwo}{1} \setminus \graphname{\examplegraphone}{1}
\end{equation}
we denote the restriction of \(\typemorph[\examplegraphtwo]{}{}\) to \(\graphname{\examplegraphtwo}{2} \setminus \graphname{\examplegraphone}{2}\) by \(\typemorph[\examplegraphtwo]{}{} \setminus \typemorph[\examplegraphone]{}{} : \graphname{\examplegraphtwo}{2} \setminus \graphname{\examplegraphone}{2} \to \graphname{\examplegraphtwo}{1} \setminus \graphname{\examplegraphone}{1}\) thus we can describe, in this case, \(\typemorph[\examplegraphtwo]{}{}\) as well as \(\typemorph[\examplegraphfour]{}{}\) by the sum of two pairs of mappings

\begin{center}
\begin{tikzpicture}[on grid,node distance=25mm]

\node[el-math] (l)										{\graphname{\examplegraphthree}{1}};
\node[el-math] (sf)	[right=32mm of l]					{\graphname{\examplegraphfour}{1} = \graphname{\examplegraphthree}{1} + \graphname{\examplegraphtwo}{1} \setminus \graphname{\examplegraphone}{1}};
\node[el-math] (s)	[right=80mm of sf]					{\graphname{\examplegraphtwo}{1} = \graphname{\examplegraphone}{1}+\graphname{\examplegraphtwo}{1} \setminus \graphname{\examplegraphone}{1}};
\node[el-math] (i)	[below of=l]						{\graphname{\examplegraphthree}{2}};
\node[el-math] (ug)	[below of=sf]						{\graphname{\examplegraphfour}{2} = \graphname{\examplegraphthree}{2} + \graphname{\examplegraphtwo}{2} \setminus \graphname{\examplegraphone}{2}};
\node[el-math] (si)	[below of=s]						{\graphname{\examplegraphtwo}{2} = \graphname{\examplegraphone}{2}+\graphname{\examplegraphtwo}{2} \setminus \graphname{\examplegraphone}{2}};
\node[el-math] (eq)	[below right=12mm and 13mm of l]	{\pb};
\node[el-math] (eq)	[below right=12mm and 40mm of sf]	{\pb};

\draw[map]		(l)		to node[la-math]		(mu1)	{\morphone^*_1}																																			(sf);
\draw[incmapr]	(s)		to node[la-math,above]	(id1)	{\morphtwo_1^* = \morphtwo_1+id_{\graphname{\examplegraphtwo}{1} \setminus \graphname{\examplegraphone}{1}}}											(sf);
\draw[map]		(i) 	to node[la-math,left]	(in1)	{\typemorph[\examplegraphthree]{}{}}																													(l);
\draw[map]		(ug)	to node[la-math,right]	(in2)	{\typemorph[\examplegraphfour]{}{} = \typemorph[\examplegraphthree]{}{} + \typemorph[\examplegraphtwo]{}{}\setminus \typemorph[\examplegraphone]{}{}}	(sf);
\draw[map]		(si)	to node[la-math]		(in3)	{\typemorph[\examplegraphtwo]{}{} = \typemorph[\examplegraphone]{}{} + \typemorph[\examplegraphtwo]{}{}\setminus \typemorph[\examplegraphone]{}{}}		(s);
\draw[map]		(i)		to node[la-math,below]	(mu2)	{\morphone_2^*}																																			(ug);
\draw[incmapr]	(si)	to node[la-math,below]	(id2)	{\morphtwo_2^* = \morphtwo_2 + id_{\graphname{\examplegraphtwo}{2} \setminus \graphname{\examplegraphone}{2}}}											(ug);
	
\end{tikzpicture}
\end{center}

\subsubsection{Pushouts for chains with equal depth}
\label{subsubsec:pushout-chain-equal-depth}

For each level \(0 \leq \indexone < \chaindepthone\) we construct the pushout and get
\begin{center}
\begin{tikzpicture}[on grid,node distance=35mm]

\node[el-math] (g0)											{\graphname{\examplegraphone}{0}};
\node[el-math] (h0)		[right of=g0]						{\graphname{\examplegraphtwo}{0}};
\node[el-math] (gi)		[below of=g0]						{\graphname{\examplegraphone}{\indexone}};
\node[el-math] (hi)		[right of=gi]						{\graphname{\examplegraphtwo}{\indexone}};
\node[el-math] (k0)		[below right=15mm and 15mm of g0]	{\graphname{\examplegraphthree}{0}};
\node[el-math] (p0)		[right of=k0]						{\graphname{\examplegraphfour}{0}};
\node[el-math] (kfi)	[below of=k0]						{\graphname{\examplegraphthree}{\maplevelone(\indexone)}};
\node[el-math] (pfi)	[right of=kfi]						{\graphname{\examplegraphfour}{\maplevelone(\indexone)}};
\node[el-math] (po1)	[below right=8mm and 25mm of g0]	{\po};
\node[el-math] (po2)	[below of=po1]						{\po};

\draw[incmapl]	(g0) to node[la-math]				(mg0h0)	{\morphone_0}									(h0);
\draw[incmapl]	(gi) to node[la-math,pos=.7]		(mgihi)	{\morphone_\indexone}							(hi);
\draw[incmapr]	(gi) to node[la-math]				(mgig0)	{\typemorph[\examplegraphone]{\indexone}{0}}	(g0);
\draw[incmapr]	(hi) to node[la-math,pos=.3,right]	(mhih0)	{\typemorph[\examplegraphtwo]{\indexone}{0}}	(h0);

\draw[incmapl]	(k0) to node[la-math,pos=.3,below]	(mk0p0)	{\morphone^*_{\maplevelone(0)}}												(p0);
\draw[incmapl]	(kfi) to node[la-math,below]		(mkfpf)	{\morphone^*_{\maplevelone(\indexone)}}										(pfi);
\draw[incmapr]	(kfi) to node[la-math,pos=.7]		(mkfk0)	{\typemorph[\examplegraphthree]{\maplevelone(\indexone)}{\maplevelone(0)}}	(k0);
\draw[mapdots]	(pfi) to node[la-math]				(mpfp0)	{}																			(p0);

\draw[map]	(g0) to node[la-math]	(mg0k0)	{\morphtwo_0}			(k0);
\draw[map]	(h0) to node[la-math]	(mh0p0)	{\morphtwo^*_0}			(p0);
\draw[map]	(gi) to node[la-math]	(mgikf)	{\morphtwo_\indexone}	(kfi);
\draw[map]	(hi) to node[la-math]	(mhipf)	{\morphtwo^*_\indexone}	(pfi);

\end{tikzpicture}
\end{center}

Since the bottom square is pushout we get \(\typemorph[\examplegraphfour]{\maplevelone(\indexone)}{\maplevelone(0)} : \graphname{\examplegraphfour}{\maplevelone(\indexone)} \to \graphname{\examplegraphfour}{0}\) that makes the square commute.
Since the top square is van Kampen the front and right square become pullbacks as well.
Pullbacks preserve monomorphisms thus \(\typemorph[\examplegraphfour]{\maplevelone(\indexone)}{\maplevelone(0)}\) becomes a monomorphism.
The bottom square is a pushout thus \(\morphone^*_{\maplevelone(\indexone)}\) and \(\morphtwo^*_\indexone\) are jointly epi.
So, what we get is a jointly-epi-mono factorisation of the pair \((\typemorph[\examplegraphthree]{\maplevelone(\indexone)}{\maplevelone(0)} ; \morphone^*_{\maplevelone(0)} , \typemorph[\examplegraphtwo]{\indexone}{0} ; \morphtwo^*_0)\) of graph homomorphisms
\begin{center}
\begin{tikzpicture}[on grid,node distance=25mm]

\node[el-math] (k0)						{\graphname{\examplegraphthree}{0}};
\node[el-math] (kfi)	[below of=k0]	{\graphname{\examplegraphthree}{\maplevelone(\indexone)}};
\node[el-math] (p0)		[right of=k0]	{\graphname{\examplegraphfour}{0}};
\node[el-math] (pfi)	[below of=p0]	{\graphname{\examplegraphfour}{\maplevelone(\indexone)}};
\node[el-math] (h0)		[right of=p0]	{\graphname{\examplegraphtwo}{0}};
\node[el-math] (hi)		[below of=h0]	{\graphname{\examplegraphtwo}{\indexone}};

\draw[incmapl]	(hi) to node[la-math]	(miii0)	{\typemorph[\examplegraphtwo]{\indexone}{0}}								(h0);
\draw[map]		(pfi) to node[la-math]	(mdfdo)	{\typemorph[\examplegraphfour]{\maplevelone(\indexone)}{\maplevelone(0)}}	(p0);
\draw[incmapl]	(kfi) to node[la-math]	(msfs0)	{\typemorph[\examplegraphthree]{\maplevelone(\indexone)}{\maplevelone(0)}}	(k0);

\draw[incmapl]	(k0) to node[la-math]		(mk0p0)	{\morphone^*_{\maplevelone(0)}}			(p0);
\draw[incmapl]	(kfi) to node[la-math]		(mkfpf)	{\morphone^*_{\maplevelone(\indexone)}}	(pfi);
\draw[incmapr]	(h0) to node[la-math,above]	(mh0p0)	{\morphtwo^*_0}							(p0);
\draw[incmapr]	(hi) to node[la-math,above]	(mhipf)	{\morphtwo^*_\indexone}					(pfi);

\end{tikzpicture}
\end{center}

Jointly-epi-mono factorisations are unique up to isomorphism thus we can choose the union of images \(\graphname{\examplegraphfour}{\maplevelone(\indexone)} = \morphone^*_{\maplevelone(0)}(\typemorph[\examplegraphthree]{\maplevelone(\indexone)}{\maplevelone(0)}(\graphname{\examplegraphthree}{\maplevelone(\indexone)})) \cup \morphtwo^*_0(\typemorph[\examplegraphtwo]{\indexone}{0}(\graphname{\examplegraphtwo}{\indexone}))\) without loss of generality.
In such a way, \(\typemorph[\examplegraphfour]{\maplevelone(\indexone)}{\maplevelone(0)}\) becomes an inclusion homomorphism.

The rest of this section depends on the special pushout construction in category \cat{Graph} for a span with one inclusion graph homomorphism and a necessary result about the structure of special mediating morphisms.

As described in Section~\ref{subsubsec:pushout-inclusion-graph-homomorphisms}, we construct the corresponding pushout of graph homomorphisms (in category \cat{Graph}) for each level \(\indexone \in [\chaindepthone]\) as follows.

\begin{center}
\begin{tikzpicture}[on grid,node distance=25mm]

\node[el-math] (gi)										{\graphname{\examplegraphone}{\indexone}};
\node[el-math] (hi)	[right=40mm of gi]					{\graphname{\examplegraphtwo}{\indexone}};
\node[el-math] (ki)	[below of=gi]						{\graphname{\examplegraphthree}{\maplevelone(\indexone)}};
\node[el-math] (pi)	[below of=hi]						{\graphname{\examplegraphfour}{\maplevelone(\indexone)} := \graphname{\examplegraphtwo}{\maplevelone(\indexone)} + \graphname{\examplegraphthree}{\indexone} \setminus \graphname{\examplegraphone}{\indexone}};
\node[el-math] (po) [below right=13mm and 20mm of gi]	{\po};

\draw[incmapl] (gi) to	node[la-math]	(ph)	{\morphone_{\indexone}}					(hi);
\draw[incmapl] (ki) to	node[la-math]	(phs)	{\morphone^*_{\maplevelone(\indexone)}}	(pi);
\draw[map] (gi) to	node[la-math,left]	(ps)	{\morphtwo_{\indexone}}					(ki);
\draw[map] (hi) to	node[la-math]		(pss)	{\morphtwo^*_{\indexone} := \morphtwo_{\indexone} + id_{\graphname{\examplegraphtwo}{\indexone} \setminus \graphname{\examplegraphone}{\indexone}}}	(pi);

\end{tikzpicture}
\end{center}

We consider any level \(1 \leq \indexone \leq \chaindepthone\) together with the base level \(0\). Since \(\graphname{\examplegraphtwo}{\indexone} \sqsubseteq \graphname{\examplegraphtwo}{0}\), condition~(\ref{eq:li-l0-equivalence}) implies \(\graphname{\examplegraphtwo}{\indexone} \setminus \graphname{\examplegraphone}{\indexone} \sqsubseteq \graphname{\examplegraphtwo}{0} \setminus \graphname{\examplegraphone}{0}\) thus condition~(\ref{eq:type-h-minus-g-condition}) in Section~\ref{subsubsec:pushout-inclusion-graph-homomorphisms} is satisfied and we get a cospan of pullbacks.

\begin{center}
\begin{tikzpicture}[on grid,node distance=25mm]

\node[el-math] (k0)						{\graphname{\examplegraphthree}{} = \graphname{\examplegraphthree}{0}};
\node[el-math] (p0)	[right=40mm of k0]	{\graphname{\examplegraphfour}{} := \graphname{\examplegraphfour}{0}};
\node[el-math] (h0)	[right=40mm of p0]	{\graphname{\examplegraphtwo}{} = \graphname{\examplegraphtwo}{0}};
\node[el-math] (kf)	[below of=k0]		{\graphname{\examplegraphthree}{\maplevelone(\indexone)}};
\node[el-math] (pf)	[below of=p0]		{\graphname{\examplegraphfour}{\maplevelone(\indexone)}};
\node[el-math] (hi)	[below of=h0]		{\graphname{\examplegraphtwo}{\indexone}};

\draw[incmapl]	(k0) to node[la-math]				(mk0p0)	{\morphone^* := \morphone^*_{0}}				(p0);
\draw[map]		(h0) to node[la-math,above]			(mh0p0)	{\morphtwo^* := \morphtwo^*_{0}}				(p0);
\draw[incmapl]	(kf) to node[la-math,left]			(mkfk0)	{\typemorph[\examplegraphthree]{\indexone}{0}}	(k0);
\draw[map]		(pf) to node[la-math]				(mpfp0)	{\typemorph[\examplegraphfour]{\indexone}{0} := \typemorph[\examplegraphthree]{\indexone}{0} + \typemorph[\examplegraphtwo]{\indexone}{0} \setminus \typemorph[\examplegraphone]{\indexone}{0}}	(p0);
\draw[incmapl]	(hi) to node[la-math,midway,right]	(mhih0)	{\typemorph[\examplegraphtwo]{\indexone}{0} := \typemorph[\examplegraphone]{\indexone}{0} + \typemorph[\examplegraphtwo]{\indexone}{0} \setminus \typemorph[\examplegraphone]{\indexone}{0}}	(h0);
\draw[incmapl]	(kf) to node[la-math,below]			(mu2)	{\morphone^*_{\maplevelone(\indexone)}}			(pf);
\draw[map]		(hi) to node[la-math,below]			(id2)	{\morphtwo^*_{\indexone}}						(pf);

\end{tikzpicture}
\end{center}

The sequence \([\graphname{\examplegraphfour}{\maplevelone(\chaindepthone)}, \graphname{\examplegraphfour}{\maplevelone(\chaindepthone-1)}, \dots, \graphname{\examplegraphfour}{\maplevelone(1)}, \graphname{\examplegraphfour}{0}]\) defines, according to Corollary~\ref{lem:inclusion-chain}, an inclusion chain that we denote by \(\chainname[\downarrow\maplevelone]{\examplegraphfour}\).
To show that the family of inclusion graph homomorphisms \((\morphone^*_{\maplevelone(\indexone)}: \graphname{\examplegraphthree}{\maplevelone(\indexone)} \hookrightarrow \graphname{\examplegraphfour}{\maplevelone(\indexone)}  \mid \indexone \in [\chaindepthone])\) defines a graph chain morphism \((\morphone^*_{\downarrow\maplevelone}, id_{[\chaindepthone]}) : \chainname[\downarrow\maplevelone]{\examplegraphthree} \hookrightarrow \chainname[\downarrow\maplevelone]{\examplegraphfour}\) we have to show, according to Definition~\ref{def:graph-chain-morphism}, that we have for any \(0 \leq \indexone < \indextwo \leq \chaindepthone\) a commutative diagram with a pullback as follows.

\begin{center}
\begin{tikzpicture}[on grid,node distance=40mm]

\def\vd{15mm}

\node[el-math]	(sfi)						{\graphname{\examplegraphthree}{\maplevelone(\indexone)}};
\node[el-math]	(dfi)	[right of=sfi]		{\graphname{\examplegraphfour}{\maplevelone(\indexone)}};
\node[el-math]	(sint)	[below=\vd of sfi]	{\graphname{\examplegraphthree}{\maplevelone(\indextwo)} \cap \graphname{\examplegraphthree}{\maplevelone(\indexone)}};
\node[el-math]	(dint)	[right of=sint]		{\graphname{\examplegraphfour}{\maplevelone(\indextwo)} \cap {\graphname{\examplegraphfour}{\maplevelone(\indexone)}}};
\node[el-math]	(sfj)	[below=\vd of sint]	{\graphname{\examplegraphthree}{\maplevelone(\indextwo)}};
\node[el-math]	(dfj)	[right of=sfj]		{\graphname{\examplegraphfour}{\maplevelone(\indextwo)}};

\draw[map]		(sfi)	to	node [la-math]		(sigfi)	{\morphone^*_{\maplevelone(\indexone)}}	(dfi);
\draw[incmapl]	(sint)	to	node [la-math]		(inint)	{\sqsubseteq}							(dint);
\draw[map]		(sfj)	to	node [la-math]		(sigjs)	{\morphone^*_{\maplevelone(\indextwo)}}	(dfj);
\draw[incmapl]	(sint)	to	node [la-math]		(insfi)	{\sqsubseteq}							(sfi);
\draw[incmapr]	(sint)	to	node [la-math,left]	(insfj)	{\sqsubseteq}							(sfj);
\draw[incmapl]	(dint)	to	node [la-math]		(indfi)	{\sqsubseteq}							(dfi);
\draw[incmapr]	(dint)	to	node [la-math,left]	(indfj)	{\sqsubseteq}							(dfj);

\node[el-math]	(pb1)	[below right=6mm and 20mm of sfi]	{\pb};
\node[el-math]	(pb2)	[below right=6mm and 20mm of sint]	{\pb};

\end{tikzpicture}
\end{center}

The intersection of two subgraphs of a given graph, however, is a pullback.
This fact together with our level-wise pushout construction provides for any \(0 \leq \indexone < \indextwo \leq \chaindepthone\) a commutative cube with pullbacks in the bottom, in the front and in the two side faces.

\begin{center}
\begin{tikzpicture}[on grid,node distance=35mm]

\node[el-math] (s0)											{\graphname{\examplegraphthree}{0} =\graphname{\examplegraphthree}{}};
\node[el-math] (d0)		[right of=s0]						{\graphname{\examplegraphfour}{0} = \graphname{\examplegraphfour}{}};
\node[el-math] (sfi)	[below of=s0]						{\graphname{\examplegraphthree}{\maplevelone(\indexone)}};
\node[el-math] (dfi)	[right of=sfi]						{\graphname{\examplegraphfour}{\maplevelone(\indexone)}};
\node[el-math] (sfj)	[below right=15mm and 15mm of s0]	{\graphname{\examplegraphthree}{\maplevelone(\indextwo)}};
\node[el-math] (dfj)	[right of=sfj]						{\graphname{\examplegraphfour}{\maplevelone(\indextwo)}};
\node[el-math] (sfji)	[below of=sfj]						{\graphname{\examplegraphthree}{\maplevelone(\indextwo)} \cap \graphname{\examplegraphthree}{\maplevelone(\indexone)}};
\node[el-math] (dfji)	[right of=sfji]						{\graphname{\examplegraphfour}{\maplevelone(\indextwo)} \cap \graphname{\examplegraphfour}{\maplevelone(\indexone)}};

\draw[incmapl] (s0) to	node[la-math]			(msd0)	{\morphone^*_0 = \morphone^*}												(d0);
\draw[incmapl] (sfi) to	node[la-math,pos=.7]	(msdi)	{\morphone^*_{\maplevelone(\indexone)}}										(dfi);
\draw[incmapr] (sfi) to	node[la-math]			(msi0)	{\typemorph[\examplegraphthree]{\maplevelone(\indexone)}{\maplevelone(0)}}	(s0);
\draw[incmapr] (dfi) to	node[la-math,pos=.3]	(mdi0)	{\typemorph[\examplegraphfour]{\maplevelone(\indexone)}{\maplevelone(0)}}	(d0);

\draw[incmapl] (sfj) to		node[la-math,pos=.3]	(msdj)		{\morphone^*_{\maplevelone(\indextwo)}}	(dfj);
\draw[mapdots] (sfji) to	node[la-math]			(msdij)		{}										(dfji);
\draw[incmapr] (sfji) to	node[la-math]			(msfjij)	{}										(sfj);
\draw[incmapr] (dfji) to	node[la-math]			(mdfjij)	{}										(dfj);

\draw[incmapr] (sfj) to		node[la-math,pos=.3,above right]	(msj0)		{\typemorph[\examplegraphthree]{\maplevelone(\indextwo)}{\maplevelone(0)}}	(s0);
\draw[incmapr] (dfj) to		node[la-math,pos=.3,above right]	(mdj0)		{\typemorph[\examplegraphfour]{\maplevelone(\indextwo)}{\maplevelone(0)}}	(d0);
\draw[incmapr] (sfji) to	node[la-math]						(msfjii)	{}																			(sfi);
\draw[incmapr] (dfji) to	node[la-math]						(mdfjii)	{}																			(dfi);

\end{tikzpicture}
\end{center}

The unique homomorphism from \(\graphname{\examplegraphthree}{\maplevelone(\indextwo)} \cap \graphname{\examplegraphthree}{\maplevelone(\indexone)}\) to \(\graphname{\examplegraphfour}{\maplevelone(\indextwo)} \cap \graphname{\examplegraphfour}{\maplevelone(\indexone)}\) that completes the square is obtained by the universal property of the right face.
The resulting commutative squares in the top and in the back are also pullbacks due to composition and decomposition of pullbacks.
This shows that we have indeed a chain morphism \((\morphone^*_{\downarrow \maplevelone}, id_{[\chaindepthone]}) : \chainname[\downarrow\maplevelone]{\examplegraphthree} \hookrightarrow \chainname[\downarrow\maplevelone]{\examplegraphfour}\).

Analogously, it can be shown that the family \((\morphtwo^*_{\indexone}: \graphname{\examplegraphtwo}{\indexone} \to \graphname{\examplegraphfour}{\maplevelone(\indexone)}  \mid \indexone \in [\chaindepthone])\) establishes a chain morphism \((\morphtwo^*, id_{[\chaindepthone]}) : \chainname{\examplegraphtwo} \to \chainname[\downarrow\maplevelone]{\examplegraphfour}\).
Since the resulting commutative square (1) of chain morphisms in Figure~\ref{fig:two-po-construct-d} is obtained by level-wise pushout constructions, it is straight forward to show that (1) becomes indeed a pushout in \cat{Chain}.

\subsubsection{Pushout by extension}
\label{subsubsec:pushout-by-extension}

To obtain an inclusion chain \(\chainname{\examplegraphfour}\) of length \(\chaindepthtwo+1\) we fill the gaps in the sequence \([\graphname{\examplegraphfour}{\maplevelone(\chaindepthone)}, \graphname{\examplegraphfour}{\maplevelone(\chaindepthone-1)}, \dots, \graphname{\examplegraphfour}{\maplevelone(1)}, \graphname{\examplegraphfour}{0}]\) of subgraphs of \(\graphname{\examplegraphfour}{} = \graphname{\examplegraphfour}{0}\), built in Section~\ref{subsubsec:pushout-chain-equal-depth}, by corresponding subgraphs of \(\examplegraphthree\) from the sequence \([\graphname{\examplegraphthree}{\chaindepthtwo}, \graphname{\examplegraphthree}{\chaindepthtwo-1}, \dots, \graphname{\examplegraphthree}{1}, \graphname{\examplegraphthree}{0}]\).
For any \(\exampleindexone \in \maplevelone([\chaindepthone])\) we denote by \(\maplevelone^{-}(\exampleindexone)\) the unique index in \([\chaindepthone]\) such that \(\maplevelone(\maplevelone^-(\exampleindexone)) = \exampleindexone\).
We define the sequence \(\tc{\examplegraphfour}\) of subgraphs of \(\graphname{\examplegraphfour}{} = \graphname{\examplegraphfour}{0}\) as follows.
\begin{equation*}
	\graphname{\examplegraphfour}{\exampleindexone} := \begin{cases}
		\graphname{\examplegraphfour}{\exampleindexone} &\text{if \(\exampleindexone \in \maplevelone([\chaindepthone])\)}\\
		\graphname{\examplegraphthree}{\exampleindexone} &\text{if \(\exampleindexone \in [\chaindepthtwo] \setminus \maplevelone([\chaindepthone])\)}
	\end{cases}
\end{equation*}
and denote by \(\chainname{\examplegraphfour} = \chain{\examplegraphfour}{\chaindepthtwo}\) the corresponding inclusion chain according to Corollary~\ref{lem:inclusion-chain}.
The family \(\overline{id} = (id_{\graphname{\examplegraphfour}{\maplevelone(\indexone)}}: \graphname{\examplegraphfour}{\maplevelone(\indexone)} \to \graphname{\examplegraphfour}{\maplevelone(\indexone)} \mid \indexone \in [\chaindepthone])\) of graph homomorphisms defines trivially a chain morphism \((\overline{id},\maplevelone) : \chainname[\downarrow\maplevelone]{\examplegraphfour} \to \chainname{\examplegraphfour}\).

It remains to show that the family \(\morphone = (\morphone_{\exampleindexone}: \graphname{\examplegraphthree}{\exampleindexone} \to \graphname{\examplegraphfour}{\exampleindexone} \mid a \in [\chaindepthtwo])\) of graph homomorphisms defined by
\begin{equation*}
	\morphone_{\exampleindexone} := \begin{cases}
		\morphone_{\exampleindexone} : \graphname{\examplegraphthree}{\exampleindexone} \hookrightarrow \graphname{\examplegraphfour}{\exampleindexone} &\mbox{if } a \in \maplevelone([\chaindepthone])\\
		id_{\graphname{\examplegraphthree}{\exampleindexone}} : \graphname{\examplegraphthree}{\exampleindexone} \to \graphname{\examplegraphfour}{\exampleindexone} = \graphname{\examplegraphthree}{\exampleindexone} &\mbox{if } a \in [\chaindepthtwo] \setminus \maplevelone([\chaindepthone])
	\end{cases}
\end{equation*}
establishes a chain morphism \((\morphone, id_{[\chaindepthtwo]}) : \chainname{\examplegraphthree} \to \chainname{\examplegraphfour}\).

That is, due to Definition~\ref{def:graph-chain-morphism}, we have to show that for arbitrary \({0 \leq \exampleindexone < \exampleindextwo \leq \chaindepthtwo}\) we do have a double pullback square
\begin{center}
\begin{tikzpicture}[on grid,node distance=30mm]

\node[el-math] (sa)									{\graphname{\examplegraphthree}{\exampleindexone}};
\node[el-math] (da)		[right of=sa]				{\graphname{\examplegraphfour}{\exampleindexone}};
\node[el-math] (sab)	[below of=sa]				{\graphname{\examplegraphthree}{\exampleindexone} \cap \graphname{\examplegraphthree}{\exampleindextwo}};
\node[el-math] (dab)	[right of=sab]				{\graphname{\examplegraphfour}{\exampleindexone} \cap \graphname{\examplegraphfour}{\exampleindextwo}};
\node[el-math] (sb)		[below of=sab]				{\graphname{\examplegraphthree}{\exampleindextwo}};
\node[el-math] (db)		[right of=sb]				{\graphname{\examplegraphfour}{\exampleindextwo}};
\node[el-math] (po1)	[below right=21mm of sa]	{\pb};
\node[el-math] (po2)	[below of=po1]				{\pb};

\draw[incmapl] (sa) to	node[la-math]	(msada)	{\morphone_\exampleindexone}						(da);
\draw[incmapl] (sab) to	node[la-math]	(msdab)	{\morphone_{\exampleindexone\mid\exampleindextwo}}	(dab);
\draw[incmapl] (sb) to	node[la-math]	(msbdb)	{\morphone_\exampleindextwo}						(db);

\draw[incmapl] (sab) to	node[la-math,left]	(msaba)	{}	(sa);
\draw[incmapr] (dab) to	node[la-math]		(mdaba)	{}	(da);
\draw[incmapr] (sab) to	node[la-math,left]	(msabb)	{}	(sb);
\draw[incmapl] (dab) to	node[la-math]		(mdabb)	{}	(db);

\end{tikzpicture}
\end{center}

In other words, we have to show that for all \(0 \leq \exampleindexone < \exampleindextwo \leq \chaindepthtwo\)
\begin{equation}\label{eq:po-by-extension-goal}
\graphname{\examplegraphthree}{\exampleindexone} \cap \graphname{\examplegraphfour}{\exampleindexone} \cap \graphname{\examplegraphfour}{\exampleindextwo} = \graphname{\examplegraphthree}{\exampleindexone} \cap \graphname{\examplegraphthree}{\exampleindextwo}
\mbox{ and }
\graphname{\examplegraphthree}{\exampleindextwo} \cap \graphname{\examplegraphfour}{\exampleindexone} \cap \graphname{\examplegraphfour}{\exampleindextwo} = \graphname{\examplegraphthree}{\exampleindexone} \cap \graphname{\examplegraphthree}{\exampleindextwo}
\end{equation}

\begin{description}
	\item[case 1:] \(\exampleindexone, \exampleindextwo \in [\chaindepthtwo] \setminus \maplevelone([\chaindepthone])\).
	In this case we have \(\graphname{\examplegraphfour}{\exampleindexone} = \graphname{\examplegraphthree}{\exampleindexone}\) and \(\graphname{\examplegraphfour}{\exampleindextwo} = \graphname{\examplegraphthree}{\exampleindextwo}\), thus condition~(\ref{eq:po-by-extension-goal}) is trivially satisfied.
	\item[case 2:] \(\exampleindexone, \exampleindextwo \in \maplevelone([\chaindepthone])\).
	This case has been shown in Section~\ref{subsubsec:pushout-chain-equal-depth} where \(\indextwo = \maplevelone^-(\exampleindextwo)\) and \(\indexone = \maplevelone^-(\exampleindexone)\).
	\item[case 3:] \(\exampleindexone \in [\chaindepthtwo] \setminus \maplevelone([\chaindepthone]), \exampleindextwo \in \maplevelone([\chaindepthone])\).
	In this case we have \(\graphname{\examplegraphfour}{\exampleindexone} = \graphname{\examplegraphthree}{\exampleindexone}\) and \(\graphname{\examplegraphfour}{\exampleindextwo} = \graphname{\examplegraphthree}{\exampleindextwo} + \graphname{\rulegraphmiddle}{\maplevelone^-(\exampleindextwo)} \setminus \graphname{\rulegraphleft}{\maplevelone^-(\exampleindextwo)}\) where \(\graphname{\rulegraphmiddle}{\maplevelone^-(\exampleindextwo)} \setminus \graphname{\rulegraphleft}{\maplevelone^-(\exampleindextwo)} = \graphname{\rulegraphmiddle}{\maplevelone^-(\exampleindextwo)} \setminus \graphname{\rulegraphleft}{0}\) due to condition~\ref{eq:li-l0-equivalence}.
	We have \(\graphname{\examplegraphfour}{\exampleindexone} \cap \graphname{\examplegraphfour}{\exampleindextwo} = \graphname{\examplegraphthree}{\exampleindexone} \cap \graphname{\examplegraphthree}{\exampleindextwo}\) since \(\graphname{\examplegraphthree}{\exampleindexone} \sqsubseteq \graphname{\examplegraphthree}{0}\) and \(\graphname{\rulegraphmiddle}{\maplevelone^-(\exampleindextwo)} \setminus \graphname{\rulegraphleft}{\maplevelone^-(\exampleindextwo)} \sqsubseteq \graphname{\rulegraphmiddle}{0} \setminus \graphname{\rulegraphleft}{0}\) are disjoint.
	Thus condition~(\ref{eq:po-by-extension-goal}) is satisfied.
	\item[case 4:] \(\exampleindexone \in \maplevelone([\chaindepthone]), \exampleindextwo \in [\chaindepthtwo] \setminus \maplevelone([\chaindepthone])\).
	In this case we have \(\graphname{\examplegraphfour}{\exampleindexone} = \graphname{\examplegraphthree}{\exampleindexone} +  \graphname{\rulegraphmiddle}{\maplevelone^-(\exampleindexone)} \setminus \graphname{\rulegraphleft}{\maplevelone^-(\exampleindexone)}\) and \(\graphname{\examplegraphfour}{\exampleindextwo} = \graphname{\examplegraphthree}{\exampleindextwo}\) such that \(\graphname{\examplegraphfour}{\exampleindexone} \cap \graphname{\examplegraphfour}{\exampleindextwo} = \graphname{\examplegraphthree}{\exampleindexone} \cap \graphname{\examplegraphthree}{\exampleindextwo}\) since \(\graphname{\examplegraphthree}{\exampleindextwo}\) and \(\graphname{\rulegraphmiddle}{\maplevelone^-(\exampleindexone)} \setminus \graphname{\rulegraphleft}{\maplevelone^-(\exampleindexone)}\) are disjoint, thus condition~(\ref{eq:po-by-extension-goal}) is satisfied.
\end{description}

\subsubsection{Proofs of pushout property for MCMTs}
\label{subsubsec:proofs-pushout-mcmts}

The pushout that is applied in the matching step of a rule, according to Figure~\ref{fig:multilevel-coupled-rule-formal}, is decomposed in two sequential pushouts, as depicted in Figure~\ref{fig:two-po-construct-d}.
The proofs that both (1) and (2) are pushouts in \cat{Chain} are as follows.

\begin{proof}[Pushout property of square (1) in Figure~\ref{fig:two-po-construct-d}]
\label{proof:pushout-property-square-1}

Take an arbitrary graph chain \(\chainname{\variablegraphone} = \chain{\variablegraphone}{\variabledepthone}\) with graph chain morphisms \(\morphone = (\morphone_\indexone : \graphname{\hierarchygraphleft}{\maplevelone(\indexone)} \to \graphname{\variablegraphone}{\mapleveltwo(\indexone)} \mid \indexone \in [\chaindepthone])\) and \(\morphtwo = (\morphtwo_\indexone : \graphname{\rulegraphmiddle}{\indexone} \to \graphname{\variablegraphone}{\maplevelthree(\indexone)} \mid \indexone \in [\chaindepthone])\) such that

\begin{equation}\label{eq:po-1-commutativity}
(\mapruletohierarchyleft , id_{[\chaindepthone]}) ; (\morphone , \mapleveltwo) = (\maprulelefttomiddle , id_{[\chaindepthone]}) ; (\morphtwo,\maplevelthree)
\end{equation}

\begin{center}
\begin{tikzpicture}[on grid,node distance=35mm]

\node[el-math] (l)										{\chainname{\rulegraphleft}};
\node[el-math] (sf)	[right of=l]						{\chainname[\downarrow\maplevelone]{\hierarchygraphleft}};
\node[el-math] (i)	[below of=l]						{\chainname{\rulegraphmiddle}};
\node[el-math] (df)	[below of=sf]						{\chainname[\downarrow\maplevelone]{\hierarchygraphmiddle}};
\node[el-math] (x)	[below right=15mm and 15mm of df]	{\chainname{\variablegraphone}};

\draw[map] (l) to	node[la-math]	(mlsf)					{(\mapruletohierarchyleft,id_{[\chaindepthone]})}								(sf);
\draw[map] (i) to	node[la-math]	(midf)					{(\mapruletohierarchymiddle,id_{[\chaindepthone]})}								(df);
\draw[map] (l) to	node[la-math,left,pos=.7]	(mli)		{(\maprulelefttomiddle,id_{[\chaindepthone]})}									(i);
\draw[map] (sf) to	node[la-math,left,pos=.7]	(msfdf)		{(\maphierarchylefttomiddle_{\downarrow\maplevelone},id_{[\chaindepthone]})}	(df);

\draw[mapdots] (df) to	node[la-math,below left]	(mdfx)	{(\morphthree,\mapleveltwo)}	(x);

\draw[map,bend left] (sf) to	node[la-math]				(msfx)	{(\morphone,\mapleveltwo)}		(x);
\draw[map,bend right] (i) to	node[la-math,below left]	(mix)	{(\morphtwo,\maplevelthree)}	(x);

\node[el-math] (1) [below right=25mm of l]	{(1)};

\end{tikzpicture}
\end{center}

We have to define a chain morphism from \(\chainname[\downarrow\maplevelone]{\hierarchygraphmiddle}\) to \(\chainname{\variablegraphone}\) that makes the two triangles commute.
Due to Definition~\ref{def:composition-graph-chain-morphisms}, condition~\ref{eq:po-1-commutativity} entails that \(\mapleveltwo = \maplevelthree\).
So, we have to define a family \(\morphthree = (\morphthree_\indexone : \graphname{\hierarchygraphmiddle}{\maplevelone(\indexone)} \to \graphname{\variablegraphone}{\mapleveltwo(\indexone)} \mid \indexone \in [\chaindepthone])\).
Square (1) in this diagram is constructed by level-wise pushouts: for each \(0 \leq \indexone \leq \chaindepthone\) we construct \(\morphthree_\indexone\) as a mediating morphism \(\morphthree_\indexone : \graphname{\hierarchygraphmiddle}{\maplevelone(\indexone)} \to \graphname{\variablegraphone}{\mapleveltwo(\indexone)}\) such that

\begin{equation}\label{eq:po-1-mediating-morphism}
\maphierarchylefttomiddle_{\maplevelone(\indexone)};\morphthree_\indexone = \morphone_\indexone\mbox{ and }\mapruletohierarchymiddle_\indexone;\morphthree_\indexone = \morphtwo_\indexone
\end{equation}

\begin{center}
\begin{tikzpicture}[on grid,node distance=35mm]

\node[el-math] (l)										{\graphname{\rulegraphleft}{\indexone}};
\node[el-math] (sf)	[right of=l]						{\graphname{\hierarchygraphleft}{\maplevelone(\indexone)}};
\node[el-math] (i)	[below of=l]						{\graphname{\rulegraphmiddle}{\indexone}};
\node[el-math] (df)	[below of=sf]						{\graphname{\hierarchygraphmiddle}{\maplevelone(\indexone)}};
\node[el-math] (x)	[below right=15mm and 15mm of df]	{\graphname{\variablegraphone}{\mapleveltwo(\indexone)}};

\draw[map] (l) to	node[la-math]		(mlsf)	{\mapruletohierarchyleft_\indexone}						(sf);
\draw[map] (i) to	node[la-math]		(midf)	{\mapruletohierarchymiddle_\indexone}					(df);
\draw[map] (l) to	node[la-math,left]	(mli)	{\maprulelefttomiddle_\indexone}						(i);
\draw[map] (sf) to	node[la-math,left]	(msfdf)	{\maphierarchylefttomiddle_{\maplevelone(\indexone)}}	(df);

\draw[map] (df) to	node[la-math]	(mdfx)	{\morphthree_\indexone}	(x);

\draw[map,bend left] (sf) to	node[la-math]	(msfx)	{\morphone_\indexone}	(x);
\draw[map,bend right] (i) to	node[la-math]	(mix)	{\morphtwo_\indexone}	(x);

\node[el-math] (po) [below right=25mm of l]		{\po};
\node[el-math] (e1) [above right=7mm of mix]	{=};
\node[el-math] (e2) [below left=12mm of msfx]	{=};

\end{tikzpicture}
\end{center}

This construction ensures \((\maphierarchylefttomiddle_{\downarrow\maplevelone},id_{[\chaindepthone]});(\morphthree,\mapleveltwo) = (\morphone,\mapleveltwo)\) and \((\mapruletohierarchymiddle,id_{[\chaindepthone]});(\morphthree,\maplevelthree) = (\morphtwo,\maplevelthree)\) (recall \(\mapleveltwo = \maplevelthree\)) as required.

It remains only to show that \((\morphthree,\mapleveltwo)\) establishes indeed a graph chain morphism from \(\chainname[\downarrow\maplevelone]{\hierarchygraphmiddle}\) to \(\chainname{\variablegraphone}\).
That is, for all \(0 \leq \indexone < \indextwo \leq \chaindepthone\) we have to show that there exists a graph homomorphism \(\morphthree_{\indextwo\mid\indexone} : \graphname{\hierarchygraphmiddle}{\maplevelone(\indextwo)} \cap \graphname{\hierarchygraphmiddle}{\maplevelone(\indexone)} \to \domain{\typemorph[\variablegraphone]{\mapleveltwo(\indextwo)}{\mapleveltwo(\indexone)}}\) such that the following diagram commutes

\begin{center}
\begin{tikzpicture}[on grid,node distance=30mm]

\node[el-math] (di)						{\graphname{\hierarchygraphmiddle}{\maplevelone(\indexone)}};
\node[el-math] (xi)		[right of=di]	{\graphname{\variablegraphone}{\mapleveltwo(\indexone)}};
\node[el-math] (dji)	[below of=di]	{\graphname{\hierarchygraphmiddle}{\maplevelone(\indextwo)} \cap \graphname{\hierarchygraphmiddle}{\maplevelone(\indexone)}};
\node[el-math] (dtx)	[right of=dji]	{\domain{\typemorph[\variablegraphone]{\mapleveltwo(\indextwo)}{\mapleveltwo(\indexone)}}};
\node[el-math] (dj)		[below of=dji]	{\graphname{\hierarchygraphmiddle}{\maplevelone(\indextwo)}};
\node[el-math] (xj)		[right of=dj]	{\graphname{\variablegraphone}{\mapleveltwo(\indextwo)}};

\draw[incmapl]	(di) to node[la-math]	(mdixi)		{\morphthree_\indexone}						(xi);
\draw[mapdots]	(dji) to node[la-math]	(mdjidtx)	{\morphthree_{\indextwo \mid \indexone}}	(dtx);
\draw[incmapl]	(dj) to node[la-math]	(mdjxj)		{\morphthree_\indextwo}						(xj);

\draw[incmapl]	(dji) to node[la-math]			(mdjidi)	{\sqsubseteq}																		(di);
\draw[map]		(dtx) to node[la-math,right]	(mdtxxi)	{\typemorph[\variablegraphone]{\mapleveltwo(\indextwo)}{\mapleveltwo(\indexone)}}	(xi);
\draw[incmapr]	(dji) to node[la-math,left]		(inlijlj)	{\sqsubseteq}																		(dj);
\draw[incmapl]	(dtx) to node[la-math,right]	(iniijij)	{\sqsubseteq}																		(xj);

\end{tikzpicture}
\end{center}

We consider the following diagram

\begin{center}
\begin{tikzpicture}[on grid,node distance=30mm]

\node[el-math] (lji)										{\graphname{\rulegraphleft}{\indextwo} \cap \graphname{\rulegraphleft}{\indexone}};
\node[el-math] (sji)	[right of=lji]						{\graphname{\hierarchygraphleft}{\maplevelone(\indextwo)} \cap \graphname{\hierarchygraphleft}{\maplevelone(\indexone)}};
\node[el-math] (lj)		[below of=lji]						{\graphname{\rulegraphleft}{\indextwo}};
\node[el-math] (sj)		[right of=lj]						{\graphname{\hierarchygraphleft}{\indextwo}};
\node[el-math] (iji)	[below right=15mm and 15mm of lji]	{\graphname{\rulegraphmiddle}{\indextwo} \cap \graphname{\rulegraphmiddle}{\indexone}};
\node[el-math] (dji)	[right of=iji]						{\graphname{\hierarchygraphmiddle}{\maplevelone(\indextwo)} \cap \graphname{\hierarchygraphmiddle}{\maplevelone(\indexone)}};
\node[el-math] (dtx)	[right of=dji]						{\domain{\typemorph[\variablegraphone]{\mapleveltwo(\indextwo)}{\mapleveltwo(\indexone)}}};
\node[el-math] (ij)		[below of=iji]						{\graphname{\rulegraphmiddle}{\indextwo}};
\node[el-math] (dj)		[right of=ij]						{\graphname{\hierarchygraphmiddle}{\indextwo}};
\node[el-math] (xj)		[right of=dj]						{\graphname{\variablegraphone}{\mapleveltwo(\indextwo)}};

\draw[map]		(lji) to node[la-math]			(mlsji)	{\mapruletohierarchyleft_{\indextwo\mid\indexone}}	(sji);
\draw[map]		(lj) to node[la-math,pos=.7]	(mljsj)	{\mapruletohierarchyleft_\indextwo}					(sj);
\draw[incmapl]	(lji) to node[la-math,pos=.7]	(mljij)	{\sqsubseteq}										(lj);
\draw[incmapl]	(sji) to node[la-math,pos=.7]	(msjij)	{\sqsubseteq}										(sj);

\draw[map]		(iji) to node[la-math,pos=.3]	(midji)	{\mapruletohierarchymiddle_{\indextwo\mid\indexone}}	(dji);
\draw[map]		(ij) to node[la-math,below]		(mijdj)	{\mapruletohierarchymiddle_\indextwo}					(dj);
\draw[incmapl]	(iji) to node[la-math,pos=.7]	(mijij)	{\sqsubseteq}											(ij);
\draw[incmapl]	(dji) to node[la-math,pos=.7]	(mdjij)	{\sqsubseteq}											(dj);

\draw[map]	(lji) to node[la-math]				(mlisf)	{\maprulelefttomiddle_{\indextwo\mid\indexone}}										(iji);
\draw[map]	(sji) to node[la-math]				(miidf)	{\maphierarchylefttomiddle_{\maplevelone(\indextwo) \mid \maplevelone(\indexone)}}	(dji);
\draw[map]	(lj) to node[la-math,below left]	(ml0s0)	{\maprulelefttomiddle_\indextwo}													(ij);
\draw[map]	(sj) to node[la-math,below left]	(mi0d0)	{\maphierarchylefttomiddle_{\maplevelone(\indextwo)}}								(dj);

\draw[mapdots]	(dji) to node[la-math]			(mdjidtx)	{\morphthree_{\indextwo\mid\indexone}}	(dtx);
\draw[map]		(dj) to node[la-math,below]		(mdjxj)		{\morphthree_\indextwo}					(xj);
\draw[incmapl]	(dtx) to node[la-math,pos=.7]	(mdtxxj)	{\sqsubseteq}							(xj);

\draw[map,bend right] (iji) to	node[la-math,below,pos=.8]	(mijidtx)	{\morphtwo_{\indextwo\mid\indexone}}	(dtx);
\draw[map,bend right] (ij) to	node[la-math,below,pos=.8]	(mijxj)		{\morphtwo_\indextwo}					(xj);
\draw[map,bend left] (sji) to	node[la-math]				(msjidtx)	{\morphone_{\indextwo\mid\indexone}}	(dtx);
\draw[map,bend left] (sj) to	node[la-math,below,pos=.7]	(msjxj)		{\morphone_\indextwo}					(xj);

\end{tikzpicture}
\end{center}

The bottom square of the cube is a pushout, by construction.
Furthermore, we have shown that all the side faces of the cube are pullbacks, thus the top face becomes a pushout.
Since we have \(\mapruletohierarchyleft_{\indextwo\mid\indexone};\morphone_{\indextwo\mid\indexone} = \maprulelefttomiddle_{\indextwo\mid\indexone};\morphtwo_{\indextwo\mid\indexone}\) there exists a unique \(\morphthree_{\indextwo\mid\indexone} : \graphname{\hierarchygraphmiddle}{\maplevelone(\indextwo)} \cap \graphname{\hierarchygraphmiddle}{\maplevelone(\indexone)} \to \domain{\typemorph[\variablegraphone]{\mapleveltwo(\indextwo)}{\mapleveltwo(\indexone)}}\) such that

\begin{equation}\label{eq:po-1-unique-morphism}
\maphierarchylefttomiddle_{\maplevelone(\indextwo)\mid\maplevelone(\indexone)};\morphthree_{\indextwo\mid\indexone} = \morphone_{\indextwo\mid\indexone} \mbox{ and } \mapruletohierarchymiddle_{\indextwo\mid\indexone};\morphthree_{\indextwo\mid\indexone} = \morphtwo_{\indextwo\mid\indexone}
\end{equation}

By our constructions and assumptions we get

\begin{alignat*}{3}
\maphierarchylefttomiddle_{\maplevelone(\indextwo)\mid\maplevelone(\indexone)};\morphthree_{\indextwo\mid\indexone};\sqsubseteq \quad & =
		& \quad \morphone_{\indextwo\mid\indexone};\sqsubseteq \\
\quad & =
		& \quad \sqsubseteq;\morphone_\indextwo \\
\quad & =
		& \quad \sqsubseteq;\maphierarchylefttomiddle_{\maplevelone(\indextwo)};\morphthree_\indextwo \\
\quad & =
		& \quad \maphierarchylefttomiddle_{\maplevelone(\indextwo)\mid\maplevelone(\indexone)};\sqsubseteq;\morphthree_\indextwo
\end{alignat*}
and, analogously
\begin{alignat*}{3}
\mapruletohierarchymiddle_{\indextwo\mid\indexone};\morphthree_{\indextwo\mid\indexone};\sqsubseteq \quad & =
		& \quad \mapruletohierarchymiddle_{\indextwo\mid\indexone};\sqsubseteq;\morphthree_\indextwo
\end{alignat*}

Since \(\maphierarchylefttomiddle_{\maplevelone(\indextwo)\mid\maplevelone(\indexone)}\) and \(\mapruletohierarchymiddle_{\indextwo\mid\indexone}\) are jointly surjective this shows \(\morphthree_{\indextwo\mid\indexone};\sqsubseteq = \sqsubseteq;\morphthree_\indextwo\).

Analogously, we obtain also \(\sqsubseteq ; \morphthree_\indexone = \morphthree_{\indextwo\mid\indexone} ; \typemorph[\variablegraphone]{\mapleveltwo(\indextwo)}{\mapleveltwo(\indexone)}\) as required.
\qed
\end{proof}

\begin{proof}[Pushout property of square (2) in Figure~\ref{fig:two-po-construct-d}]
\label{proof:pushout-property-square-2}

Let be given an arbitrary graph chain \(\chainname{\variablegraphone} = \chain{\variablegraphone}{\variabledepthone}\) with \(\variabledepthone \geq \chaindepthtwo \geq \chaindepthone\) and graph chain morphisms \((\morphone,\mapleveltwo) : \chainname{\hierarchygraphleft} \to \chainname{\variablegraphone}\) with \(\mapleveltwo : [\chaindepthtwo] \to [\variabledepthone]\), \(\morphone = (\morphone_\indexone : \graphname{\hierarchygraphleft}{\indexone} \to \graphname{\variablegraphone}{\mapleveltwo(\indexone)} \mid \indexone \in [\chaindepthtwo])\), \((\morphtwo,\maplevelthree) : \chainname[\downarrow\maplevelone]{\hierarchygraphmiddle} \to \chainname{\variablegraphone}\) with \(\maplevelthree : [\chaindepthone] \to [\variabledepthone]\), \(\morphtwo = (\morphtwo_\indextwo : \graphname{\hierarchygraphmiddle}{\maplevelone(\indextwo)} \to \graphname{\variablegraphone}{\maplevelthree(\indextwo)} \mid \indextwo \in [\chaindepthone])\) such that
\begin{equation}\label{eq:po-2-commutativity}
(\overline{id}^{\chainname{\hierarchygraphleft}}_{\downarrow\maplevelone},\maplevelone);(\morphone,\mapleveltwo) = (\maphierarchylefttomiddle_{\downarrow\maplevelone},id_{[\chaindepthone]});(\morphtwo,\maplevelthree)
\end{equation}

\begin{center}
\begin{tikzpicture}[on grid,node distance=35mm]

\node[el-math] (sf)										{\chainname[\downarrow\maplevelone]{\hierarchygraphleft}};
\node[el-math] (s)	[right of=sf]						{\chainname{\hierarchygraphleft}};
\node[el-math] (df)	[below of=sf]						{\chainname[\downarrow\maplevelone]{\hierarchygraphmiddle}};
\node[el-math] (d)	[below of=s]						{\chainname{\hierarchygraphmiddle}};
\node[el-math] (x)	[below right=15mm and 15mm of d]	{\chainname{\variablegraphone}};

\draw[map] (sf) to	node[la-math]				(mlsf)	{(\overline{id}^{\chainname{\hierarchygraphleft}}_{\downarrow\maplevelone},\maplevelone)}	(s);
\draw[map] (df) to	node[la-math]				(midf)	{(\overline{id}^{\chainname{\hierarchygraphmiddle}}_{\downarrow\maplevelone},\maplevelone)}	(d);
\draw[map] (sf) to	node[la-math,left,pos=.7]	(mli)	{(\maphierarchylefttomiddle_{\downarrow\maplevelone},id_{[\chaindepthone]})}				(df);
\draw[map] (s) to	node[la-math,left,pos=.7]	(msfdf)	{(\maphierarchylefttomiddle,id_{[\chaindepthtwo]})}											(d);

\draw[mapdots] (d) to	node[la-math,below left]	(mdx)	{(\morphthree,\mapleveltwo)}	(x);

\draw[map,bend left] (s) to		node[la-math]				(msx)	{(\morphone,\mapleveltwo)}		(x);
\draw[map,bend right] (df) to	node[la-math,below left]	(mdfx)	{(\morphtwo,\maplevelthree)}	(x);

\node[el-math] (2) [below right=25mm of sf]	{(2)};

\end{tikzpicture}
\end{center}

Due to the definition of composition of graph chain morphism (see Definition~\ref{eq:type-compatibility-rule-maps-2}) this means that, first,
\begin{equation}\label{eq:po-2-maps-compatibility}
\maplevelone;\mapleveltwo = \maplevelthree
\end{equation}

This means that we have to construct a graph chain morphism \((\morphthree,\mapleveltwo) : \chainname{\hierarchygraphmiddle} \to \chainname{\variablegraphone}\) with \(\morphthree = (\morphthree_\exampleindexone : \graphname{\hierarchygraphmiddle}{\exampleindexone} \to \graphname{\variablegraphone}{\mapleveltwo(\exampleindexone)} \mid \exampleindexone \in [\chaindepthtwo])\) such that 
\begin{equation}\label{eq:po-2-k-definition}
(\maphierarchylefttomiddle,id_{[\chaindepthtwo]});(\morphthree,\mapleveltwo) = (\morphone,\mapleveltwo)
\mbox{ and }
(\overline{id}^{\chainname{\hierarchygraphmiddle}}_{\downarrow\maplevelone},\maplevelone);(\morphthree,\mapleveltwo) = (\morphtwo,\maplevelthree)
\end{equation}

\begin{description}
	\item[case 1:] \(\exampleindexone \in \maplevelone([\chaindepthone])\).
	In this case we have \(\graphname{\hierarchygraphmiddle}{\exampleindexone} = \graphname{\hierarchygraphleft}{\exampleindexone} + \graphname{\rulegraphmiddle}{\maplevelone^-(\exampleindexone)} \setminus \graphname{\rulegraphleft}{\maplevelone^-(\exampleindexone)}\) and due to the construction of \(\chainname{\hierarchygraphmiddle}\) we do have a trivial pushout square. thus \(\morphthree_\exampleindexone\) is uniquely defined by
	\begin{equation}\label{eq:po-2-k-construction-1}
	\morphthree_\exampleindexone := \morphtwo_{\maplevelone^-(\exampleindexone)} : \graphname{\hierarchygraphmiddle}{\exampleindexone} \to \graphname{\variablegraphone}{\mapleveltwo(\exampleindexone)}
	\end{equation}
	Keep in mind that \(\maplevelthree(\maplevelone^-(\exampleindexone)) = \mapleveltwo(\exampleindexone)\) according to~\ref{eq:po-2-maps-compatibility}.
	\item[case 2:] \(\exampleindexone \in [\chaindepthtwo] \setminus \maplevelone([\chaindepthone])\).
	We have \(\graphname{\hierarchygraphmiddle}{\exampleindexone} = \graphname{\hierarchygraphleft}{\exampleindexone}\) and only the second condition in~\ref{eq:po-2-k-definition} is relevant, thus \(\morphthree_\exampleindexone\) is uniquely defined by
	\begin{equation}\label{eq:po-2-k-construction-2}
	\morphthree_\exampleindexone := \morphone_\exampleindexone : \graphname{\hierarchygraphleft}{\exampleindexone} \to \graphname{\variablegraphone}{\mapleveltwo(\exampleindexone)}
	\end{equation}
	\begin{center}
	\begin{tikzpicture}[on grid,node distance=35mm]
	
	\node[el-math] (s)					{\graphname{\hierarchygraphleft}{\exampleindexone}};
	\node[el-math] (d)	[below of=s]	{\graphname{\hierarchygraphmiddle}{\exampleindexone}=\graphname{\hierarchygraphleft}{\exampleindexone}};
	\node[el-math] (x)	[right of=d]	{\graphname{\variablegraphone}{\mapleveltwo(\exampleindexone)}};
	
	\draw[map] 				(s) to	node[la-math]	(mlsf)	{id_{\graphname{\hierarchygraphleft}{\exampleindexone}}}	(d);
	\draw[mapdots] 			(d)	to	node[la-math]	(mdx)	{\morphthree_\exampleindexone}								(x);
	\draw[map,bend left]	(s) to	node[la-math]	(mdfx)	{\morphone_\exampleindexone}								(x);
	
	\end{tikzpicture}
	\end{center}
	After we have constructed the only family \(\morphthree = (\morphthree_\exampleindexone : \graphname{\hierarchygraphmiddle}{\exampleindexone} \to \graphname{\variablegraphone}{\mapleveltwo(\exampleindexone)} \mid \exampleindexone \in [\chaindepthtwo])\) satisfying condition~\ref{eq:po-2-k-definition}, it remains to show that the pair \((\morphthree,\mapleveltwo)\) establishes a graph chain morphism \((\morphthree,\mapleveltwo) : \chainname{\hierarchygraphmiddle} \to \chainname{\variablegraphone}\).
	Due to Definition~\ref{def:graph-chain-morphism}, we have to show that for all \(0 \leq \exampleindexone < \exampleindextwo \leq \chaindepthtwo\) there exists a graph homomorphism \(\morphthree_{\exampleindextwo\mid\exampleindexone} : \graphname{\hierarchygraphmiddle}{\exampleindexone} \cap \graphname{\hierarchygraphmiddle}{\exampleindextwo} \to \domain{\typemorph[\variablegraphone]{\mapleveltwo(\exampleindextwo)}{\mapleveltwo(\exampleindexone)}}\) such that the following diagram commutes
	\begin{center}
	\begin{tikzpicture}[on grid,node distance=30mm]
	
	\node[el-math] (da)						{\graphname{\hierarchygraphmiddle}{\exampleindexone}};
	\node[el-math] (xa)		[right of=da]	{\graphname{\variablegraphone}{\exampleindexone}};
	\node[el-math] (dab)	[below of=da]	{\graphname{\hierarchygraphmiddle}{\exampleindexone} \cap \graphname{\hierarchygraphmiddle}{\exampleindextwo}};
	\node[el-math] (dtx)	[right of=dab]	{\domain{\typemorph[\variablegraphone]{\mapleveltwo(\exampleindextwo)}{\mapleveltwo(\exampleindexone)}}};
	\node[el-math] (db)		[below of=dab]	{\graphname{\hierarchygraphmiddle}{\exampleindextwo}};
	\node[el-math] (xb)		[right of=db]	{\graphname{\variablegraphone}{\exampleindextwo}};
	
	\draw[map]	(da) to node[la-math]	(mdixi)		{\morphthree_\exampleindexone}							(xa);
	\draw[map]	(dab) to node[la-math]	(mdjidtx)	{\morphthree_{\exampleindextwo \mid \exampleindexone}}	(dtx);
	\draw[map]	(db) to node[la-math]	(mdjxj)		{\morphthree_\exampleindextwo}							(xb);
	
	\draw[incmapl]	(dab) to node[la-math]			(mdabda)	{\sqsubseteq}																					(da);
	\draw[map]		(dtx) to node[la-math,right]	(mdtxxa)	{\typemorph[\variablegraphone]{\mapleveltwo(\exampleindextwo)}{\mapleveltwo(\exampleindexone)}}	(xa);
	\draw[incmapr]	(dab) to node[la-math,left]		(mdabdb)	{\sqsubseteq}																					(db);
	\draw[incmapl]	(dtx) to node[la-math,right]	(mdtxxb)	{\sqsubseteq}																					(xb);
	
	\node[el-math]	(e1)	[below right=21mm of da]	{=};
	\node[el-math]	(e2)	[below of=e1]				{=};
	
	\end{tikzpicture}
	\end{center}
	We show this by case distinction:
	\begin{description}
		\item[case 2a:] \(\exampleindexone,\exampleindextwo \in [\chaindepthtwo] \setminus \maplevelone([\chaindepthone])\).
		In this case, we have \(\graphname{\hierarchygraphmiddle}{\exampleindexone} = \graphname{\hierarchygraphleft}{\exampleindexone}\), \(\graphname{\hierarchygraphmiddle}{\exampleindextwo} = \graphname{\hierarchygraphleft}{\exampleindextwo}\) and \(\morphthree_\exampleindexone = \morphone_\exampleindexone, \morphthree_\exampleindextwo = \morphone_\exampleindextwo\) thus \(\morphthree_{\exampleindextwo\mid\exampleindexone} := \morphone_{\exampleindextwo\mid\exampleindexone}\) makes the diagram commute since \((\morphone,\mapleveltwo)\) is a graph chain morphism.
		\item[case 2b:] \(\exampleindexone,\exampleindextwo \in \maplevelone([\chaindepthone])\).
		In this case, we have \(\morphthree_\exampleindexone = \morphtwo_{\maplevelone^-(\exampleindexone)}\) and \(\morphthree_\exampleindextwo = \morphtwo_{\maplevelone^-(\exampleindextwo)}\) thus \(\morphthree_{\exampleindextwo\mid\exampleindexone} := \morphtwo_{\maplevelone^-(\exampleindextwo)\mid\maplevelone^-(\exampleindexone)}\) makes the diagram commute since \((\morphtwo,\maplevelthree)\) is a graph chain morphism.
		\item[case 2c:] \(\exampleindexone \in [\chaindepthtwo] \setminus \maplevelone([\chaindepthone]), \exampleindextwo \in \maplevelone([\chaindepthone])\).
		In this case, we have \(\graphname{\hierarchygraphmiddle}{\exampleindexone} = \graphname{\hierarchygraphleft}{\exampleindexone}\), and inclusion \(\maphierarchylefttomiddle_\exampleindextwo : \graphname{\hierarchygraphleft}{\exampleindextwo} \hookrightarrow \graphname{\hierarchygraphmiddle}{\exampleindextwo}\) and \(\graphname{\hierarchygraphmiddle}{\exampleindexone} \cap \graphname{\hierarchygraphmiddle}{\exampleindextwo} = \graphname{\hierarchygraphleft}{\exampleindexone} \cap \graphname{\hierarchygraphleft}{\exampleindextwo}\). 
		\begin{center}
		\begin{tikzpicture}[on grid,node distance=30mm]
		
		\node[el-math] (da)									{\graphname{\hierarchygraphmiddle}{\exampleindexone} = \graphname{\hierarchygraphleft}{\exampleindexone}};
		\node[el-math] (xa)		[right=40mm of da]			{\graphname{\variablegraphone}{\mapleveltwo(\exampleindexone)}};
		\node[el-math] (dab)	[below of=da]				{\graphname{\hierarchygraphmiddle}{\exampleindexone} \cap \graphname{\hierarchygraphmiddle}{\exampleindextwo} = \graphname{\hierarchygraphleft}{\exampleindexone} \cap \graphname{\hierarchygraphleft}{\exampleindextwo}};
		\node[el-math] (dtx)	[right=40mm of dab]			{\domain{\typemorph[\variablegraphone]{\mapleveltwo(\exampleindextwo)}{\mapleveltwo(\exampleindexone)}}};
		\node[el-math] (db)		[below of=dab]				{\graphname{\hierarchygraphmiddle}{\exampleindextwo}};
		\node[el-math] (xb)		[right=40mm of db]			{\graphname{\variablegraphone}{\mapleveltwo(\exampleindextwo)}};
		\node[el-math] (sb)		[below right=22mm of dab]	{\graphname{\hierarchygraphleft}{\exampleindextwo}};
		
		\draw[map]	(da) to node[la-math]		(mdaxa)		{\morphthree_\exampleindexone = \morphone_\exampleindexone}						(xa);
		\draw[map]	(dab) to node[la-math]		(mdabdtx)	{\morphone_{\exampleindextwo\mid\exampleindexone}}								(dtx);
		\draw[map]	(db) to node[la-math,below]	(mdbxb)		{\morphthree_\exampleindextwo = \morphtwo_{\maplevelone^-(\exampleindextwo)}}	(xb);
		
		\draw[incmapl]	(dab) to node[la-math]			(mdabda)	{\sqsubseteq}																					(da);
		\draw[map]		(dtx) to node[la-math,right]	(mdtxxa)	{\typemorph[\variablegraphone]{\mapleveltwo(\exampleindextwo)}{\mapleveltwo(\exampleindexone)}}	(xa);
		\draw[incmapr]	(dab) to node[la-math,left]		(mdabdb)	{\sqsubseteq}																					(db);
		\draw[incmapl]	(dtx) to node[la-math,right]	(mdtxxb)	{\sqsubseteq}																					(xb);
		
		\draw[incmapl]	(dab) to node[la-math]	(mdabsb)	{}												(sb);
		\draw[incmapl]	(sb) to node[la-math]	(msbdb)		{\maphierarchylefttomiddle_\exampleindextwo}	(db);
		\draw[map]		(sb) to node[la-math]	(msbxb)		{\morphone_\exampleindextwo}					(xb);
		
		\end{tikzpicture}
		\end{center}
		Moreover, we have \(\morphthree_\exampleindexone = \morphone_\exampleindexone\), \(\morphthree_\exampleindextwo = \morphtwo_{\maplevelone^-(\exampleindextwo)}\) and \(\morphthree_{\exampleindextwo\mid\exampleindexone} := \morphone_{\exampleindextwo\mid\exampleindexone}\) that makes the upper square and the lower right triangle commute since \((\morphone,\mapleveltwo)\) is a graph chain morphism.
		The lower left triangle of inclusions commutes trivially, while the remaining triangle commutes by definition of \(\morphthree_\exampleindextwo\).
		This shows that also the lower square commutes, as required.
		\item[case 2d:] \(\exampleindexone \in\maplevelone([\chaindepthone]), \exampleindextwo \in [\chaindepthtwo] \setminus  \maplevelone([\chaindepthone])\).
		In this case, we have \(\graphname{\hierarchygraphmiddle}{\exampleindextwo} = \graphname{\hierarchygraphleft}{\exampleindextwo}\), an inclusion \(\maphierarchylefttomiddle_\exampleindexone : \graphname{\hierarchygraphleft}{\exampleindexone} \hookrightarrow \graphname{\hierarchygraphmiddle}{\exampleindexone}\) and \(\graphname{\hierarchygraphmiddle}{\exampleindexone} \cap \graphname{\hierarchygraphmiddle}{\exampleindextwo} = \graphname{\hierarchygraphleft}{\exampleindexone} \cap \graphname{\hierarchygraphleft}{\exampleindextwo}\).
		\begin{center}
		\begin{tikzpicture}[on grid,node distance=30mm]
		
		\node[el-math] (da)									{\graphname{\hierarchygraphmiddle}{\exampleindexone}};
		\node[el-math] (xa)		[right=40mm of da]			{\graphname{\variablegraphone}{\mapleveltwo(\exampleindexone)}};
		\node[el-math] (dab)	[below of=da]				{\graphname{\hierarchygraphmiddle}{\exampleindexone} \cap \graphname{\hierarchygraphmiddle}{\exampleindextwo} = \graphname{\hierarchygraphleft}{\exampleindexone} \cap \graphname{\hierarchygraphleft}{\exampleindextwo}};
		\node[el-math] (dtx)	[right=40mm of dab]			{\domain{\typemorph[\variablegraphone]{\mapleveltwo(\exampleindextwo)}{\mapleveltwo(\exampleindexone)}}};
		\node[el-math] (db)		[below of=dab]				{\graphname{\hierarchygraphmiddle}{\exampleindextwo} = \graphname{\hierarchygraphleft}{\exampleindextwo}};
		\node[el-math] (xb)		[right=40mm of db]			{\graphname{\variablegraphone}{\mapleveltwo(\exampleindextwo)}};
		\node[el-math] (sa)		[below right=22mm of da]	{\graphname{\hierarchygraphleft}{\exampleindexone}};
		
		\draw[map]	(da) to node[la-math]		(mdaxa)		{\morphthree_\exampleindexone = \morphtwo_{\maplevelone^-(\exampleindexone)}}	(xa);
		\draw[map]	(dab) to node[la-math]		(mdabdtx)	{\morphone_{\exampleindextwo\mid\exampleindexone}}								(dtx);
		\draw[map]	(db) to node[la-math,below]	(mdbxb)		{\morphthree_\exampleindextwo = \morphone_\exampleindextwo}						(xb);
		
		\draw[incmapl]	(dab) to node[la-math]			(mdabda)	{\sqsubseteq}					(da);
		\draw[map]		(dtx) to node[la-math,right]	(mdtxxa)	{\typemorph[\variablegraphone]{\mapleveltwo(\exampleindextwo)}{\mapleveltwo(\exampleindexone)}}	(xa);
		\draw[incmapr]	(dab) to node[la-math,left]		(mdabdb)	{\sqsubseteq}					(db);
		\draw[incmapl]	(dtx) to node[la-math,right]	(mdtxxb)	{\sqsubseteq}					(xb);
		
		\draw[incmapr]	(sa) to node[la-math]	(msada)		{\maphierarchylefttomiddle_\exampleindexone}	(da);
		\draw[incmapl]	(sa) to node[la-math]	(msadab)	{}												(dab);
		\draw[map]		(sa) to node[la-math]	(msaxa)		{\morphone_\exampleindexone}					(xa);
		
		\end{tikzpicture}
		\end{center}
		Moreover, we have \(\morphthree_\exampleindextwo = \morphone_\exampleindextwo\), \(\morphthree_\exampleindexone = \morphtwo_{\maplevelone^-(\exampleindexone)}\) (recall \(\mapleveltwo(\exampleindexone) = \maplevelthree(\maplevelone^-(\exampleindexone))\)) and \(\morphthree_{\exampleindextwo \mid \exampleindexone} := \morphone_{\exampleindextwo\mid\exampleindexone}\) that makes the lower square and the upper right triangle commute since \((\morphone, \mapleveltwo)\) is a graph chain morphism.
		The upper left triangle of inclusions commutes trivially, while the remaining triangle commutes by definition of \(\morphthree_\exampleindexone\).
	\end{description}
\end{description}
\qed
\end{proof}

Composing the two pushouts (1) and (2) we obtain a pushout in \cat{Chain} that provides us, finally, with a chain morphism from \(\chainname{\hierarchygraphmiddle}\) to \(\chainname{\hierarchygraph}\) that materialises the required multilevel typing of the constructed graph \(\graphname{\hierarchygraphmiddle}{}\):

\begin{center}
\begin{tikzpicture}[on grid,node distance=30mm]

\node[el-math] (l)										{\chainname{\rulegraphleft}};
\node[el-math] (i)	[right of=l]						{\chainname{\rulegraphmiddle}};
\node[el-math] (mm)	[above of=i]						{\chainname{\rulegraph}};
\node[el-math] (s)	[below right=10mm and 25mm of l]	{\chainname{\hierarchygraphleft}};
\node[el-math] (d)	[right of=s]						{\chainname{\hierarchygraphmiddle}};
\node[el-math] (tg)	[above of=d]						{\chainname{\hierarchygraph}};

\draw[map]		(l) to node[la-math,below left]		(mf)	{(\mapruletohierarchyleft,\maplevelone)}						(s);
\draw[map] 		(l) to node[la-math]				(li)	{(\maprulelefttomiddle,id_{[\chaindepthone]})}					(i);
\draw[map] 		(s) to node[la-math,below]			(sd)	{(\maphierarchylefttomiddle,id_{[\chaindepthtwo]})}				(d);
\draw[mapdots]	(d) to node[la-math,right,pos=.3]	(dt)	{(\chainmorph{\hierarchygraphmiddle},id_{[\chaindepthtwo]})}	(tg);
\draw[map]		(i) to node[la-math,pos=.3]			(im)	{(\chainmorph{\rulegraphmiddle},id_{[\chaindepthone]})}			(mm);
\draw[map]		(s) to node[la-math,pos=.9]			(st)	{(\chainmorph{\hierarchygraphleft},id_{[\chaindepthtwo]})}		(tg);
\draw[map]		(l) to node[la-math]				(lm)	{(\chainmorph{\rulegraphleft},id_{[\chaindepthone]})}			(mm);
\draw[map]		(i) to node[la-math]				(df)	{(\mapruletohierarchymiddle,\maplevelone)}						(d);
\draw[map]		(mm) to	node[la-math]				(bf)	{(\mapruletohierarchybinding,\maplevelone)}						(tg);

\end{tikzpicture}
\end{center}

The left triangle commutes due to compatible typing of the rule.
The roof square commutes since the match \((\mapruletohierarchyleft , \maplevelone)\) is type compatible.
This construction gives us \((\mapruletohierarchyleft , \maplevelone) ; (\chainmorph{\hierarchygraphleft},id_{[\chaindepthtwo]}) = (\maprulelefttomiddle , id_{[\chaindepthone]});(\chainmorph{\rulegraphmiddle} , id_{[\chaindepthone]}) ; (\mapruletohierarchybinding , \maplevelone)\), thus the pushout universal property of the back of the square gives a unique chain morphism \((\chainmorph{\hierarchygraphmiddle} , id_{[\chaindepthtwo]}) : \chainname{\hierarchygraphmiddle} \to \chainname{\hierarchygraph}\) such that the two desired typing compatibilities \((\maphierarchylefttomiddle , id_{[\chaindepthtwo]}) ; (\chainmorph{\hierarchygraphmiddle} , id_{[\chaindepthtwo]}) = (\chainmorph{\hierarchygraphleft} , id_{[\chaindepthtwo]})\) and \((\mapruletohierarchymiddle , \maplevelone);(\chainmorph{\hierarchygraphmiddle} , id_{[\chaindepthtwo]}) = (\chainmorph{\rulegraphmiddle} , id_{[\chaindepthone]}) ; (\mapruletohierarchybinding , \maplevelone)\) are satisfied.

\begin{remark}
Based in Theorem 2 in~\cite{wolter2015topoi}, we can show that square (1) is a pushout even if \(\maprulelefttomiddle : \graphname{\rulegraphleft}{} \to \graphname{\rulegraphmiddle}{}\) is an arbitrary graph homomorphism.
However, square (2) may not be a pushout in this case, thus we need to restrict to inclusions \(\maprulelefttomiddle : \graphname{\rulegraphleft}{} \hookrightarrow \graphname{\rulegraphmiddle}{}\).
\end{remark}

\subsection{Pullbacks in the category \cat{Chain}}
\label{subsec:pullbacks-category-chain}

Following a similar way of reasoning to Section~\ref{subsec:pushouts-category-chain}, the second step in the MCMT rule application should go as follows.

We construct a final pullback complement
\begin{center}
\begin{tikzpicture}[on grid,node distance=25mm]

\node[el-math] (i)							{\graphname{\rulegraphmiddle}{}};
\node[el-math] (r)	[right of=i]			{\graphname{\rulegraphright}{}};
\node[el-math] (d)	[below of=i]			{\graphname{\hierarchygraphmiddle}{}};
\node[el-math] (t)	[below of=r]			{\graphname{\hierarchygraphright}{}};
\node[el-math] (pb)	[below right=18mm of i]	{\fpbc};

\draw[incmapr]	(r) to node[la-math,above]	(mr)	{\maprulerighttomiddle}			(i);
\draw[incmapr]	(t) to node[la-math,above]	(mt)	{\maphierarchyrighttomiddle}	(d);
\draw[map]		(i) to node[la-math,left]	(md)	{\mapruletohierarchymiddle}		(d);
\draw[map]		(r) to node[la-math]		(mn)	{\mapruletohierarchyright}		(t);

\end{tikzpicture}
\end{center}

We reduce the typing of \(\graphname{\hierarchygraphmiddle}{}\), i.e.\ the inclusion chain \(\chainname{\hierarchygraphmiddle} = \chain{\hierarchygraphmiddle}{\chaindepthtwo}\), along \(\maphierarchyrighttomiddle\), due to Lemma~\ref{lem:inclusion-chains-establish-chain-morphism}, and obtain an inclusion chain \(\chainname{\hierarchygraphright} = \chain{\hierarchygraphright}{\chaindepthtwo}\) that is a closed subchain of \(\chainname{\hierarchygraphmiddle}\).
The multilevel typing \(\chainmorph{\hierarchygraphright}\) of \(\graphname{\hierarchygraphright}{}\) is borrowed from \(\chainmorph{\hierarchygraphmiddle}\) and this gives us exactly the intended type compatibility of \(\maphierarchyrighttomiddle : \graphname{\hierarchygraphright}{} \to \graphname{\hierarchygraphmiddle}{}\).

\begin{figure}[ht!]
\begin{center}
	\begin{minipage}{.4\linewidth}
	\centering
	\begin{tikzpicture}[on grid,node distance=25mm]
	
	\node[el-math] (tg)						{\graphname{\hierarchygraph}{\indexone}};
	\node[el-math] (d0) [below left of=tg]	{\graphname{\hierarchygraphmiddle}{0} = \graphname{\hierarchygraphmiddle}{}};
	\node[el-math] (t0) [below right of=tg]	{\graphname{\hierarchygraphright}{0} = \graphname{\hierarchygraphright}{}};
	\node[el-math] (di)	[below of=d0]		{\graphname{\hierarchygraphmiddle}{\indexone}};
	\node[el-math] (ti)	[below of=t0]		{\graphname{\hierarchygraphright}{\indexone}};
	
	\draw[incmapr]	(t0) to node[la-math,above]			(mt0d0)	{\maprulerighttomiddle_{0} = \maprulerighttomiddle}	(d0);
	\draw[incmapr]	(ti) to node[la-math,above]			(mtidi)	{\maprulerighttomiddle_{\indexone}}					(di);
	\draw[incmapl]	(di) to node[la-math]				(mdid0)	{}													(d0);
	\draw[incmapl]	(ti) to node[la-math]				(mtit0)	{}													(t0);
	\draw[map]		(d0) to node[la-math]				(mditg)	{\chainmorph[\indexone]{\hierarchygraphmiddle}}		(tg);
	\draw[mapdots]	(t0) to node[la-math,above right]	(mtitg)	{\chainmorph[\indexone]{\hierarchygraphright}}		(tg);
		
	\draw[map,bend left=75]			(di.west) to node[la-math]			(mditg)			{\chainmorph[\indexone]{\hierarchygraphmiddle}}	(tg.west);
	\draw[mapdots,bend right=75]	(ti) to node[la-math,above right]	(mtitg.east)	{\chainmorph[\indexone]{\hierarchygraphright}}	(tg.east);
	
	\node[el-math] (pb)	[below=14mm of mt0d0]	{\pb};
	
	\end{tikzpicture}
	\end{minipage}
	\quad
	\begin{minipage}{.4\linewidth}
	\centering
	\begin{tikzpicture}[on grid,node distance=25mm]
		
	\node[el-math] (tg)						{\graphname{\hierarchygraph}{\indexone}};
	\node[el-math] (d)	[below left of=tg]	{\graphname{\hierarchygraphmiddle}{}};
	\node[el-math] (t)	[below right of=tg]	{\graphname{\hierarchygraphright}{}};
	\node[el-math] (e1)	[below=10mm of tg]	{=};
	
	\draw[partmap]	(d) to node[la-math]				(tl)	{\chainmorph[\indexone]{\hierarchygraphmiddle}}	(tg);
	\draw[partmap]	(t) to node[la-math,above right]	(ti)	{\chainmorph[\indexone]{\hierarchygraphright}}	(tg);
	\draw[incmapr]	(t) to node[la-math,below]			(lid)	{\maprulerighttomiddle}							(d);
		
	\end{tikzpicture}
	\end{minipage}
\end{center}
\end{figure}

The remaining graph chain morphism \((\mapruletohierarchyright,\maplevelone) : \chainname{\rulegraphright} \to \chainname{\hierarchygraphright}\) between chains \(\chainname{\rulegraphright} = \chain{\rulegraphright}{\chaindepthone}\) and \(\chainname{\hierarchygraphright} = \chain{\hierarchygraphright}{\chaindepthtwo}\) with \(\mapruletohierarchyright = (\mapruletohierarchyright_\indexone : \graphname{\rulegraphright}{\indexone} \to \graphname{\hierarchygraphright}{\maplevelone(\indexone)} \mid \indexone \in [\chaindepthone])\) such that
\begin{equation}\label{eq:chainmorph-r-to-t}
(\maprulerighttomiddle,id_{[\chaindepthone]});(\mapruletohierarchymiddle,\maplevelone) = (\mapruletohierarchyright,\maplevelone);(\maphierarchyrighttomiddle,id_{[\chaindepthtwo]})
\end{equation}
is simply given by pullback composition and decomposition:

\begin{center}
\begin{tikzpicture}[on grid,node distance=35mm]

\node[el-math] (i0)											{\graphname{\rulegraphmiddle}{0} = \graphname{\rulegraphmiddle}{}};
\node[el-math] (r0)		[right of=i0]						{\graphname{\rulegraphright}{0} = \graphname{\rulegraphright}{}};
\node[el-math] (ii)		[below of=i0]						{\graphname{\rulegraphmiddle}{\indexone}};
\node[el-math] (ri)		[right of=ii]						{\graphname{\rulegraphright}{\indexone}};
\node[el-math] (d0)		[below right=15mm and 15mm of i0]	{\graphname{\hierarchygraphmiddle}{0} = \graphname{\hierarchygraphmiddle}{}};
\node[el-math] (t0)		[right of=d0]						{\graphname{\hierarchygraphright}{0} = \graphname{\hierarchygraphright}{}};
\node[el-math] (dfi)	[below of=d0]						{\graphname{\hierarchygraphmiddle}{\maplevelone(\indexone)}};
\node[el-math] (tfi)	[right of=dfi]						{\graphname{\hierarchygraphright}{\maplevelone(\indexone)}};

\draw[incmapr] (r0) to	node[la-math,above]			(mr0i0)	{\maprulerighttomiddle_0 = \maprulerighttomiddle}	(i0);
\draw[incmapr] (ri) to	node[la-math,above,pos=.3]	(mriii)	{\maprulerighttomiddle_\indexone}					(ii);
\draw[incmapr] (ii) to	node[la-math]				(msj0)	{}													(i0);
\draw[incmapr] (ri) to	node[la-math]				(mdj0)	{}													(r0);

\draw[incmapr] (t0) to	node[la-math,above,pos=.7]	(mt0d0)		{\maphierarchyrighttomiddle_0 = \maphierarchyrighttomiddle}	(d0);
\draw[incmapr] (tfi) to	node[la-math,above]			(mtfidfi)	{\maphierarchyrighttomiddle_\indexone}						(dfi);
\draw[incmapr] (dfi) to	node[la-math]				(mdfid0)	{}															(d0);
\draw[incmapr] (tfi) to	node[la-math]				(mtfit0)	{}															(t0);

\draw[map] (i0) to	node[la-math]	(mi0d0)		{\mapruletohierarchymiddle_0 = \mapruletohierarchymiddle}	(d0);
\draw[map] (r0) to	node[la-math]	(mr0t0)		{\mapruletohierarchyright_0 = \mapruletohierarchyright}		(t0);
\draw[map] (ii) to	node[la-math]	(miidfi)	{\mapruletohierarchymiddle_\indexone}						(dfi);
\draw[map] (ri) to	node[la-math]	(mritfi)	{\mapruletohierarchyright_\indexone}						(tfi);

\end{tikzpicture}
\end{center}

The back square is pullback due to the type compatibility of \(\maprulerighttomiddle : \graphname{\rulegraphright}{} \to \graphname{\rulegraphmiddle}{}\).
The left square has been shown to be a pullback due to the pushout construction in the first step.
The bottom square is constructed as a final pullback complement and the front square is pullback due to the construction of \(\maphierarchyrighttomiddle_\indexone\).
The diagonal square from the inclusion \(\graphname{\rulegraphright}{\indexone} \sqsubseteq \graphname{\rulegraphright}{}\) to the inclusion \(\graphname{\hierarchygraphmiddle}{\indexone} \sqsubseteq \graphname{\hierarchygraphmiddle}{}\) is a pullback due to the composition of the back pullback and the left pullback.
The decomposition of this diagonal pullback w.r.t.\ the front pullback gives us \(\mapruletohierarchyright_\indexone : \graphname{\rulegraphright}{\indexone} \to \graphname{\hierarchygraphright}{\indexone}\) making the cube commute and making the right square to a pullback as well.
According to Lemma~\ref{lem:inclusion-chains-establish-chain-morphism} the family \(\mapruletohierarchyright = (\mapruletohierarchyright_\indexone : \graphname{\rulegraphright}{\indexone} \to \graphname{\hierarchygraphright}{\maplevelone(\indexone)} \mid \indexone \in [\chaindepthone])\) of graph homomorphisms define a graph chain morphism \((\mapruletohierarchyright,\maplevelone) : \chainname{\rulegraphright} \to \chainname{\hierarchygraphright}\) establishing \(\chainname{\rulegraphright}\) as the reduct of \(\chainname{\hierarchygraphright}\) along \(\mapruletohierarchyright : \graphname{\rulegraphright}{} \to \graphname{\hierarchygraphright}{}\) where condition~\ref{eq:chainmorph-r-to-t} is simply satisfied by construction.

The proof of the pullback property in the category \cat{Chain} in the construction of our MCMT rules is planned as future work as part of our further research concerning this category.

\section{MCMTs for Three Different Dimensions}
\label{sec:mcmts-three-dimensions}

In Section~\ref{sec:three-dimensions} we present the three different dimensions which compose our framework.
We also include a hierarchy from each of those dimensions in order to illustrate our approach.
For the application dimension, where the ``main'' languages reside, we use the Robolang hierarchy.
For the supplementary dimension, which contains languages that are used to double-type elements in application hierarchies, we present the LTL hierarchy.
And for the data type dimension, we include an excerpt of the only hierarchy it contains, where we define data types and how they relate to each other.

MCMTs are oblivious to the dimension where their target hierarchy resides.
However, for the sake of completeness, we include in this section MCMT rules which are applied in each of the three example hierarchies aforementioned.
So in Section~\ref{subsec:mcmts-robolang} we present the remaining MCMT rules for the Robolang example, in Section~\ref{subsec:mcmts-ltl} we show two illustrative examples for LTL and in Section~\ref{subsec:data-type-dimension} we include an example rule for the data type hierarchy.

\subsection{MCMTs for Robolang}
\label{subsec:mcmts-robolang}

Using the Robolang example, we explain in Section~\ref{sec:why-mcmt} the advantages of using MCMTs with respect to reusability and shortly presented a comparison to two-level and multilevel model transformation rules.
Here we will show the remaining MCMT rules which together define the behaviour of autonomous robots which directly or indirectly conform to \elementname{robolang}.

These rules belong to two different kinds: \emph{behavioural} rules and \emph{environmental} rules.
In the first group, we include those rules which specify how the autonomous robots should carry own the execution of their own workflow, by firing transitions in order to move from one task to the next.
The second group contains the rules which we use to simulate changes in the environment of the robot.
This separation is useful since the environmental rules are used during simulation only.
However, the behavioural rules are used both for simulation and for deployment in the actual robots and real-world execution, where the environment does not need to be simulated.
The idea of specifying explicitly the coordination of such sets of rules is discussed as future work in Section~\ref{sec:future-work}.

\subsubsection{Behavioural MCMT rules}
\label{subsubsec:robolang-behavioural-mcmts}

\begin{figure}[tb]
	\centering
	\includegraphics{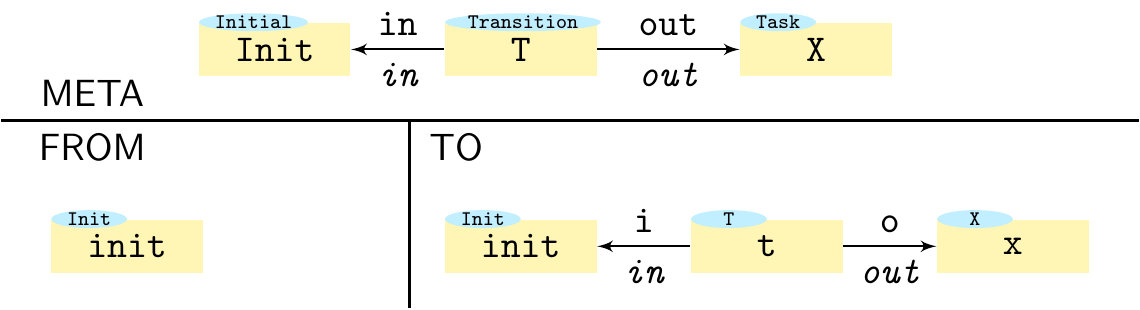}
	\caption{Rule \emph{Start}: the initial transition is fired}
	\label{fig:robolang-rule-start}
\end{figure}

The first rule in this group, shown in Figure~\ref{fig:robolang-rule-start}, is the first rule to be fired at the start of the simulation, hence it is called \emph{Start}.
This rule deletes the instance of the initial task to which the flow is initialised (recall Figure~\ref{fig:robolang-multilevel-hierarchy}(d)) and fires the connected transition.
In our examples, this is the only case where we use an input-free transition which can be fired immediately.

The next rule in this group, \emph{Fire Transition}, has already been presented in Figure~\ref{fig:robolang-rule-fire-transition} as part of the motivation for using MCMTs.
This is the main behavioural rule in the sense that it allows the workflow simulation to evolve, by allowing the robot to react to environmental changes (inputs).

\begin{figure}[tb]
	\centering
	\includegraphics{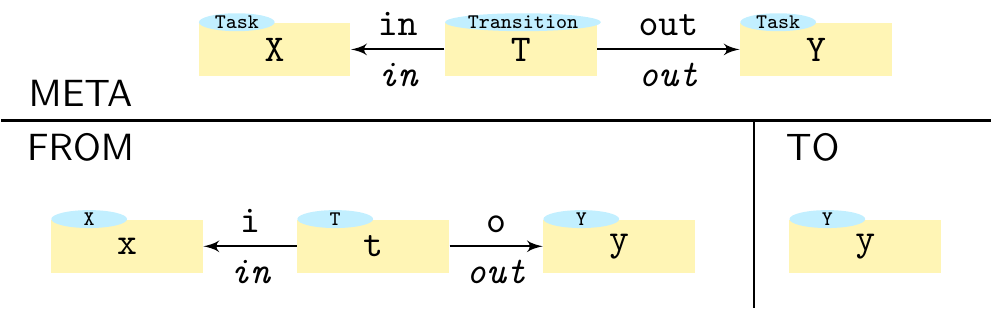}
	\caption{Rule \emph{Delete Task}: a finished task is removed}
	\label{fig:robolang-rule-delete-task}
\end{figure}

Finally, note that the \emph{Fire Transition} rule does not remove the previous instance of task from the workflow after spawning the new one, so one last rule is required to do so.
Hence, the last rule in this group, called \emph{Delete Task} and depicted in Figure~\ref{fig:robolang-rule-delete-task}, is responsible for doing precisely that.

An alternative to the use of this rule consists of ``merging'' \emph{Fire Transition} and \emph{Delete Task}, so that the next task is created and the previous one deleted in a single step.
However, this rule does not create the intermediate instance where the transition is being fired, which reduces the possible atomic propositions that we can define when combining with temporal verification properties, as we presented in Section~\ref{subsec:supplementary-dimension}.

\subsubsection{Environmental MCMT rules}
\label{subsubsec:robolang-environmental-mcmts}

This second group handles the simulation of environmental conditions for the simulation of autonomous robots.
Given the simplicity of Robolang, this semantics only require managing the appearance of instances of inputs in the running instance.
In order to do that, we created three MCMT rules, which we present in the following.

\begin{figure}[tb]
	\centering
	\includegraphics{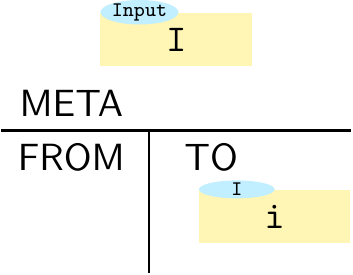}
	\caption{Rule \emph{Insert Input}: an input is created}
	\label{fig:robolang-rule-insert-input}
\end{figure}

Figure~\ref{fig:robolang-rule-insert-input} shows a simple MCMT rule called \emph{Insert Input} which creates an input nondeterministically and without constraints in the running instance.
The appearance of this instance of input may not cause any effect on the execution.
Note that the FROM block is empty since the current state of the execution does not affect which input appears.

The rule \emph{Insert Input} can be helpful to generate environmental conditions not expected by the workflow, which can uncover design flaws.
However, sometimes it may be desirable to introduce an input which will have an immediate and expected effect on the simulation.
That is, an instance of input which can cause one of the transitions coming out of the current task to be fired.
Hence, we provide a more constrained version of \emph{Insert Input}: \emph{Insert Effective Input}.

The rule \emph{Insert Effective Input}, depicted in Figure~\ref{fig:robolang-rule-fire-transition-multilevel}, creates an input that will cause the currently running task to finish and the next one to start, by firing the transition between them.
That is, its appearance will cause an effect on the execution.
Note that, although this rule is restricted to create an input that causes a direct effect, it is still non-deterministic, since we cannot know in advance which input instance will be created in case of several existing candidates (several outgoing transitions from the current task with different associated inputs).
If desired, more specific versions of this rule can be created for one specific instance of \elementname{Input}, like \elementname{Edge} or \elementname{Obstacle}.

\begin{figure}[tb]
	\centering
	\includegraphics{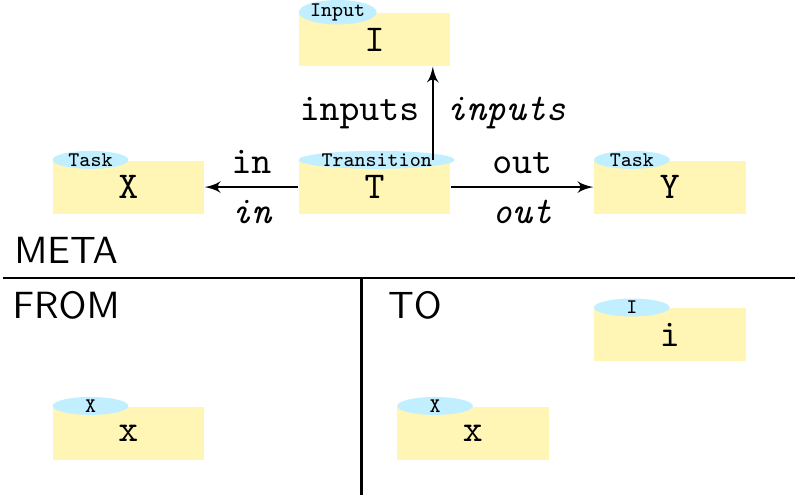}
	\caption{Rule \emph{Insert Effective Input}: a relevant input is created}
	\label{fig:robolang-rule-insert-effective-input}
\end{figure}

\begin{figure}[tb]
	\centering
	\includegraphics{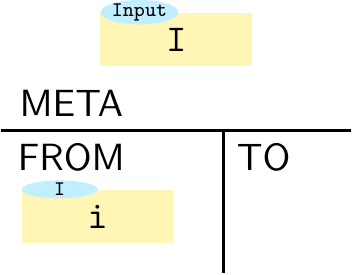}
	\caption{Rule \emph{Delete Input}: an existing input is removed}
	\label{fig:robolang-rule-delete-input}
\end{figure}

Finally, the inputs created with the two previous rules may disappear from the running instance (e.g.\ the robot may back off from an obstacle and not detect it any longer).
Therefore, the rule \emph{Delete Input}, presented in Figure~\ref{fig:robolang-rule-delete-input}, removes a previously existing instance of input from the simulation.

\subsection{MCMTs for LTL}
\label{subsec:mcmts-ltl}

The LTL supplementary hierarchy used as an example in Section~\ref{subsec:supplementary-dimension} is used for Runtime Verification~\cite{leucker2009briefaccount}.
In order to evaluate temporal properties represented as models we can make use of term-rewriting techniques, some of them known as unrolling~\cite{leucker2011rv}.
Such rewriting can be performed with MCMT rules, which will do two main types of operations.
In the first case, the unrolling operations ``split'' the temporal operators into two parts: one to be evaluated right away, and one to be evaluated in the next point in time.
Then, we match the atomic propositions in the temporal properties against the model with the corresponding state of the execution, represented as a running instance (e.g.\ the model \elementname{robot\_1\_run\_1} from Figure~\ref{fig:robolang-multilevel-hierarchy}(d)) by using a subgraph matching algorithm like the one described in Section~\ref{subsec:graph-matching}.
And in the second case, after replacing the atomic propositions by instances of \elementname{True} or \elementname{False}, another set of operations which simplify the formula by applying boolean equivalences and removing some temporal operators can be applied.

Since a comprehensive list of all MCMT rules for unrolling would be too dense and provide little insight about our approach, we do include two rules for illustrative purposes: one which performs an unrolling operation and one which removes a temporal operator.
Note that these rules use a new construction available on the MCMT syntax: constants.
A constant is a node or an arrow which is matched in a more restrictive way than the ones we have used in previous MCMT rules, which can be called variables.
When an element is defined as a constant, it must match an element with exactly the same name in the target hierarchy, in addition to the rest of conditions for matching related to graph homomorphisms and typing relations.
This is an useful construction when an MCMT rule is related to elements of a certain type in a model, but not to other elements of the same type in that model, and we need a mechanism to filter the relevant ones.
Since the name of a node is unique within a model, this name is enough to uniquely identify the node, and therefore defining also the type of a constant node is redundant.
For an arrow, the name is also enough to identify it uniquely as long as its source node is a constant too, but we allow for constant arrows with variable source nodes since such a construction has proven useful in our case studies.
As a consequence, constants can be identified in the graphical notation of MCMTs easily: constant nodes lack the blue ellipse where their type is specified and constant arrows lack the label in italics which indicates their type.
Constants in the META block can be used as types for other elements in the levels below in the same way as variables.
Actually, declaring a variable where its type is not explicitly written (like we do in the rules in Section~\ref{subsec:mcmts-robolang}) is equivalent to defining that type as a constant in the level above the variable, which yields equivalent rules but with a more verbose specification.
More examples of constants in MCMT rules can be found in the PLS case study in Section~\ref{subsec:pls}.

The first rule implements the unrolling operation of the binary operator ``until'', represented in LTL syntax as \(\LTLuntil\) and in our LTL model, in Figure~\ref{fig:ltl-supplementary-model} as \elementname{U}.
The process of unrolling is described in the literature (e.g.\ in~\cite{leucker2011rv}) with the next formula:

\[\phi \LTLuntil \psi \equiv \psi \vee (\phi \wedge \LTLnext (\phi \LTLuntil \psi))\]

\begin{figure}[ht]
	\centering
	\includegraphics{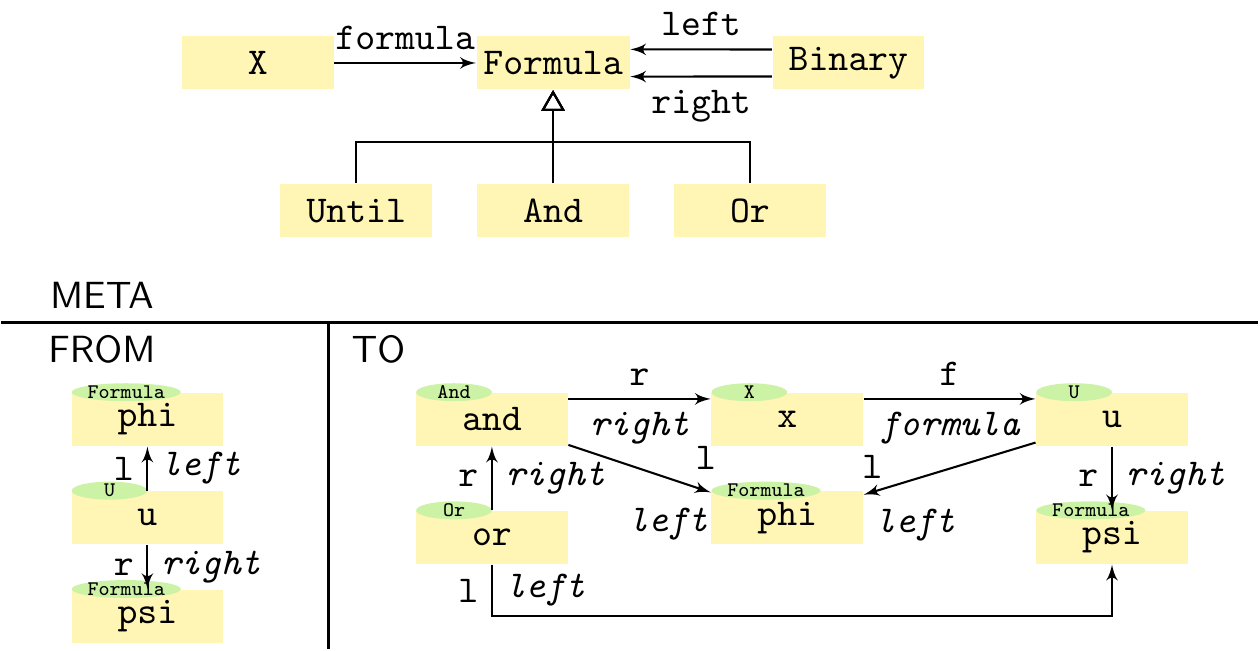}
	\caption{Rule \emph{Unroll Until}: the unrolling is applied to the \(\LTLuntil\) operator}
	\label{fig:ltl-rule-unroll-until}
\end{figure}

We display the MCMT rule \emph{Unroll Until}, which specifies the unrolling operation for \elementname{U} in Figure~\ref{fig:ltl-rule-unroll-until}.
It preserves the \elementname{phi} and \elementname{psi} elements, and adds the corresponding operators to create the right-hand side of the equivalence in the previous formula.

\begin{figure}[ht]
	\centering
	\includegraphics{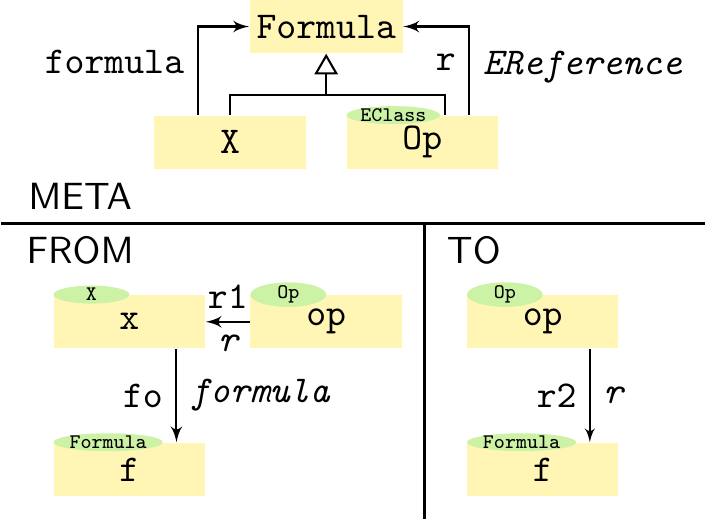}
	\caption{Rule \emph{Delete Next}: the operator \(\LTLnext\) is removed, keeping its subformula}
	\label{fig:ltl-rule-delete-next}
\end{figure}

The second rule we display is \emph{Delete Next}, in Figure~\ref{fig:ltl-rule-delete-next}.
This rule removes the ``next'' operator, commonly represented as \(\LTLnext\), from the model and replaces it by its subformula.
Using LTL syntax, this means replacing a (sub)formula \(\LTLnext(\phi)\) by just \(\phi\).
This rewriting is applied to all appearances of the next operator (\elementname{X} in our model) before the resulting formula from one evaluation is applied and evaluated in the next snapshot in the sequence (or \emph{letter} in the \emph{word}, in LTL jargon).

\subsection{MCMTs for data types}
\label{subsec:mcmts-data-types}

\begin{figure}[ht!]
	\centering
	\includegraphics{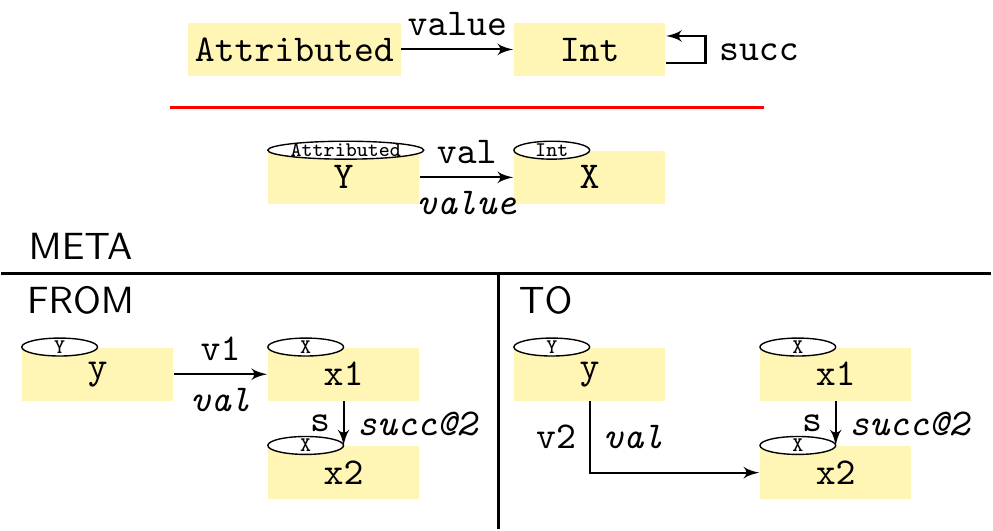}
	\caption{Rule \emph{Increment}: an integer value is replaced by its successor}
	\label{fig:data-type-rule-increment}
\end{figure}

The models presented in Section~\ref{subsec:data-type-dimension}, which we define for the representation of data types and their values from an abstract point of view in our framework, are actually infinite and not used in practice.
However, we can still represent the behaviour of the operations among data types and their values using MCMT rules, and use them as a formal description of how data types relate to each other and how the operations performed with them behave.
That is, such operations can be defined in a manner which is consistent with the rest of the framework, providing an interface from the actual implementation of the data types.
To finish this chapter, we include an example of such a rule in Figure~\ref{fig:data-type-rule-increment}.

The MCMT rule \emph{Increment} displays the behaviour of the \elementname{succ} operation defined for \elementname{Int}.
That is, the operator that increases the value of an integer by one.
The META part matches any element \element{Y}{Attributed} connected to an attribute \element{X}{Int} by an instance of the \elementname{value} reference.
Note that \elementname{Int} and \elementname{succ} are defined as constants in the first level of the META block to ensure that they match against the elements with the same name in the data type hierarchy.
The existence of the \elementname{value} edge in a level not depicted is made implicit by the specification of \element{val}{value} in the second level of the META block.
In the FROM block, if we find an instance \element{y}{Y} which instantiates the attribute \elementname{X} to a value \elementname{x1} which has \elementname{x2} as successor (since they are connected by \element{s}{succ}), the value of the attribute is changed to the value of the successor in the FROM block, by replacing the instance of \elementname{val} which connects them.

%% file: thesis/04-tooling.tex
\chapter{Tool Support with MultEcore}
\label{chap:tooling}

In this chapter, we focus on the tooling aspects of our approach.
We divide the contents on two sections.
Section~\ref{sec:modelling-tooling} presents the core of the framework: a set of Eclipse-based plugins which realise multilevel modelling within the Eclipse Modelling Framework, according to the syntax and definitions from Chapter~\ref{chap:mlm}, including a graphical, Sirius-based editor.
Then, Section~\ref{sec:mcmt-tooling} introduces our textual editor for multilevel coupled model transformation rules, according to the concepts presented in Chapter~\ref{chap:mcmt}.
All tools are available at the MultEcore website~\cite{multecore}.

\section{MLM Tooling}
\label{sec:modelling-tooling}

\begin{figure}[ht]
	\centering
	\includegraphics{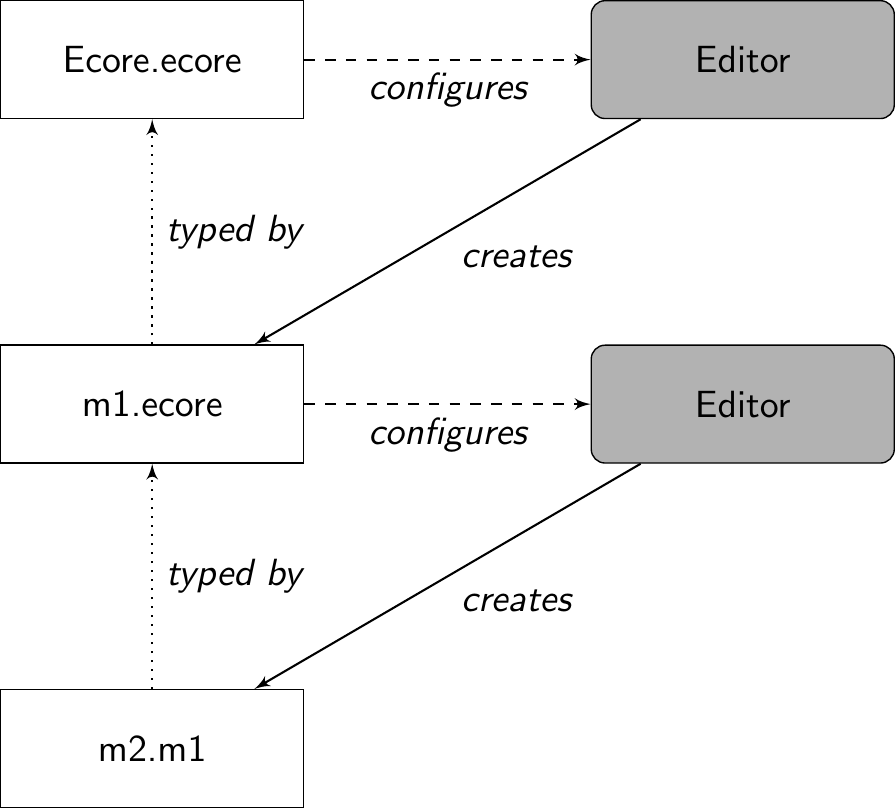}
	\caption{An overview of the modelling process in EMF}
	\label{fig:emf-modelling-process}
\end{figure}

The central MultEcore tool is a set of Eclipse plugins which aim at combining the best from traditional (fixed-level) and multilevel modelling: the mature tool ecosystem and familiarity of the former, and the expressiveness and flexibility of the latter~\cite{multecore,macias2016multecore,macias2017challenge}.
The first part is achieved by providing full compatibility and integration with EMF.
Using MultEcore, model designers can create multilevel hierarchies while still keeping all the advantages they get from the Eclipse Modelling Framework (EMF).
Hence, one of the goals for the tooling aspect of our approach is keeping full compatibility with EMF.
This way, the user does not depend on a set of \emph{closed-garden} tools where the lack of proper documentation and maintainability is common.
In order to explain how we achieve such compatibility, let us first examine how the basic modelling process is carried out in the EMF, using Figure~\ref{fig:emf-modelling-process} as reference.
To avoid confusion, notice that the relation \emph{typed by} is equivalent to \textit{instance of}, and that we employ the terms \emph{metamodel} and \emph{instance} to clarify the role that a model is playing (as metalanguage for the definition of other models, or as one of these latter models, respectively), and the word \emph{model} when we do not want to make such distinction.

\begin{enumerate}
	\item The user starts with the EMF self-defining meta-metamodel \texttt{Ecore.ecore} at her disposal.
	\item She uses an editor compatible with Ecore (which can be graphical, tree-based or textual) to create her own metamodel, which we call \elementname{m1.ecore} in the figure.
	This metamodel is typed by \elementname{Ecore.ecore}.
	The typing relationship means that the elements in this model are instances of Ecore elements, such as \elementname{EClass} and \elementname{EReference}.
	\item Once the user has created a custom model, she can use it to create a second editor.
	\item The user can now build instances of the custom metamodel \elementname{m1.ecore} using this second editor.
	The usual convention in EMF is that this instance has a file extension which is the same as the name of its metamodel.
	Hence the name \texttt{m2.m1} in Figure~\ref{fig:emf-modelling-process}.
	This extension has no effect on the serialization of the instance, which always uses the XMI format.
\end{enumerate}

\begin{figure}[ht]
	\centering
	\includegraphics{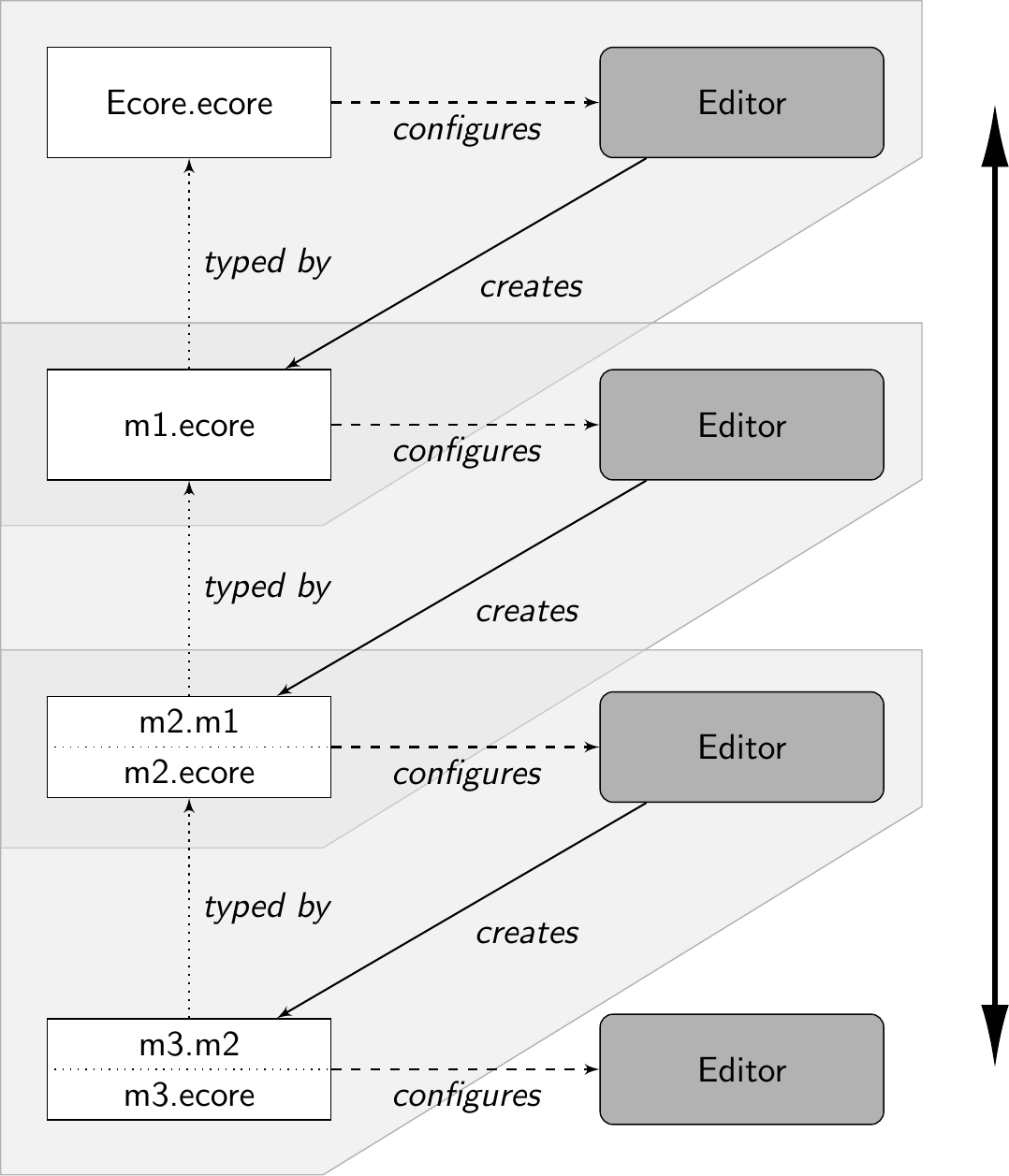}
	\caption{Sliding window: repeated, bidirectional application of two-level cascading}
	\label{fig:sliding-window}
\end{figure}

At this point, the EMF hierarchy reaches its limit: it is not possible to repeat again the process of creating an editor from the bottommost model (XMI instance).
In other words, it is not possible to create a new model which uses \elementname{m2.m1} as metamodel and is therefore typed by it.
If the XMI instance \elementname{m2.m1} could be represented as a model (in Ecore format), the process of creating an \elementname{m3.m2} instance could be carried out in the same way as the two last steps of the process described above.
MultEcore allows the user to do exactly that.
Let us assume that there is a way to exchange the representation of any model between Ecore and XMI, so that they convey the same information, and can act both as a metamodel (in the former case, to create new instances) or as an instance of another metamodel (in the latter case).
If such a thing is possible, the process illustrated in Figure~\ref{fig:emf-modelling-process} does not need to stop at \elementname{m2.m1}.
As Figure~\ref{fig:sliding-window} shows, \elementname{m2.m1} is transformer into its metamodel version: \elementname{m2.ecore}, and can now play the same role as \elementname{m1.ecore} did in Figure~\ref{fig:emf-modelling-process}.
It can be used to generate an editor, which in turns allows the \elementname{m3.m2}.
This new model can be exchanged into its Ecore version too (\elementname{m3.ecore}), allowing the user to repeat the process indefinitely ``downwards'' in the hierarchy.

It is also desirable that the process of exchange of representations can be also done ``upwards'' in the hierarchy, so that an instance (XMI) which was turned in a metamodel (Ecore) can be transformed back into its XMI representation and manipulated as an instance of its metamodel.
That is, the process of exchange between representations must be \emph{repeatable} and \emph{bidirectional}.
To achieve this, we implemented in MultEcore the concept of \emph{sliding window}, shown in Figure~\ref{fig:sliding-window}.
This concept extends the idea of \emph{two-level cascading}~\cite{atkinson2005concepts} (also referred to as \emph{promotion transformation}~\cite{delara2015dsmml}) in order to make it repeatable and bidirectional.
Repeatable implies that, instead of applying the two-level cascading just once, we allow to repeat it every time the user requires to create a new model as an instance of a previously existing instance, by transforming the latter into a model that can be instantiated.
In EMF, this transformation required generating an Ecore metamodel from an XMI instance.
Bidirectional implies that the process of cascading can also be reversed while preserving all the information already existing in all models.
That is, re-generating an XMI instance from an Ecore metamodel.

Since EMF can only manage two levels (one Ecore metamodel and its XMI instance) at once, we define our window as the two levels that a user is manipulating at an specific point in time, which are only a fragment of the whole hierarchy (to which EMF is oblivious).
This window is shown in Figure~\ref{fig:sliding-window}, as a background trapezium shape which comprises these two levels and the editor that is used to manipulate the instance based on the information from the metamodel.
When the user slides the window upwards or downwards to manage different parts of the hierarchy, the representations of each level need to change accordingly.
We overcome the two-level limitation of EMF by creating an intermediate, level-agnostic, graph-based representation of a model, independent of its level, original format and typing relations.
This representation can be generated from both an XMI instance and an Ecore model, and can be used to regenerate both of them.
As a consequence, the user can get either the model or instance representation of a model in any level arbitrarily.

\begin{figure}[ht]
	\centering
	\includegraphics[width=\textwidth]{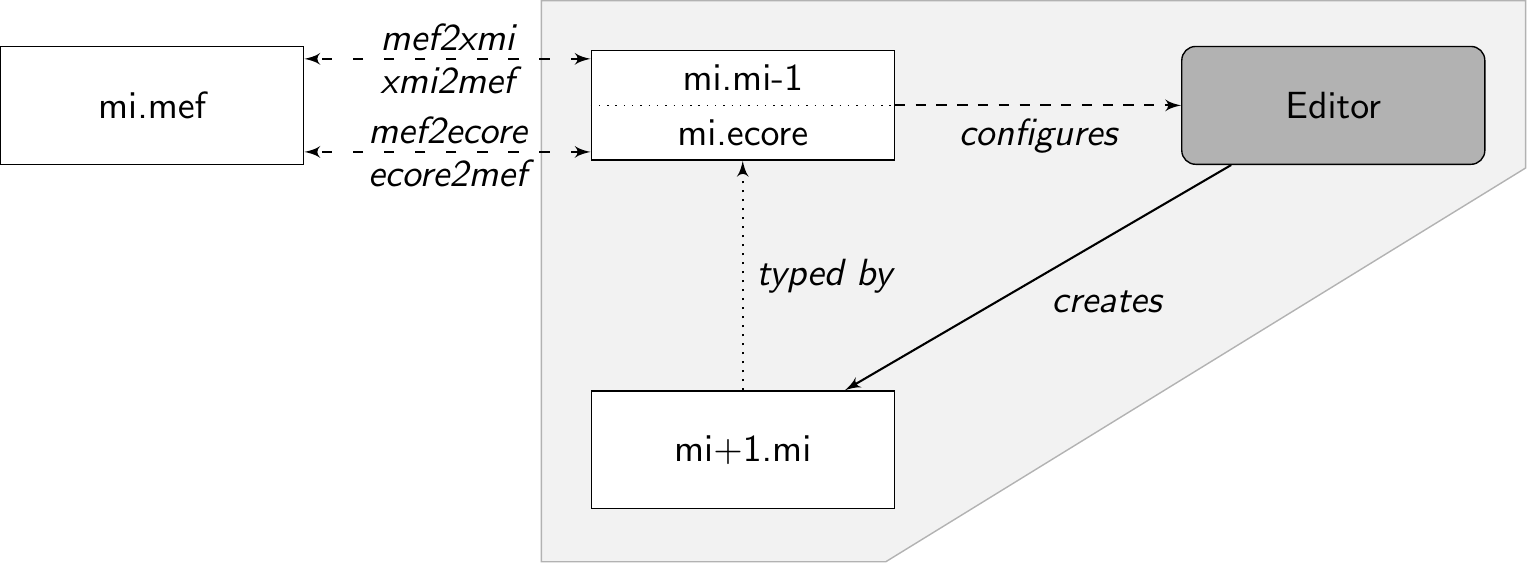}
	\caption{Implementation of the sliding window using MEF files} 
	\label{fig:sliding-window-mef-format}
\end{figure}

Figure~\ref{fig:sliding-window-mef-format} displays how our intermediate model representation is connected to the concept of sliding window by switching among the formats aforementioned, given an arbitrary metamodel.
We use an XML representation called MultEcore Format, and abbreviated MEF, for our intermediate representation.
In this abstract representation, models are considered to be attributed-graph based.
Therefore, they only consist of \textit{elements} and \textit{relations} (equivalent to \texttt{EClass} and \texttt{EReference} in EMF).
In addition, we store information such as the name, type, and potency of each element.

We can see in Figure~\ref{fig:sliding-window-mef-format} how, for an arbitrary model \elementname{mi}, we can manipulate both its Ecore representation (\elementname{mi.ecore}) and its XMI representation (\elementname{m1.mi-1}) by exchanging their information with our custom-made file (\elementname{mi.mef}).
This synchronisation among the three representations is achieved by four transformations
We defined two translations for each representation:

\begin{itemize}
	\item \textit{mef2xmi}, to get the instance representation of a level;
	\item \textit{xmi2mef}, to reflect the changes from the instance into the MEF file;
	\item \textit{mef2ecore}, to generate the metamodel representation of a level; and
	\item \textit{ecore2mef}, to reflect the changes from the metamodel into the MEF file.
\end{itemize}

If our window slides down one level, the instance previously at the bottom of the window needs to be transformed into an Ecore model.
On the contrary, sliding the window up one level requires to transform the model on top into an instance, and generating a model from the level above.
This way, the user can (1) edit the instance \elementname{mi.mi-1} and then slide down the window by switching to its metamodel representation \elementname{mi.ecore}, which allows to create or edit an instance \elementname{mi+1.mi}; or (2) get the instance representation of the current Ecore model to edit it, which implies sliding up the window.

It is worth pointing out that the execution of \textit{mef2xmi} for a model \elementname{mi} triggers the execution of \textit{mef2ecore} for the model \elementname{mi-1}.
The four translations are executed in a user-transparent fashion and their execution times are negligible.
This way, the user is only aware of the existence of several levels and the possibility of getting the model or the instance representations for each of them.
Any change on the generated models is automatically reflected in the corresponding MEF file every time an Ecore or XMI file is updated and saved.

Moreover, the compatibility of our approach with conventional EMF allows to insert any pre-existing Ecore model into the ontological stack by just specifying the typing relations with the adjacent levels.
Thus, all the already existing two-level tools designed for standard EMF can be used seamlessly with our hierarchies.
Furthermore, this compatibility provides all the modelling capabilities of Ecore by default, such as abstract classes, specialization (inheritance) and multiplicity for references and attributes.

It is also important to note that, in the actual tool implementation, explicit level numbers are replaced by references from a model to its immediate metamodels, since the numbers are only needed in our approach for the formal constructions.
In the hypothetical case where these numbers were introduced and needed to be updated when new levels are added on top of them, traversing the hierarchy and updating them would be automatic and almost trivial.

\subsection{MultEcore Eclipse environment}
\label{subsec:multecore-eclipse-environment}

\begin{figure}[ht]
	\centering
	\includegraphics[width=\textwidth]{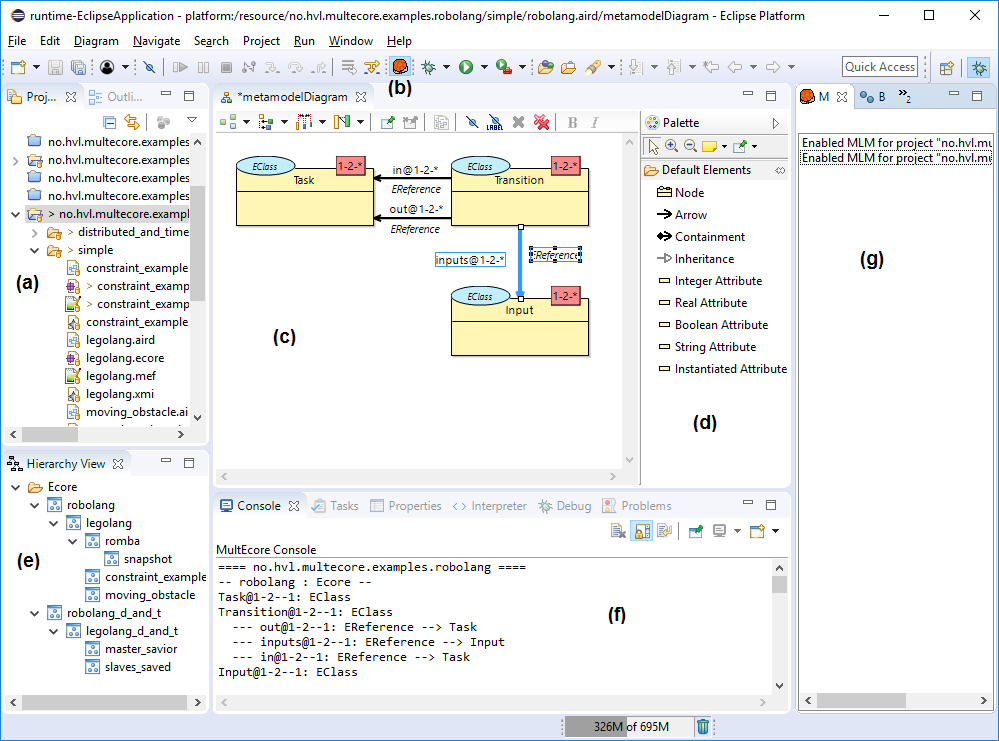}
	\caption{Modelling environment for MultEcore in Eclipse} 
	\label{fig:multecore-screenshot-general}
\end{figure}

A normal configuration of the views and functionalities offered by the MultEcore tool may look like the screenshot from Figure~\ref{fig:multecore-screenshot-general}.

Following the aim of being as transparent as possible, MultEcore works by selecting a project in Eclipse's default Project Explorer, displayed in Figure~\ref{fig:multecore-screenshot-general}(a), and simply clicking the only button which we add to Eclipse's toolbars, shown in Figure~\ref{fig:multecore-screenshot-general}(b).
This toggle-like button enables (or disables) the multilevel features of a project, by scanning all Ecore, XMI and MEF files in the project and creating a runtime version of the hierarchy.
From the moment MLM is enabled in a project, MultEcore will keep track of the changes to any Ecore or XMI file, and react accordingly to keep the whole hierarchy updated.
We do not, for now, support repairing or co-evolution operations within MultEcore.

Although the Project Explorer and the MultEcore button are the only requirements for the tool to work properly, we created a complementary set of views which can facilitate the creation and management of multilevel hierarchies.
We present them in the following.

\subsubsection{Graphical editor}
\label{subsubsec:graphical-editor}

\begin{figure}[ht]
	\centering
	\includegraphics[width=\textwidth]{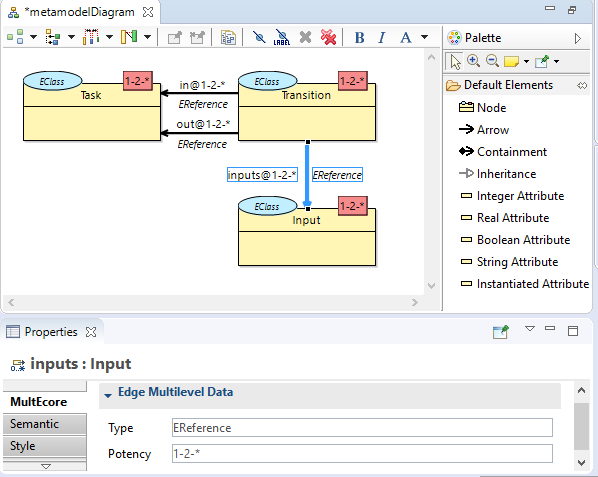}
	\caption{Graphical editor in MultEcore} 
	\label{fig:multecore-screenshot-editor}
\end{figure}

MultEcore provides a custom graphical editor implemented in Eclipse Sirius~\cite{sirius}, designed specifically for multilevel modelling.
This editor (also called \emph{viewpoint} in Sirius' terminology) has been used to generate the example hierarchies displayed in Sections~\ref{sec:three-dimensions} and~\ref{sec:case-studies}.
Figure~\ref{fig:multecore-screenshot-editor} shows a closer look at this editor from Figure~\ref{fig:multecore-screenshot-general}(c), with the \elementname{robolang} language in the editing window.

All of the modelling aspects and the full graphical syntax used in the rest of this thesis are supported in MultEcore, including the type of a node as a blue ellipse, the type of an arrow as a label in italics, the potency of nodes in a red rectangle and so on.
The screenshot also gives a closer look at the palette from Figure~\ref{fig:multecore-screenshot-general}(d) which can be used to create new elements, and the additions to Eclipse's Properties View, where a new tab called ``MultEcore'' is shown with multilevel-specific information like the type and potency of the selected edge.
Apart from that, the names (labels) of most elements can also be edited directly in the editor, by selecting them and typing directly the new value.

\subsubsection{Hierarchy View}
\label{subsubsec:hierarchy-view}

\begin{figure}[ht]
	\centering
	\includegraphics{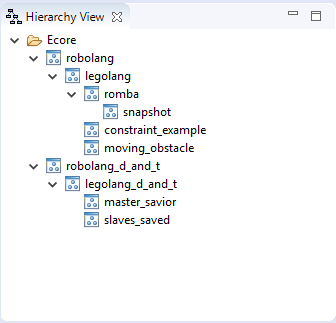}
	\caption{Hierarchy View in MultEcore} 
	\label{fig:multecore-screenshot-hierarchy}
\end{figure}

Figure~\ref{fig:multecore-screenshot-hierarchy} shows MultEcore's hierarchy view, depicted also in Figure~\ref{fig:multecore-screenshot-general}(e).
This window lists all the models in a project, using a tree viewer.
This means that the root of the hierarchy, displayed as a folder, is always Ecore, and its first-level children are the models in level \(1\) which are direct instances of Ecore (e.g.\ \elementname{robolang}).
These elements may have children of their own, representing their direct instances, and so on (e.g.\ \elementname{legolang} is a direct instance of \elementname{robolang}).

If MLM is not enabled in the selected project in the Project Explorer, this view will show a message stating precisely that.
Otherwise, double-clicking in any of the displayed models (except for Ecore) will open the graphical editor associated to that model.

\subsubsection{MultEcore Console}
\label{subsubsec:multecore-console}

\begin{figure}[ht]
	\centering
	\includegraphics[width=\textwidth]{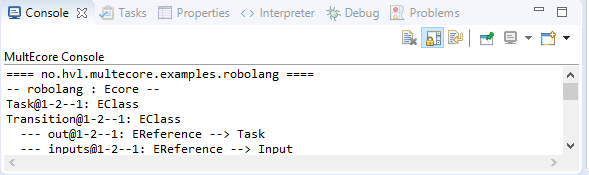}
	\caption{Custom console with MultEcore} 
	\label{fig:multecore-screenshot-console}
\end{figure}

The standard Eclipse console can be replaced by one which displays a textual version of the hierarchy, useful to see at a glance all the models on the hierarchy, their direct metamodels, and their elements and types.
This console is shown in Figure~\ref{fig:multecore-screenshot-general}(f), and more closely in Figure~\ref{fig:multecore-screenshot-console}.
This console can be selected by clicking in the drop-down list visible in the top right of the figure.

The console shows the information of the hierarchy once MLM is enabled on a project, and offers a compact, read-only representation of all the contents of the multilevel hierarchy associated to that project.
For example, Figure~\ref{fig:multecore-screenshot-console} shows the information about the Robolang hierarchy and part of the \elementname{robolang} model and its elements.

\subsubsection{MultEcore Log}
\label{subsubsec:multecore-log}

\begin{figure}[ht]
	\centering
	\includegraphics{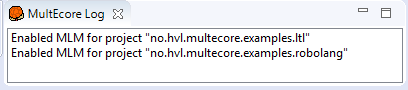}
	\caption{Custom console with MultEcore} 
	\label{fig:multecore-screenshot-log}
\end{figure}

This view was created in the early stages of the tool's development.
As Figure~\ref{fig:multecore-screenshot-log} shows, the only information that it currently shows is the projects in which MLM is enabled or disabled.

We preserve it in the current state of the tool since it can be useful to track other actions like file updates (disabled for now for being too verbose), MCMT proliferation (see Section~\ref{sec:mcmt-tooling}) or future extensions to the tool.

\section{MCMT Tooling}
\label{sec:mcmt-tooling}

We present in this section our current results towards the full support for MCMT rule specification, proliferation and execution. 
Specifically, we introduce in Section~\ref{subsec:textual-dsml} our textual language for the specification of MCMTs, equivalent to the graphical representation used in Sections~\ref{subsec:mcmts-robolang} and~\ref{sec:case-studies}, and its associated editor.
We also present, in Section~\ref{subsec:proliferation}, a proliferation algorithm that generates two-level MTs out of given set of MCMTs, which can then be used on conventional model transformation engines.
This algorithm depends on two other algorithms, which are presented in Sections~\ref{subsec:matching-algorithm} and~\ref{subsec:graph-matching}.

\subsection{Textual DSML for MCMTs}
\label{subsec:textual-dsml}

\begin{figure}
	\centering
	\includegraphics[width=\textwidth]{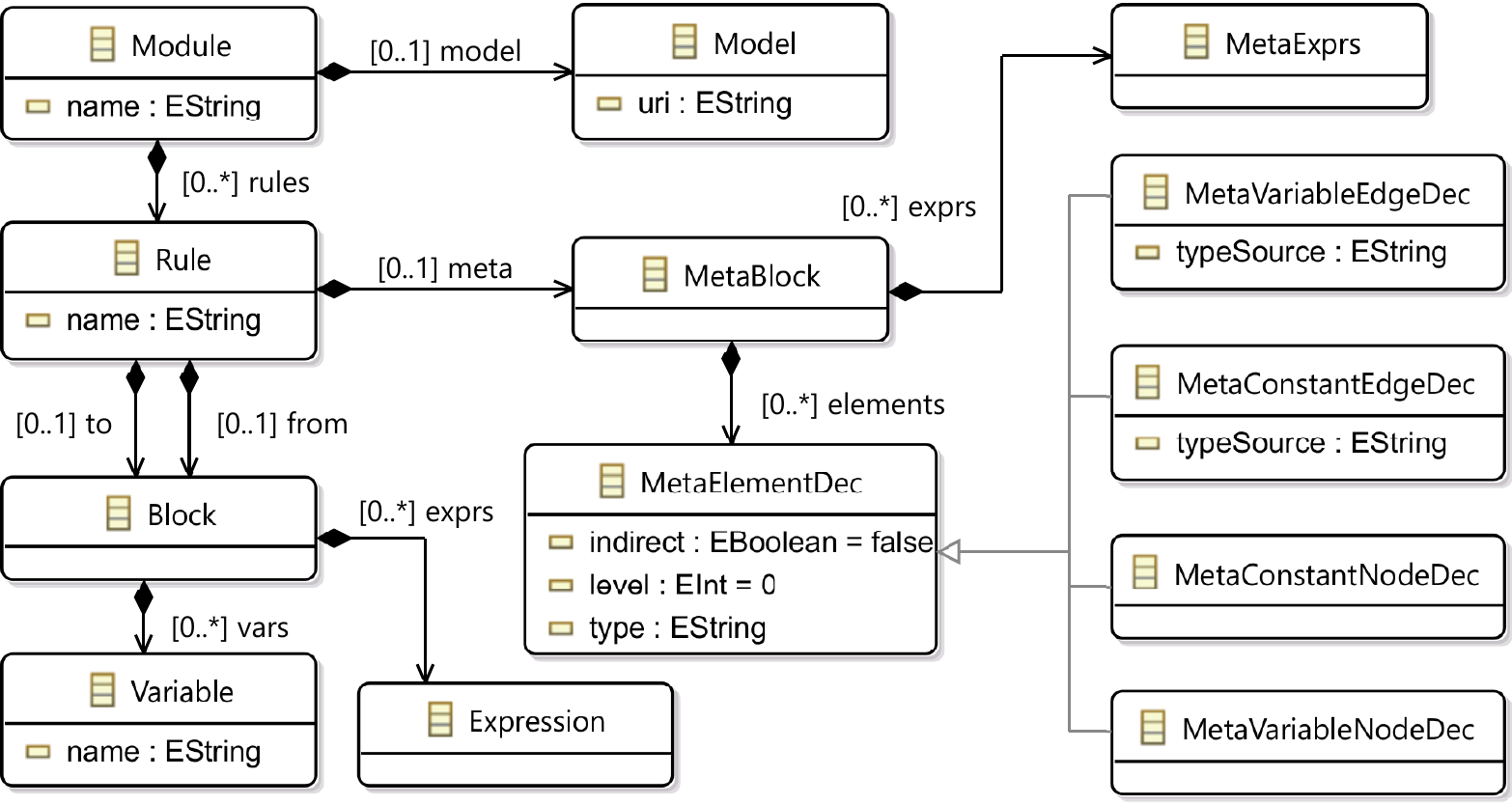}
	\caption{Fragment of the metamodel defining the abstract syntax of our textual DSML}
	\label{fig:textual-dsml-metamodel}
\end{figure}

The abstract syntax of our DSML for MCMTs is defined as an Ecore metamodel, partially depicted in Figure~\ref{fig:textual-dsml-metamodel}.
This metamodel has been used to create an editor using Xtext~\cite{bettini2016implementing} which provides syntax highlighting, error detection, code completion and outline visualization.

This DSML provides the specification of modules containing a collection of multilevel coupled transformation rules defined independently from a hierarchy, but that will be matched against one during the proliferation process (see Section~\ref{subsec:proliferation}).
That is, the MCMTs are applied to a single hierarchy in each execution.

\begin{figure}
	\centering
	\includegraphics[width=\textwidth]{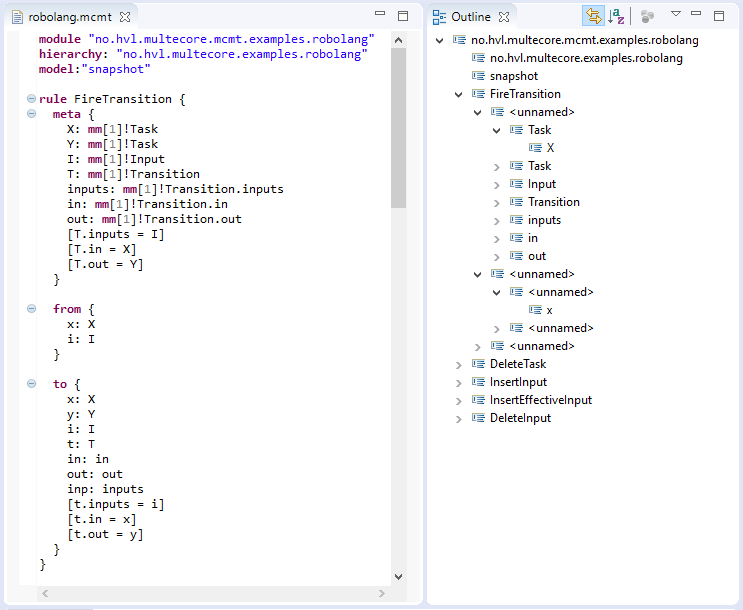}
	\caption{Screenshot of the MCMT editor}
	\label{fig:mcmt-editor-screenshot}
\end{figure}

In order to briefly present this DSML, we use the textual representation of the MCMT rule \emph{Fire Transition}, already introduced in Section~\ref{sec:why-mcmt}.
This textual version is shown in Figure~\ref{fig:mcmt-editor-screenshot}, from a screenshot of our Eclipse-based editor.
As stated in Section~\ref{sec:why-mcmt}, \textsf{Rule} elements are structured into three organizational components, namely META, FROM and TO, named in lowercase in the textual syntax.
These blocks contain graph pattern declarations, plus expressions that relate the elements to each other.
The \textsf{meta} block must contain a valid, non-empty pattern, but the \textsf{from} and \textsf{to} blocks may be empty.
In the \textsf{meta} block, a pattern is specified by means of \textsf{MetaElementDec}, used to declare both nodes and edges, which can in turn be constant or not.
Also, this block may contain assignment expressions, used in this example to specify the structural relationships between the declared nodes by means of the declared edges.
In the example, \textsf{X} and \textsf{Y} are declared as node variables, while \textsf{in} and \textsf{out} are declared as a edge variables.
Specifying any of these as a constant would just require to add the character \textsf{\$} as a suffix to their type.
Note that the type names of \textsf{meta} block elements are suffixed by the keyword \textsf{mm} followed by the level (index) at which these types is actually declared, starting from \(1\) and increasing downwards.
Note that this editor assumes that Ecore is at level \(0\) and does not have to be explicitly given.

In the \textsf{from} and \textsf{to} blocks we can define patterns according to the \textsf{MetaElementDec} previously declared in the \textsf{meta} part.
In the example, the \textsf{from} block of the rule defines a pattern consisting of just two variables \textsf{x} and \textsf{i}, while its \textsf{to} block comprises seven variable declarations and three assignment expressions among them.
This pattern mirrors precisely the one depicted in Figure~\ref{fig:robolang-rule-fire-transition}.

\subsection{Proliferation process}
\label{subsec:proliferation}

As mentioned in Section~\ref{sec:why-mcmt}, an MCMT rule can generically, but precisely, replace a number of two-level rules which are applicable to just one given source model.
For matching an MCMT in a modelling hierarchy and executing it, we can consider two different approaches.

A first approximation requires the implementation of a custom execution engine that tackles MCMTs and multilevel modelling stacks.
This kind of approach would offer a more seamless execution of MCMTs in an integrated environment, also facilitating possible extensions and eliminating the necessity of back and forth translation of hierarchies and MCMTs to existing transformation engines.
However, the cost of implementing such a custom engine from scratch, including the re-implementation of basic functionalities which already exist in transformation engines, would be too high.
Moreover, it might lock the transformation engine to a single MLM-tool.

As an alternative, it is possible to perform a pre-processing step that can automatically generate the operationally equivalent two-level rules which, while correct, are in general too numerous and cumbersome to specify by hand.
This approach can be combined with other well-established, well-understood and optimised transformation engines like Henshin~\cite{arendt2010henshin}, Groove~\cite{rensink2003groove,ghamarian2012groove}, ETL~\cite{kolovos2008epsilon} or ATL~\cite{jouault2008atl}.
In other words, this process of automatic proliferation results in a more flexible implementation, in which we do not even strictly depend on a particular transformation engine.
Besides, the process of adapting both the target hierarchy and the MCMTs to a two-level setting can be done automatically, and indeed needs to be performed only once before running the whole scenario.

Specifically, our automatic proliferation procedure consists in the generation of two-level model transformation rules for every possible match of the variables defined in an MCMT.
The goal of this process is to replace the variability in the types of the elements in the FROM and TO blocks,  which cannot be interpreted by mainstream model transformation engines, by specific types, so that we can generate automatically an executable set of rules for any two-level engine.

Once specified, an MCMT is run by specifying in which model it will be applied, as in any model transformation engine.
The specification of the input model, which is part of a multilevel hierarchy, uniquely identifies its branch in the hierarchy tree.
That is, we can ignore at execution time the same-level sibling models of the input one as well as the sibling models of any metamodel of the input model in the levels above.
Furthermore, in the exceptional case that the input model is not in the bottommost level of the hierarchy, all models in the levels below it can be ignored for the execution.
As a consequence of these simplifications, the relevant part of the hierarchy will always be a stack of models, without any branches.
Hence, we use hereafter the term \emph{stack} instead of \emph{hierarchy}.

The way to replace the variable types of the MCMT by specific ones, is by looking at the META part of the MCMT rule and finding all possible matches of its variables into the types defined in the modelling stack.
Looking back to Figure~\ref{fig:multilevel-coupled-rule-formal}, this process will establish all the possible maps \(\mapruletohierarchybinding_\exampleindexone\), and generate one two-level rule for each valid combination that involves one map per level in the META part, which is represented as the sequence of graphs \(\graphname{\rulegraph}{\exampleindexone}\).
Once proliferated, it is necessary to perform a second matching process \(\mapruletohierarchyleft\) between the FROM graph \(\graphname{\rulegraphleft}{}\) and the input model \(\graphname{\hierarchygraphleft}{}\), in order to actually apply the now proliferated rule.
Such operation will be performed by the selected two-level engine, so both \(\graphname{\rulegraphleft}{}\) and \(\graphname{\hierarchygraphleft}{}\) are excluded from the proliferation algorithm.

The core of our MCMT proliferation mechanism is a matching algorithm between the meta part of an MCMT rule and the stack of metamodels of the input model. 
We describe our matching algorithm in the following sections.

\subsection{Matching algorithm}
\label{subsec:matching-algorithm}

The algorithm that generates the list of all possible values of the variables that yield valid two-level MTs is shown in Algorithm~\ref{alg:matching}.
It consists of a recursive function with four input parameters and an input/output parameter.

\begin{algorithm}[tb]
	\footnotesize
    \caption{Matching algorithm}
    \label{alg:matching}
    \begin{algorithmic}[1] 
      \Procedure{Match}{$MM,TG,\mathit{mmLevel},\mathit{tgLevel},matches$}
        \If{$\mathit{mmLevel} = MM.size$}
          \State \textbf{return} $true$ \Comment{End of pattern reached}
        \EndIf
        \State $\mathit{found} \gets false$
        \While{$\mathit{tgLevel}<TG.size$} \Comment{Every level in hierarchy}
          \State $\mathit{maps \gets graphMatch(MM[mmLevel],TG[tgLevel],matches[matches.size - 1])}$ \label{alg:matching:graphmatch}
          \ForAll {$m \in maps$}
            \State $size \gets matches.size$
            \If{$size > 0\ \textbf{and}\ \mathit{mmLevel} > 0$}
              \State $\mathit{currentMatch} \gets matches[size - 1]$ \label{alg:matching:current-1}
              \State $matches[size - 1] \gets \mathit{currentMatch} \cup m$ \label{alg:matching:union-m}
              \State $\mathit{found} \gets \mathit{found}\ \textbf{or}\ match(MM,TG,\mathit{mmLevel}+1,\mathit{tgLevel}+1,matches)$
              \State $matches[matches.size] \gets \mathit{currentMatch}$ \label{alg:matching:current-2}
            \Else
              \State $matches[size] \gets m$
              \State $\mathit{found} \gets \mathit{found}\ \textbf{or}\ match(MM,TG,\mathit{mmLevel}+1,\mathit{tgLevel}+1,matches)$
            \EndIf
          \EndFor
          \State $\mathit{tgLevel} \gets \mathit{tgLevel} + 1$
        \EndWhile
        \If{$\mathit{mmLevel} > 0$}
          \State $matches[matches.size-1] \gets \emptyset$  \Comment{Remove incomplete match}
        \EndIf
        \State \textbf{return} $\mathit{found}$ \label{alg:matching:return}
      \EndProcedure
    \end{algorithmic}
\end{algorithm}

The first two input parameters, \algstyle{MM} and \algstyle{TG}, represent all the metalevels of an MCMT depicted in Figure~\ref{fig:multilevel-coupled-rule-formal} as \(\graphname{\rulegraph}{\exampleindexone}\) and \(\graphname{\hierarchygraph}{\exampleindextwo}\).
Hence, they represent the META part of an MCMT (the typing chain \algstyle{MM}) and the levels above the one were the MCMT will be applied (\algstyle{TG}).
That is, \algstyle{MM} represents the pattern that the algorithm tries to match and \algstyle{TG} represent the stack against which the algorithm tries to match \algstyle{MM}.
Both inputs are accessed level by level, in all valid combinations, as the algorithm progresses.
This way, it establishes all possible \emph{bindings} between pairs of levels that conform a full \emph{match}, and therefore result in a proliferated rule.

The next two parameters, \algstyle{mmLevel} and \algstyle{tgLevel}, are the indexes used to indicate the current levels that are being accessed in \algstyle{MM} and \algstyle{TG}, respectively.
In the first call, both have value zero, which will increase as the algorithm makes recursive calls.
Since we number the levels of both the \algstyle{MM} and \algstyle{TG} increasing downwards, the matches will be generated in a top-down manner.
This decision is not arbitrary: the match between two levels depends, among other things, of the types of their elements; since those types may be other elements in levels above, these need to be matched first, so that we can check that the types of any two elements are compatible.
That is, the type in the pattern has been previously mapped to the type in the stack.

The last parameter, \algstyle{matches}, contains the list of all valid matches found by the algorithm.
It is initially empty, and it is required to work as an input/output variable (by reference) so that the algorithm can modify it in each recursive call and, afterwards, propagate the result of such calls to provide the final list.
For the same reason described in the previous paragraph, the information of the maps established so far for any match needs to be passed to \algstyle{graphMatch} so that it can be taken into account to produce only those maps that are consistent with the maps of the types.

The algorithm depends on an auxiliary function \algstyle{graphMatch} (line~\ref{alg:matching:graphmatch}) that calculates the possible matches between two models.
This auxiliary algorithm is described in Section~\ref{subsec:graph-matching}.

The function returns a boolean value that indicates whether the algorithm has found valid matches or not.
In the latter case, the rule would not be applicable and, consequently, would not be proliferated.
This value is also used by the algorithm to filter all potential matches that are discarded, in case their level-wise list of bindings could not be completed.

The algorithm works as follows.
First, the base case of the recursion can be triggered if the \algstyle{mmLevel} has reached the size of the pattern of the rule (\algstyle{MM}).
This would indicate that the algorithm has found a sequence of bindings all the way down to the last level of \algstyle{MM}, which constitute a full match.
If that is not the case, the first time that a pair of levels is explored as part of one sequence of matches, the flag \algstyle{found} is set to false.

The core of the algorithm is a while loop that attempts to match the current \algstyle{mmLevel} with all the remaining \algstyle{tgLevel}s.
The first value is the one which was passed to the function in the call, which will increase by 1 in each iteration.
On each iteration, the list of maps that are possible to establish for \algstyle{mmLevel} and \algstyle{tgLevel} are calculated with the call to \algstyle{graphMatch}.
This function takes the two levels of \algstyle{MM} and \algstyle{TG} indicated by the indexes, and gives back a list of maps from all the elements in the pattern into elements of the multilevel stack.
This results are stored in the \algstyle{maps} variable, and then iterated in the for loop that performs the recursive calls in order to complete the match.

Inside the for loop, the current match \algstyle{m} is stored in \algstyle{matches}, and the recursive call is made afterwards.
However, a case distinction is required.
In those cases where both the size of the \algstyle{matches} variable and \algstyle{mmLevel} are not zero, these values indicate that the algorithm has already found some maps for the current match.
In that case, such information is stored in the last position of the list (\algstyle{matches[size-1]}), and \algstyle{m} is simply added to that position (line~\ref{alg:matching:union-m}).
Furthermore, in case more than one match can be generated from the current sequence of maps and other values of \algstyle{m}, that current information is temporarily stored in \algstyle{currentMatch} and duplicated in the last position of \algstyle{matches} after the recursive call (lines~\ref{alg:matching:current-1} and~\ref{alg:matching:current-2}).
In the cases where the \algstyle{size} of \algstyle{matches} is zero (empty list) or \algstyle{mmLevel} is zero, there is no partial information to be preserved before storing \algstyle{m} and making the recursive call, so \algstyle{m} is simply added to the tail of the list, since it is the first map of the sequence.
In both cases, the recursive call is performed in the same way, and its return value is used to indicate if the match was completed after the current map or not.
The result, stored in \algstyle{found}, will be then propagated to the calling function in the same way (line~\ref{alg:matching:return}).

After all iterations of a loop are completed, and due to the duplication described above when \algstyle{mmLevel} and \algstyle{size} are not zero, the last partial result that was not completed is taken away from the list of matches.

Note that \algstyle{mmLevel} and \algstyle{tgLevel} are increased in such a way that it always holds that \(\algstyle{mmLevel} \leq \algstyle{tgLevel}\).
Moreover, every time that \algstyle{mmLevel} is increased, the same happens to \algstyle{tgLevel}.
This ensures that the sequence of graph matches are consistent with their depiction in Figure~\ref{fig:multilevel-coupled-rule-formal}, where two \(\mapruletohierarchybinding_\exampleindexone\) cannot have the same source or target, or ``cross'' each other.

To sum up, the algorithm is first called with the \algstyle{MM} specified in the META part of the MCMT, the \algstyle{TG} that represent the multilevel stack on top of the model to be transformed, both \algstyle{mmLevel} and \algstyle{tgLevel} initialized at zero and \algstyle{matches} initialized empty.
It will recursively explore all valid combination of maps that cover all the levels in the pattern, and return those in \algstyle{matches}, plus a boolean value set to false if no full matches were found.

\begin{figure}[ht]
	\centering
	\includegraphics[width=\textwidth]{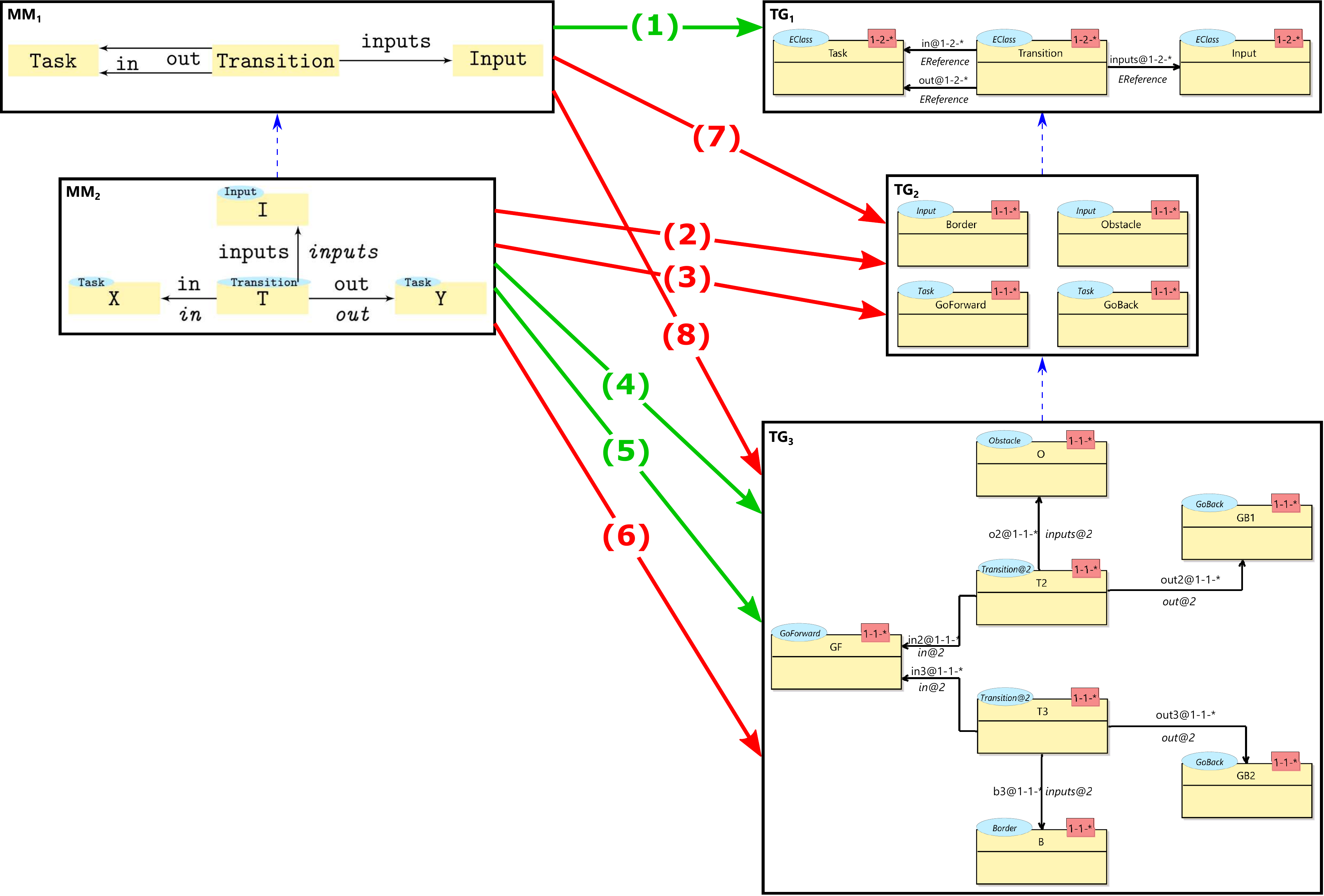}
	\caption{Example of application of the matching algorithm to the rule \emph{Fire Transition}}
	\label{fig:proliferation-example-overview}
\end{figure}

Figure~\ref{fig:proliferation-example-overview} displays an example of the application of Algorithm~\ref{alg:matching} to the MCMT rule \emph{Fire Transition}, which we use as a running example in Chapter~\ref{chap:mcmt} and depict in Figure~\ref{fig:robolang-rule-fire-transition}.
Following the top-down mechanism we already described, the two \(\graphname{\rulegraph}{\exampleindexone}\) representing the META part of the rule are matched against the multilevel stack in top of the model where the rule is applied.
The rule \emph{Fire Transition} is applied in this example to the model \elementname{robot\_1\_run\_1} in Figure~\ref{fig:robolang-multilevel-hierarchy}(d).
Hence, the stack of models \(\graphname{\hierarchygraph}{\exampleindextwo}\) consists of \elementname{robolang}, \elementname{legolang} and \elementname{robot\_1}, from Figure~\ref{fig:robolang-multilevel-hierarchy}(a), (b) and (c), respectively.
For simplicity reasons, the potencies of all elements, as well as the multiplicities of relations, are excluded from the diagram, since they are not specified and hence make no difference in the execution of the algorithm.
Moreover, \(\graphname{\hierarchygraph}{2}\) and \(\graphname{\hierarchygraph}{3}\) contain only fragments from \elementname{legolang} and \elementname{robot\_1} in order to simplify the example.

In the Figure, we represent how the algorithm tries to match every level in the rule META block with every model in the typing chain of the model being transformed.
Valid matches are represented in green, and failed attempts in red.
After completing the algorithm, in this example, the resulting list of matches would contain the information shown in Table~\ref{tab:proliferation-results}.
With the full models instead of the fragments used in the example the list of results would contain seven matches: as many as transitions in \elementname{robot\_1\_run\_1}.

\begin{table}[tb]
\centering
\caption{Matches found in the the proliferation example}
\label{tab:proliferation-results}
\begin{tabular}{|c|l||c|l|}
	
	\hline
	\multicolumn{2}{|c||}{\bfseries Match 1}				& \multicolumn{2}{c||}{\bfseries Match 2}				\\
	\hline
	\multirow{6}{*}{MM1 - TG1}	& Task - Task				& \multirow{6}{*}{MM1 - TG1} & Task - Task				\\
								& Transition - Transition	& 							 & Transition - Transition	\\
								& Input - Input				& 							 & Input - Input			\\
								& in - in					& 							 & in - in					\\
								& out - out					& 							 & out - out				\\
								& inputs - inputs			& 							 & inputs - inputs			\\
	\hline
	\multirow{7}{*}{MM2 - TG3}	& X - GF					& \multirow{7}{*}{MM2 - TG3} & X - GF					\\
								& Y - GB1					& 							 & Y - GB2					\\
								& T - T2					& 							 & T - T3					\\
								& I - Obstacle				& 							 & I - Edge					\\
								& in - in2					& 							 & in - in3					\\
								& out - out2				& 							 & out - out3				\\
								& inputs - o2				& 							 & inputs - b3				\\
	\hline
\end{tabular}
\end{table}

\begin{figure}
    \begin{minipage}{0.45\textwidth}
		\centering
		\includegraphics[width=\linewidth]{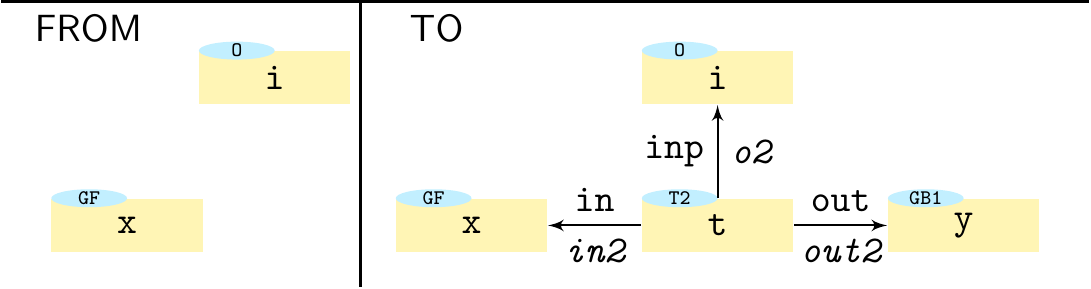}
		\caption{Proliferation of \emph{Fire Transition} for the first match}
		\label{fig:proliferation-example-result-robolang-1}
    \end{minipage}\hfill
    \begin{minipage}{0.45\textwidth}
		\centering
		\includegraphics[width=\linewidth]{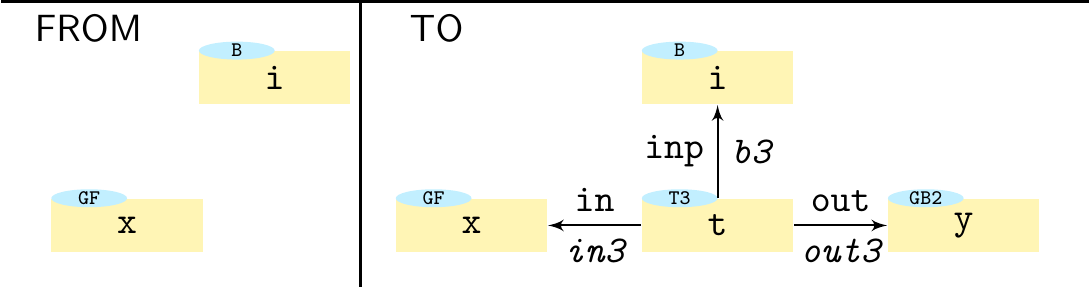}
		\caption{Proliferation of \emph{Fire Transition} for the second match}
		\label{fig:proliferation-example-result-robolang-2}
	\end{minipage}
\end{figure}

Replacing the variable types used in the FROM and TO blocks of the MCMT rule \emph{Fire Transition} by their matches, and removing the META block, we get the two-level rules which encode the same behaviour as the original 
As desired, we get two resulting rules as a result from the proliferation algorithm, depicted in Figures~\ref{fig:proliferation-example-result-robolang-1} and~\ref{fig:proliferation-example-result-robolang-2}.
These two rules are exactly the same ones we specify manually in Figures~\ref{fig:robolang-rule-fire-transition-2-level-1} and~\ref{fig:robolang-rule-fire-transition-2-level-2}, just with different names for the variables which do not affect their semantics.

\subsection{Graph matching}
\label{subsec:graph-matching}

The match between a graph defined in the META part of an MCMT an a subgraph of one of the models in the multilevel stack is done by means of graph homomorphisms, plus some restrictions.
The algorithm for graph matching is a modification of the Ullman algorithm~\cite{ullmann1976algorithm}, as proposed in~\cite{saltz2013fast}.
Basically, we take into account some modelling aspects in order to adapt the process from pure graphs to modelling.

Recall that nodes within a graph are uniquely identified by their name, which act as unique identifiers.
As for arrows, names can be reused as long as the source or target nodes (at least one) are different.
Moreover, both nodes and arrows are fundamentally defined by their types.
Hence, in order to match any element, it is required that the one in the pattern and its counterpart in the hierarchy have matching types.
This restriction does not conflict with the possibility of having variable types.
See for example the nodes \element{x}{X} and \element{x}{Y}.

As shown in Section~\ref{subsec:proliferation}, the proliferation algorithm works in a top-down manner, so that before matching any element, its type has already been matched, and it is possible to use that information for the current match.

For arrows, where multiplicity is part of their definition, the lower and upper bounds can also be used for matching purposes, taking into account that a more restrictive multiplicity in the pattern will match a less restrictive one in the hierarchy.
For example, a pattern arrow with multiplicity \elementname{1..2} will match a hierarchy arrow with multiplicity \elementname{0..3} if the rest of the matching conditions are met.

In addition, the notion of potency can also be used in the match if required, although it is ignored if not specified.
In a similar manner to arrow multiplicity, the less restrictive ones in the hierarchy (broader intervals) can be matched by more restrictive patterns (narrower intervals).

Lastly, the distinction between variables and constants, already presented in Section~\ref{subsec:mcmts-ltl}, also influences the graph matching algorithm.
In this case, the name of the element is taken into account for the matching.
For nodes, the algorithm must find a corresponding node with all the restrictions aforementioned, plus an equal name, in order to get a successful match.
To properly identify one specific arrow, we need to identify its source as a constant too, at least.
Otherwise, it is possible to match different arrows with the same name and whose source nodes have the same type.

\begin{figure}[ht]
	\centering
	\includegraphics[width=\textwidth]{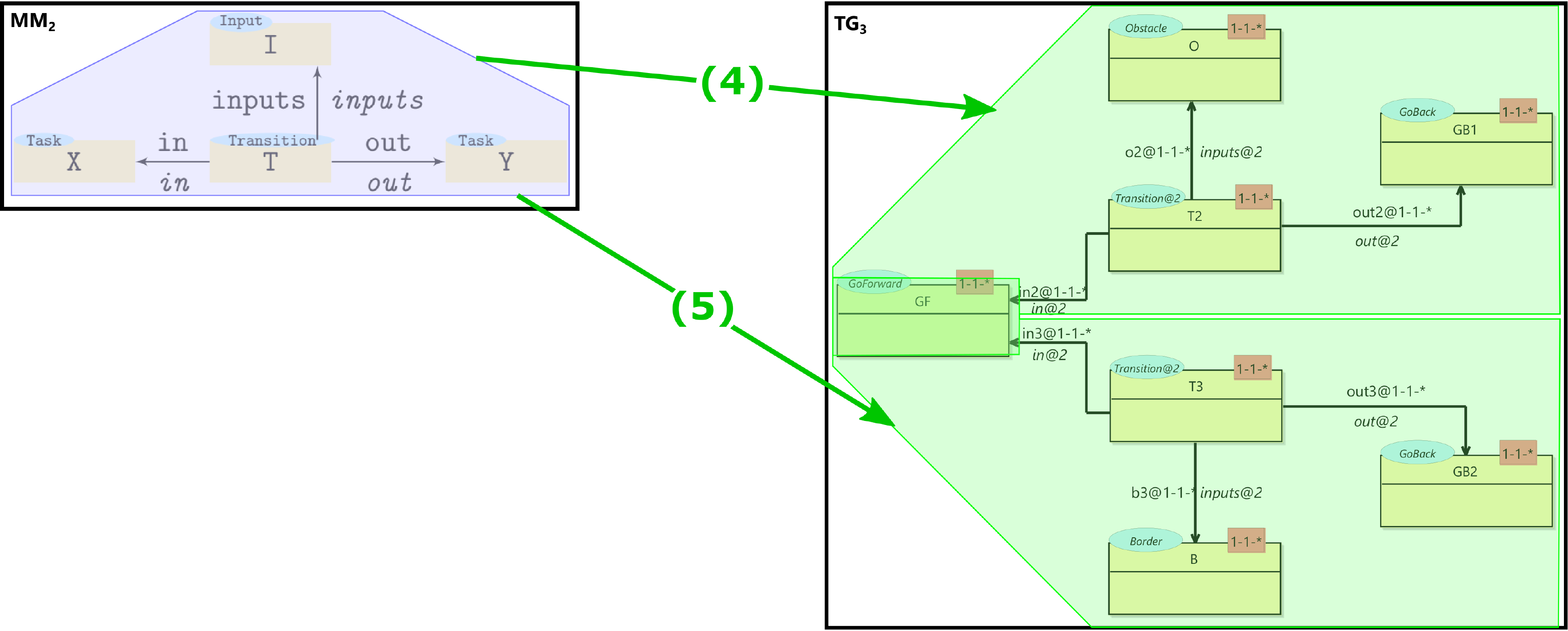}
	\caption{Detail of the graph matching process for the rule \emph{Fire Transition}}
	\label{fig:proliferation-example-detail}
\end{figure}

Figure~\ref{fig:proliferation-example-detail} displays a fragment of Figure~\ref{fig:proliferation-example-overview}, in which, after matching \(\graphname{\rulegraph}{1}\) to \(\graphname{\hierarchygraph}{1}\), the proliferation algorithm proceeds to find matches between \(\graphname{\rulegraph}{2}\) to \(\graphname{\hierarchygraph}{3}\).
As depicted, the pattern (blue) can match two different subgraphs (green), resulting in two different proliferated rules.
This can be achieved since the proliferation process has already matched, before a recursive call, all the elements in \(\graphname{\rulegraph}{1}\) to their homonyms in \(\graphname{\hierarchygraph}{1}\).

%% file: thesis/05-validation.tex
\chapter{Validation}
\label{chap:validation}

In this section, we present the methods used to validate three different aspects of our approach.
First, we introduce the main case studies created with MultEcore in Section~\ref{sec:case-studies}, which allows us to test the expressive capabilities of our approach, both for the construction of DSMLs (syntax) and the specification of their behaviour (semantics).
Second, in Section~\ref{sec:empirical-evaluation} we present an experiment which compares an early version of our tool MultEcore with other MLM tools, and its updated results with a more recent version of our tool.
We discuss how these results have influenced the development of the tool to ensure its suitability to the demands of the MLM community.
We also evaluate our own case studies in a similar manner.
And third, we use some of the examples in these case studies to check that our constructions in the category \cat{Chain} yield the results we require in Section~\ref{sec:examples-constructions-chain}.

\section{Case Studies}
\label{sec:case-studies}

This section presents several examples of multilevel hierarchies and MCMTs.
All of them serve the purposes of illustrating our proposal and validating our approach to multilevel modelling.
Section~\ref{subsec:petri-nets} contains an early approach to semantics reuse for Petri Nets using MLM, without explicit MCMT-based behaviour.
Section~\ref{subsec:extended-robolang} contains an extended version of the running example Robolang, that we use in this thesis.
This version of the DSML and its MCMT-based semantics can be used to specify complex systems with different agents (robots), explicit timing and message passing.
Similarly, Section~\ref{subsec:ltl} defines a supplementary hierarchy with an extended version of LTL which includes the concepts of distribution and explicit time.
In Section~\ref{subsec:pls} we present our multilevel version of a well-known case study from the domain of Product Line Systems, which can be enhanced with the small supplementary hierarchy presented in Section~\ref{subsec:counter} to include counters of the processed elements inside a product line.
Finally, Section~\ref{subsec:bicycle-challenge} details the solution created with MultEcore to a challenge proposed in the context of the MULTI workshops~\cite{multievents}.

Given the large size of some of the models which we include in this section, several multilevel hierarchies cannot be depicted in a single figure while keeping them readable.
In such cases, we show first an overview of the structure of the hierarchy, followed by figures showing the contents of each of the models which conform it.

\subsection{Petri Nets}
\label{subsec:petri-nets}

\begin{figure}[ht!]
	\centering
	\includegraphics[width=\textwidth]{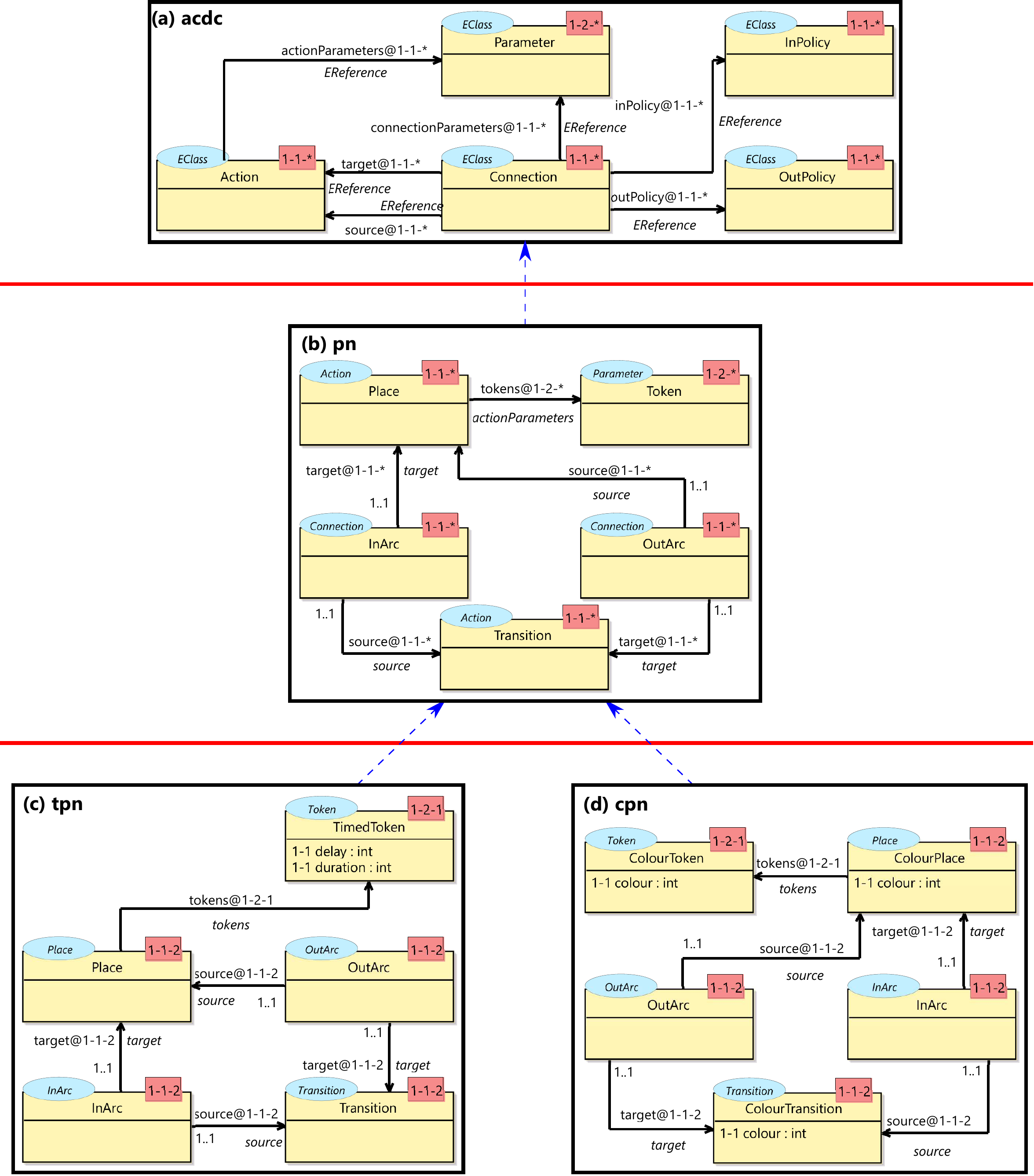}
	\caption{Multilevel hierarchy for the Petri Nets case study}
	\label{fig:petri-nets-multilevel-hierarchy}
\end{figure}

This is an early example of multilevel hierarchy which served us to develop the ideas presented in this thesis.
An early version of it can be found in~\cite{macias2016multecore}.
This initial example approaches the idea of reusing semantics by organising languages in a multilevel hierarchy based on their similarity and degree of abstraction.
Such languages are represented as models and relate to each other using typing relations.
The hierarchy, shown in Figure~\ref{fig:petri-nets-multilevel-hierarchy}, contains four models organised in three levels.

The model in level \(1\) contains the basic concepts of workflow-like languages, such as Petri Nets, BPMN and state machines.
Its name, \elementname{acdc}, is the acronym of \elementname{Action-Connection Domain Concepts}, and contains the node \elementname{Action} as an abstraction of concepts like Petri Net's places, BPMN's activities or states in a state machine.
These are connected by \elementname{Connections}, and the language does not limit the number of actions that can be in its \elementname{source} and \elementname{target}.
The policies to traverse a connection can be specified by instantiating \elementname{InPolicy} and \elementname{OutPolicy}, for example, using \emph{and} to indicate that all actions in the source of a connection must be \emph{active} for the connection to be fired.
Furthermore, the concept of passing information from one step of the workflow to the next can be specified using \elementname{Parameter}, which can be part of both actions and connections, as reflected by the arrows \elementname{actionParameters} and \elementname{connectionParameters}.

As an example of how \elementname{acdc} can be instantiated, the model \elementname{pn} contains a basic specification of Petri Nets, including the concepts of \elementname{Place}, \elementname{Transition} and \elementname{Token} and the two types of arcs than can connect them: \elementname{InArc} for connecting a transition to a place and \elementname{OutArc} for the opposite.
This ensures that places cannot be connected directly to each other, and likewise for transitions.
The multiplicity \elementname{1..1} of the two \elementname{source} and the two \elementname{target} relations also ensures that every arc has exactly one source of one target (recall that, unless depicted, all relations have the default multiplicity \elementname{0..*}).
There are two ways in which an instance of Petri Net could be executed: either the instance specifies the structure and initial state of the system, and then evolves to simulate the execution through model transformations, or it is further instantiated into a replica of itself that is then executed.
We use both kinds of paradigms for defining the behavioural semantics in other case studies in this thesis.
For example, the Robolang multilevel hierarchy that we use as running example in this thesis uses the latter (re-instantiate and use that instance for execution), whereas the PLS example we present later in this section uses the former (the bottom level is at the same time a specific instance and represents the running state of the system).
Since the Petri Net case study is used to explore the different possibilities of our framework, it allows for both.
The potency \elementname{1-2-*} of token comes from this fact, since it allows to create direct instances of token either one or two levels below, whichever we use as running instance.
The depth is still unbounded since we still allow for refinement of the concept of token in other kinds of Petri Nets.
For example, Timed Petri Nets and Coloured Petri Nets.

In the left-hand side, the model \elementname{tpn} refines tokens to add time constraints.
The node \elementname{TimedToken} contains two integer attributes for the \elementname{delay} and \elementname{duration}, in time units, that the passing of a token can have.
As in the basic Petri Net model, the potency of the token allows for both modes of execution, but here the depth is \elementname{1} since we do not want to allow further refinements of the concept.
When an instance of a timed token is created, it is expected to be in the running instance, and therefore instantiate its duration and delay with specific values.
To allow that the rest of the concepts are instantiated in the same two levels, the depth of the rest of the elements is \elementname{2}.
Also, the refined \elementname{source} and \elementname{target} relations in this model maintain the same multiplicity as the original ones.

In a similar manner, the model \elementname{cpn} refines token, place and transition, to add an integer attribute that indicates their colour.
The potency of \elementname{ColourToken} is the same as the one of \elementname{TimedToken} for the same reasons aforementioned.
In the case of \elementname{ColourPlace} and \elementname{ColourTransition}, we keep the door open to instantiate them in the same way as the rest of the concepts in the model, hence the potency \elementname{1-1-2}.

One criticism of this example is the amount of repetition in the models, where concepts like \elementname{Place}, or the arcs (\elementname{InArc}, \elementname{OutArc}) are redefined via instantiation without adding new information (attributes or references).
However, there are more advanced ways to use potency than the one this example illustrates.
For example, the Robolang language used in previous chapters contains elements which exploit potency in order to avoid repetitions, like the node \elementname{Transition} and its associated arrows \elementname{in} and \elementname{out}.
Other case studies included in this section contain additional examples of such techniques.

\subsection{Extended Robolang}
\label{subsec:extended-robolang}

The Robolang language, presented originally in~\cite{macias2018mlm} is used in this thesis as a running example in Chapters~\ref{chap:mlm} and~\ref{chap:mcmt}.
We present here an extended version of it.
This version enriches the original language, and allows to specify different agents, each with its own, independent workflow, as well as the messages that they send each other when firing a transition.
These messages can also act as guards of a transition so that it is only fired once the message is received.
Additionally, the concept of explicit time passing is now part of the models, and the duration of a timeout is made explicit.
This language is also designed for both simulation using MCMTs and code generation that can be deployed in simple, autonomous robots using the Lego EV3 and Arduino platforms~\cite{monk2011arduino,banzi2008arduino}.

\begin{figure}[ht!]
	\centering
	\includegraphics{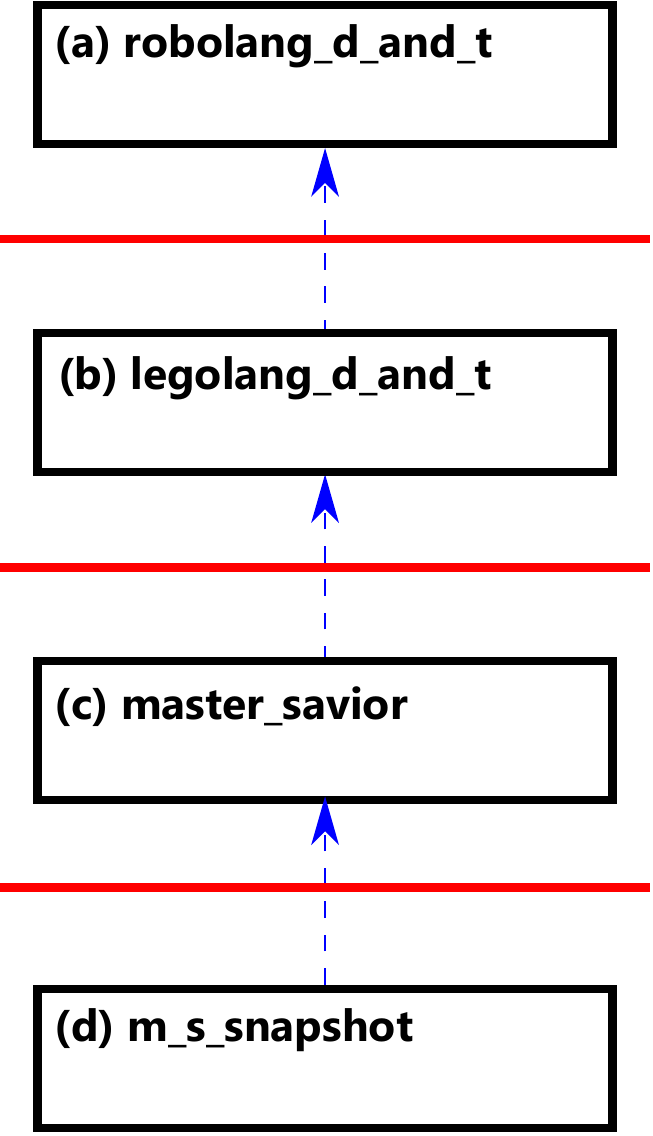}
	\caption{Multilevel hierarchy for the extended Robolang case study}
	\label{fig:extended-robolang-multilevel-hierarchy-overview}
\end{figure}

Figure~\ref{fig:extended-robolang-multilevel-hierarchy-overview} shows an overview of the hierarchy for the extended version of Robolang, including an example scenario and snapshot. 

\begin{figure}[ht!]
	\centering
	\includegraphics[width=\textwidth]{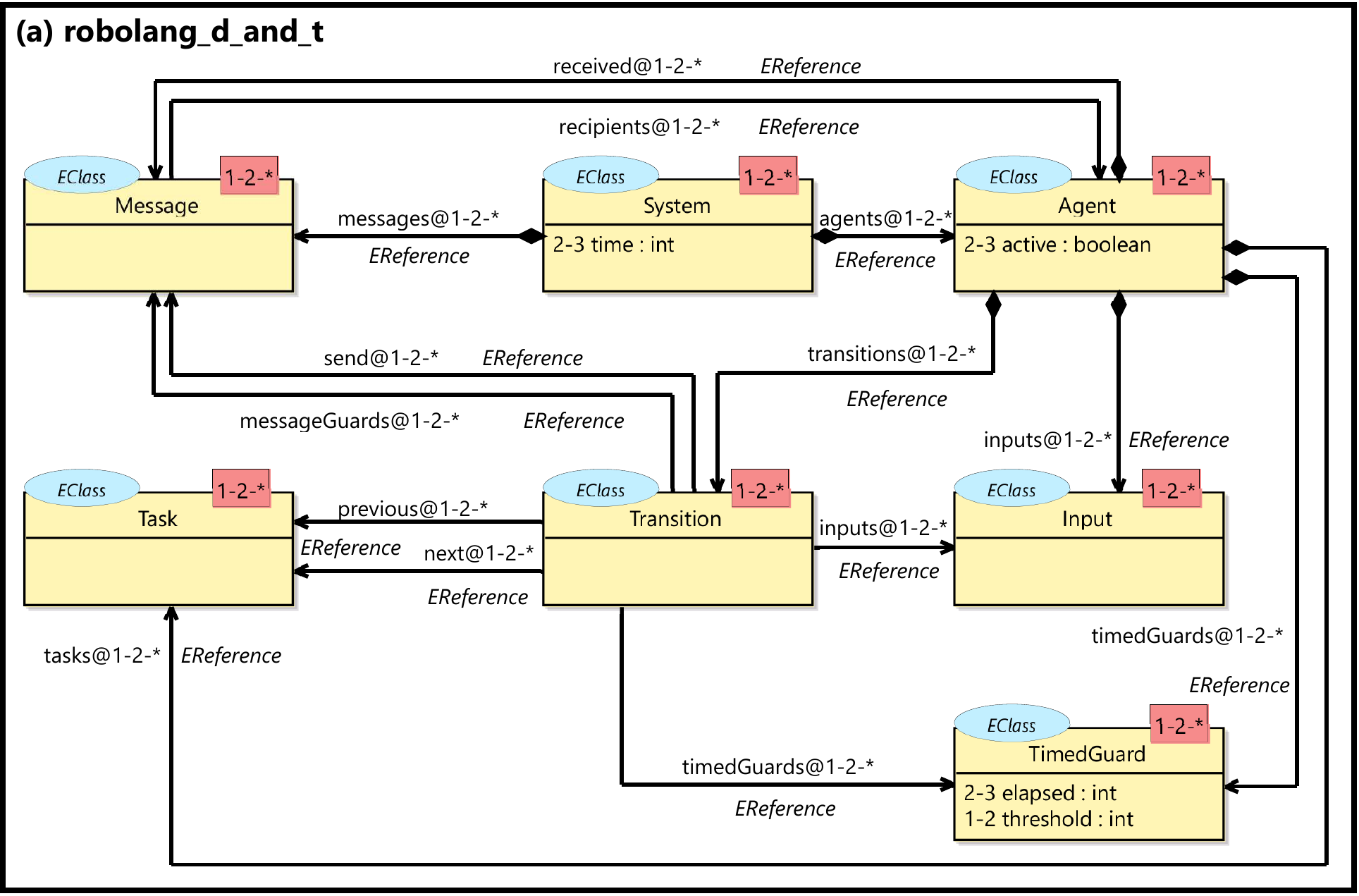}
	\caption{Extended Robolang - Level 1: Robolang}
	\label{fig:extended-robolang-level-1-robolang-d-and-t}
\end{figure}

The contents of the top-most model in Figure~\ref{fig:extended-robolang-multilevel-hierarchy-overview}(a), \elementname{robolang\_d\_and\_t}, are shown in Figure~\ref{fig:extended-robolang-level-1-robolang-d-and-t}.
This model extends the original Robolang with concepts related to distribution and time, so the name of the model is a shortened version of ``Robolang - Distributed and Timed''.
The model contains the same elements \elementname{Task}, \elementname{Transition} and \elementname{Input} as the original one, and the same references among them.
These three elements are now contained inside \elementname{Agent}, a concept used to represent each of the robots participating in the scenario.
Agents have an \elementname{active} boolean attribute which indicates if the robot has switched to a new state in the last time step or remained in its current one, e.g.\ by firing a transition or receiving a message.
Agents can also contain \elementname{TimedGuards}, which replace the concept of timeout from the original Robolang.
The two attributes inside a timed guard are meant to be instantiated in adjacent levels: when designing a specific workflow, the \elementname{threshold} attribute must be given a value which indicates the minimum amount of time that the task before a transition must be executed before the transition (which connects to the guard via the \elementname{timedGuards} arrow) can be fired.
In the level below, the \elementname{elapsed} attribute is used to keep track of the amount of time that the task has already been running.
Once this number become greater or equal than the threshold specified in the level above, the transition can be fired.
The elapsed time in timed guards must increase in the same amount as the global \elementname{time} contained in \elementname{System}, which represents the whole modelled scenario and contains all agents.
We pointed out at the beginning that this version of Robolang also allows to model messages which are sent among agents.
As shown in the figure, \elementname{Message} can be contained by either the system (if the message was just sent) or by an agent (if the message was received already).
A transition can send the messages indicated with by the \elementname{send} relation when it is fired to the agents that the message connects to, using the \elementname{recipients} relation.
When the message is received, it can trigger other transitions to which it is connected using the \elementname{messageGuard} relation.
The potencies of all elements allow to either create a specific workflow for a robot directly or, as we illustrate in the hierarchy, add an intermediate level to refine some concepts into more specific ones which are suitable for different domains.

\begin{figure}[ht!]
	\centering
	\includegraphics[width=\textwidth]{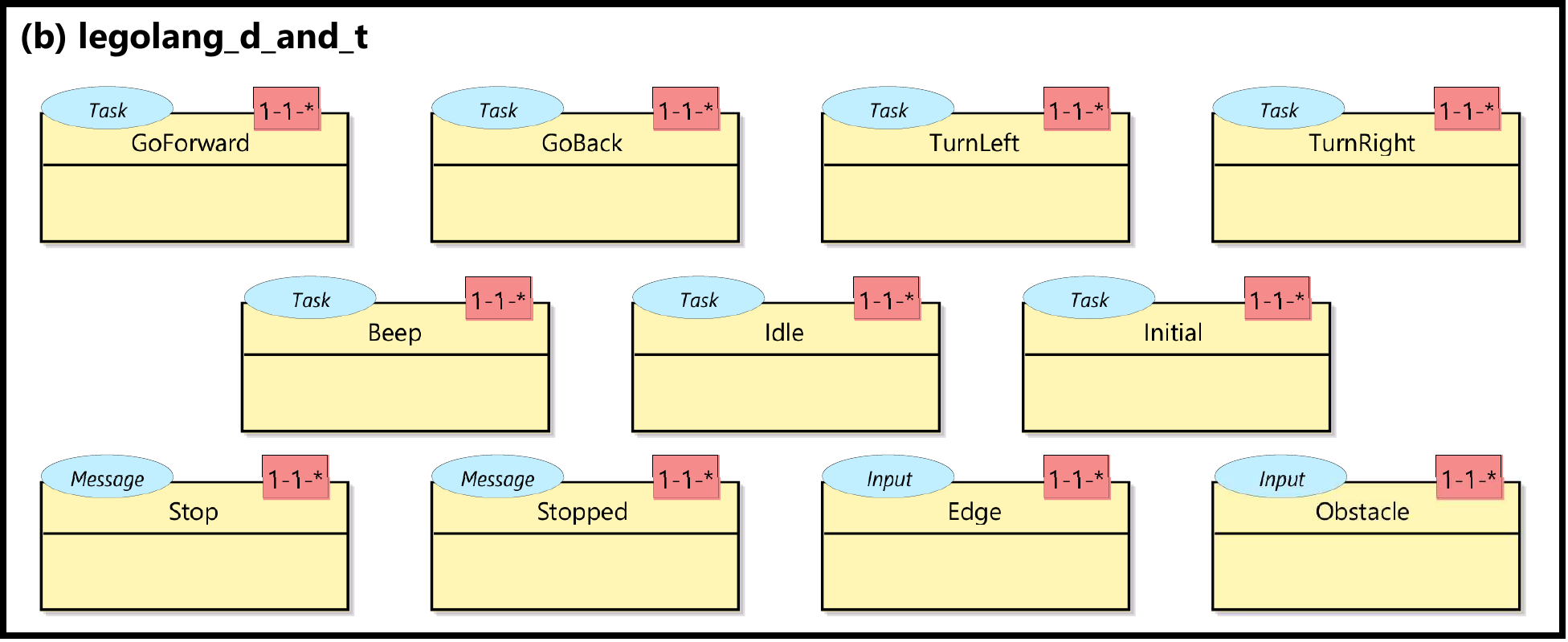}
	\caption{Extended Robolang - Level 2: Legolang}
	\label{fig:extended-robolang-level-2-legolang-d-and-t}
\end{figure}

Figure~\ref{fig:extended-robolang-level-2-legolang-d-and-t} shows in detail the model \elementname{legolang\_d\_and\_t} from Figure~\ref{fig:extended-robolang-multilevel-hierarchy-overview}(b), which contains the refinements of task, input and message into more specific ones, adapted to the domain of the simple Lego EV3 robots with which we work.
We specify here four types of movement in a flat surface for our robots: \elementname{GoForward}, \elementname{GoBack}, \elementname{TurnLeft} and \elementname{TurnRight}.
There are also a task to make the robot \elementname{Beep}, stay \elementname{Idle} and the special task \elementname{Inital} which marks the starting point of the workflow.
There are also the inputs \elementname{Edge} and \elementname{Obstacle}, same as in the original Robolang, and two types of message: \elementname{Stop}, where the sender orders the receiver to remain idle, and \elementname{Stopped}, as an acknowledgement for it.

\begin{figure}[ht!]
	\centering
	\includegraphics{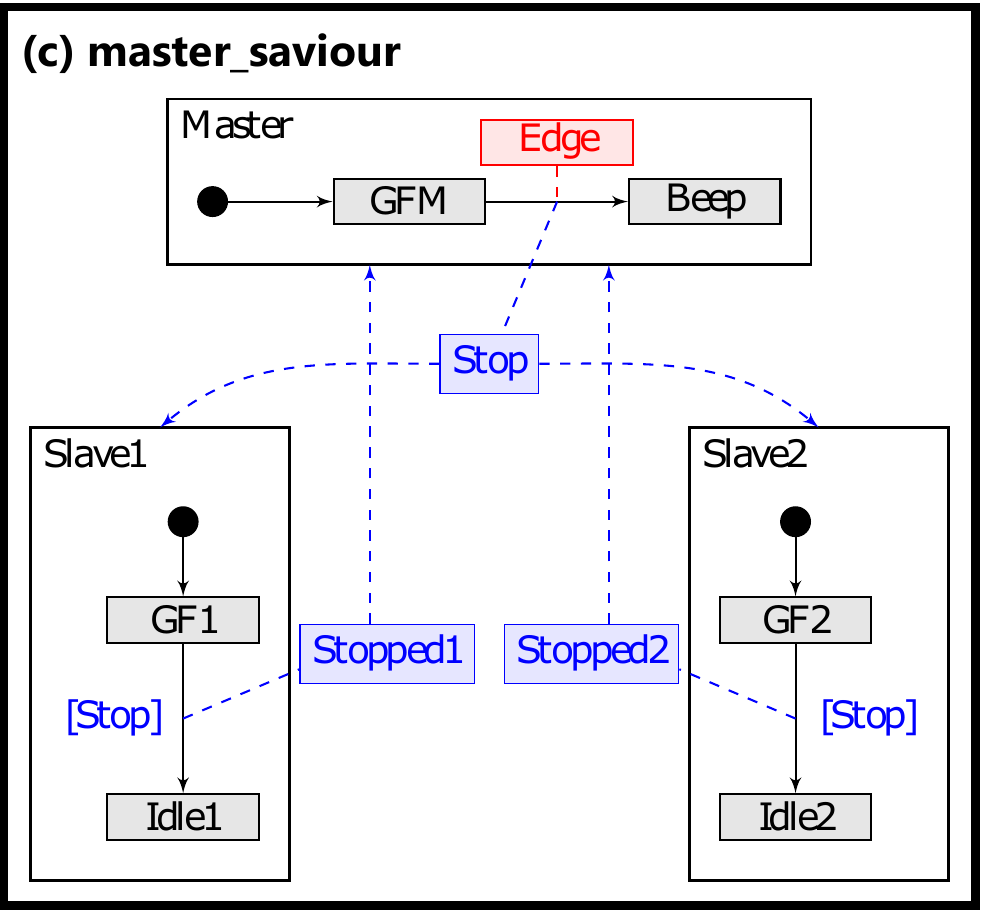}
	\caption{Model \elementname{master\_saviour} with concrete syntax}
	\label{fig:extended-robolang-level-3-master-saviour-concrete-syntax}
\end{figure}

The next model in the hierarchy, in Figure~\ref{fig:extended-robolang-multilevel-hierarchy-overview}(c), is called \elementname{master\_saviour}.
It specifies a relatively simple scenario which uses all the new concepts in this version of Robolang.
This model is difficult to understand with the abstract syntax that we use in MultEcore, as shown in Figure~\ref{fig:extended-robolang-level-3-master-saviour}.
For that reason, we also illustrate in this case study how a concrete syntax can contribute greatly to the design and understanding of DSMLs, same as the original Robolang.
In Figure~\ref{fig:extended-robolang-level-3-master-saviour-concrete-syntax}, we display the same \elementname{master\_saviour}, but this time using specific representations for the instances of different types:

\begin{figure}[ht!]
	\centering
	\includegraphics[width=\textwidth]{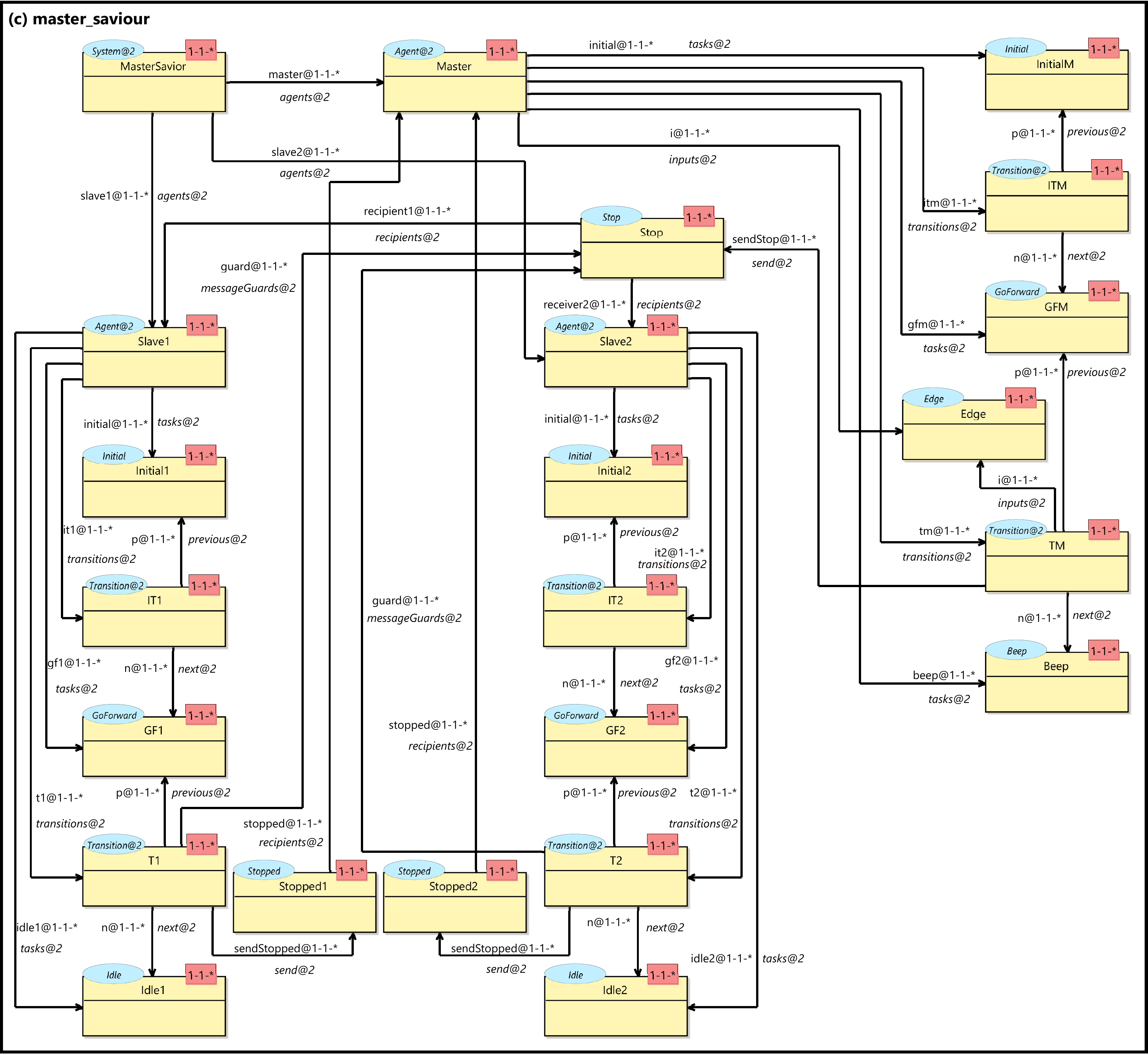}
	\caption{Extended Robolang - Level 3: Master Saviour}
	\label{fig:extended-robolang-level-3-master-saviour}
\end{figure}

\begin{itemize}
	\item The instance of system containing the other elements has no explicit representation, but simply becomes the canvas in which the rest of the graphical elements are laid out.
	If the \elementname{time} attribute is instantiated, it appears inside the canvas too.
	\item Instances of agents appear as black boxes.
	\item Instances of tasks are represented as grey rectangles with their name inside, except the initial ones, which are represented with a black dot.
	\item Instances of transitions appear as arrows connecting their previous and next tasks.
	\item Instances of inputs are depicted with red boxes with their name inside and these are connected by a dashed line to the transition they trigger.
	\item Instances of messages which are sent appear as blue boxes containing their name.
	These boxes are connected to the transition where they are sent and to their receivers by dashed blue lines.
	Whenever these messages act as guards, their name appears within square brackets as a blue label for the transition they trigger.
\end{itemize}

Keeping this representation in mind, it is now easier to understand the modelled scenario.
There are three agents, which we can imagine moving in the same direction.
One of them, the master, has a sensor able to detect obstacles, whereas the other two agents, the slaves, are blind and moving slightly behind their master.
In case there is an obstacle in the way, the master must alert the slaves so that they can stop before crashing into it, and then beep as a notification that an obstacle was found.
The slaves must just keep going forward until they receive the \elementname{Stop} message, to which they respond by sending an instance of the \elementname{Stopped} message each, and them become idle.
Note that the types of the tasks and messages are not visible in this representation, but can be checked in the abstract one, e.g.\ \elementname{GF1} is an instance of \elementname{GoForward}.

\begin{figure}[ht!]
	\centering
	\includegraphics{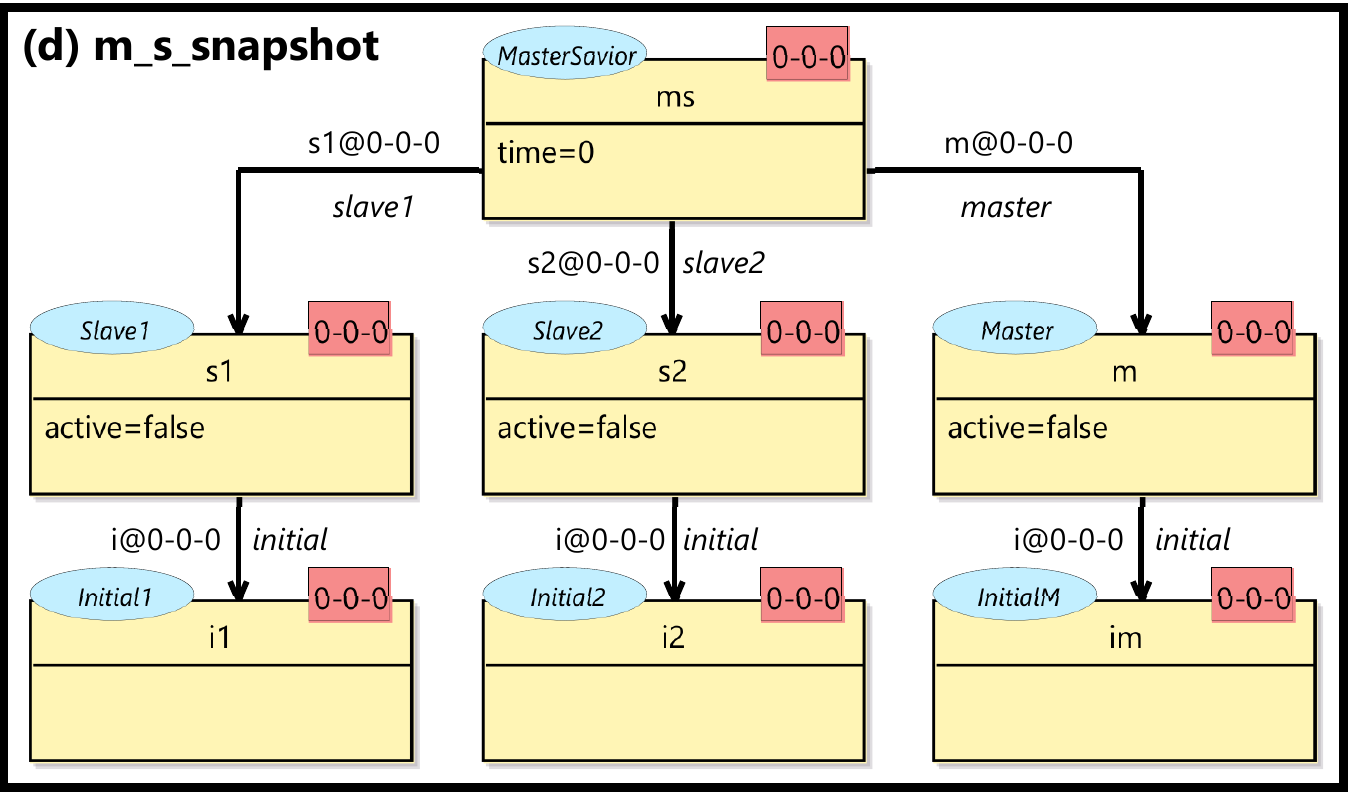}
	\caption{Extended Robolang - Level 4: Initial Snapshot}
	\label{fig:extended-robolang-level-4-m-s-snapshot}
\end{figure}

Moreover, this representation can be extended to instances of instances, so that the running ones (snapshots) in this example can be depicted in a similar manner.
That is the case with the last level of the extended Robolang hierarchy, from Figure~\ref{fig:extended-robolang-multilevel-hierarchy-overview}(d).
This model, representing the initial state of the execution, defines the three agents to be in the initial state, where their corresponding instance of \elementname{Initial} has not been discarded.
The global time of the system is initialised to zero, as shown in Figure~\ref{fig:extended-robolang-level-4-m-s-snapshot}.

\begin{figure}[ht!]
	\centering
	\includegraphics{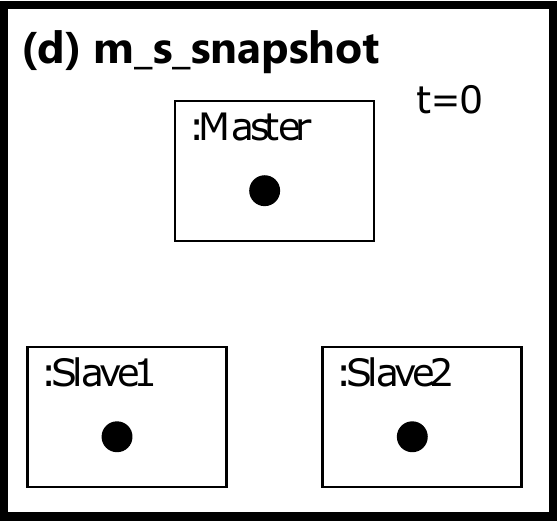}
	\caption{Model \elementname{m\_s\_snapshot} with concrete syntax}
	\label{fig:extended-robolang-level-4-m-s-snapshot-concrete-syntax}
\end{figure}

Again, the information contained in this model can be represented in a more concise and understandable way as shown in Figure~\ref{fig:extended-robolang-level-4-m-s-snapshot-concrete-syntax}, if we apply the same visualisation to elements based on their \emph{transitive} type.
For example, instances of instances of agents (that is, transitive instances of agents) are depicted as black, labelled rectangles, exactly as we did for direct instances of agents.

We also use this concrete syntax to specify the MCMT rules for the behavioural semantics of this example.
There are four different groups of rules, depending on their function, which must be coordinated in a certain way.
In Section~\ref{sec:future-work} we address the ways in which we can achieve this coordination in the future.

\subsubsection{Behavioural MCMT rules}
\label{subsubsec:extended-robolang-behavioural-mcmts}

As with the initial version of Robolang, we call the first group \emph{behavioural rules}, since they model the behaviour of the agents, i.e.\ the semantics of how agents evolve through time, executing tasks, switching to one task to the next, and interacting with messages and environmental information.
All these rules simulate agent activity, so all the rules in this group set the active boolean attribute of the agent to true.
Visually, this flag is represented with a blue \emph{active mark} in the upper-right corner of the agent.

\begin{figure}[ht!]
	\centering
	\includegraphics{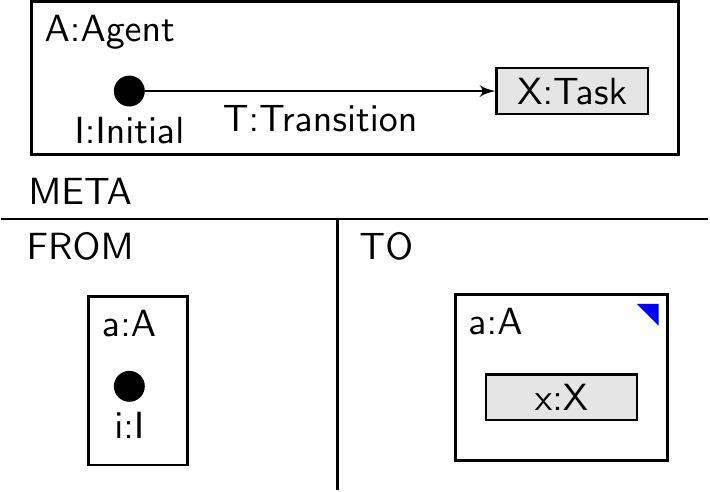}
	\caption{Rule \emph{Fire Initial Transition}: the initial transition is fired}
	\label{fig:extended-robolang-rule-fire-initial-transition}
\end{figure}

The rule \emph{Fire Initial Transition}, depicted in Figure~\ref{fig:extended-robolang-rule-fire-initial-transition} is the first rule to be executed which affects agents, since it deletes the initial task of an agent and fires the next one (connected with a transition).
This is the only case where a guard-free transition is used.

\begin{figure}[ht!]
	\centering
	\includegraphics{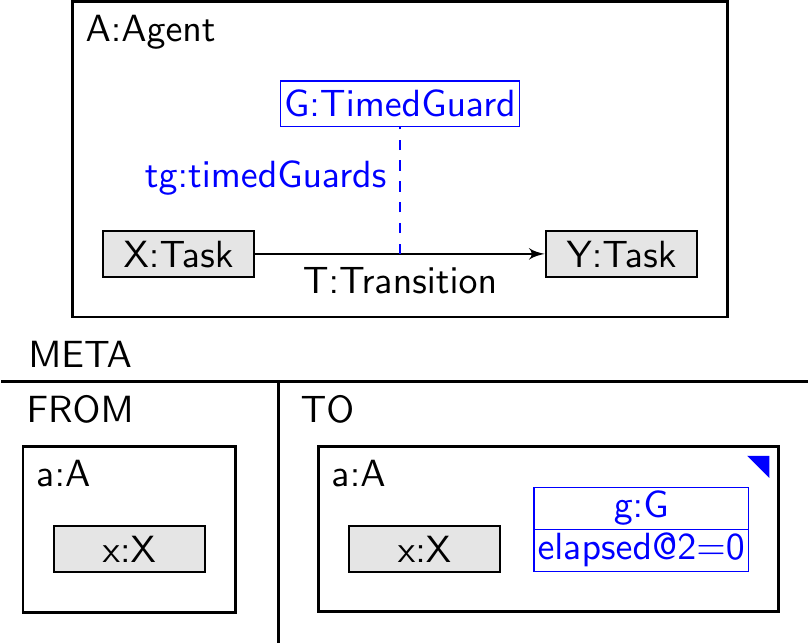}
	\caption{Rule \emph{Add Timed Guard}: the timeout counter is initialised}
	\label{fig:extended-robolang-rule-add-timed-guard}
\end{figure}

Before a timed guard can start keeping track of the elapsed time, the rule \emph{Add Timed Guard} in Figure~\ref{fig:extended-robolang-rule-add-timed-guard} is responsible for creating it, with the elapsed time initialised to zero, whenever the agent is executing a task before a transition which is triggered by a timed guard.

\begin{figure}[ht!]
	\centering
	\includegraphics[width=\textwidth]{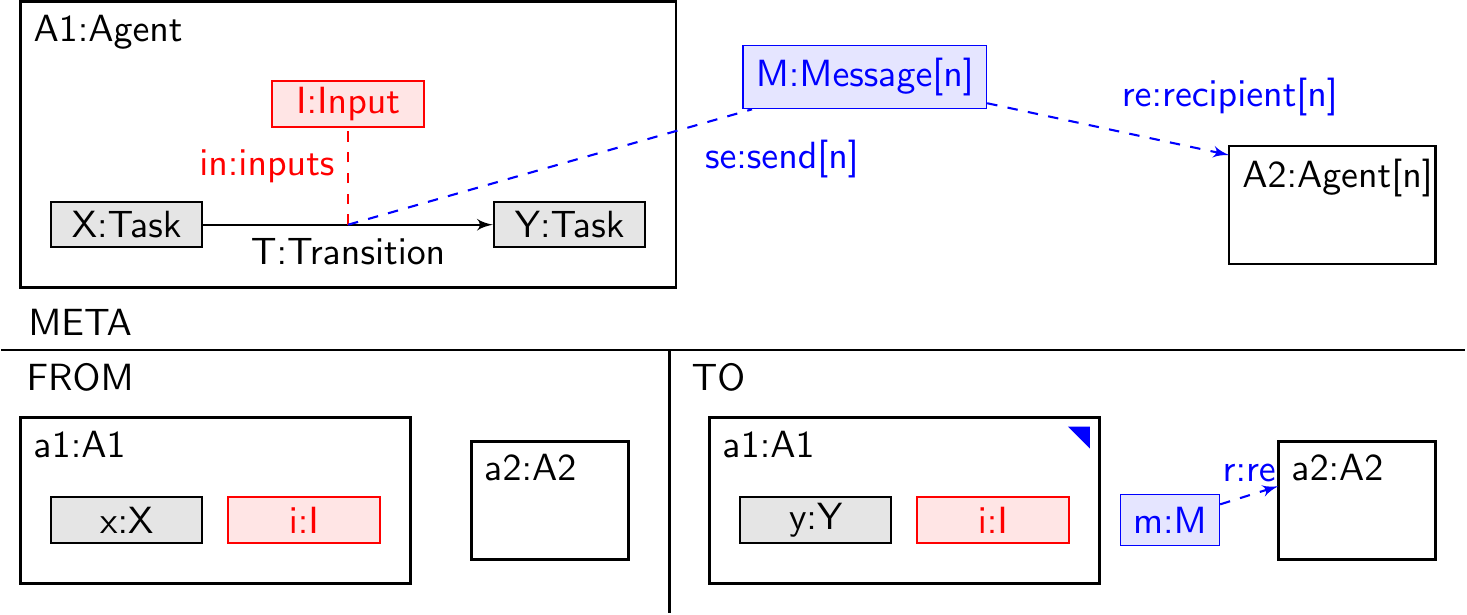}
	\caption{Rule \emph{Fire Transition with Input}: a transition gets fired by the associated input}
	\label{fig:extended-robolang-rule-fire-transition-with-input}
\end{figure}

The rule \emph{Fire Transition with Input}, displayed in Figure~\ref{fig:extended-robolang-rule-fire-transition-with-input}, allows an agent to progress by responding to environmental inputs.
If a message has to be sent during the transition, messages into the system (outside any agent) are created.
One message per receiving agent is created, with exactly one reference to the sending agent and exactly one reference to the recipient.
Note that the \elementname{[n]} multiplicity specifier can be any value between 0 and an arbitrarily large number.
That is, this rule will match scenarios in which \elementname{A1} sends any number of messages to the same number of agents.

\begin{figure}[ht!]
	\centering
	\includegraphics[width=\textwidth]{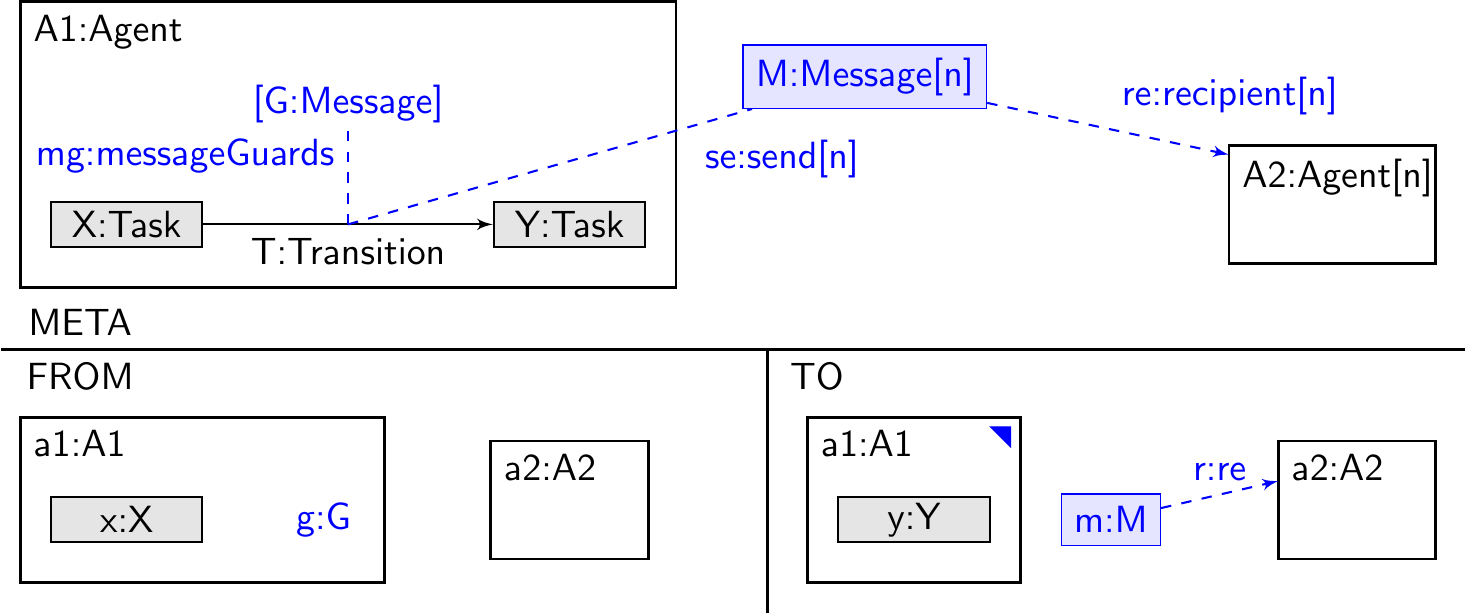}
	\caption{Rule \emph{Fire Transition with Message Guard}: a transition gets fired by the associated message upon reception}
	\label{fig:extended-robolang-rule-fire-transition-with-message-guard}
\end{figure}

The rule \emph{Fire Transition with Message Guard} is similar to the previous one, but here the transition is triggered by the reception of a message, which acts as a guard.
As opposed to the inputs in \emph{Fire Transition With Input}, guards are consumed when the transition is fired.
That means that this MCMT removes the guard from the running instance (in the TO block) once it's fired, as Figure~\ref{fig:extended-robolang-rule-fire-transition-with-message-guard} shows the graphical representation of this MCMT rule.

\begin{figure}[ht!]
	\centering
	\includegraphics[width=\textwidth]{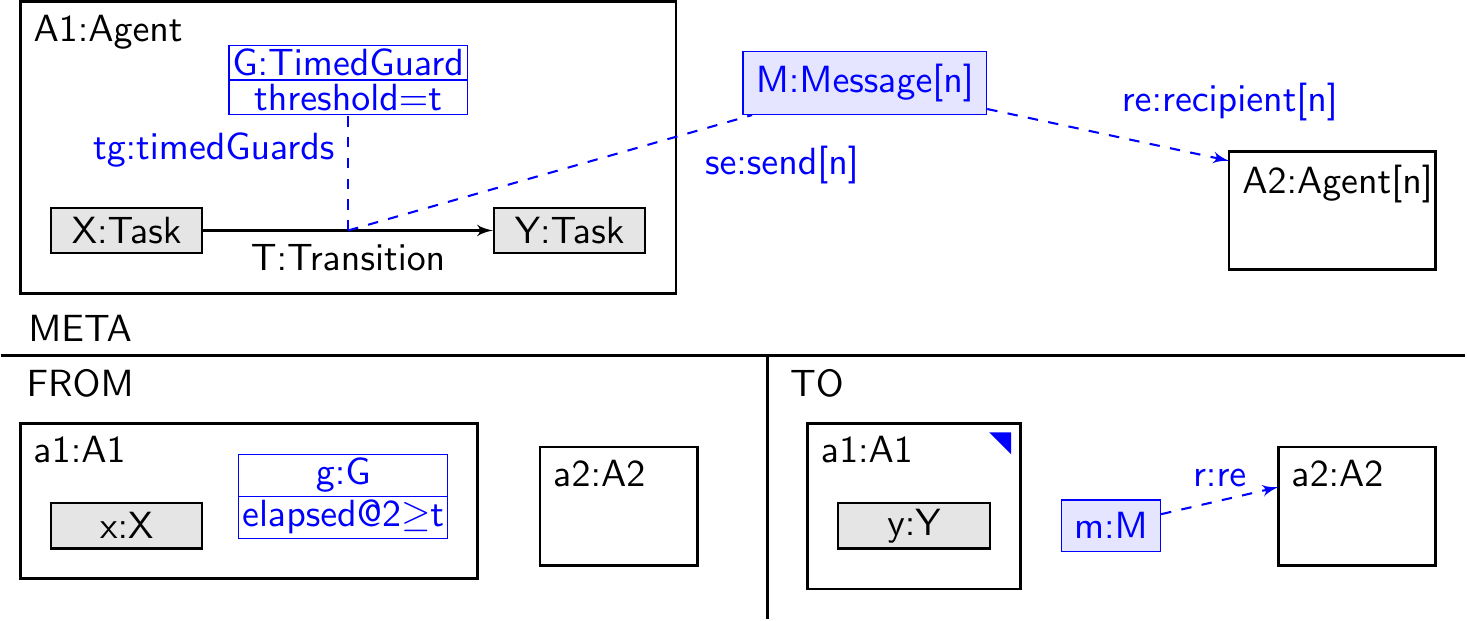}
	\caption{Rule \emph{Fire Transition with Timed Guard}: a transition gets fired by the associated timer going over threshold}
	\label{fig:extended-robolang-rule-fire-transition-with-timed-guard}
\end{figure}

In the last environmental rule, \emph{Fire Transition with Timed Guard}, a timed guard fires the transition if the elapsed time is above the threshold.
As with the previous one, this MCMT rule consumes the guard once the transition is fired.
Figure~\ref{fig:extended-robolang-rule-fire-transition-with-timed-guard} shows the graphical syntax of the rule.

Note that our modelling hierarchy does not forbid transitions from having more than one input or guard connected to them.
The semantics of this last MCMT, as well as the two previous ones, allows the transition to be triggered as soon as one of them appears.
Hence, it can be considered that the semantics of attaching several inputs/guards to a transition is that of a boolean \emph{or}.

\subsubsection{Environmental MCMT rules}
\label{subsubsec:extended-robolang-environmental-mcmts}

This second group was also present in the original version of Robolang used throughout the first chapters of this thesis, and some of its rules are agent-aware versions of the original ones.
Environmental rules are applied to simulate changes on the environment of the agents, such as a change perceived through some input from the agent's sensors or a message being received from another agent.

\begin{figure}[ht!]
	\centering
	\includegraphics{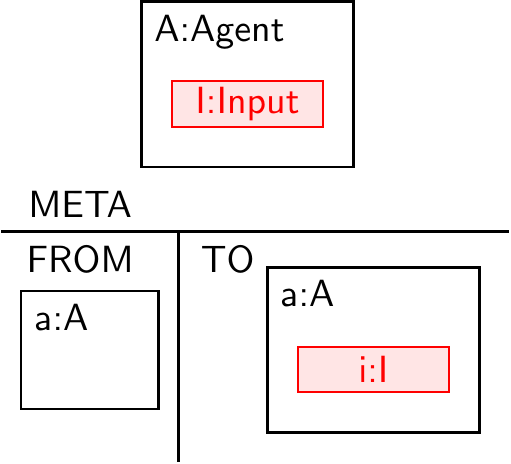}
	\caption{Rule \emph{Insert Input}: an arbitrary input is created}
	\label{fig:extended-robolang-rule-insert-input}
\end{figure}

The rule \emph{Insert Input} creates an input inside an agent, nondeterministically and without any kind of constraint.
This rule, depicted in Figure~\ref{fig:extended-robolang-rule-insert-input}, is a non-deterministic and loose version of the next rule, since the appearance of such input may not cause any effect on the execution.

\begin{figure}[ht!]
	\centering
	\includegraphics{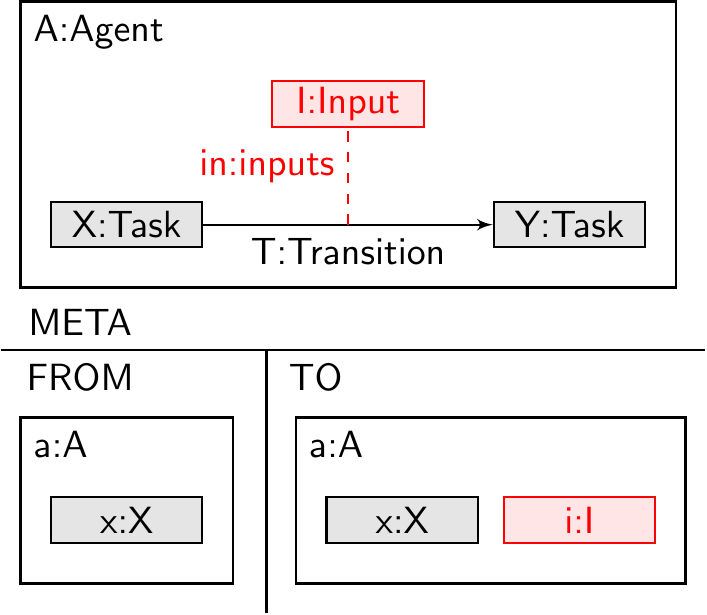}
	\caption{Rule \emph{Insert Effective Input}: a relevant input is created}
	\label{fig:extended-robolang-rule-insert-effective-input}
\end{figure}

The rule \emph{Insert Effective Input} creates an input inside an agent, ensuring that it will cause its currently running task to stop by firing one of the task's outgoing transitions.
That is, the introduced input will cause some effect on the execution.
Note that, although this rule is more restrictive than \emph{Insert Input}, it is still nondeterministic, since there may be more than one agent in which this rule can be matched.
The graphical representation of \emph{Insert Effective Input} is shown in Figure~\ref{fig:extended-robolang-rule-insert-effective-input}.

\begin{figure}[ht!]
	\centering
	\includegraphics{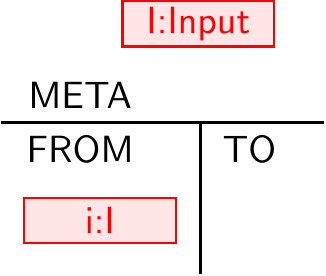}
	\caption{Rule \emph{Delete Input}: an existing input is removed}
	\label{fig:extended-robolang-rule-delete-input}
\end{figure}

Figure~\ref{fig:extended-robolang-rule-delete-input} represents the rule \emph{Delete Input}, used to remove a previously existing input inside an agent.
Semantically speaking, this rule reflects the fact that an input, previously detected by an agent, cannot be detected any more.
For instance, if an agent is moving backwards, an obstacle detected in front of it should eventually disappear.
Note that it is not necessary to identify the containing agent, since the previous rules always ensure that inputs are created inside agents.

\begin{figure}[ht!]
	\centering
	\includegraphics{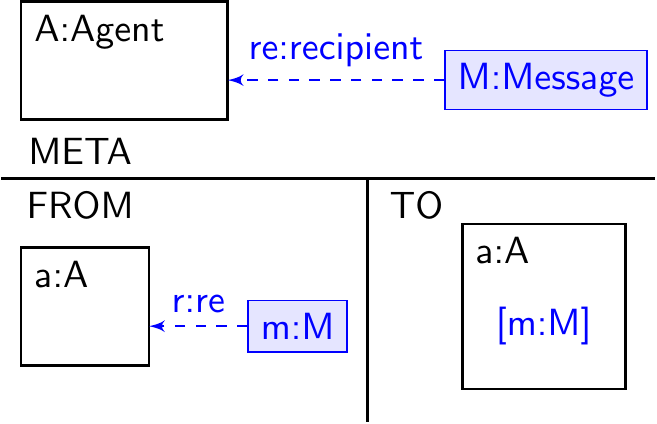}
	\caption{Rule \emph{Receive Message}: a message is received by an agent}
	\label{fig:extended-robolang-rule-receive-message}
\end{figure}

If a message was previously sent to an agent, but not received by it yet, the rule \emph{Receive Message} can be applied to simulate the reception of such message.
This rule, shown in Figure~\ref{fig:extended-robolang-rule-receive-message}, changes the containment relation from the outer system to the internal containment inside an agent, by replacing one relation from the other.
After the execution of this rule, the message will act as a guard, and hence its graphical representation is changed.

\subsubsection{Time-stepping rule}
\label{subsubsec:extended-robolang-time-stepping-mcmt}

\begin{figure}[ht!]
	\centering
	\includegraphics{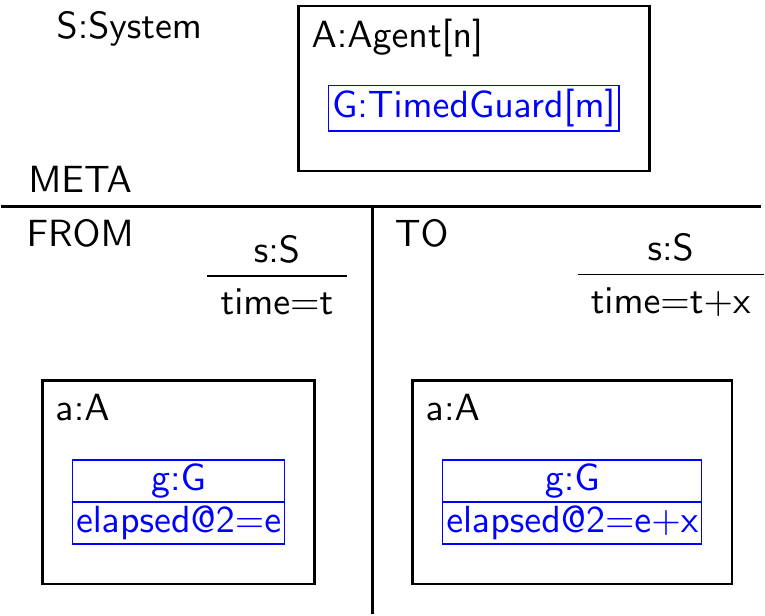}
	\caption{Rule \emph{Step Time}: the global clock advances}
	\label{fig:extended-robolang-rule-step-time}
\end{figure}

This next rule needs to be in its own group since it must be executed only at a specific point in each step of the execution.
The rule \emph{Step Time}, displayed in Figure~\ref{fig:extended-robolang-rule-step-time}, represents the passage of time forward for a given amount of time units, which is specified in the \elementname{x} parameter.
The rule then increases the global clock of the system and the elapsed time attribute on the timed guards of each agent.
This is the only rule with a parameter presented in this thesis, but that only represents syntactic sugar to modifying the rule manually or having several copies with different values for \elementname{x}.

Similarly, the next rule is also on its own group.

\subsubsection{Mark-deleting rule}
\label{subsubsec:extended-robolang-mark-deleting-mcmt}

\begin{figure}[ht!]
	\centering
	\includegraphics{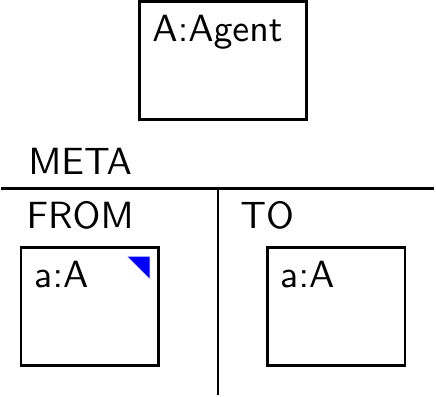}
	\caption{Rule \emph{Remove Active Mark}: the attribute is set to false}
	\label{fig:extended-robolang-rule-delete-active-mark}
\end{figure}

Figure~\ref{fig:extended-robolang-rule-delete-active-mark} shows the rule \emph{Remove Active Mark}, which simply removes the active marks (setting the attribute value to false) introduced by the behavioural rules from all the agents that have executed some behaviour.

\subsubsection{Coordination of MCMT rules}
\label{subsubsec:extended-robolang-coordination}

All the MCMT rules aforementioned must be coordinated with certain priorities in order for them to yield the expected result.
This process of coordination consists simply of a number of layers or priorities, to which each rule belongs.
Inside a group, we can either choose to run a single matching rule from that layer, or either sequentially run rules from the same layer until no more rules match.
Notice that this coordination can be modified to suit other needs, so the one presented here can be used as an example.

The coordination for our example scenario is as follows:

\begin{enumerate}
	\item The first layer of rules is conformed by the behavioural rules which specify the agents' behaviour.
	From this first layer, as many rules as possible must be run until no further matching rules are found, thus allowing all agents to progress by reacting to changes in the environment or incoming messages.
	\item The second layer contains the environmental rules, from which just one rule can be applied.
	If more than one rule matches the current snapshot, the user can manually select (hence guiding the execution), or the system will non-deterministically choose one of them.
	\item The third layer is only formed by the rule \emph{Step Time}, which is run only once per iteration.
	\item Finally, the rule \emph{Delete Active Mark}  is executed as many times as possible, so that it ensures that all active marks are removed.
\end{enumerate}

Once the last layer is reached, one new snapshot has been generated, and the process starts from the first layer again.

\subsection{Distributed and timed linear temporal logic}
\label{subsec:ltl}

As a consequence of the extension of the original Robolang presented in Section~\ref{subsec:extended-robolang}, we also extend the original LTL supplementary hierarchy, which we presented in Section~\ref{subsec:supplementary-dimension}, in order to exploit the additional features in the extended Robolang, namely the distribution of tasks among agents, the communication via messages and the existence of explicit time.
We present a new version of Linear Temporal Logic which adds: (1) distributed evaluation of properties, (2) usage of explicit time constraints and (3) usage of a 4-valued boolean algebra which accounts for uncertain evaluations of the temporal properties due to their distributed nature.

We apply simultaneously three different modifications of LTL to achieve a language capable of managing distribution and explicit time.
First, LTL can be extended to timed LTL, abbreviated as TLTL~\cite{raskin1999tltl}, to express real-time timing constraints.
In TLTL, it is required that each symbol is paired with a time stamp, which running instances of extended Robolang are capable of doing by using the global clock.
Second, LTL can be extended to be applied to asynchronous distributed systems, e.g.\ with the Distributed Temporal Logic, abbreviated DTL~\cite{scheffel2014dtl}, where the evaluation of fragments of the property can be delegated to the agents, which also fits the extended version of Robolang.
And third, when applying runtime verification, it is not always possible to reach a final verdict of true/false for the evaluation immediately.
Hence, we also use a 4-valued boolean logic for the evaluation of the verdict considering the trace until now, using the values ``presumably true'' and ``presumably false'' until the final true/false verdict is reached~\cite{leucker2011rv}.
Due to all these modifications, we name this new temporal logic as DTLTL4.

\begin{figure}[ht!]
	\centering
	\includegraphics[width=\textwidth]{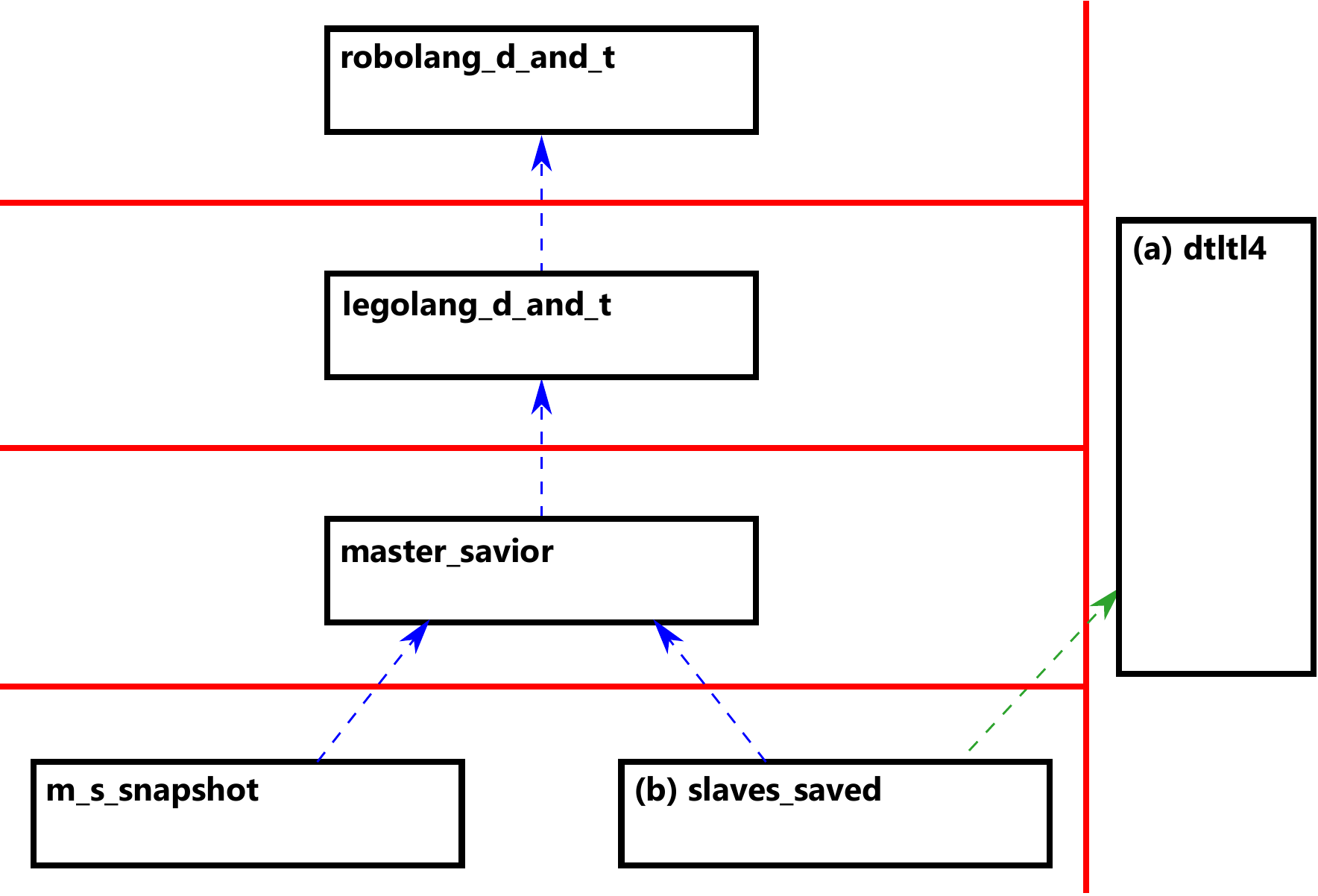}
	\caption{Multilevel supplementary hierarchy for 4-valued, timed, distributed LTL}
	\label{fig:dtltl-multilevel-hierarchy-overview}
\end{figure}

Figure~\ref{fig:dtltl-multilevel-hierarchy-overview} shows how the DTLTL4 supplementary hierarchy is combined with that of extended Robolang (recall Figure~\ref{fig:extended-robolang-multilevel-hierarchy-overview}) in order to create a temporal property.
This is the same structure that we construct with the original hierarchies for Robolang and LTL, as depicted in Section~\ref{subsec:supplementary-dimension}, Figure~\ref{fig:ltl-multilevel-hierarchy-outline}).

\begin{figure}[ht!]
	\centering
	\includegraphics[width=\textwidth]{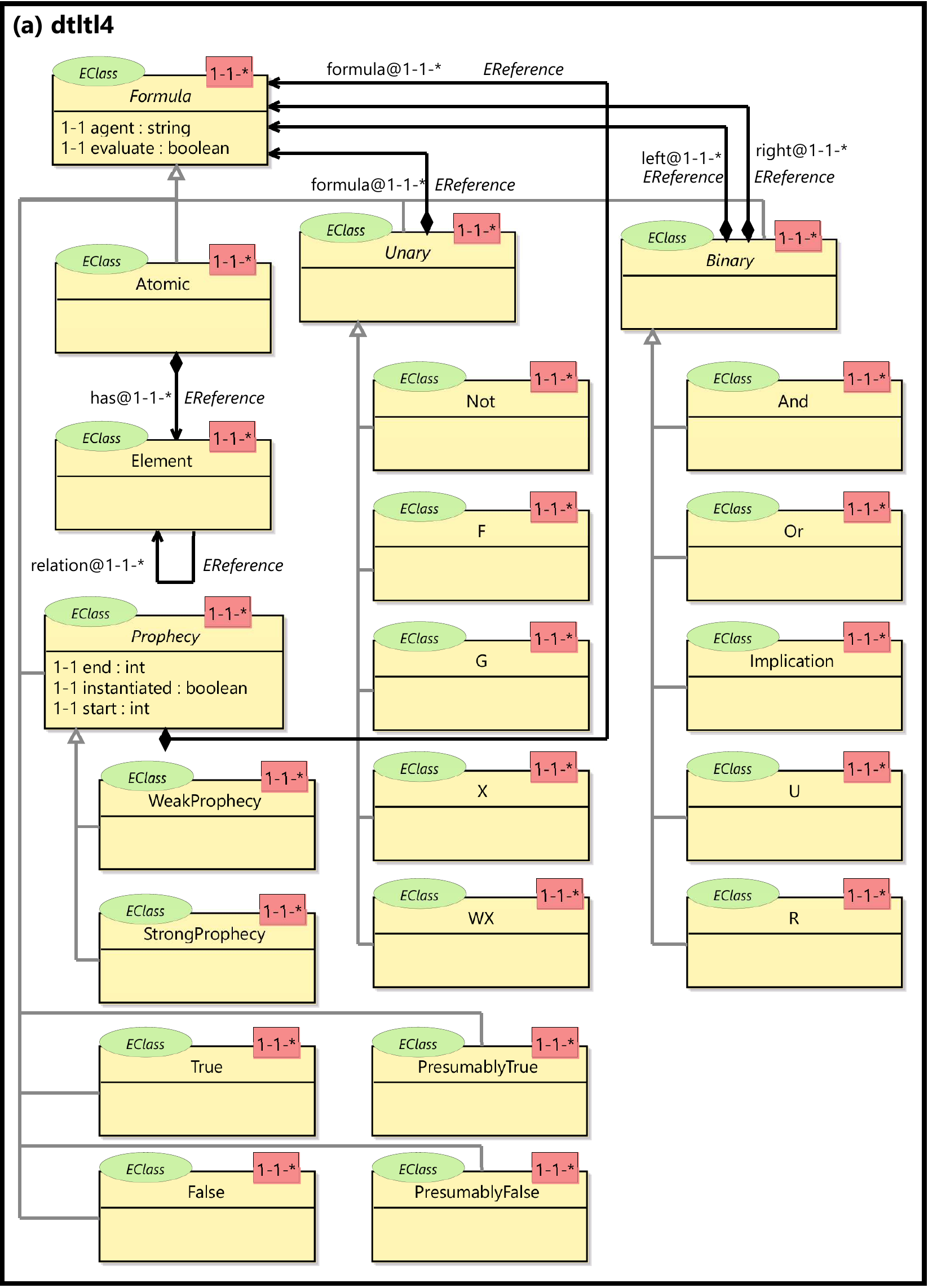}
	\caption{Distributed, timed, 4-valued LTL supplementary model}
	\label{fig:dtltl-level-1-dtltl4}
\end{figure}

The contents of the model in Figure~\ref{fig:dtltl-multilevel-hierarchy-overview}(a), where we define the structure of the DTLTL4 language, are shown in Figure~\ref{fig:dtltl-level-1-dtltl4}.
This extends LTL with new prophecy operators for the explicit-time constraints, which are evaluated to true if and only if the sub-property they contain is fulfilled between the two time boundaries specified by the values of \elementname{start} and \elementname{end}.
The \elementname{instantiated} flag is used during the evaluation process to mark when these boundaries start to be updated.
The detailed semantics of the prophecy operators are described in~\cite{raskin1999tltl}.
Furthermore, the \elementname{Formula} element now contains an attribute which specifies the id of the agent where the (sub)formula must be evaluated.
Finally, the two additional boolean values \elementname{PresumablyTrue} and \elementname{PresumablyFalse} are used to get the full DTLTL4 specification.

\begin{figure}[ht!]
	\centering
	\includegraphics[width=\textwidth]{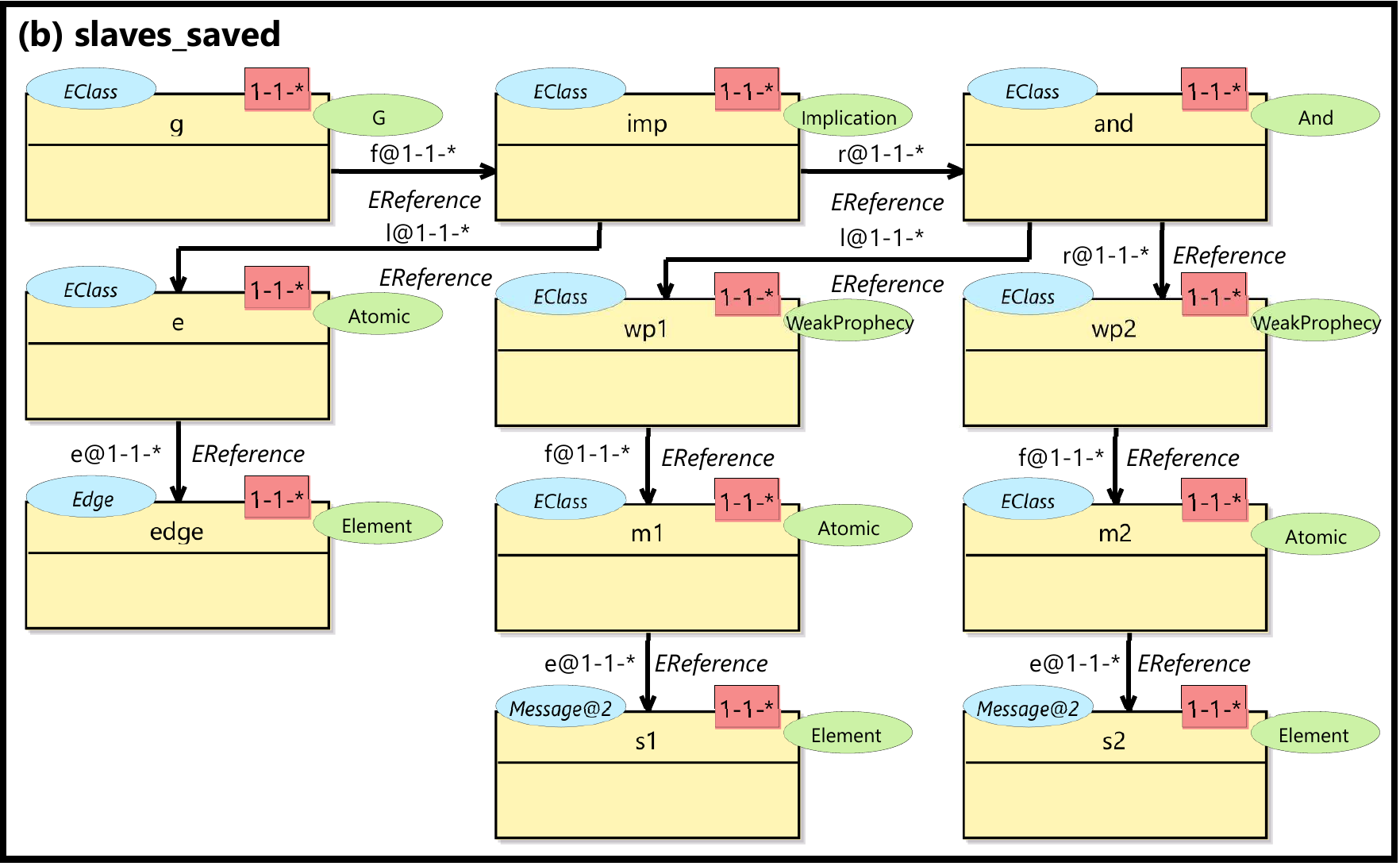}
	\caption{A DTLTL4 property for the \elementname{master\_saviour} scenario}
	\label{fig:dtltl-property-slaves-saved}
\end{figure}

The example temporal property created with DTLT4, placed in the hierarchy in Figure~\ref{fig:dtltl-multilevel-hierarchy-overview}(b), is depicted in detail in Figure~\ref{fig:dtltl-property-slaves-saved}.
This property can be written using the textual notations of DTL and TLTL as follows:

\[@_{\text{Master}} \LTLglobally (e \LTLimp \LTLprophecy{0,3} @_{\text{Slave1}} s1 \LTLand \LTLprophecy{0,3} @_{\text{Slave2}} s2)\]

The property means that, once the master finds the obstacle, both slaves must stop within three seconds.
The master evaluates the property, including the \elementname{e} atomic proposition which represents the detection of an edge by the master, but delegates to each slave the evaluation of their respective atomic propositions \elementname{s1} and \elementname{s2}.
These atomic propositions are evaluated to true when the message from the master commanding the slaves to stop is received.
In Figure~\ref{fig:dtltl-property-slaves-saved} we show the \emph{Slaves saved} property, which is satisfied when the slaves are notified in time about an edge lying ahead that they cannot detect, so that they stop before reaching the edge and falling off. 

As it was the case with the original version of LTL, the MCMT rules for DTLTL4 are too numerous and uninteresting to display them in this thesis.
However, we can reuse the same two example rules we included in Section~\ref{subsec:mcmts-ltl} for the unrolling of the \elementname{U} operator and the simplification of the subformulas with the \elementname{X} operator.

\subsection{Product line systems}
\label{subsec:pls}

This example is an adaptation into multilevel modelling of a well-known example within MDSE, used for example in~\cite{cabot2008analysing} and~\cite{rivera2009formal}.
In that example, inheritance is used to refine abstract types like ``Machine'' into more specific instances of machines, like an ``Assembler'' that put pieces together in order to manufacture the final product.
This example provides a good opportunity to compare our approach, and MLM in general, against traditional model-driven techniques, when they are applied to DSML development, in a similar manner as we did with Robolang in Sections~\ref{sec:why-mlm} and~\ref{sec:why-mcmt}.
Here, we just present the example hierarchy and its associated MCMT rules, as another example of modelling in our approach and using MultEcore.

\begin{figure}[ht!]
	\centering
	\includegraphics[width=\textwidth]{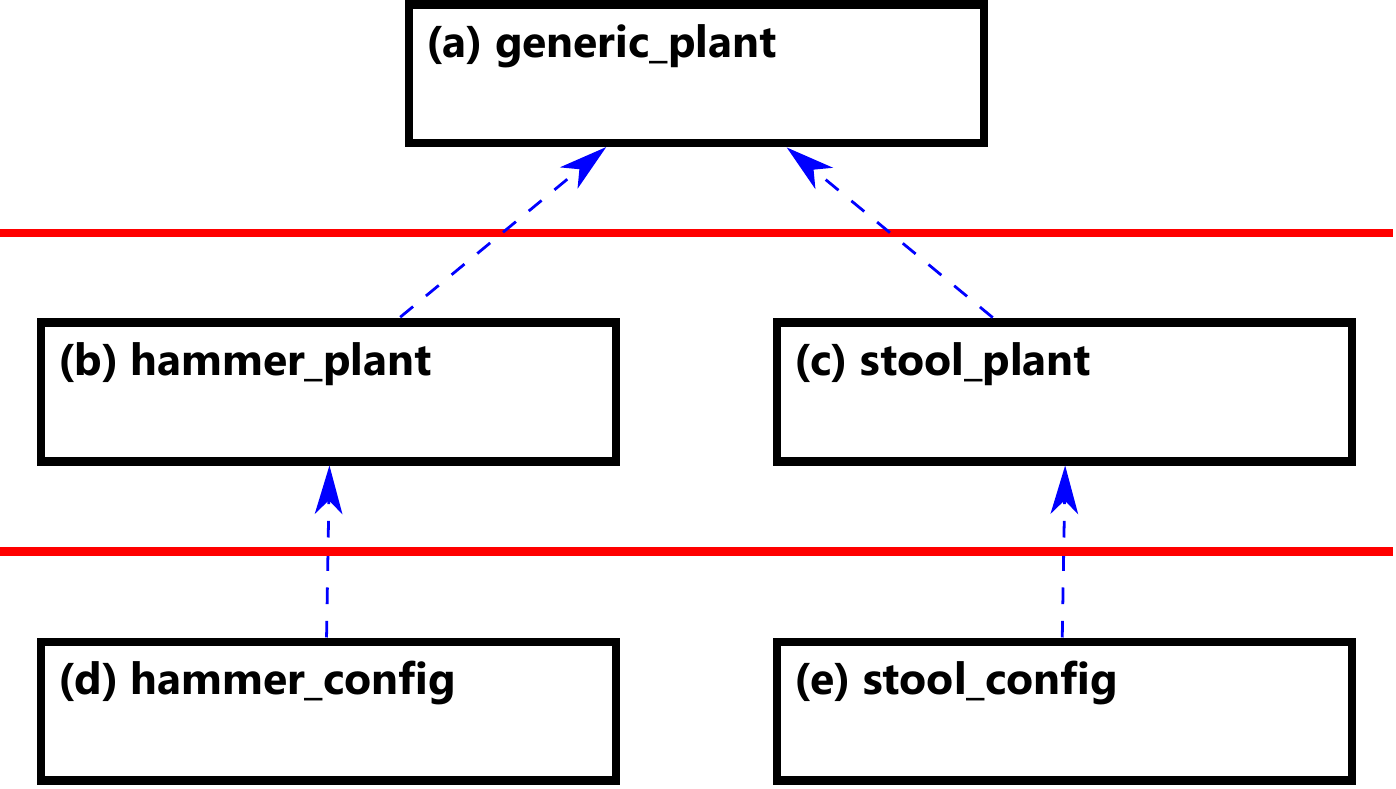}
	\caption{Multilevel hierarchy for the PLS case study}
	\label{fig:pls-multilevel-hierarchy-overview}
\end{figure}

Figure~\ref{fig:pls-multilevel-hierarchy-overview} shows the overview of the multilevel hierarchy which we created to improve the flexibility and conciseness of the original PLS example.

\begin{figure}[ht!]
	\centering
	\includegraphics[width=\textwidth]{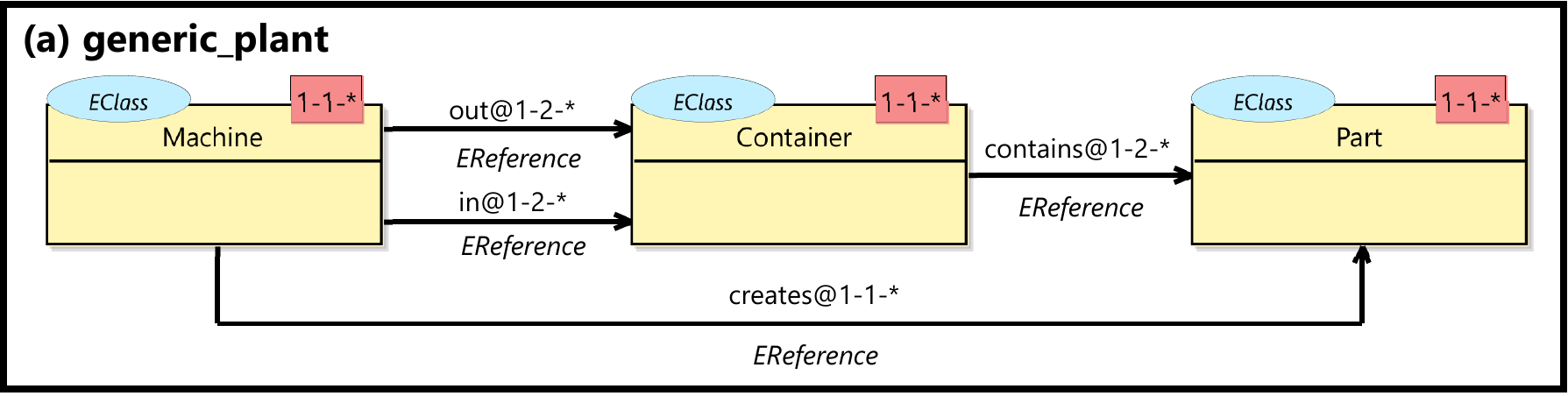}
	\caption{PLS - Level 1: Generic Plant}
	\label{fig:pls-level-1-generic-plant}
\end{figure}

The hierarchy contains a top model, called \elementname{generic\_plant} in Figure~\ref{fig:pls-multilevel-hierarchy-overview}(a), which is depicted in Figure~\ref{fig:pls-level-1-generic-plant}.
In this model we define abstract concepts related to product lines that manufacture physical objects.
\elementname{Machine} defines any device that can create, modify or combine objects, which are represented by the concept \elementname{Part}.
In the first case, we indicate the relation from a generator-like machine to the part it generates with the \elementname{creates} relation.
In order for parts to be transported between machines or to be stored, they can be inside \elementname{Containers}, and this relation is expressed by the \elementname{contains} relation.
All machines may have containers to take parts from or to leave manufactured ones in.
These two relations are identified with the \elementname{in} and \elementname{out} relations, respectively.

\begin{figure}[ht!]
	\centering
	\includegraphics[width=\textwidth]{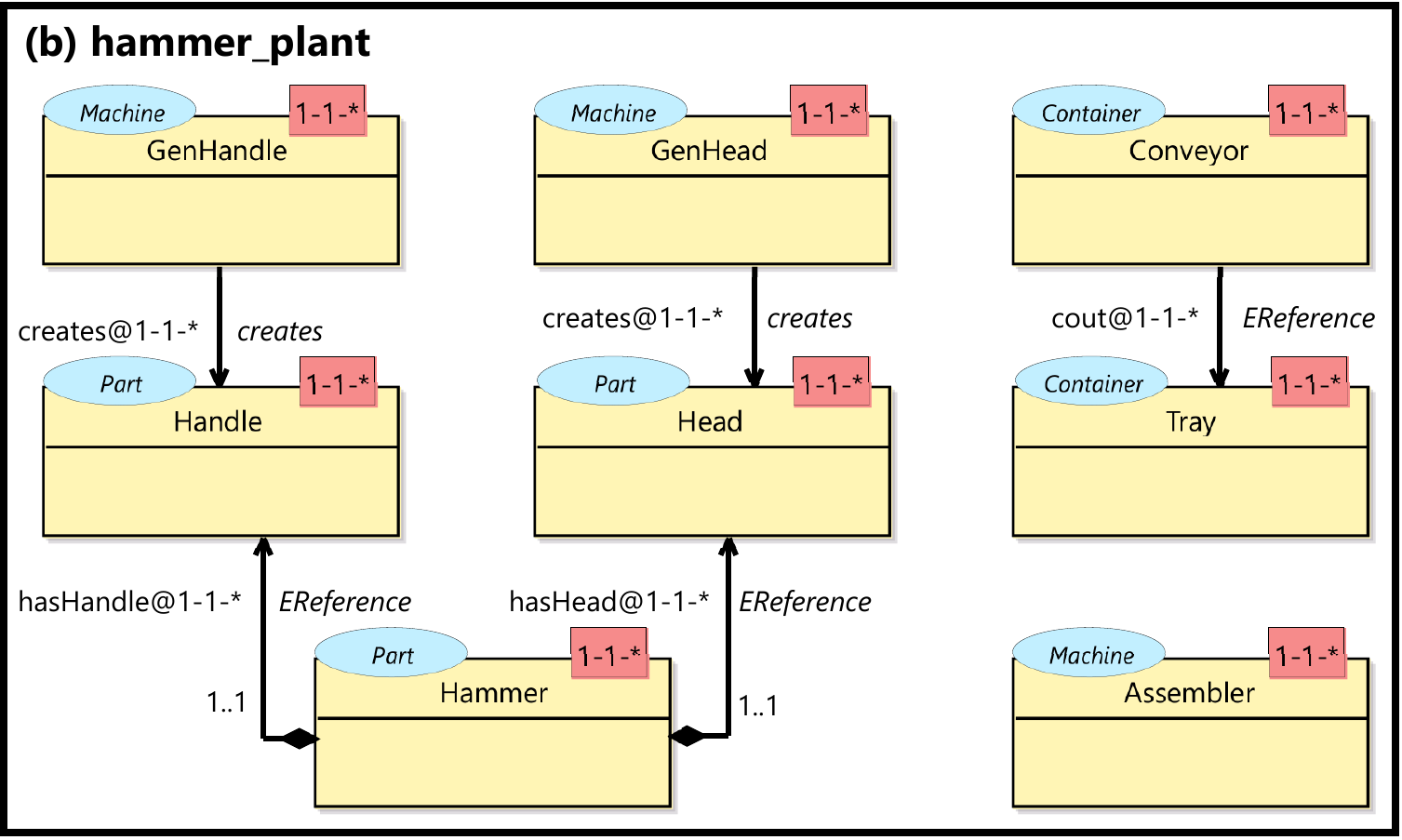}
	\caption{PLS - Level 2: Hammer Plant}
	\label{fig:pls-level-2-hammer-plant}
\end{figure}

The second level of the hierarchy contains two models, that define concepts related to specific types of plants.
The first of these models, \elementname{hammer\_plant} from Figure~\ref{fig:pls-multilevel-hierarchy-overview}(b), defines an instance of plant where the final product \elementname{Hammer} is created by combining one \elementname{Handle} and one \elementname{Head}, as shown in Figure~\ref{fig:pls-level-2-hammer-plant}.
Both \elementname{hasHandle} and \elementname{hasHead} relations express this fact, and their \elementname{1..1} multiplicity states that a \elementname{Hammer} must be created out of exactly one \elementname{Head} and one \elementname{Handle}.
The type of these two relations is \elementname{EReference} (from \elementname{Ecore}), since there is no relation defined for parts in \elementname{generic\_plant}, because the concept of assembling parts is too specific to be located in \elementname{generic\_plant}.
Note that thanks to the use of potency, we can define \elementname{hasHandle} and \elementname{hasHead} without forcing \elementname{Part} to have a relation with itself in \elementname{generic\_plant} in order to use it as their type, since it is not desirable that a specific detail in a lower level (\elementname{hammer\_plant}) affects the specification of an upper one (\elementname{generic\_plant}).
In the \elementname{hammer\_plant} model we also define three types of machines.
First, \elementname{GenHandle} and \elementname{GenHead}, that create the corresponding parts, indicated by the two \elementname{creates} arrows.
And secondly, \elementname{Assembler}, that generates hammers by assembling the corresponding parts.
Finally, the model contains two specific instances of \elementname{Container}, namely \elementname{Conveyor} and \elementname{Tray}.
The \elementname{cout} arrow between them indicates that a \elementname{Conveyor} must always be connected to a \elementname{Tray}.
As explained before, this relation can be defined without requiring a loop arrow in \elementname{Container}, so that the separation in levels of abstraction is maintained.

\begin{figure}[ht!]
	\centering
	\includegraphics[width=\textwidth]{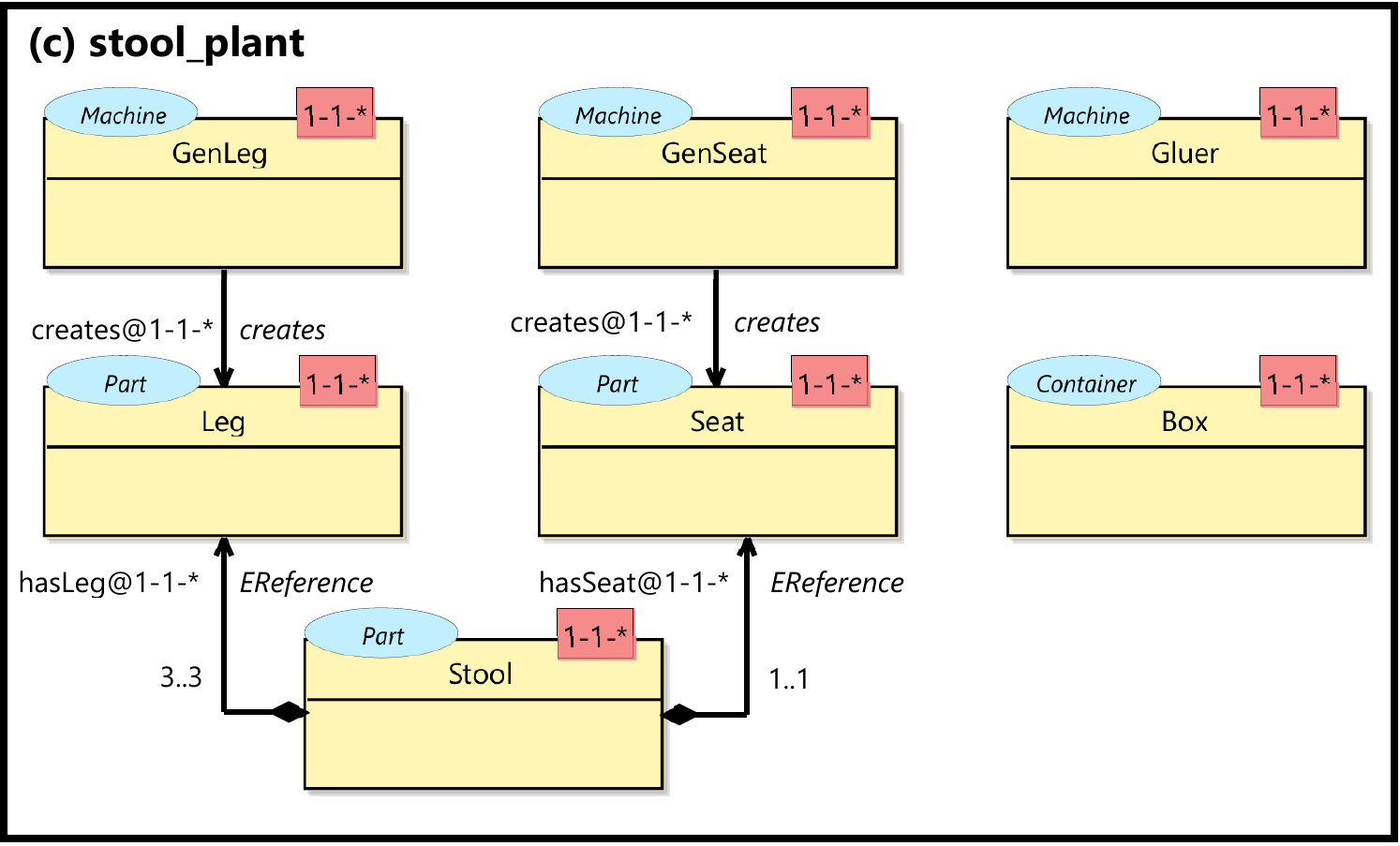}
	\caption{PLS - Level 2: Stool Plant}
	\label{fig:pls-level-2-stool-plant}
\end{figure}

The other model in the second level, in Figure~\ref{fig:pls-multilevel-hierarchy-overview}(c), is depicted in Figure~\ref{fig:pls-level-2-stool-plant}.
Here, we specify another kind of manufacturing plant for stools.
In this model, the relations between \elementname{Stool}, \elementname{Leg} and \elementname{Seat} resembles those of \elementname{Hammer}, \elementname{Handle} and \elementname{Head}.
The multiplicity of the arrow between the first two indicates that a \elementname{Stool} must have exactly three \elementname{Legs}.
Two different types of machine, \elementname{GenLeg} and \elementname{GenSeat}, manufacture \elementname{Legs} and \elementname{Seats}, respectively.
The remaining one, \elementname{Gluer}, takes three \elementname{Legs} and a \elementname{Seat} and creates a \elementname{Stool} out of them.
Finally, the only container defined for this kind of plant is \elementname{Box}.

\begin{figure}[ht!]
	\centering
	\includegraphics[width=\textwidth]{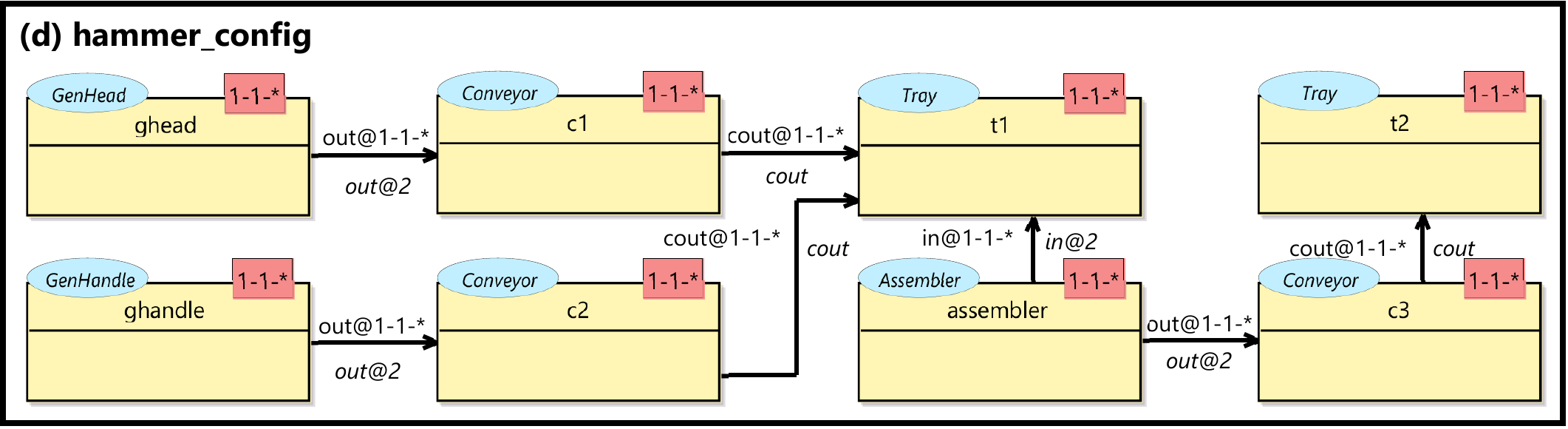}
	\caption{PLS - Level 3: A possible configuration of a hammer-manufacturing plant}
	\label{fig:pls-level-3-hammer-config}
\end{figure}

\begin{figure}[ht!]
	\centering
	\includegraphics[width=\textwidth]{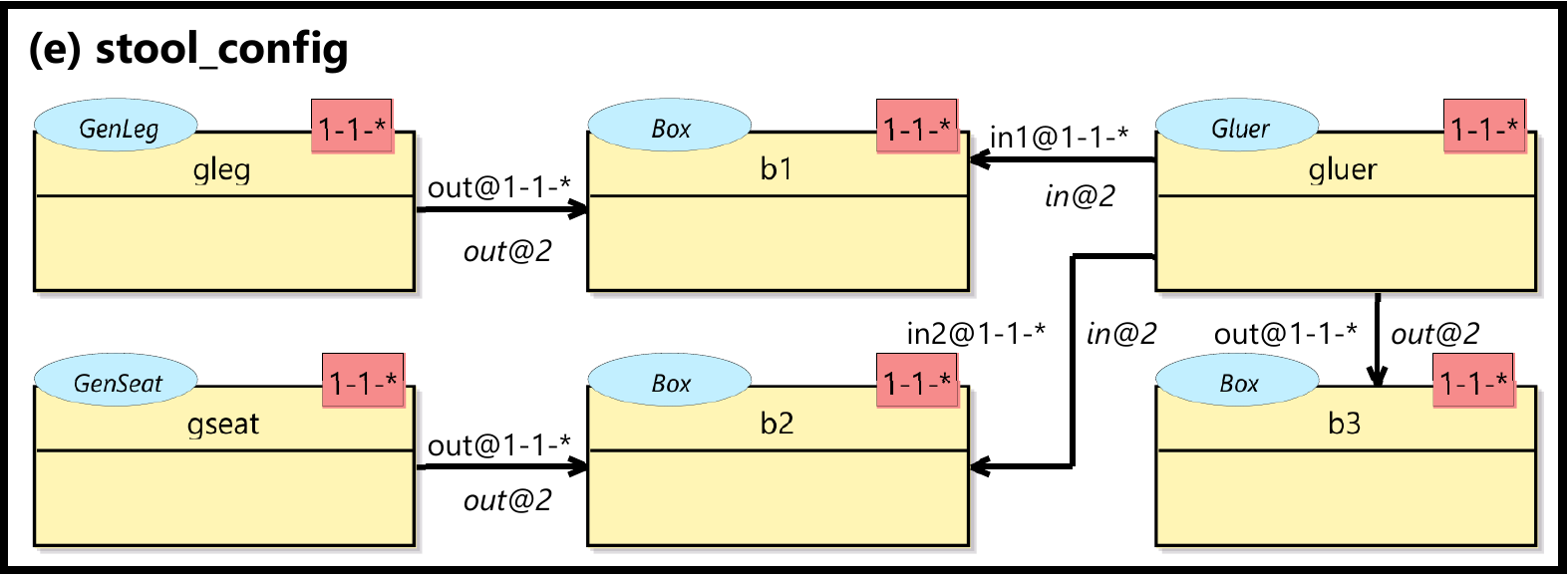}
	\caption{PLS - Level 3: A possible configuration of a stool-manufacturing plant}
	\label{fig:pls-level-3-stool-config}
\end{figure}

The two models at the bottom of the hierarchy, in Figures~\ref{fig:pls-multilevel-hierarchy-overview}(d) and~\ref{fig:pls-multilevel-hierarchy-overview}(e), represent specific configurations of a hammer PLS (\elementname{hammer\_config}, in Figure~\ref{fig:pls-level-3-hammer-config}) and a stool PLS (\elementname{stool\_config}, in Figure~\ref{fig:pls-level-3-stool-config}).
They contain specific instances of the concepts defined in the level above, organized to specify correct product lines, in which parts get transferred from generator machines to machines that combine them to obtain the final manufactured products.
Note that, again, the use of potencies enables the creation of instances of the \elementname{out} relation defined two levels above between \elementname{Machine} and \elementname{Container}, without requiring a type defined in the intermediate level of \elementname{hammer\_plant} and \elementname{stool\_plant}.

Another interesting remark about this example, when comparing our MLM version of PLS against the original, fixed-level ones is that we are capable of reusing the concepts defined in \elementname{generic\_plant} in both of its instances, but we keep the concepts related to hammer manufacturing from those related to stool manufacturing.
This is not possible in the original version of the example, where the addition of the stool-related concepts would pollute the original metamodel with additional classes and references.
This effect gets magnified if we keep creating new kinds of plants, making the metamodel more difficult to maintain, understand and extend.

The MCMT rules we define for this example are capable of exploiting the MLM features of our hierarchy without losing specificness or causing unwanted side effects, and we also use them to illustrate the benefits of our approach.
We use them to define the behaviour of any plant which is typed directly or indirectly by \elementname{generic\_plant}.

\begin{figure}[ht]
	\centering
	\includegraphics{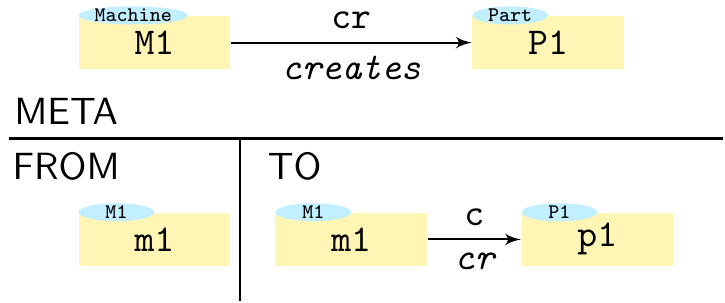}
	\caption{Rule \emph{Create Part}: a machine creates a part}
	\label{fig:pls-rule-create-part}
\end{figure}

The first rule is \emph{CreatePart}, shown in Figure~\ref{fig:pls-rule-create-part}, which is used to create parts by the corresponding generator machines.
This rule can be applied, as note above, on any PLS typed by \elementname{generic\_plant}, since the variable \elementname{P1} can match any of the parts (in our PLS hierarchy those are \elementname{Head}, \elementname{Handle}, \elementname{Seat} and \elementname{Leg}) and the variable \elementname{M1} can match any of the generator machines (correspondingly, \elementname{GenHead}, \elementname{GenHandle}, \elementname{GenSeat} and \elementname{GenLeg}).
That is, \emph{Create Part} can only match the generators of parts and the parts they generate, since we require \elementname{M1} in the META block to have a \elementname{creates} relation to the \elementname{P1}.
If the META is matched correctly, the proliferated rules can be applied to any instance of a generator in order to generate the appropriate instance of part, without any side effects.

\begin{figure}[tb]
	\centering
	\includegraphics[width=\textwidth]{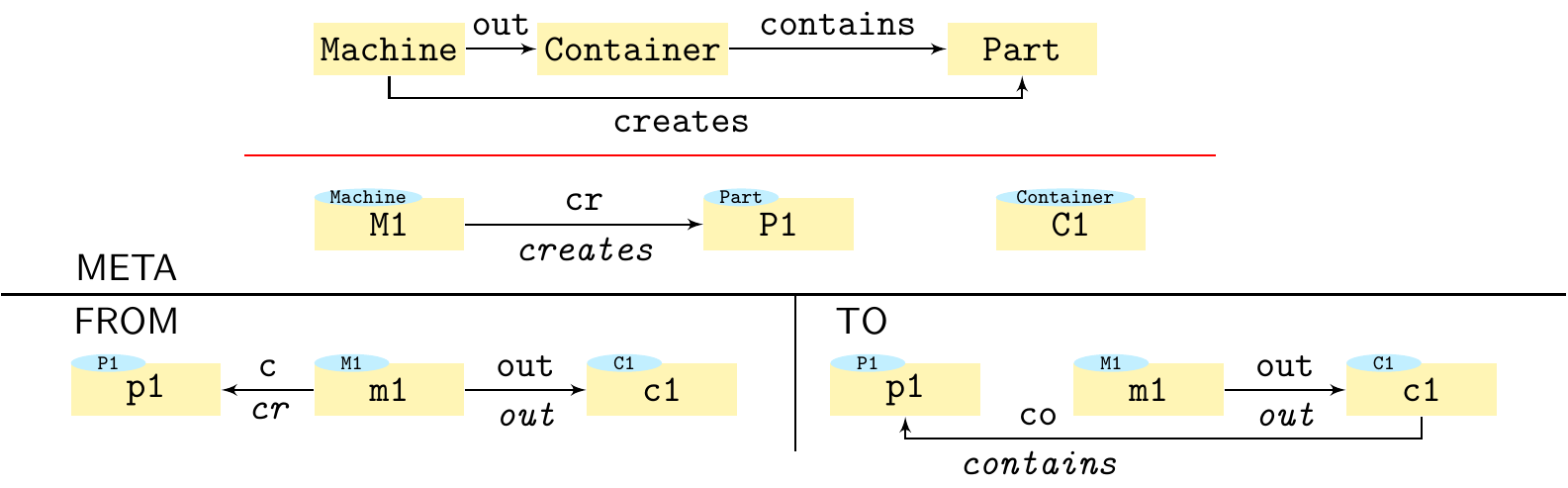}
	\caption{Rule \emph{Send Part Out}: a part leaves the machine that created it}
	\label{fig:pls-rule-send-part-out}
\end{figure}

Another rule is \emph{Send Part Out}, shown in Figure~\ref{fig:pls-rule-send-part-out}, used for moving a created part from its generator into the output container of this machine.
This rule displays a richer META block than \emph{Create Part}, in which we need to identify elements from two different models.
At the top level, we mirror part of \elementname{generic\_plant}, defining elements like \elementname{out} and \elementname{contains}, that are used directly as types in the FROM and TO blocks.
These elements are defined as constants, meaning that the name of the pattern element must match an element with the same name in the typing chain.

\begin{figure}
	\centering
	\includegraphics[width=\textwidth]{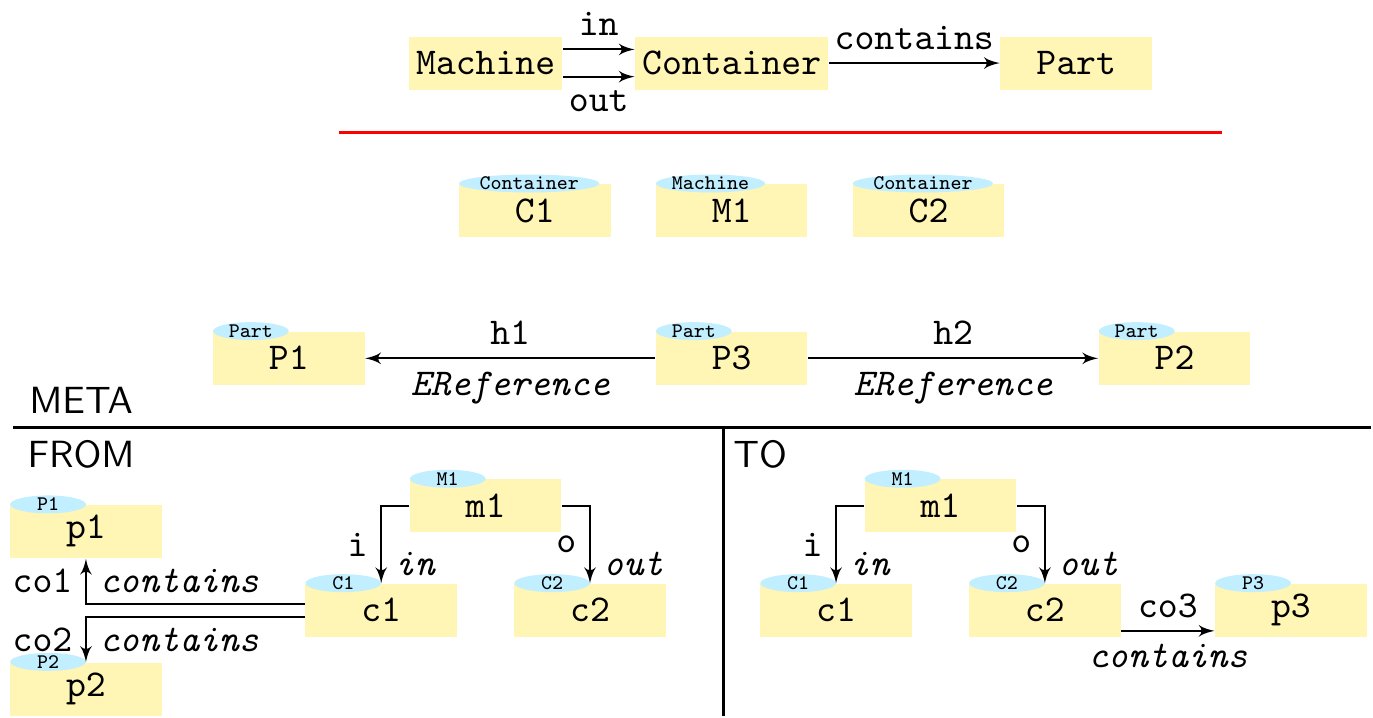}
	\caption{Rule \emph{Assemble}: a machine assembles two parts into a new one}
	\label{fig:pls-rule-assemble}
\end{figure}

Once the component parts have reached the right containers, it is possible to combine them to create final products (parts), like \elementname{hammer} or \elementname{stool}.
The \emph{Assemble} rule does exactly that: it assembles two parts into a different part, as can be seen in Figure~\ref{fig:pls-rule-assemble}.
It requires the resulting part \elementname{p3} to be a part that can consist of, or be built from, parts \elementname{p1} and \elementname{p2}.
One particularity of this rule is that the META block defines several instances of the same type, but different from each other.
In such a way, we ensure that, for example, \elementname{P1} is different from \elementname{P2}, but both are instances of \elementname{Part}.

The semantics of matching makes this rule applicable for a variable number of parts which are assembled into one part.
To apply this rule we use the cardinality information of relations \elementname{h1} and \elementname{h2}, as long as the upper bound is different from \elementname{*}.
The cardinalities of the matches of \elementname{h1} and \elementname{h2} in the corresponding model will indicate the number of instances of \elementname{P1} and \elementname{P2} that should appear in the FROM block.
This information is then used to replicate the instances \elementname{p1} and \elementname{p2} accordingly, together with all their related references (\elementname{co1} and \elementname{co2}, in this case).
That is, this rule fits both the \elementname{Gluer} and the \elementname{Assembler} functionalities of the two sample plants (for hammers and stools).
For example, this process gives us the semantics that, in order to build an instance of \elementname{Stool}, we require three instances of \elementname{Leg} and one instance of \elementname{Seat}.
We can also achieve this result by using explicit multiplicities in the rule, with a similar technique as we use in Section~\ref{subsubsec:extended-robolang-environmental-mcmts}.
In case that the cardinalities allow for several values, i.e., the minimum and maximum values are different, one copy of the rule will be generated for each possible value.

\begin{figure}
	\centering
	\includegraphics[width=\textwidth]{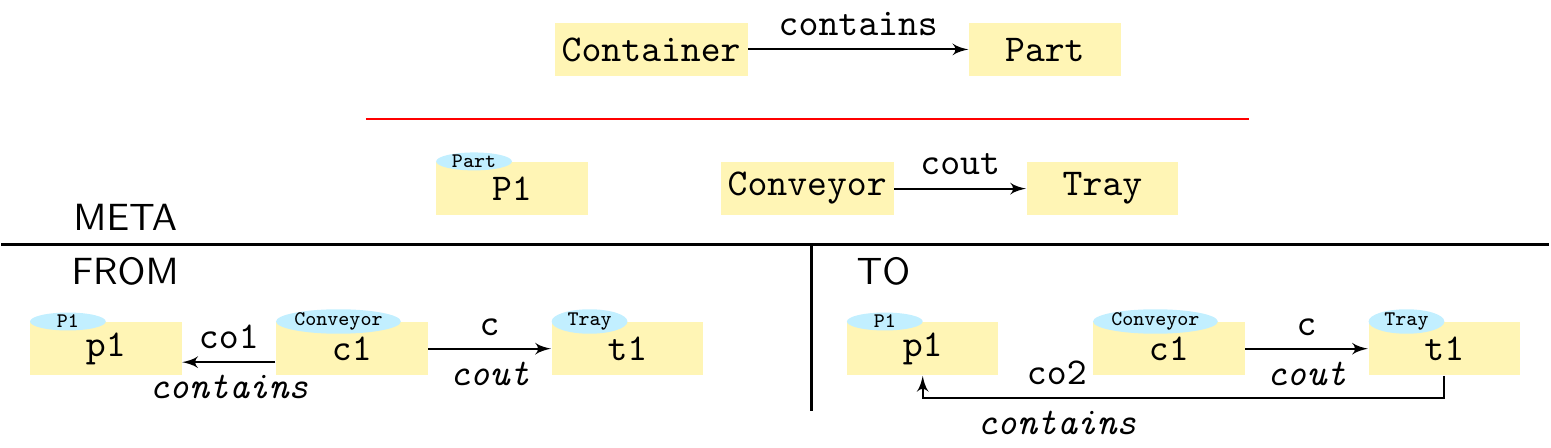}
	\caption{Rule \emph{Transfer Part}: a part is transferred from a conveyor into a tray}
	\label{fig:pls-rule-transfer-part}
\end{figure}

The last rule is used to specify the transfer of the assembled part from the assembler's output conveyor into a tray.
This behaviour is achieved by applying the rule \emph{Transfer Part} displayed in Figure~\ref{fig:pls-rule-transfer-part}.
Note that in this rule we want to define the behaviour related to the elements \elementname{Conveyor} and \elementname{Tray}, which are related to \elementname{hammer\_plant} only.
Thus, this rule cannot (and should not) be applied in the stool-related branch of the hierarchy, and is a good example of how MCMT rules allow us to be precise if needed.

\subsection{Counter}
\label{subsec:counter}

\begin{figure}[ht!]
	\centering
	\includegraphics{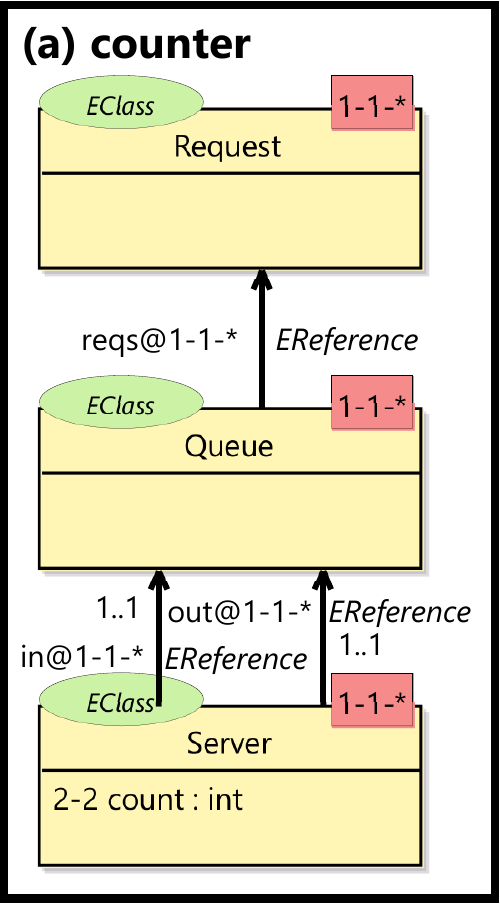}
	\caption{Model for a simple counter}
	\label{fig:counter-level-1-counter}
\end{figure}

This simple hierarchy was created as a proof of concept for the idea of supplementary typing in the context of the PLS case study, also inspired by the example in~\cite{rivera2009formal}.
The hierarchy only has one level apart from Ecore, which is displayed in Figure~\ref{fig:counter-level-1-counter}.
One of the nodes represents the concept of \elementname{Request}.
Requests are sent to a \elementname{Server} which has exactly one \elementname{Queue} for the ones it receives (\elementname{in}), and exactly another one for those it sends back (\elementname{out}).
These concepts can be used to double-type any kind of element which has input and output channels and sends and receives the same type of elements through them.
Once the double types are applied, the \elementname{counter} attribute can be instantiated in order to keep track of the amount of elements which go through the instance of \elementname{Server}.
So, it could be easily used to add counting capabilities to the instances of \elementname{Assembler} or \elementname{Gluer} in the PLS example from Section~\ref{subsec:pls}, which we do not depict due to its simplicity.

This early example does not have its behaviour specified as MCMTs, but it served us to identify the need of future work in amalgamation of MCMTs (see Section~\ref{sec:future-work}), since whichever rule is used to increase the counter must be matched and executed \emph{at the same time} as the rule \emph{Assemble}, so that an instance of \elementname{Assembler} puts together all the required parts and counts in one single rewriting (likewise for instances of \elementname{Gluer}).

\subsection{The bicycle challenge}
\label{subsec:bicycle-challenge}

This case study was presented in~\cite{macias2017challenge} as a solution for the bicycle challenge~\cite{bicyclechallenge2017}, proposed in the context of the MULTI workshop series~\cite{multievents}.
For the sake of self-containment, the main description of the challenge is reproduced here with minimal corrections.
The requirements for the solution and the evaluation criteria have been omitted.

\begin{quote}
A configuration is a physical artefact that is composed of components.
A component may be composed of other components or of basic parts.
There is a difference between the type of a component and its instances.
A component has a weight.
A bicycle is built of components like frame, a handle bar, two wheels\dots
A bicycle component is a component.
A frame, a fork, a wheel, etc. are bicycle components.
Frames and forks exist in various colours.
Every frame has a unique serial number.
Front wheel and rear wheel must have the same size.
Each bicycle has a purchase price and a sales price.
There are different types of bicycles for different purposes such as race, mountains, city\dots
A mountain bike or a city bike may have a suspension.
A mountain bike make have a rear suspension.
That is not the case for city bikes.
A racing fork does not have a suspension.
It does not have a mud mount either.
A racing bike is not suited for tough terrains.
A racing bike is suited for races.
It can be used in cities, too.
Racing frames are specified by top tube, down tube, and seat tube length.
A racing bike can be certified by the UCI.
A racing frame is made of steel, aluminium, or carbon.
A pro race bike is certified by the UCI.
A pro race frame is made of aluminium or carbon.
A pro racing bike has a minimum weight of 5200 gr.
A carbon frame type allows for carbon or aluminium wheel types only.
``Challenger A2-X'' is a pro racer for tall cyclists.
The regular sales price is 4999.00.
Some exemplars are sold for a lower price.
It is equipped with a Rocket-A1-XL pro race frame.
The Rocket-A1-XL has a weight of 920.0 gr.
A sales manager may be interested in the average sales price of all exemplars of a certain model.
He may also be interested in the average sales price of all mountain bikes, all racing bikes etc.
\end{quote}

\begin{figure}
	\centering
	\includegraphics[width=\textwidth]{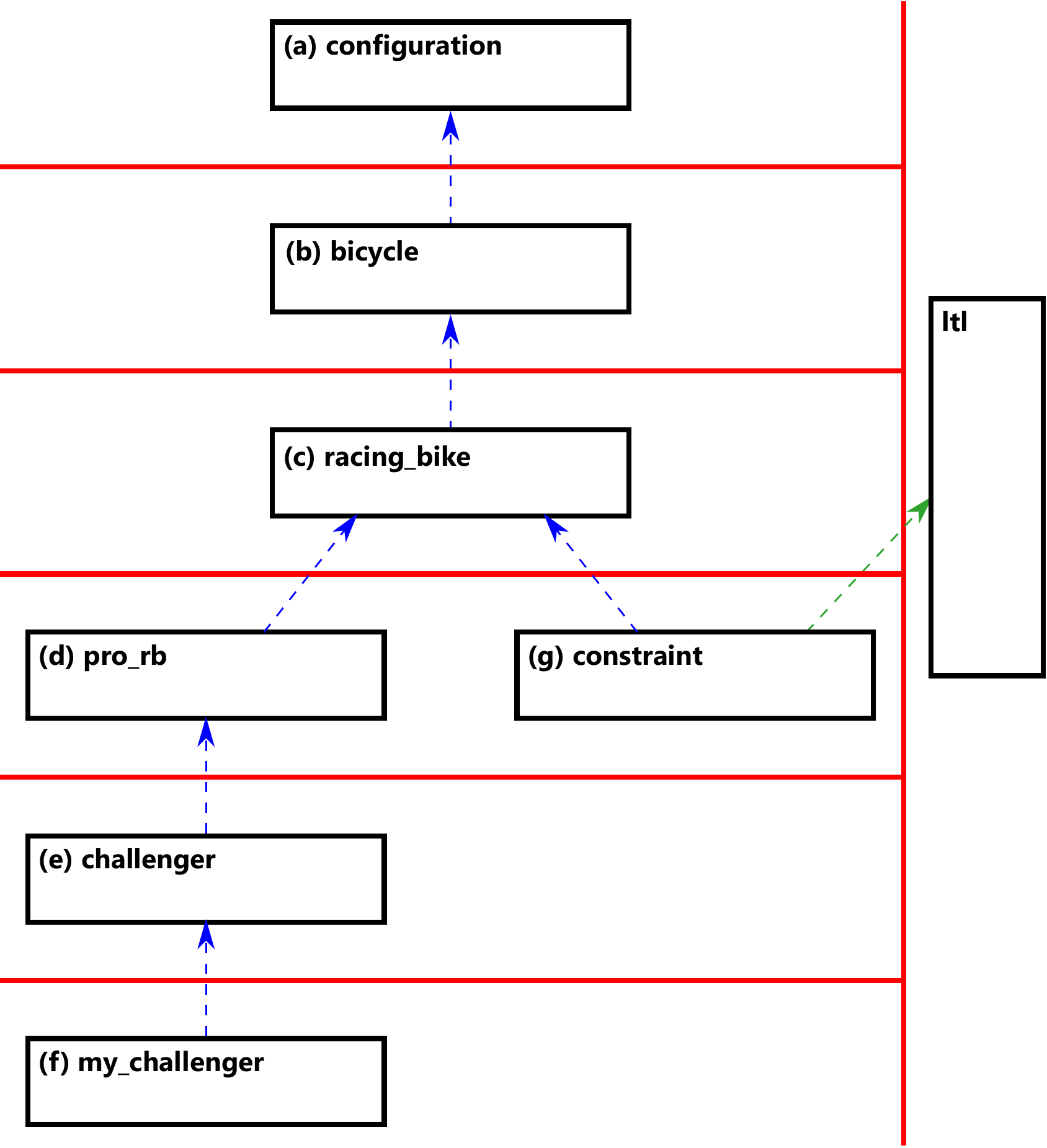}
	\caption{Multilevel hierarchy for the Bicycle case study} 
	\label{fig:bicycle-multilevel-hierarchy-overview}
\end{figure}

The outline of the hierarchy we present as a solution for this challenge is shown in Figure~\ref{fig:bicycle-multilevel-hierarchy-overview}.
Given the nature of the challenge, most of the solution only requires a stack of models in the application dimension, but we do reuse the LTL language used in this thesis in Chapters~\ref{chap:mlm} and~\ref{chap:mcmt} in order to specify a required constraint.

\begin{figure}
	\centering
	\includegraphics{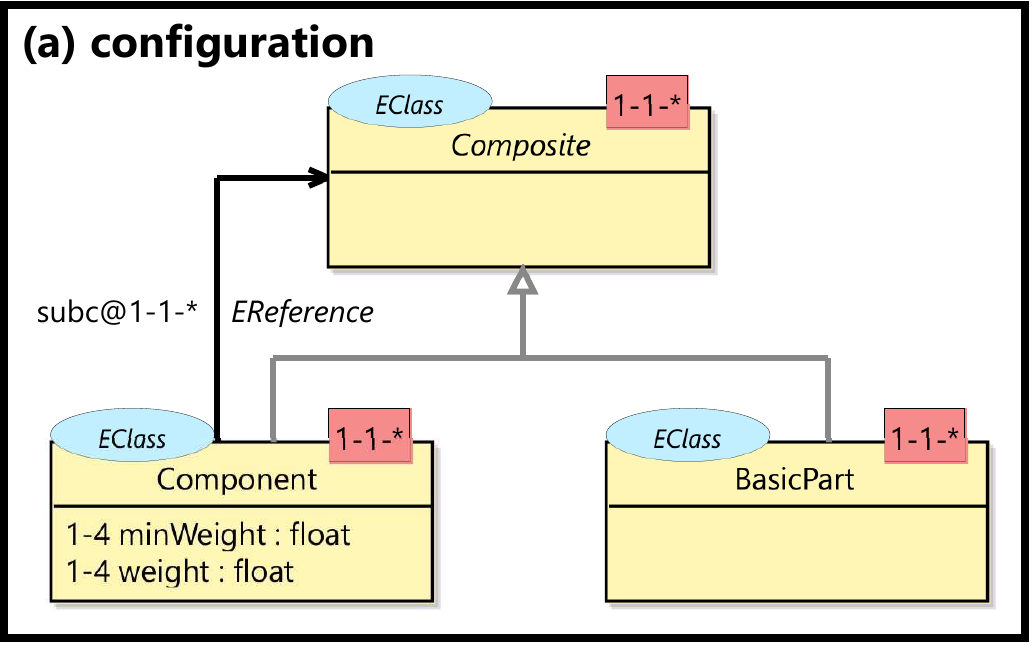}
	\caption{Bicycle challenge - Level 1: Configuration} 
	\label{fig:bicycle-level-1-configuration}
\end{figure}

The only model in the first level of the application hierarchy, in Figure~\ref{fig:bicycle-multilevel-hierarchy-overview}(a) is called \elementname{configuration}.
This level resembles the traditional object-oriented Composite pattern~\cite{gamma1994design}.
Hence, the model exploits inheritance to achieve the pattern, as displayed in Figure~\ref{fig:bicycle-level-1-configuration}.

Apart from the nodes themselves, the description mentions that components may have a weight and a minimum weight, included as attributes.
The minimum weight attribute has potency \elementname{1--3}, indicating that it can be instantiated directly in the level immediately below, two levels below or three levels below (levels 2, 3 and 4, respectively).
This attempts to fulfil the requirement that the knowledge about the domain must be located as high as possible.
In case we want to add more levels to our hierarchy, we may need to update this potency to a higher number.
Also, the sentence ``There is a difference between the type of a component and its instances'' seems to clearly point out a separation between this level and the next one.

\begin{figure}
	\centering
	\includegraphics{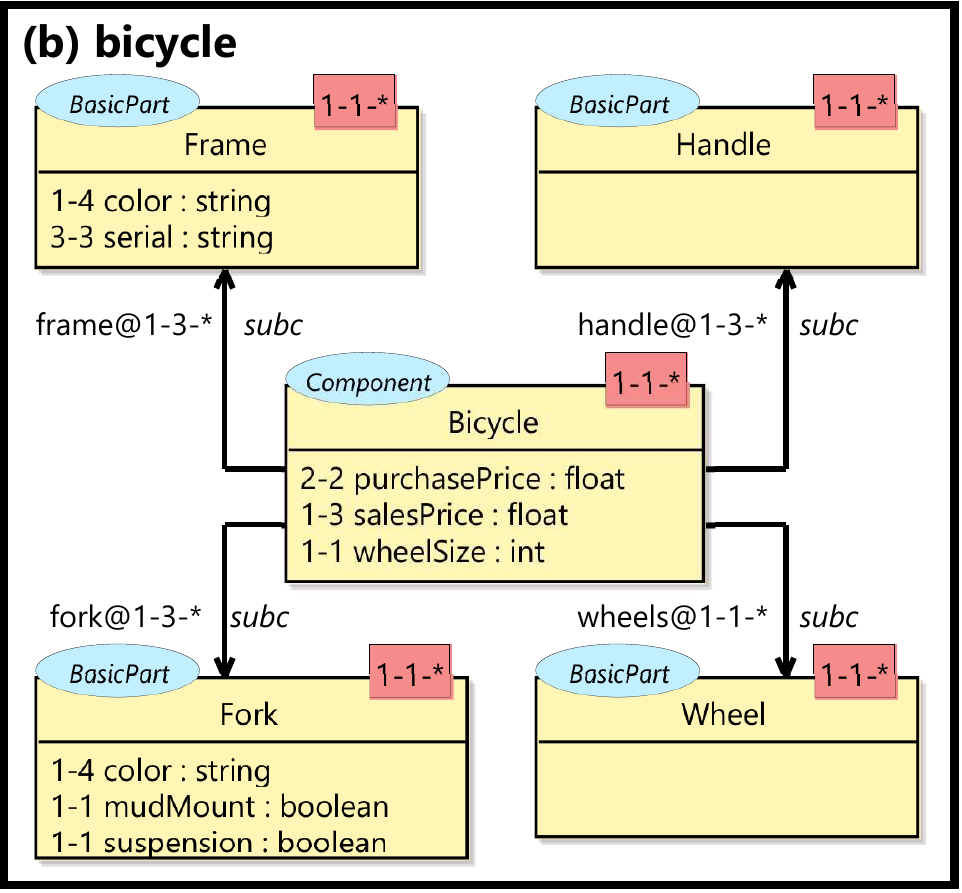}
	\caption{Bicycle challenge - Level 2: Bicycle} 
	\label{fig:bicycle-level-2-bicycle}
\end{figure}

The next model on the hierarchy, located in Figure~\ref{fig:bicycle-multilevel-hierarchy-overview}(b) contains the abstract description of a bicycle and its parts (see Figure~\ref{fig:bicycle-level-2-bicycle}).
Most of the requirements from the description are quite straightforward to model.
The most relevant design decision is giving cardinality \elementname{0..1} to \elementname{purchasePrice} , since some instances of it do not specify it.
It has a potency of \elementname{2--2} since there are still two levels to instantiate it below.
Also, the fact that both wheels must have the same size has been solved by the attribute \elementname{wheelSize} in the class Bicycle hence forcing all wheels to be of the same size.
This design decision can be replaced by the more complex (but arguably more semantically adequate) solution of duplicating the attribute in both wheels and creating a constraint that ensures that they have the same value.

One remarkable usage of potency in this level is that all the relations (except for \elementname{wheels}) are defined with potency \elementname{1--3} so that they do not need to be redefined in the intermediate levels while they can be directly instantiated in the lower levels.
The other interesting use of our range-like potency is for both \elementname{color} attributes.
We chose \elementname{1--4} for it since we believe it is a nice example of flexibility, so that a colour can be specified at the upper levels if that component is made with that colour, but we allow for the lower instances to redefine it, so that a particular bike can differ from its original colours.

The fact that frames have a unique serial number can be easily provided by exploiting Ecore's ID feature.

\begin{figure}
	\centering
	\includegraphics{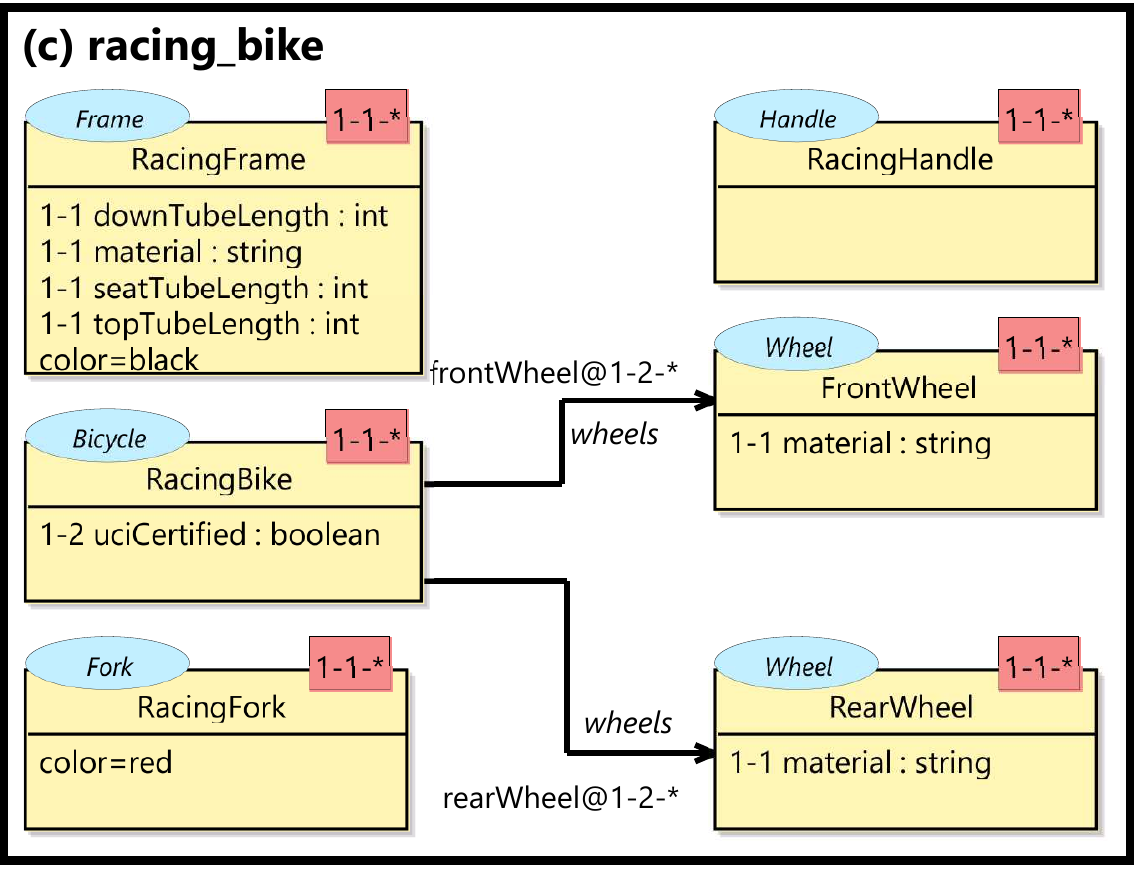}
	\caption{Bicycle challenge - Level 3: Racing Bicycle} 
	\label{fig:bicycle-level-3-racing-bike}
\end{figure}

The next model, \elementname{racing\_bike} in Figure~\ref{fig:bicycle-multilevel-hierarchy-overview}(c), simply instantiates some attributes previously defined, and specifies new ones like the lengths of the different tubes for the frame, as Figure~\ref{fig:bicycle-level-3-racing-bike} shows.
The fact that some bicycles are suitable to certain environments did not look suitable for explicitly modelling, since maintaining a list of them is cumbersome.
At most, we consider that a simple string attribute with the description could be added.

\begin{figure}
	\centering
	\includegraphics{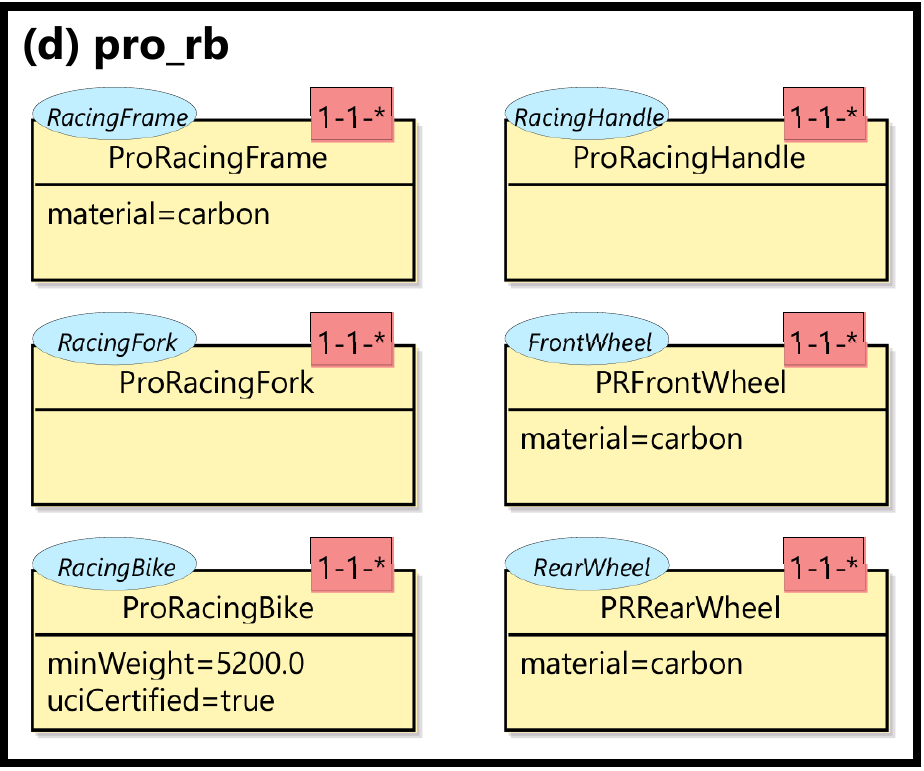}
	\caption{Bicycle challenge - Level 4: Pro Racing Bicycle} 
	\label{fig:bicycle-level-4-pro-rb}
\end{figure}

The model \elementname{pro\_rb} in Figure~\ref{fig:bicycle-multilevel-hierarchy-overview}(d), is quite simple, and just instantiates the attributes previously defined (see Figure~\ref{fig:bicycle-level-4-pro-rb}).

\begin{figure}
	\centering
	\includegraphics{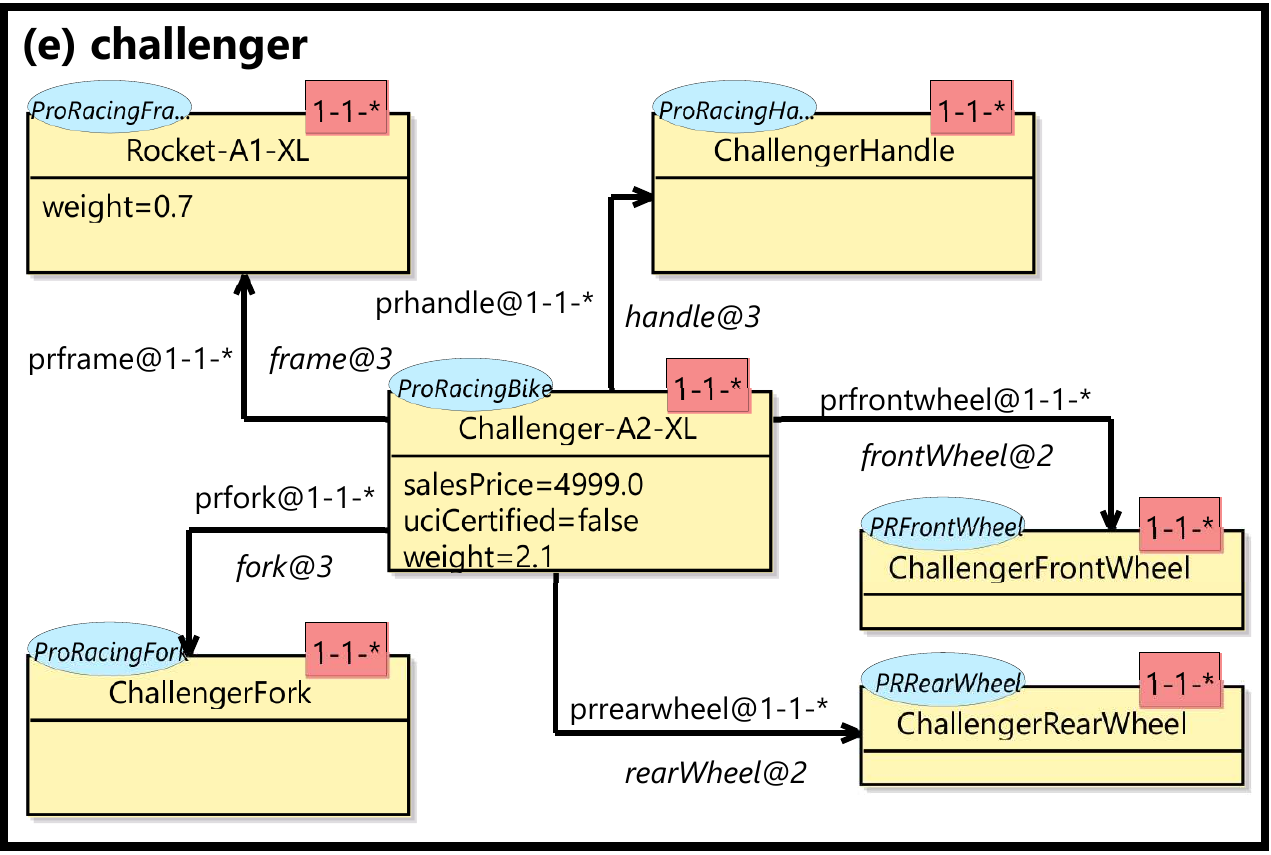}
	\caption{Bicycle challenge - Level 5: Challenger A2-XL} 
	\label{fig:bicycle-level-5-challenger}
\end{figure}

The last level required by the description is the model of a specific brand of bicycle, called \elementname{challenger} in Figure~\ref{fig:bicycle-multilevel-hierarchy-overview}(e), where we can see the instantiation of three attributes, \elementname{weight} (for both frame and bicycle) and \elementname{salesPrice}, defined at higher levels.
The model is depicted in Figure~\ref{fig:bicycle-level-5-challenger}. 
Since \elementname{weight} is defined at level 1 together with \elementname{minWeight}, it is possible to also define a cross-level constraint where we forbid an actual weight to be less than the minimum weight.
Specific instances of this model will instantiate the frame serial ID which will differentiate specific bicycles from each other.

\begin{figure}
	\centering
	\includegraphics{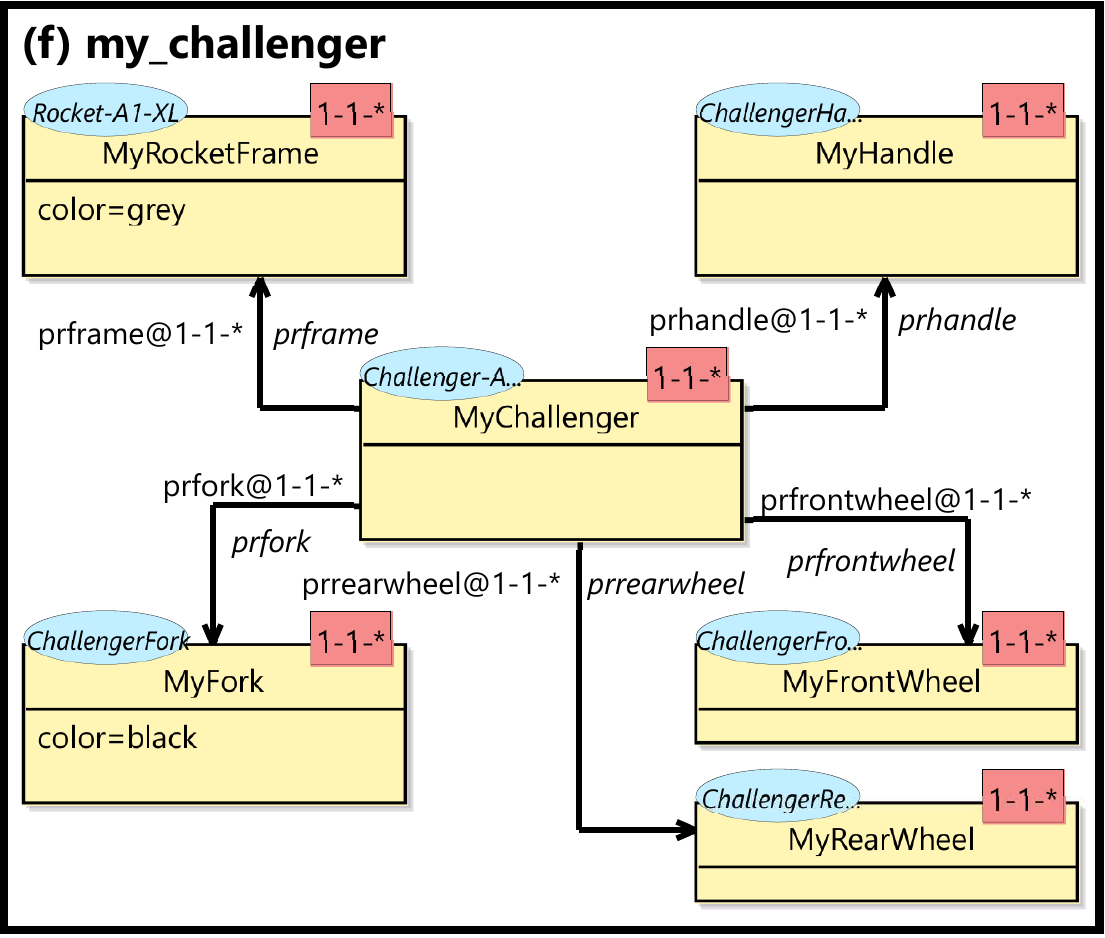}
	\caption{Bicycle challenge - Level 6: My Challenger} 
	\label{fig:bicycle-level-6-my-challenger}
\end{figure}

The previous model defines a particular model of a bike, but probably more than one bike of that model is available.
If the modeller wishes to represent this information, along with some specific differences between this bike an the rest of the Challenger A2-XL bikes, an instance of the previous model can be specified, like the model \elementname{my\_challenger} in Figure~\ref{fig:bicycle-multilevel-hierarchy-overview}(f), that we depict in Figure~\ref{fig:bicycle-level-6-my-challenger}.

Here, the particular colours to which the bike was painted (or re-painted) can be specified, as displayed in \elementname{MyRocketFrame} and \elementname{MyFork}.

\begin{figure}
	\centering
	\includegraphics[width=\textwidth]{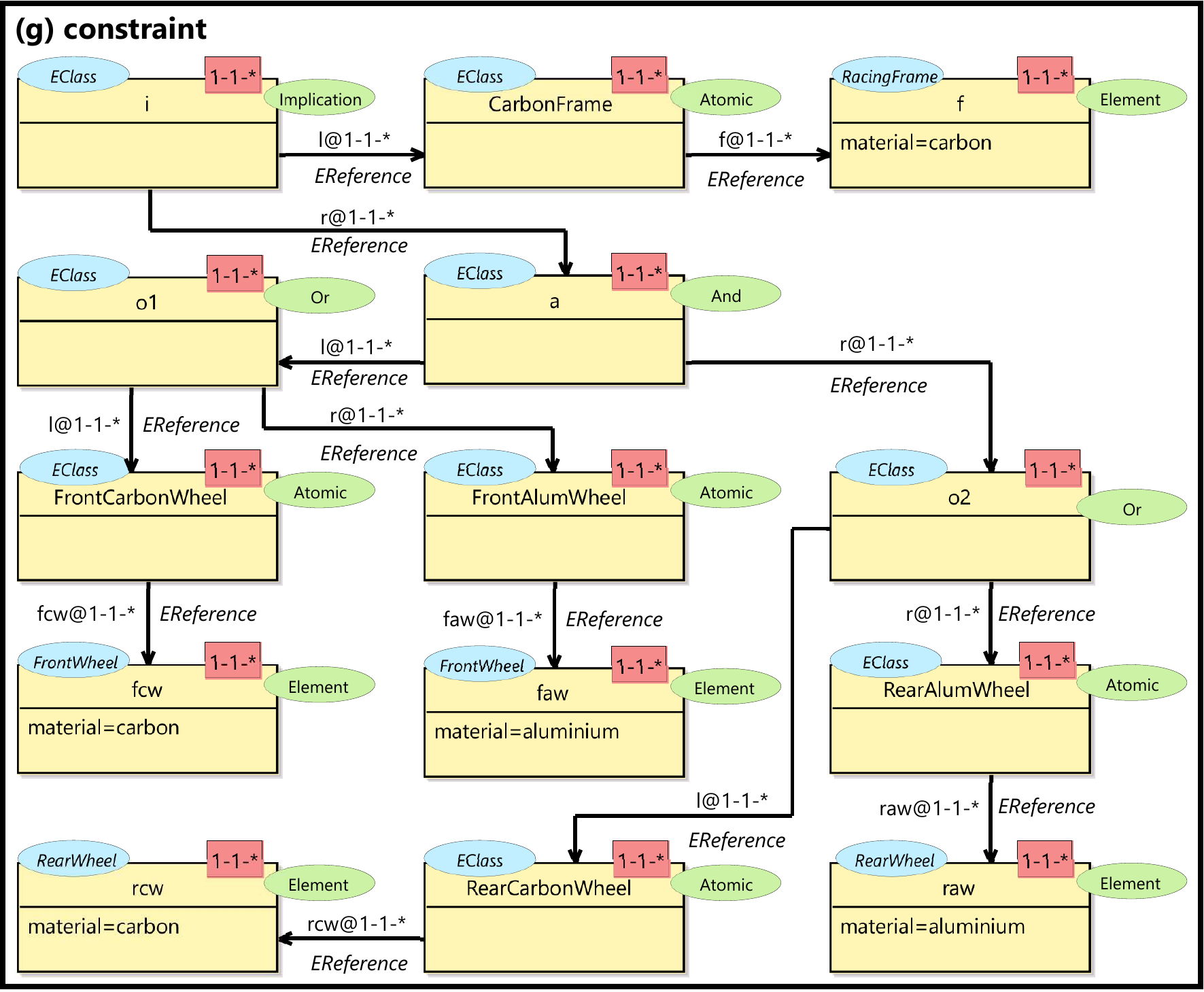}
	\caption{Bicycle challenge - Level 4: Required constraint on materials} 
	\label{fig:bicycle-level-4-constraint}
\end{figure}

The only part of the challenge that we have not discussed so far is the restriction on the material of the wheels: ``A carbon frame type allows for carbon or aluminium wheel types only''.
This requirement is addressed in the following, by defining a constraint using supplementary typing.
We actually use the same hierarchy for LTL presented in~\ref{subsec:supplementary-dimension} to avoid introducing a new one, although we do not use the temporal operators of the language in this case.
Figure~\ref{fig:bicycle-level-4-constraint} depicts the \elementname{constraint} model located in Figure~\ref{fig:bicycle-multilevel-hierarchy-overview}(g), where we define an implication with an atomic proposition the left-hand side and a conjunction of disjunctions in the right-hand side.
For the sake of simplicity, we assume that an instance of the model \elementname{racing\_bike} where this constraint can be checked will contain the elements for defining a single bike, and that the constraint is only meant to be applied in direct instances of \elementname{racing\_bike}.

The atomic proposition in the left-hand side \elementname{CarbonFrame} defines an instance \element{f}{RacingFrame} made of carbon, by instantiating the \elementname{material} attribute defined in \elementname{RacingFrame}, from model \elementname{racing\_bike}.
If a match of this element is found, the left-hand side of the implication is evaluated to true.
Similarly, the right-hand side of the implication contains four atomic propositions where we define the four possible combinations of the two allowed materials (carbon and aluminium) for the two wheels of a racing bike (front and rear).
These atomic propositions are connected with instances of the \elementname{And} and \elementname{Or} operators so that the whole right-hand side is evaluated to true if both wheels are made of either carbon or aluminium, in any combination.
The elements contained within the atomic propositions instantiate the \elementname{material} attribute to express this.

If we use the same names for the atomic propositions, the textual version of the property with boolean logic syntax is \(\mathit{CarbonFrame} \to ((\mathit{FrontCarbonWheel} \vee \mathit{FrontAlumWheel}) \wedge (\mathit{RearCarbonWheel} \vee \mathit{RearAlumWheel}))\).

\section{Experimental Evaluation of MultEcore}
\label{sec:empirical-evaluation}

The tool MultEcore has been developed from scratch in parallel to the work presented in this thesis, based on the principles of our approach presented in Chapters~\ref{chap:mlm} and~\ref{chap:mcmt}.
In this section, we present a set of experiments using publicly available metamodels from the AtlanMod Zoos~\cite{zoos} (to avoid bias since they were developed as external case studies by researchers unrelated to us), which are refactored into multilevel hierarchies.

These experiments, which we originally presented in~\cite{macias2017rearchitecting} and later extended in~\cite{macias2018convergence}, were aimed towards the validation of Multilevel Modelling in general.
That work included our implementation of a rearchitecting tool and a metamodel, introduced in the following, together with the execution of additional experiments to evaluate the validity of MLM against two-level approaches.
While the full set of experiments is not the main focus of this section, the results obtained in some of them enable us to assess the suitability of MultEcore as a target representation for arbitrary multilevel hierarchies.

\begin{figure}[ht]
	\centering
	\includegraphics[width=\linewidth]{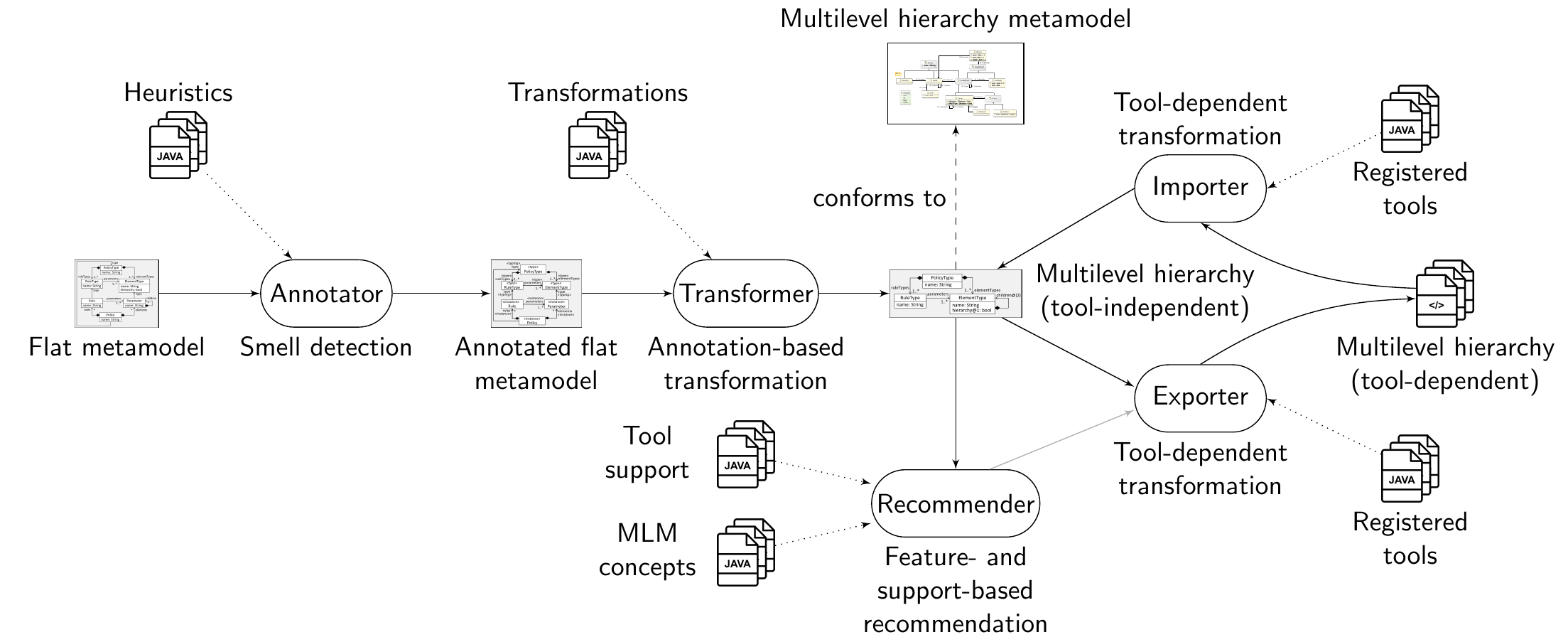}
	\caption{Overview of the modules of the Rearchitecter and their dependencies}
	\label{fig:rearchitecting-process}
\end{figure}

\begin{figure}[ht]
	\centering
	\includegraphics[width=\linewidth]{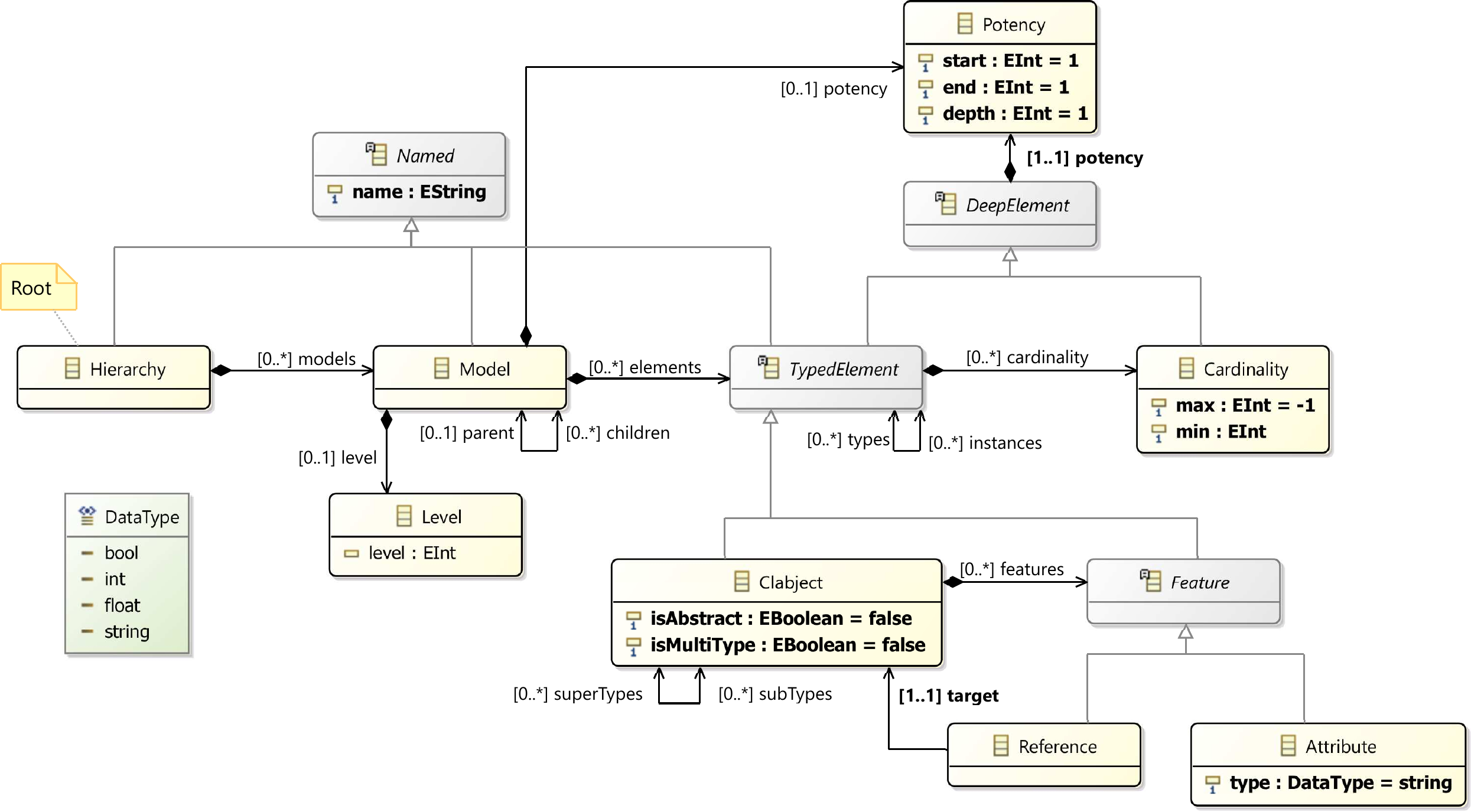}
	\caption{The tool-agnostic metamodel for representing multilevel hierarchies}
	\label{fig:multilevel-hierarchy-metamodel}
\end{figure}

The structure of the tool developed for the experiment, called \emph{Rearchitecter} is shown in Figure~\ref{fig:rearchitecting-process}.
It consists of several modules which follow a pipeline-like structure and manipulate EMF-based models which conform to our own tool-agnostic metamodel for representing multilevel hierarchies.
This metamodel, depicted in Figure~\ref{fig:multilevel-hierarchy-metamodel}, defines the concept of \emph{Hierarchy}, which contains \emph{Models}, which in turn consist of \emph{Clabjects} and \emph{References}.
A more detailed description of this metamodel can be found in~\cite{macias2018convergence}.

The modules that compose the Rearchitecter are the following:

\begin{itemize}
	\item\textbf{Annotator}
	Looks for bad design ``smells'' in Ecore metamodels, and uses EAnnotations to mark them, together with a confidence level.
	Different techniques like structural patterns, string similarity for names and semantic proximity are used in the already implemented heuristics.
	The full description of smells can be found at~\cite{delara2014whenandhow}.
	As a result of this step, we obtain the same Ecore metamodel from the input, enriched with annotations in the elements that conform a smell and a confidence level, normalised between 0 and 10.
	\item\textbf{Transformer}
	Takes an annotated Ecore metamodel as input.
	Based on the information carried by these annotations (which are obtained in the previous step and can be extended or corrected manually) different transformations are applied, in order to turn the elements in the metamodel (e.g.\ an EClass) into instances of elements from our tool-agnostic metamodel (e.g.\ an instance of \emph{Clabject}).
	The result of this process is a model representing a multilevel hierarchy that optimizes the original metamodel, and which is an instance of our custom metamodel.
	\item\textbf{Recommender}
	This module takes an instance model of the tool-agnostic metamodel, which can be generated by the previous step or by importing a hierarchy (explained below), and ranks the registered multilevel tools based on their suitability for modelling the hierarchy represented by the model.
	This ranking is calculated by counting the number of multilevel features that appear in the model, and whether the MLM tools can support them (2 points), emulate them (1 point) or do not support them (-1 point and warning).
	The ranking is both written as console output and as a CSV file; some of these results can be found online~\cite{rearchitecter}.
	This information provides an empirical way of comparing MLM tools, and their suitability for different scenarios.
	\item\textbf{Exporter}
	Since the rearchitecting tool itself does not provide any custom editors for MLM, this module allows to transform the tool-agnostic representation into a tool-dependent one.
	The nature of the generated hierarchy depends completely on the selected tool, so it is the responsibility of the contributor to respect the format used for the target representation.
	\item\textbf{Importer}
	This module is designed to increase the usefulness of the rearchitecting tool for the MLM community, by allowing the exchange of representation of multilevel hierarchies from tool-dependent to tool-independent.
	Similar to the Exporter, it is the contributor's responsibility to ensure the proper exchange of representation, both in parsing the tool-dependent representation and generating a valid tool-agnostic model instance.
	The Importer does not use any model transformation or model weaving techniques (such as~\cite{delfabro2005amma}) since it cannot be guaranteed that the tools use a metamodel for their representation format, or that it will be compatible with EMF technologies.
	This argument also applies to the Exporter.
\end{itemize}

\begin{table}[ht]
\centering
\caption{Results for the initial rearchitecting experiment: best scores in bold and cases with unsupported features in red}
\label{tab:initial-experiment-results}
\begin{tabular}{|l|c|c|c|c|c|c|c|}
	\hline
	\multirow{2}{*}{\bfseries Hierarchy}	& \multicolumn{4}{|c|}{\bfseries Size (multilevel)}	&	\multirow{2}{*}{\bfseries Melanee}	& \multirow{2}{*}{\bfseries MetaDepth}	& \multirow{2}{*}{\bfseries MultEcore}	\\
	\cline{2-5}
					& \bfseries \#M		&\bfseries \#C		& \bfseries \#R		& \bfseries \#A		&				&				&					\\
	\hline
	Sec. Policies	& 1					& 4					& 5					& 4					& \bfseries 31	& 29			& \color{red} 10	\\
	\hline
	Agate			& 1					& 64				& 118				& 81				& 398			& \bfseries 518	& 279				\\
	\hline
	CloudML			& 1					& 15				& 17 				& 26				& 111			& \bfseries 120	& 64				\\
	\hline
	CloudML-2.0		& 1					& 21				& 40				& 44				& 188			& \bfseries 214	& 112				\\
	\hline
	HAL				& 1					& 41				& 15				& 72				& 256			& \bfseries 284	& 156				\\
	\hline
\end{tabular}
\end{table}

All previous modules are designed with interfaces to facilitate extending all modules, as Figure~\ref{fig:rearchitecting-process} shows.
For example, the Annotator can receive new heuristics, or the Exporter can be extended with new tool formats.
With this implementation, we performed an initial experiment with an early version of MultEcore, in order to identify its strengths and weaknesses and guide the development process.
Five metamodels from the AtlanMod Zoos were rearchitected into their multilevel counterparts, and the Recommender evaluated the suitability of three different tools as a target representation for the hierarchy: MetaDepth~\cite{delara2010deep}, Melanee~\cite{atkinson2016melanee} and our tool MultEcore.
The scores obtained for each metamodel and tool are shown in Table~\ref{tab:initial-experiment-results}.
The columns \emph{\#M}, \emph{\#C}, \emph{\#R} and \emph{\#A}, show the size of the multilevel hierarchies (resulting from the rearchitecting process) in terms of number of metamodels, classes, references and attributes, respectively.

\begin{table}[hb]
\centering
\caption{Results for the second rearchitecting experiment: best scores in bold}
\label{tab:second-experiment-results}
\begin{tabular}{|l|c|c|c|c|c|c|c|}
	\hline
	\multirow{2}{*}{\bfseries Hierarchy}	& \multicolumn{4}{|c|}{\bfseries Size (multilevel)}	&	\multirow{2}{*}{\bfseries Melanee}	& \multirow{2}{*}{\bfseries MetaDepth}	& \multirow{2}{*}{\bfseries MultEcore}	\\
	\cline{2-5}
					& \bfseries \#M		&\bfseries \#C		& \bfseries \#R		& \bfseries \#A		&		&				&				\\
	\hline
	Sec. Policies	& 1					& 4					& 5					& 4					& 31	& 29			& \bfseries 32	\\
	\hline
	Agate			& 1					& 64				& 118				& 81				& 398	& \bfseries 518	& \bfseries 518	\\
	\hline
	CloudML			& 1					& 15				& 17 				& 26				& 111	& \bfseries 120	& \bfseries 120	\\
	\hline
	CloudML-2.0		& 1					& 21				& 40				& 44				& 188	& \bfseries 214	& \bfseries 214	\\
	\hline
	HAL				& 1					& 41				& 15				& 72				& 256	& \bfseries 284	& \bfseries 284	\\
	\hline
\end{tabular}
\end{table}

As expected, the immature state of MultEcore at the time of the first experiment caused it to perform poorly.
Analysing these results closely, we identified the most relevant features in which to focus our efforts, namely the support for potency depth (as explained in Section~\ref{subsec:potency}), the handling of attributes and attribute potencies (see Section~\ref{subsec:data-type-dimension}) and the semantics of inheritance relations among nodes (Section~\ref{subsec:inheritance-and-cardinalities}).
After an important overhaul of the tool, we performed the experiment again, updating the support of MultEcore for the different multilevel features in the Recommender.
The new scores for our tool are depicted in Table~\ref{tab:second-experiment-results}.
Here, we see how MultEcore performs after its update at least as well as the other two tools, if not better.
Therefore, these experiments give us some empirical validation that our tool is able to support most relevant multilevel features and can be used in realistic scenarios.

\begin{table}[htb]
\centering
\caption{Results for each tool after importing and recommendation: best scores in bold and cases with unsupported features in red}
\label{tab:third-experiment-results}
\begin{tabular}{|l|c|c|c|c|c|c|c|}
	\hline
	\multirow{2}{*}{\bfseries Hierarchy}	& \multicolumn{4}{|c|}{\bfseries Size (Multilevel)}	&	\multirow{2}{*}{\bfseries Melanee}	& \multirow{2}{*}{\bfseries MetaDepth}	& \multirow{2}{*}{\bfseries MultEcore}	\\
	\cline{2-5}
						& \bfseries \#M	&\bfseries \#C	& \bfseries \#R	& \bfseries \#A	&					&						&				\\
	\hline
	bicycle				& 6				& 32			& 17			& 17			& \color{red} 125	& \color{red} 119		& \bfseries 140	\\
	\hline
	datatypes			& 3				& 21			& 27			& 0				& \bfseries 96		& \bfseries 96			& \bfseries 96	\\
	\hline
	ltl					& 5				& 59			& 35			& 4				& \bfseries 196		& \bfseries 196			& \bfseries 196	\\
	\hline
	petrinets			& 4				& 20			& 21			& 3				& \color{red} 85	& \color{red} 82		& \bfseries 94	\\
	\hline
	pls					& 5				& 32			& 25			& 0				& \color{red} 105	& \color{red} 105		& \bfseries 114	\\
	\hline
	robolang			& 7				& 69			& 78			& 1				& \color{red} 215	& \color{red} 214		& \bfseries 296	\\
	\hline
\end{tabular}
\end{table}

In addition to these experiments, we also used some of our own examples to validate MultEcore.
These examples are designed to illustrate our tool's strengths.
So a way to validate our claims is to use the Recommender to rank the suitability of the three MLM tools aforementioned.
If the best-scoring tool is MultEcore, we can gain confidence that our claims are well founded.

In order to perform this experiment, we used the Importer to obtain the tool-agnostic representations of our examples and input them in the Recommender.
The results of this evaluation are summarised in Table~\ref{tab:third-experiment-results}.

We see how the scores of this experiment yields the expected results: MultEcore is the only suitable tool in the more complex examples, and ties with the other two in the simpler ones.
Consequently, we get an empirical confirmation that MultEcore possesses certain strengths which differentiate it from other MLM tools.

All the experiences presented in this section, together with some additional experiments, detailed explanations and the related code artefacts have been presented in more detail in~\cite{macias2018convergence} and made available in~\cite{rearchitecter}, as a contribution to the MLM community.

\section{Examples of Constructions in \cat{Chain}}
\label{sec:examples-constructions-chain}

In Chapters~\ref{chap:mlm} and~\ref{chap:mcmt} of this thesis, we presented the formal constructions which support our approach to multilevel modelling and multilevel model transformations, using Category Theory.
To verify that our formalisation is aligned with our requirements and intuitions, we recall in the following an example hierarchy and an MCMT rule and apply our formal constructions to them.
We then verify that the results we obtain are consistent with the expected behaviour of this constructions, and hence validate our formalisation with actual examples.
In addition, this section can help the reader understand how the formal concepts of multilevel typing, domain of definition, rule matching and application, etc. relate to each other.
Due to the complexity of the figures contained here, and the limitations imposed by page size, we made available the sources of these figures in the MultEcore website~\cite{multecore}, where the layers assigned to different maps can be disabled at will to aid their understanding.

Shifting away from the Robolang example that we use throughout this thesis, we choose here our examples from the PLS case study presented in Section~\ref{subsec:pls}.
We use the left branch of the multilevel hierarchy depicted in Figure~\ref{fig:pls-multilevel-hierarchy-overview}(a,b,d) and the MCMT \emph{Transfer Part} from Figure~\ref{fig:pls-rule-transfer-part}.
We include below a simplified version of these two hierarchies to facilitate the reading of this section.
Note that the topmost Ecore metamodel is taken into account to validate the constructions, so it appears in several of the following diagrams.

The concepts and constructions we focus on are typing chains and bindings (morphisms) between them for rule application, as shown in Figure~\ref{fig:mlm-binding-diagram-formal}, and inclusion chains and refactoring of multilevel typing into typing chain morphisms, as Figure~\ref{fig:mlm-refactoring-diagram-formal} depicts.
Versions of these two diagrams for each combination of models that we check are also depicted below.

Furthermore, we use these examples to motivate the use of inequality instead of equality in condition~\ref{eq:preserve-reflect}, which we reproduce here for convenience:
\[	\domain{\typemorph[\rulegraph]{\indextwo}{\indexone}} \sqsubseteq \mapruletohierarchybinding_{\indextwo}^{-1}(\domain{\typemorph[\hierarchygraph]{\maplevelone(\indextwo)}{\maplevelone(\indexone)}})\]
That is, we show an example where a version of this condition with \(=\) instead of \(\sqsubseteq\) is too strict, which justifies the more relaxed condition that we use.

\begin{figure}[ht]
	\centering
	\includegraphics[scale=.5]{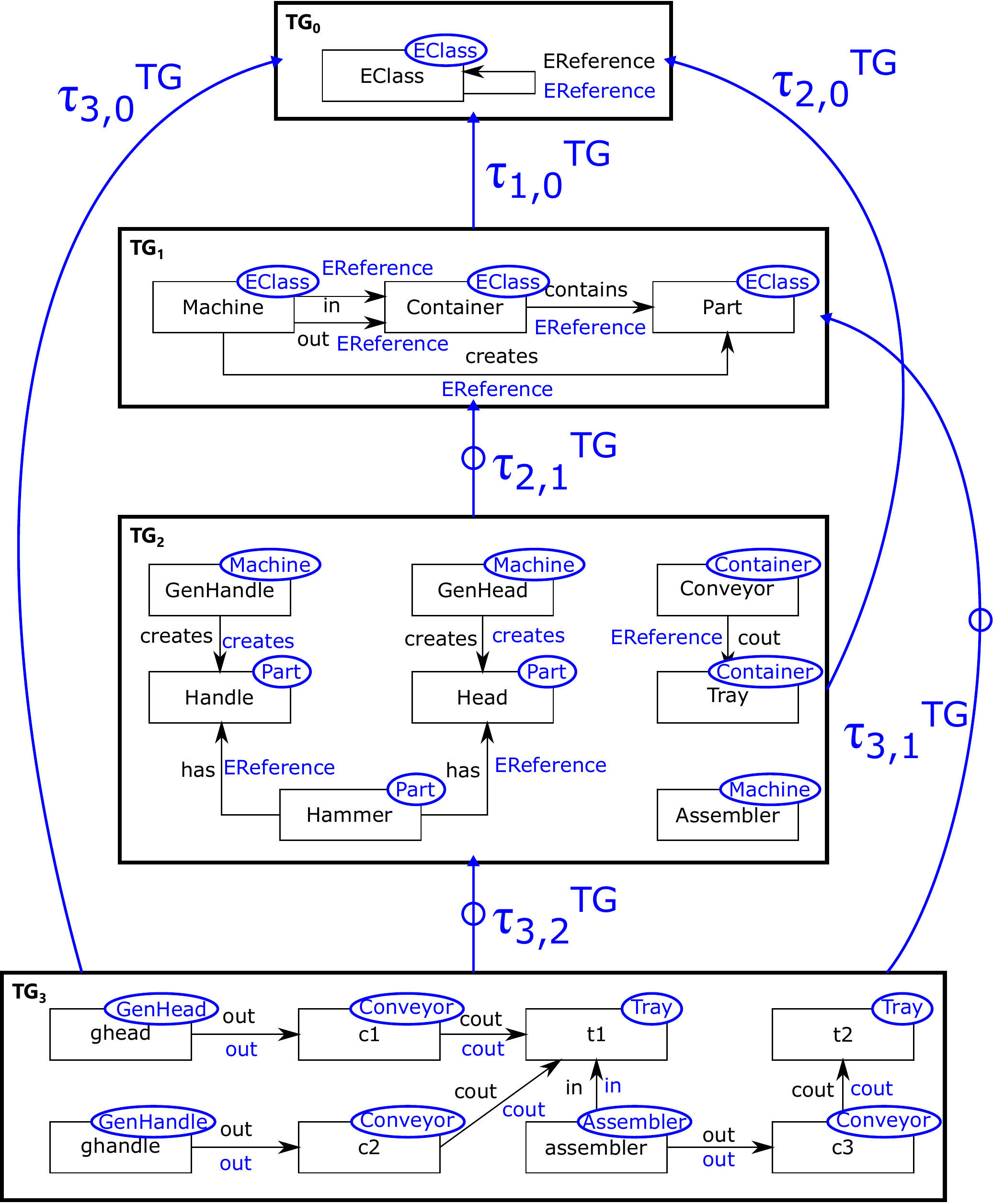}
	\caption{Simplified PLS hierarchy used for formal constructions}
	\label{fig:pls-hammer-branch-multilevel-hierarchy-formal}
\end{figure}

The hierarchy example that we are using (notation simplified here), is shown in Figure~\ref{fig:pls-hammer-branch-multilevel-hierarchy-formal}.
It consists of the typing chain \(\chainname{\hierarchygraph} = \chain{\hierarchygraph}{3}\) with a sequence \(\tc{\hierarchygraph} = [\graphname{\hierarchygraph}{3}, \graphname{\hierarchygraph}{2}, \graphname{\hierarchygraph}{1}, \graphname{\hierarchygraph}{0}]\) of  models where, as stated before, \(\graphname{\hierarchygraph}{3} = \elementname{hammer\_config}\), \(\graphname{\hierarchygraph}{2} = \elementname{hammer\_plant}\), \(\graphname{\hierarchygraph}{1} = \elementname{generic\_plant}\), and \(\graphname{\hierarchygraph}{0} = \elementname{Ecore}\). 

\begin{figure}[ht]
	\centering
	\includegraphics[scale=.5]{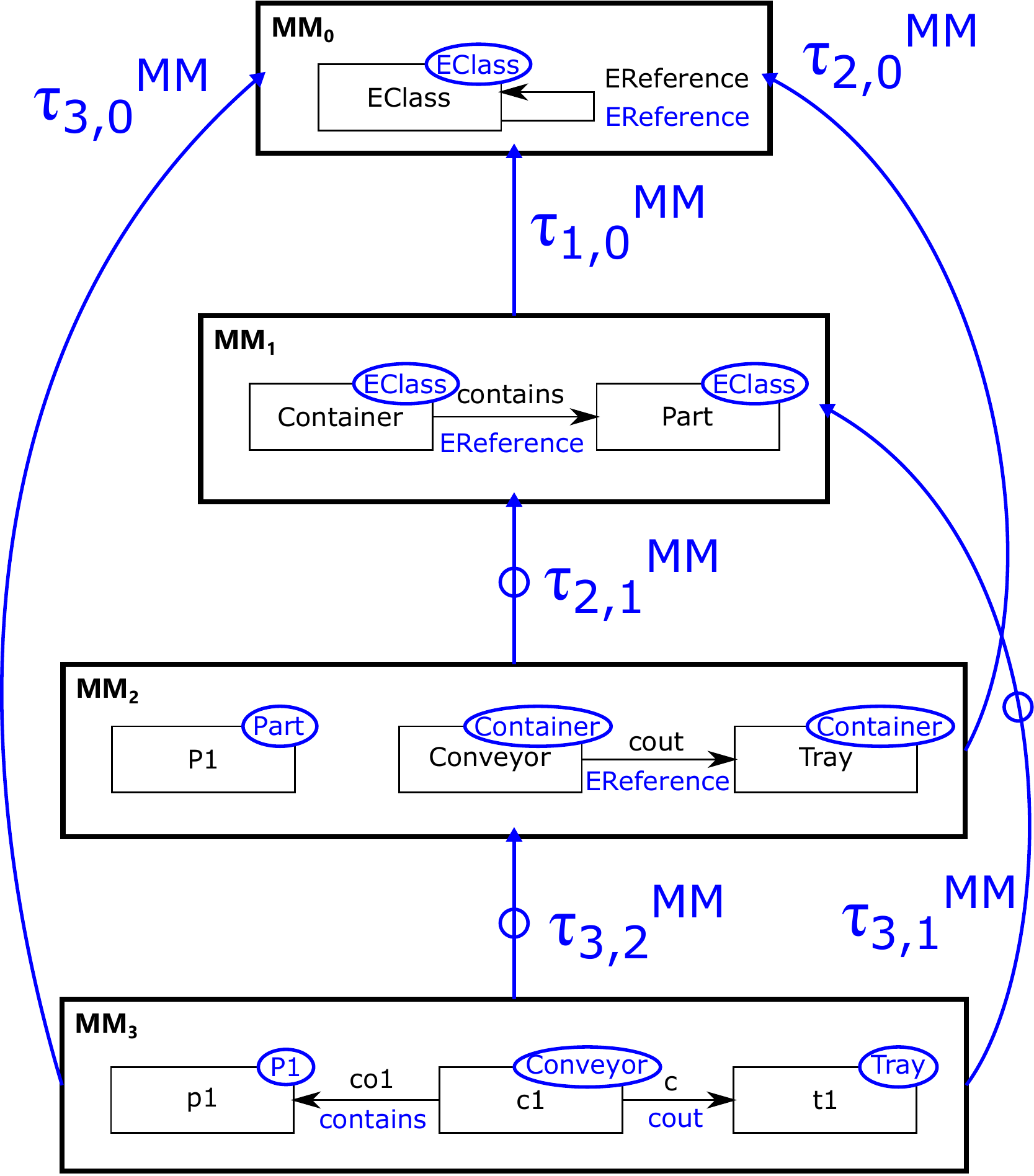}
	\caption{Simplified \emph{Transfer Part} rule hierarchy used for formal constructions}
	\label{fig:pls-mcmt-transfer-part-multilevel-hierarchy-formal}
\end{figure}

The left-hand side (FROM block, relevant for calculating rule matches) of the rule \emph{Transfer Part} is displayed in Figure~\ref{fig:pls-mcmt-transfer-part-multilevel-hierarchy-formal} with the same simplified notation.
This diagram shows the typing chain \(\chainname{\rulegraph} = \chain{\rulegraph}{3}\) with \(\tc{\rulegraph} = [\graphname{\rulegraph}{3}, \graphname{\rulegraph}{2}, \graphname{\rulegraph}{1}, \graphname{\rulegraph}{0}]\) being a sequence of models with the same simplified notation used in the previous one.

In the following we use the names \(\graphname{\hierarchygraphleft}{}\) and \(\graphname{\hierarchygraph}{\chaindepthone}\) interchangeably, depending on the context.
The same applies to \(\graphname{\rulegraphleft}{}\) and \(\graphname{\rulegraph}{\chaindepthone}\).

\subsection{Required calculations}
\label{subsec:required-calculations}

In order to check the constructions from Figures~\ref{fig:mlm-binding-diagram-formal} and~\ref{fig:mlm-refactoring-diagram-formal} and condition~\ref{eq:preserve-reflect} we first need to calculate the domain of definition between every pair of levels, for both hierarchies in Figures~\ref{fig:pls-hammer-branch-multilevel-hierarchy-formal} and~\ref{fig:pls-mcmt-transfer-part-multilevel-hierarchy-formal}.

\begin{figure}[ht!]
	\centering
	\includegraphics[scale=.5]{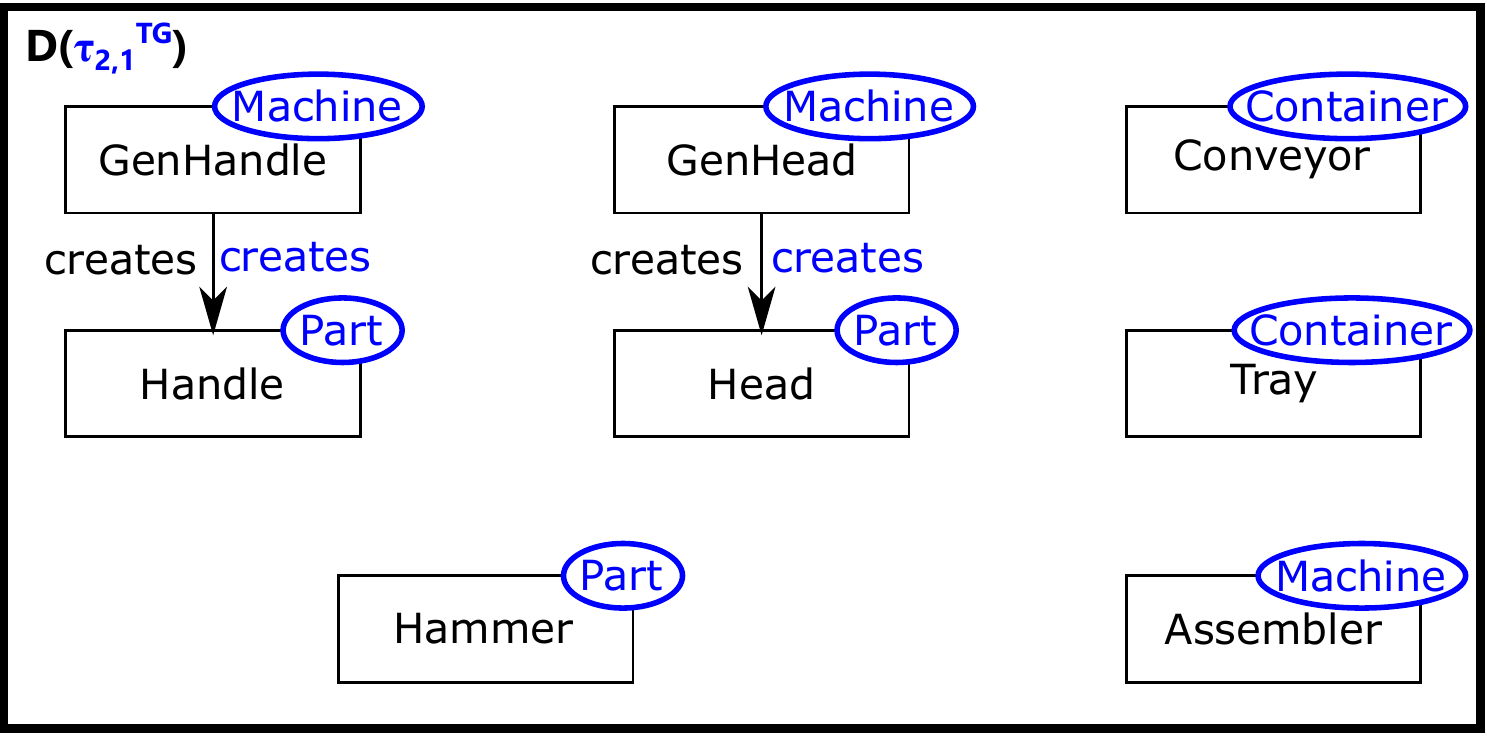}
	\caption{Visual representation of \(\domain{{\color{cdblue}\typemorph[\hierarchygraph]{2}{1}}}\)}
	\label{fig:pls-hammer-branch-formal-domain-2-1}
\end{figure}

\begin{figure}[ht!]
	\centering
	\includegraphics[scale=.5]{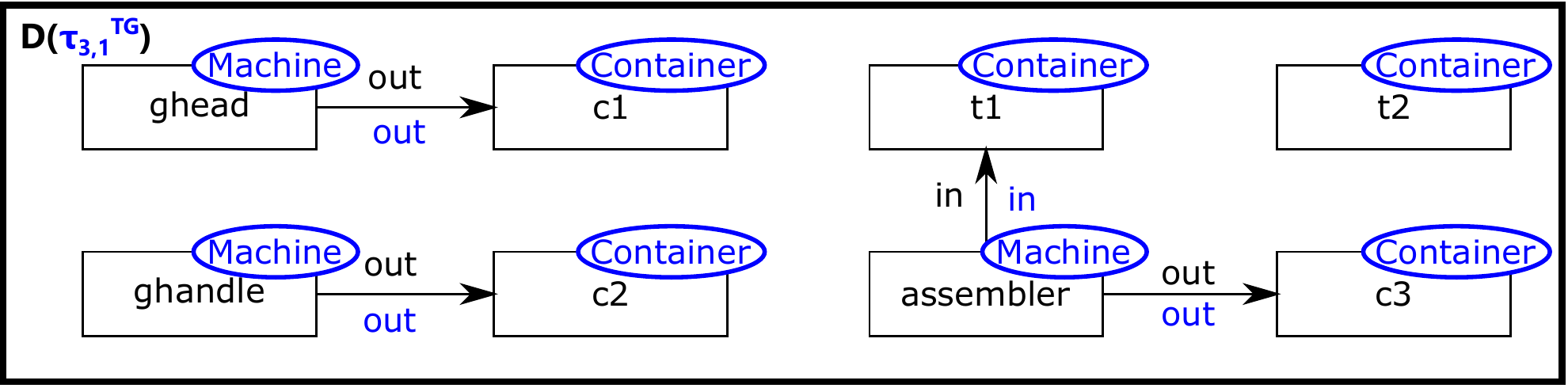}
	\caption{Visual representation of \(\domain{{\color{cdblue}\typemorph[\hierarchygraph]{3}{1}}}\)}
	\label{fig:pls-hammer-branch-formal-domain-3-1}
\end{figure}

\begin{figure}[ht!]
	\centering
	\includegraphics[scale=.5]{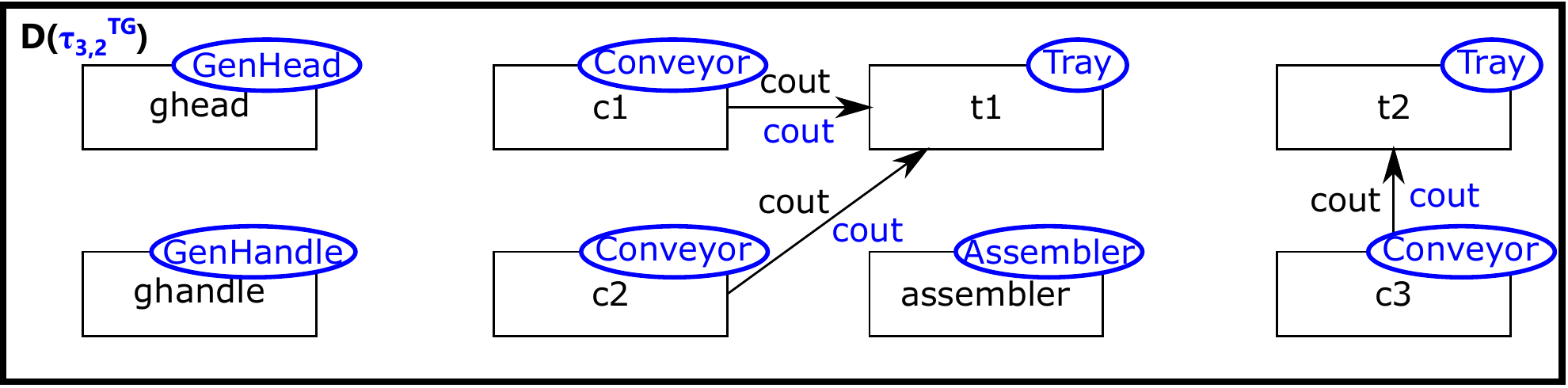}
	\caption{Visual representation of \(\domain{{\color{cdblue}\typemorph[\hierarchygraph]{3}{2}}}\)}
	\label{fig:pls-hammer-branch-formal-domain-3-2}
\end{figure}

Since \(\color{cdblue}\typemorph[\hierarchygraph]{1}{0}\), \(\color{cdblue}\typemorph[\hierarchygraph]{2}{0}\) and \(\color{cdblue}\typemorph[\hierarchygraph]{3}{0}\) are total, we get that \(\domain{{\color{cdblue}\typemorph[\hierarchygraph]{1}{0}}} = \graphname{\hierarchygraph}{1}\), \(\domain{{\color{cdblue}\typemorph[\hierarchygraph]{2}{0}}} = \graphname{\hierarchygraph}{2}\) and \(\domain{{\color{cdblue}\typemorph[\hierarchygraph]{3}{0}}} = \graphname{\hierarchygraph}{3}\).
Then, the only ones that we need to calculate are \(\domain{{\color{cdblue}\typemorph[\hierarchygraph]{2}{1}}}\), \(\domain{{\color{cdblue}\typemorph[\hierarchygraph]{3}{1}}}\) and \(\domain{{\color{cdblue}\typemorph[\hierarchygraph]{3}{2}}}\), respectively shown in Figures~\ref{fig:pls-hammer-branch-formal-domain-2-1},~\ref{fig:pls-hammer-branch-formal-domain-3-1} and~\ref{fig:pls-hammer-branch-formal-domain-3-2}.

Furthermore, we also require the inclusion chains for \(\graphname{\hierarchygraphleft}{} = \graphname{\hierarchygraph}{3}\).

As explained in Section~\ref{subsec:pls}, the model \(\graphname{\hierarchygraph}{3}= \elementname{hammer\_config}\) represents a specific configuration of a manufacturing plant for hammers, and can be modified over time via transformation rule applications.
Hence, we can consider that \elementname{hammer\_config} is the actual graph \(\graphname{\hierarchygraphleft}{}\) where a rule may be applied.
Consequently, \(\graphname{\hierarchygraphleft}{} = \graphname{\hierarchygraph}{3}\) is typed over the remaining models in its typing chain \(\chainname{\hierarchygraph'} = \chain{\hierarchygraph'}{2}\) with \(\tc{\hierarchygraph'} = [\graphname{\hierarchygraph}{2}, \graphname{\hierarchygraph}{1}, \graphname{\hierarchygraph}{0}]\).

\begin{figure}[h!]
	\centering
	\includegraphics[scale=.5]{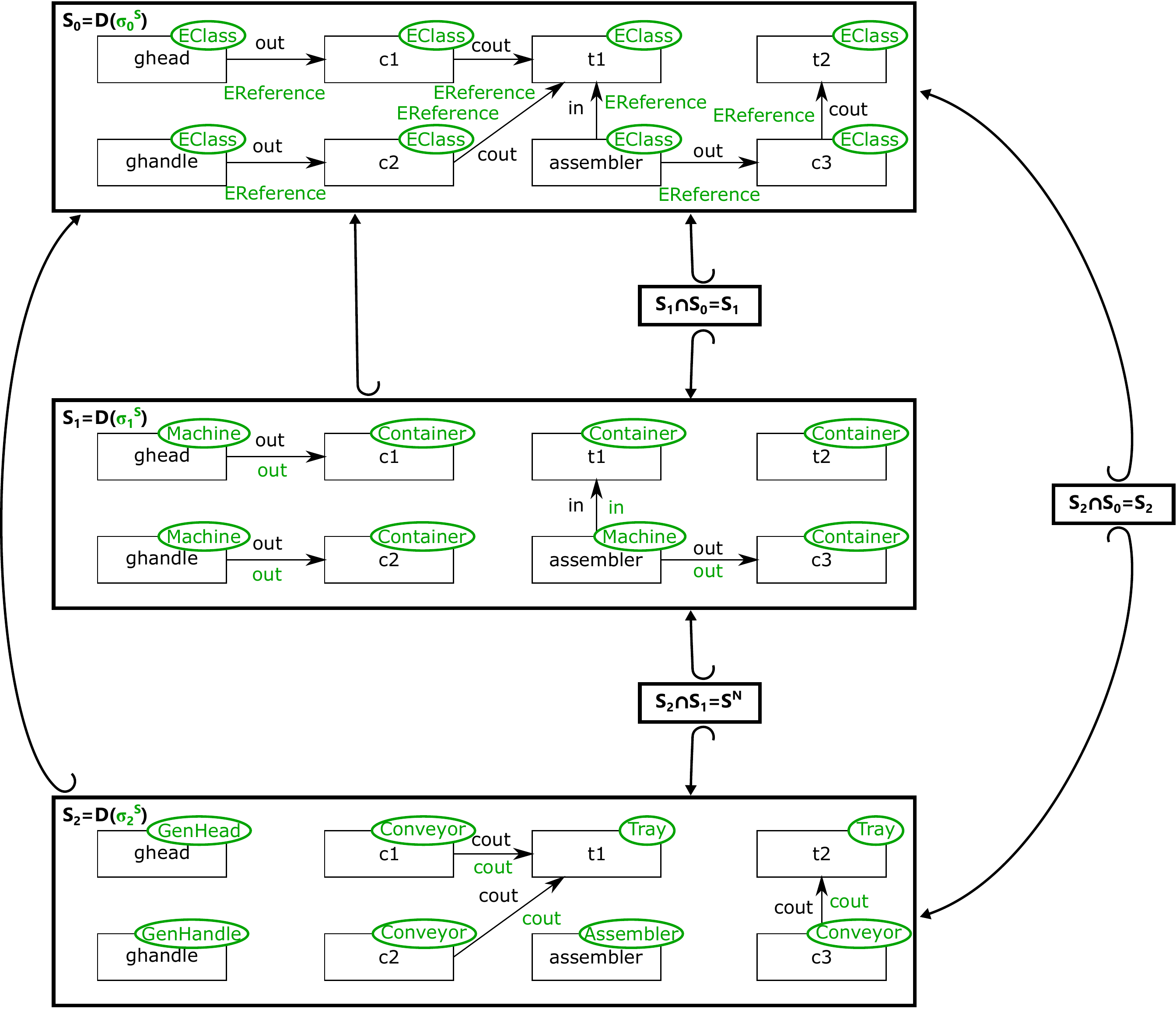}
	\caption{Inclusion chain \(\chainmorph{\hierarchygraphleft} : \graphname{\hierarchygraphleft}{} \Rightarrow \chainname{\hierarchygraph'}\)}
	\label{fig:pls-hammer-branch-formal-inclusion-chains-tg3}
\end{figure}

We can now construct the multilevel typing of \elementname{hammer\_config} as an inclusion chain \(\chainmorph{\hierarchygraphleft} : \graphname{\hierarchygraphleft}{} \Rightarrow \chainname{\hierarchygraph'}\) with \(\chainname{\hierarchygraphleft} = \chain{\hierarchygraphleft}{2}\) and \(\graphname{\hierarchygraphleft}{\indextwo} := \domain{\chainmorph[\indextwo]{\hierarchygraphleft}}\) for all \(\indextwo \in [2]\), by constructing the chain morphism \((\chainmorph{\hierarchygraphleft},id_{[2]}) : \chainname{\hierarchygraphleft} \to \chainname{\hierarchygraph'}\).
Using a simplified, colour-coded representation similar to the previous one, we represent this inclusion chain as shown in Figure~\ref{fig:pls-hammer-branch-formal-inclusion-chains-tg3}.

\begin{figure}[h!]
	\centering
	\includegraphics[scale=.5]{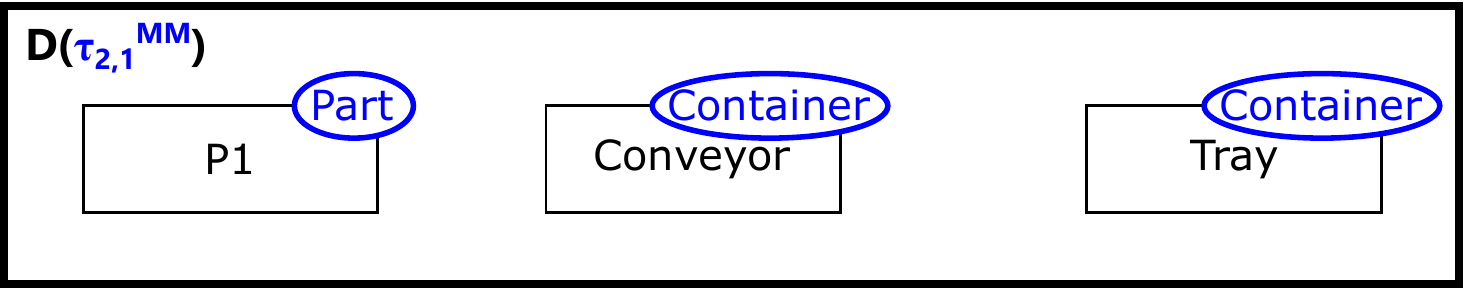}
	\caption{Visual representation of \(\domain{{\color{cdblue}\typemorph[\rulegraph]{2}{1}}}\)}
	\label{fig:pls-mcmt-transfer-part-formal-domain-2-1}
\end{figure}

\begin{figure}[h!]
	\centering
	\includegraphics[scale=.5]{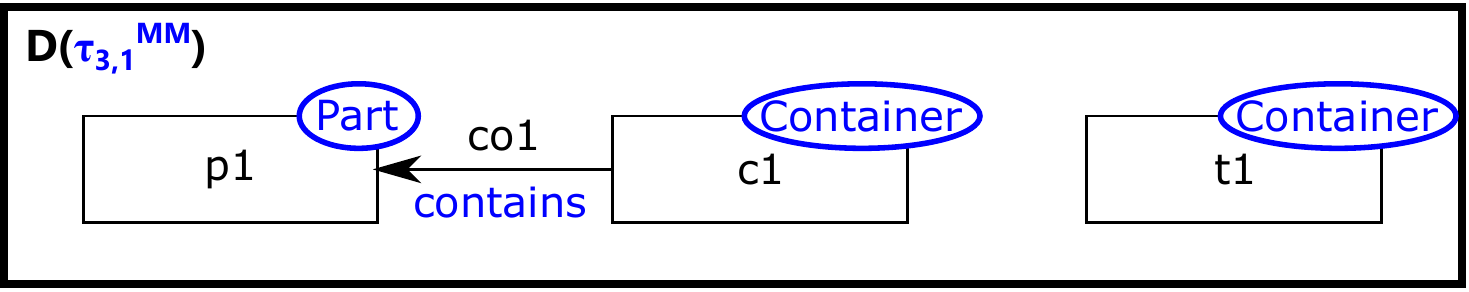}
	\caption{Visual representation of \(\domain{{\color{cdblue}\typemorph[\rulegraph]{3}{1}}}\)}
	\label{fig:pls-mcmt-transfer-part-formal-domain-3-1}
\end{figure}

\begin{figure}[h!]
	\centering
	\includegraphics[scale=.5]{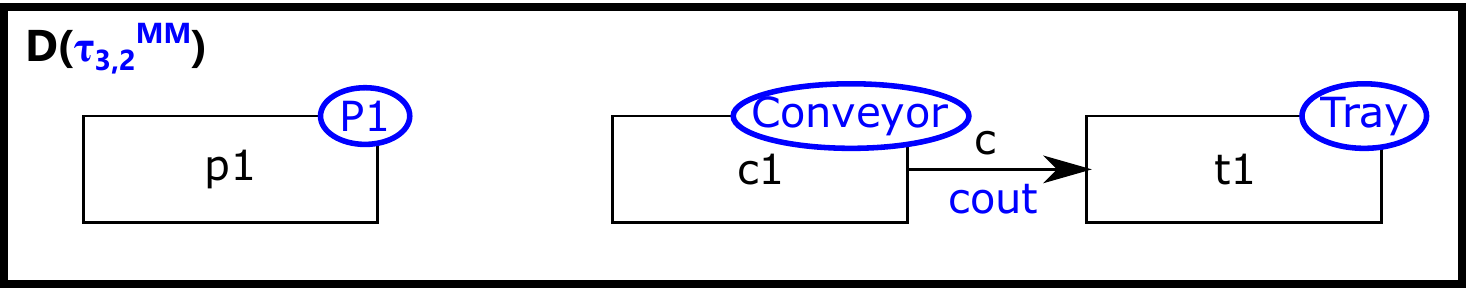}
	\caption{Visual representation of \(\domain{{\color{cdblue}\typemorph[\rulegraph]{3}{2}}}\)}
	\label{fig:pls-mcmt-transfer-part-formal-domain-3-2}
\end{figure}

For the typing chain of rule \emph{Transfer Part}, and using the same reasoning as in the previous one, we only need to calculate \(\domain{{\color{cdblue}\typemorph[\rulegraph]{2}{1}}}\), \(\domain{{\color{cdblue}\typemorph[\rulegraph]{3}{1}}}\) and \(\domain{{\color{cdblue}\typemorph[\rulegraph]{3}{2}}}\) since the rest of the domains of definition are trivial.
These three graphs are respectively shown in Figures~\ref{fig:pls-mcmt-transfer-part-formal-domain-2-1},~\ref{fig:pls-mcmt-transfer-part-formal-domain-3-1} and~\ref{fig:pls-mcmt-transfer-part-formal-domain-3-2}.

Since the FROM block of a rule is the one to be matched against the source model \(\graphname{\hierarchygraphleft}{}\), we now take for the \emph{Transfer Part} MCMT rule that \(\graphname{\rulegraphleft}{} = \graphname{\rulegraph}{3}\).
Therefore, \(\graphname{\rulegraphleft}{}\) is typed over the typing chain \(\chainname{\rulegraph'} = \chain{\rulegraph'}{2}\) with \(\tc{\rulegraph'} = [\graphname{\rulegraph}{2}, \graphname{\rulegraph}{1}, \graphname{\rulegraph}{0}]\).

\begin{figure}[h!]
	\centering
	\includegraphics[scale=.5]{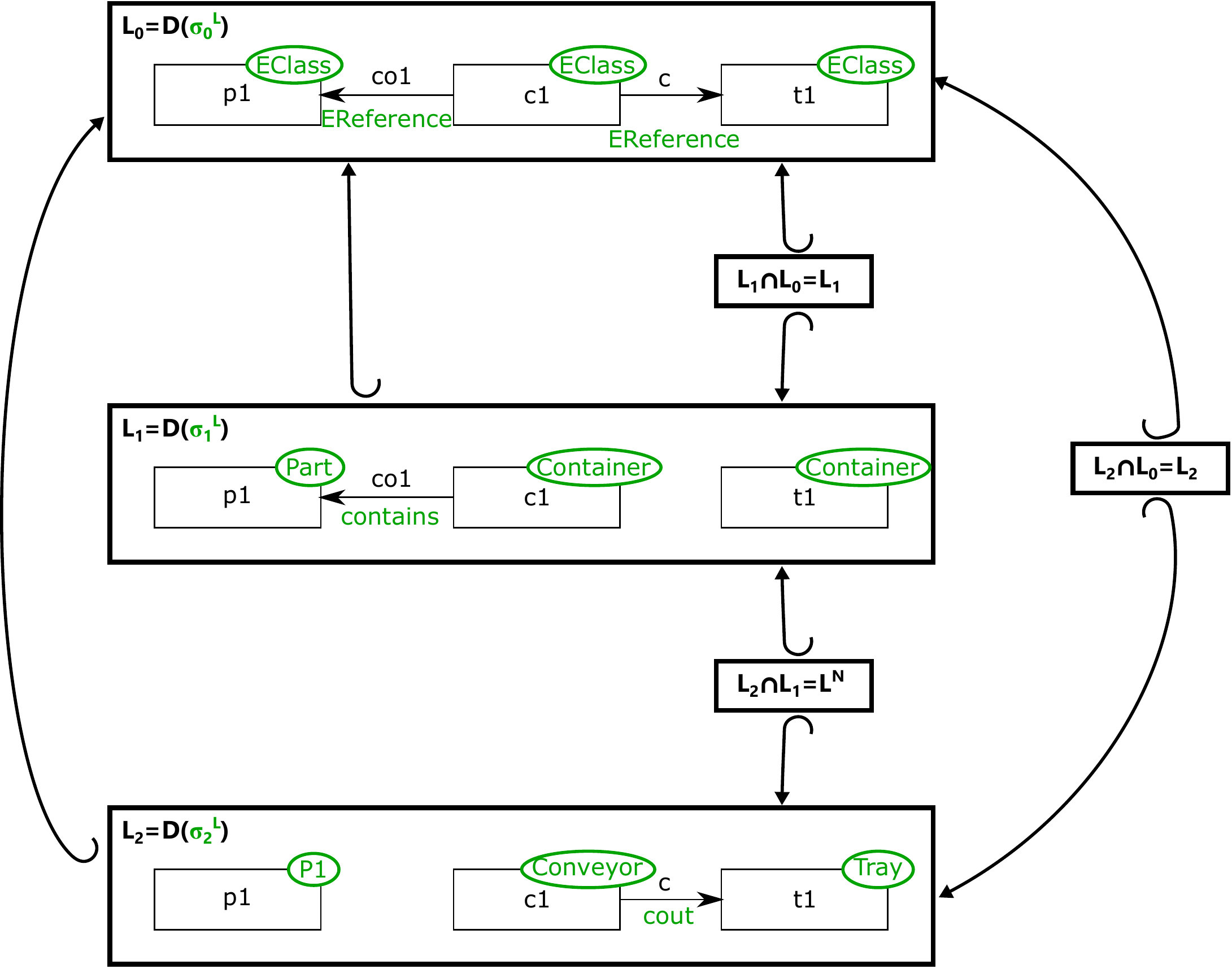}
	\caption{Inclusion chain \(\chainmorph{\rulegraphleft} : \graphname{\rulegraphleft}{} \Rightarrow \chainname{\rulegraph'}\)}
	\label{fig:pls-mcmt-transfer-part-formal-inclusion-chains-mm3}
\end{figure}

We can then construct \(\chainmorph{\rulegraphleft} : \graphname{\rulegraphleft}{} \Rightarrow \chainname{\rulegraph'}\) by considering the inclusion chain \(\chainname{\rulegraphleft} = \chain{\rulegraphleft}{2}\)  with \(\graphname{\rulegraphleft}{\indextwo} := \domain{\chainmorph[\indextwo]{\rulegraphleft}}\) for all \(\indextwo \in [2]\), and constructing the corresponding typing chain morphism \((\chainmorph{\rulegraphleft},id_{[2]}) : \chainname{\rulegraphleft} \to \chainname{\rulegraph'}\), as Figure~\ref{fig:pls-mcmt-transfer-part-formal-inclusion-chains-mm3} depicts.

\subsection{Examples of refactoring multilevel typing as typing chain morphisms}
\label{subsec:constructions-refactoring}

We can check now that the pullback condition in the diagram from Figure~\ref{fig:mlm-refactoring-diagram-formal} holds for both typing chains.
Starting with the \elementname{hammer\_config} typing chain, we need to do so for all three pairs \((\graphname{\hierarchygraphleft}{0}, \graphname{\hierarchygraphleft}{1})\), \((\graphname{\hierarchygraphleft}{0}, \graphname{\hierarchygraphleft}{2})\) and \((\graphname{\hierarchygraphleft}{1}, \graphname{\hierarchygraphleft}{2})\).
The first pair gives us a trivial pullback, as shown in Figure~\ref{fig:mlm-refactoring-diagram-formal-pls-hammer-tg2-0-1}:

\begin{figure}[H]
	\centering
	\includegraphics{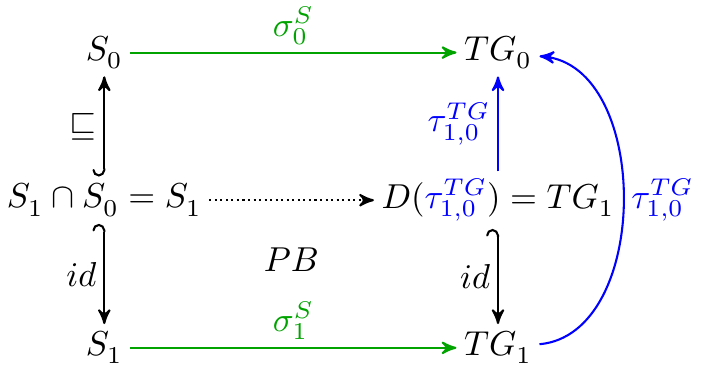}
	\caption{Pullback condition for refactoring in \((\graphname{\hierarchygraphleft}{0}, \graphname{\hierarchygraphleft}{1})\)}
	\label{fig:mlm-refactoring-diagram-formal-pls-hammer-tg2-0-1}
\end{figure}

\begin{figure}[H]
	\centering
	\includegraphics[width=.66\textheight,height=\linewidth,keepaspectratio,angle=270,origin=c]{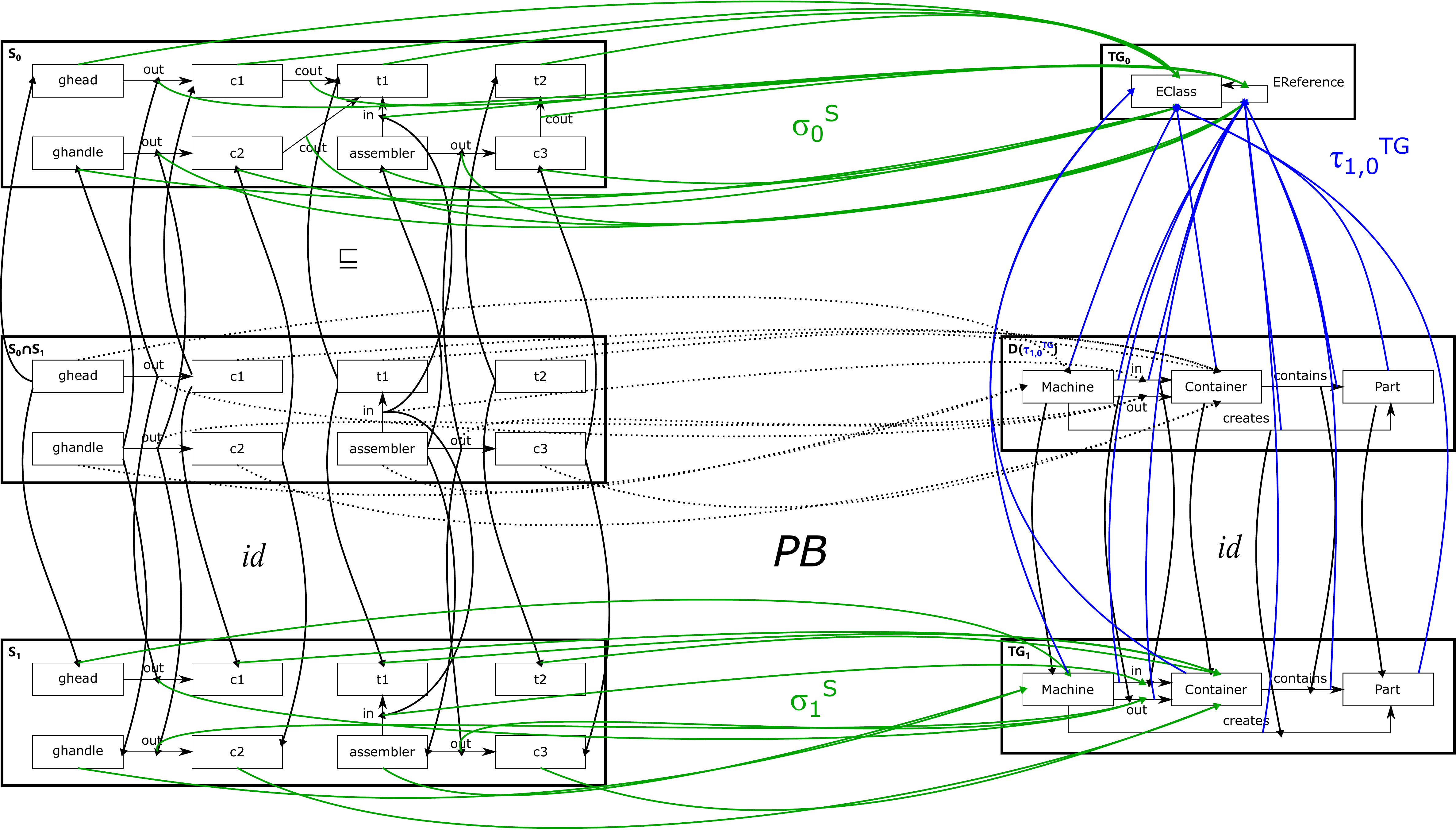}
	\caption{Detailed pullback condition in \((\graphname{\hierarchygraphleft}{0}, \graphname{\hierarchygraphleft}{1})\)}
	\label{fig:mlm-refactoring-diagram-pls-hammer-tg2-0-1}
\end{figure}

The expanded version of this diagram is depicted in Figure~\ref{fig:mlm-refactoring-diagram-pls-hammer-tg2-0-1}.

\begin{figure}[H]
	\centering
	\includegraphics{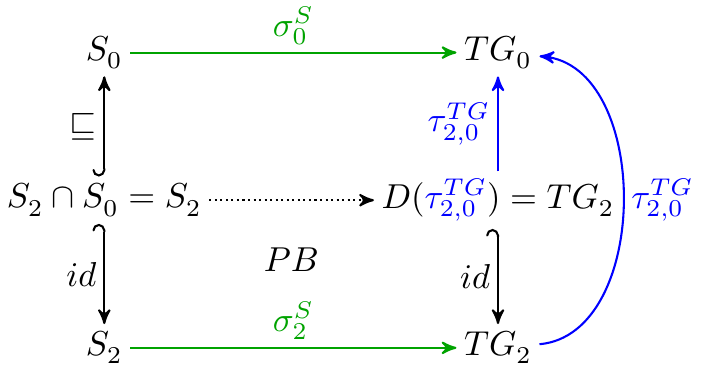}
	\caption{Pullback condition for refactoring in \((\graphname{\hierarchygraphleft}{0}, \graphname{\hierarchygraphleft}{2})\)}
	\label{fig:mlm-refactoring-diagram-formal-pls-hammer-tg3-0-2}
\end{figure}

\begin{figure}[H]
	\centering
	\includegraphics[width=.63\textheight,height=\linewidth,keepaspectratio,angle=270,origin=c]{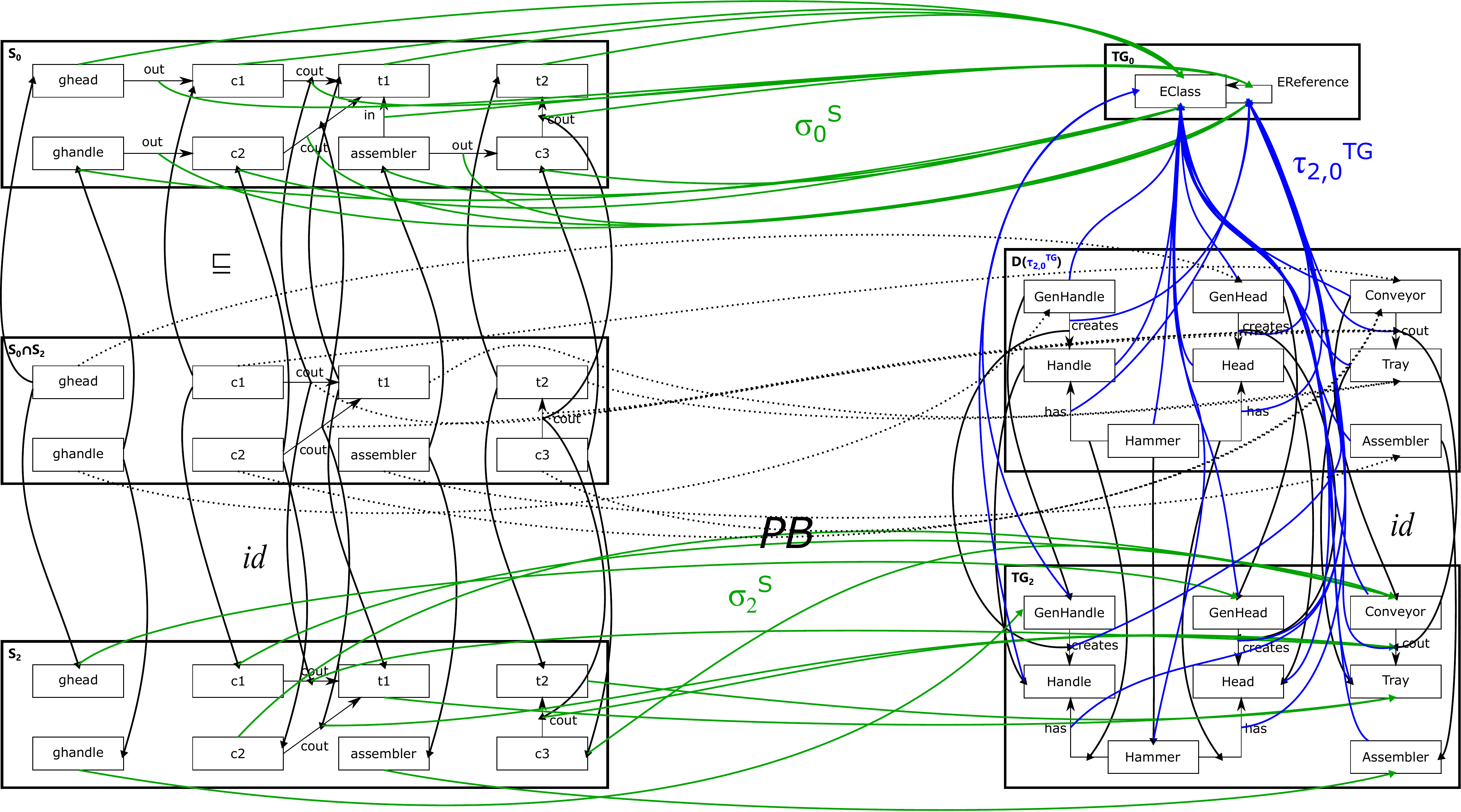}
	\caption{Detailed pullback condition in \((\graphname{\hierarchygraphleft}{0}, \graphname{\hierarchygraphleft}{2})\)}
	\label{fig:mlm-refactoring-diagram-pls-hammer-tg3-0-2}
\end{figure}

For the pair \((\graphname{\hierarchygraphleft}{0}, \graphname{\hierarchygraphleft}{2})\) we get the diagram in Figure~\ref{fig:mlm-refactoring-diagram-formal-pls-hammer-tg3-0-2} and its expanded version in Figure~\ref{fig:mlm-refactoring-diagram-pls-hammer-tg3-0-2}.

\begin{figure}[H]
	\centering
	\includegraphics{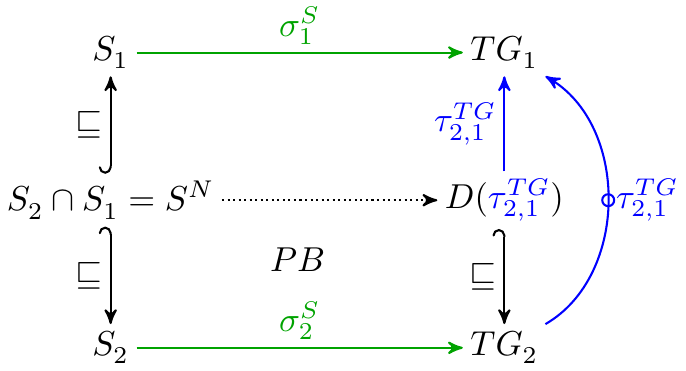}
	\caption{Pullback condition for refactoring in \((\graphname{\hierarchygraphleft}{1}, \graphname{\hierarchygraphleft}{2})\)}
	\label{fig:mlm-refactoring-diagram-formal-pls-hammer-tg3-1-2}
\end{figure}

\begin{figure}[H]
	\centering
	\includegraphics[width=.62\textheight,height=\linewidth,keepaspectratio,angle=270,origin=c]{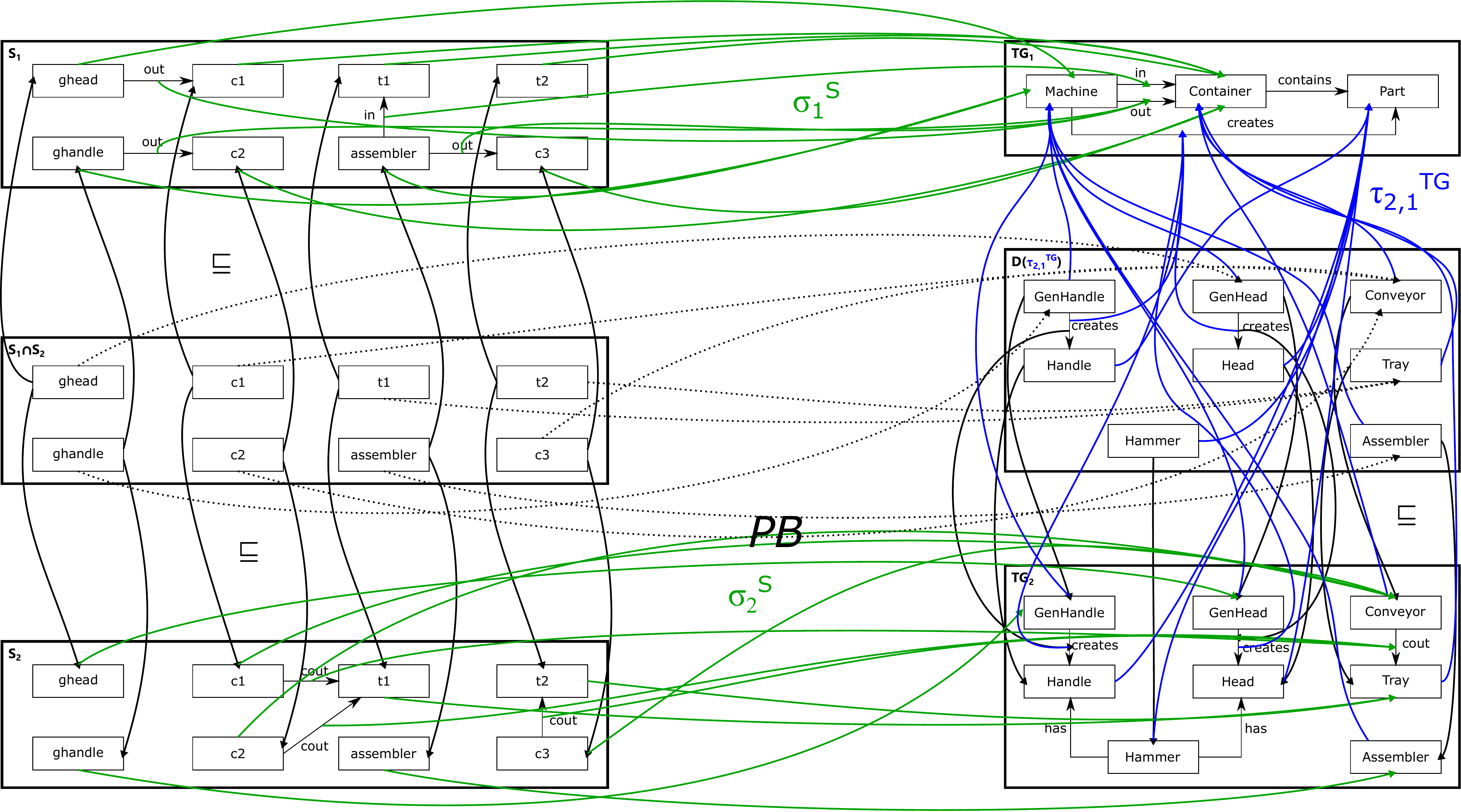}
	\caption{Detailed pullback condition in \((\graphname{\hierarchygraphleft}{1}, \graphname{\hierarchygraphleft}{2})\)}
	\label{fig:mlm-refactoring-diagram-pls-hammer-tg3-1-2}
\end{figure}

And for the pair \((\graphname{\hierarchygraphleft}{1}, \graphname{\hierarchygraphleft}{2})\) we get the diagrams in Figures~\ref{fig:mlm-refactoring-diagram-formal-pls-hammer-tg3-1-2} and~\ref{fig:mlm-refactoring-diagram-pls-hammer-tg3-1-2} (recall that \(\graphname[\graphnodes]{\hierarchygraphleft}{}\) refers to a version of \(\graphname{\hierarchygraphleft}{}\) with just the nodes).

\begin{figure}[H]
	\centering
	\includegraphics{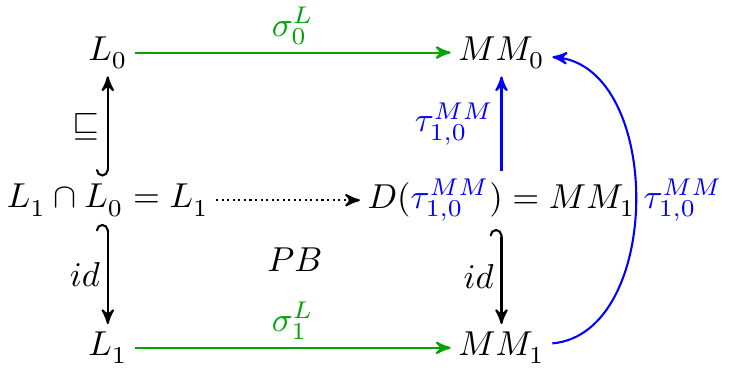}
	\caption{Pullback condition for refactoring in \((\graphname{\rulegraphleft}{0}, \graphname{\rulegraphleft}{1})\)}
	\label{fig:mlm-refactoring-diagram-formal-pls-mcmt-tranfer-part-mm2-0-1}
\end{figure}

\begin{figure}[H]
	\centering
	\includegraphics[width=.62\textheight,height=\linewidth,keepaspectratio,angle=270,origin=c]{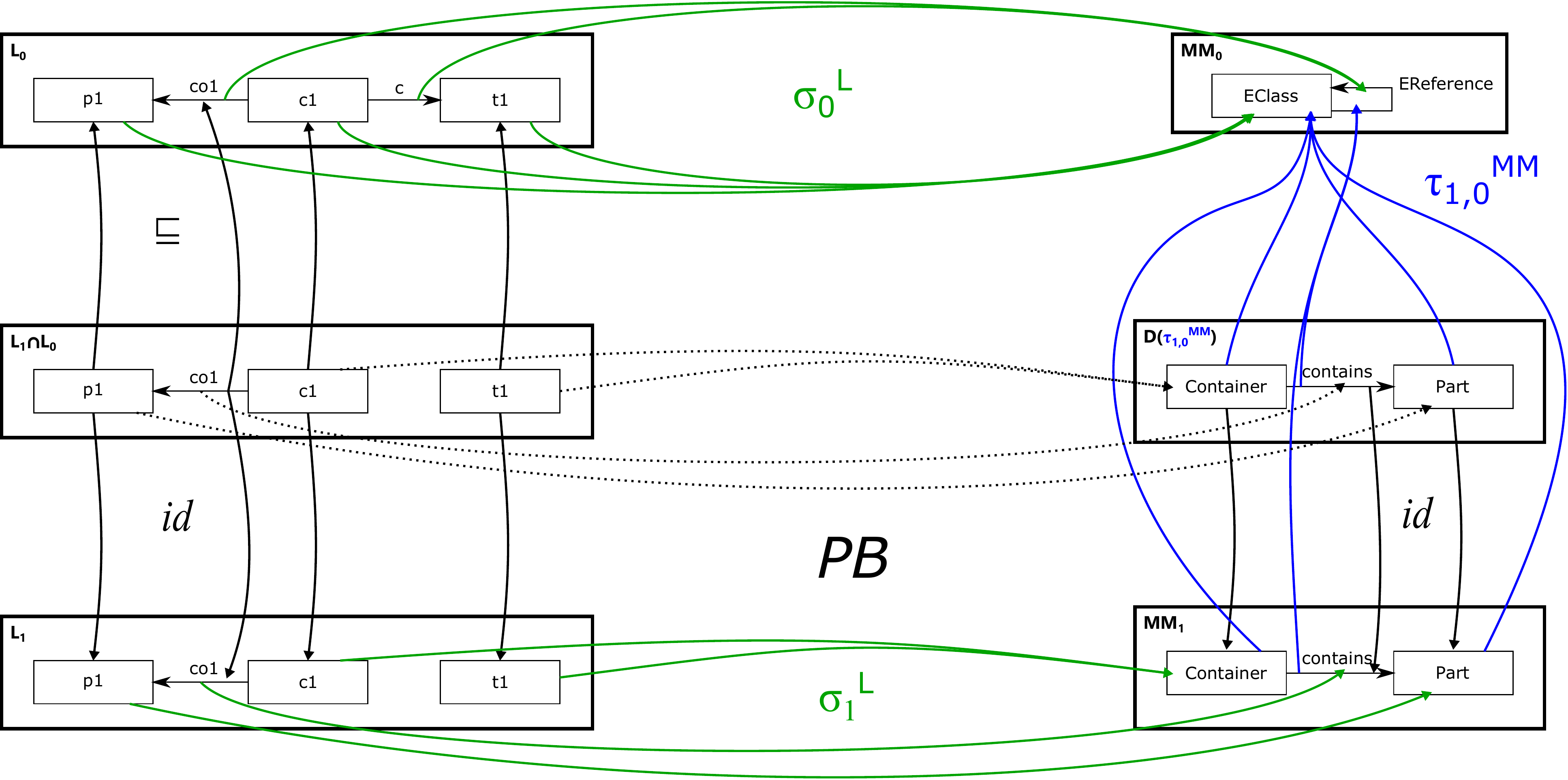}
	\caption{Detailed pullback condition in \((\graphname{\rulegraphleft}{0}, \graphname{\rulegraphleft}{1})\)}
	\label{fig:mlm-refactoring-diagram-pls-mcmt-tranfer-part-mm3-0-1}
\end{figure}

For the typing chain of \(\graphname{\rulegraph}{3}\) we need to check the condition for pairs \((\graphname{\rulegraphleft}{0}, \graphname{\rulegraphleft}{1})\), \((\graphname{\rulegraphleft}{0}, \graphname{\rulegraphleft}{2})\) and \((\graphname{\rulegraphleft}{1}, \graphname{\rulegraphleft}{2})\).
\((\graphname{\rulegraphleft}{0}, \graphname{\rulegraphleft}{1})\) yields the trivial pullback in Figures~\ref{fig:mlm-refactoring-diagram-formal-pls-mcmt-tranfer-part-mm2-0-1} and~\ref{fig:mlm-refactoring-diagram-pls-mcmt-tranfer-part-mm3-0-1}.

\begin{figure}[H]
	\centering
	\includegraphics{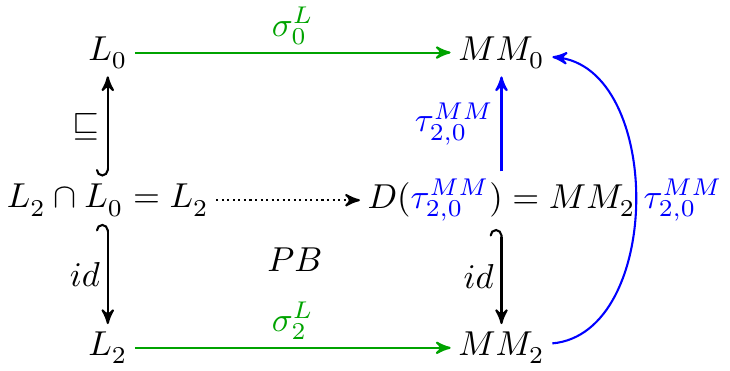}
	\caption{Pullback condition for refactoring in \((\graphname{\rulegraphleft}{0}, \graphname{\rulegraphleft}{2})\)}
	\label{fig:mlm-refactoring-diagram-formal-pls-mcmt-tranfer-part-mm3-0-2}
\end{figure}

\begin{figure}[H]
	\centering
	\includegraphics[width=.62\textheight,height=\linewidth,keepaspectratio,angle=270,origin=c]{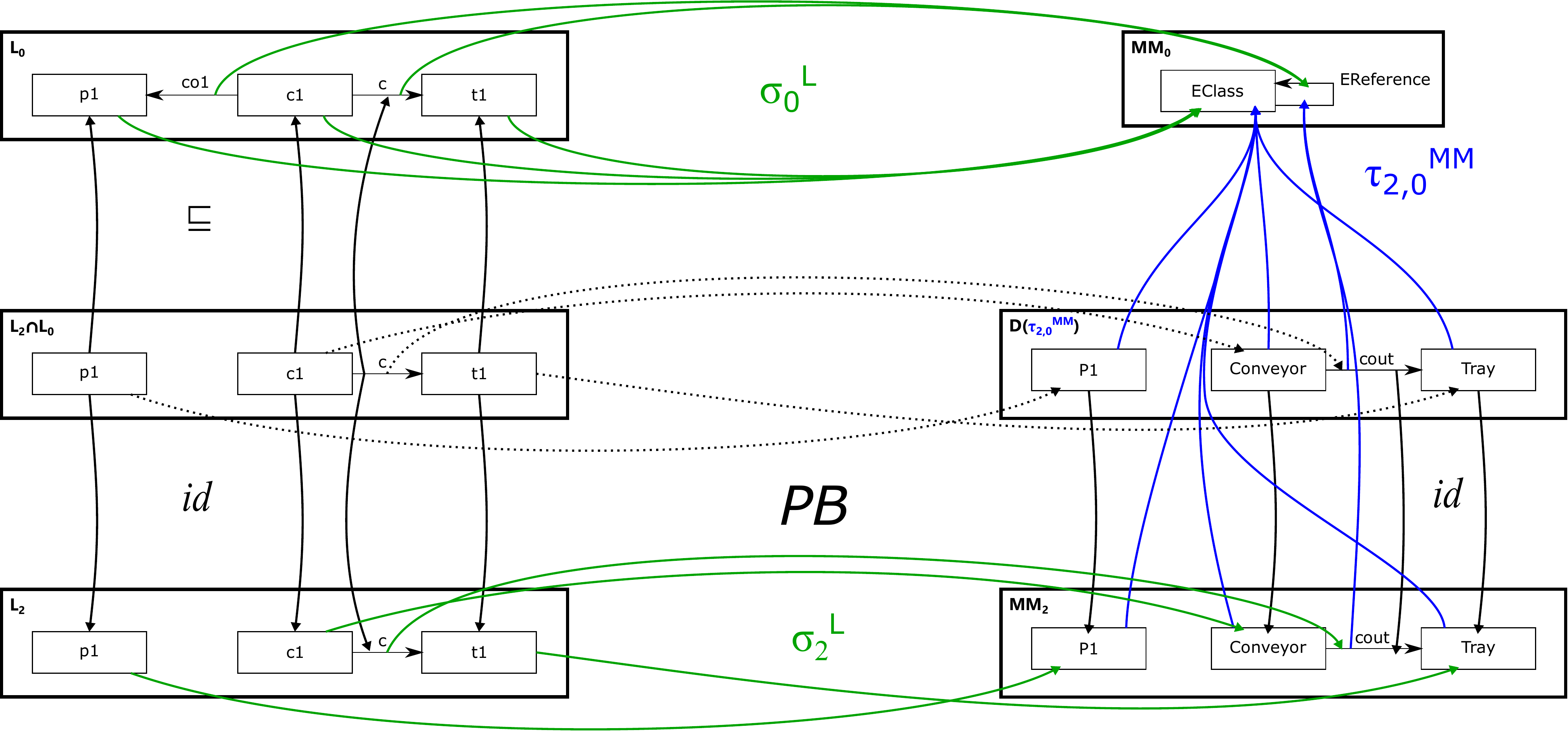}
	\caption{Detailed pullback condition in \((\graphname{\rulegraphleft}{0}, \graphname{\rulegraphleft}{2})\)}
	\label{fig:mlm-refactoring-diagram-pls-mcmt-tranfer-part-mm3-0-2}
\end{figure}

Likewise, the pair \((\graphname{\rulegraphleft}{0}, \graphname{\rulegraphleft}{2})\) gives us a trivial pullback diagram shown in Figures~\ref{fig:mlm-refactoring-diagram-formal-pls-mcmt-tranfer-part-mm3-0-2} and~\ref{fig:mlm-refactoring-diagram-pls-mcmt-tranfer-part-mm3-0-2}.

\begin{figure}[H]
	\centering
	\includegraphics{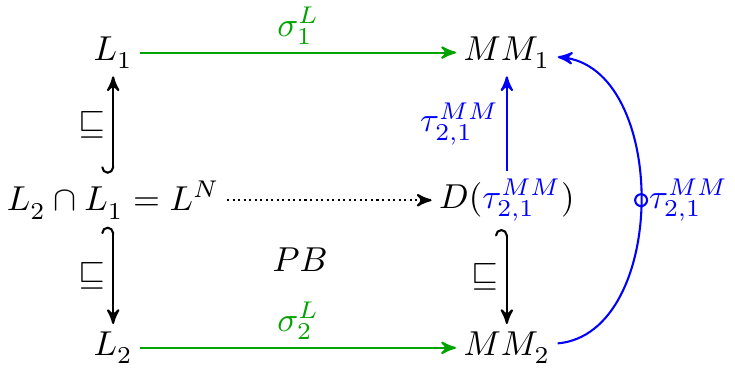}
	\caption{Pullback condition for refactoring in \((\graphname{\rulegraphleft}{1}, \graphname{\rulegraphleft}{2})\)}
	\label{fig:mlm-refactoring-diagram-formal-pls-mcmt-tranfer-part-mm3-1-2}
\end{figure}

\begin{figure}[H]
	\centering
	\includegraphics[width=.62\textheight,height=\linewidth,keepaspectratio,angle=270,origin=c]{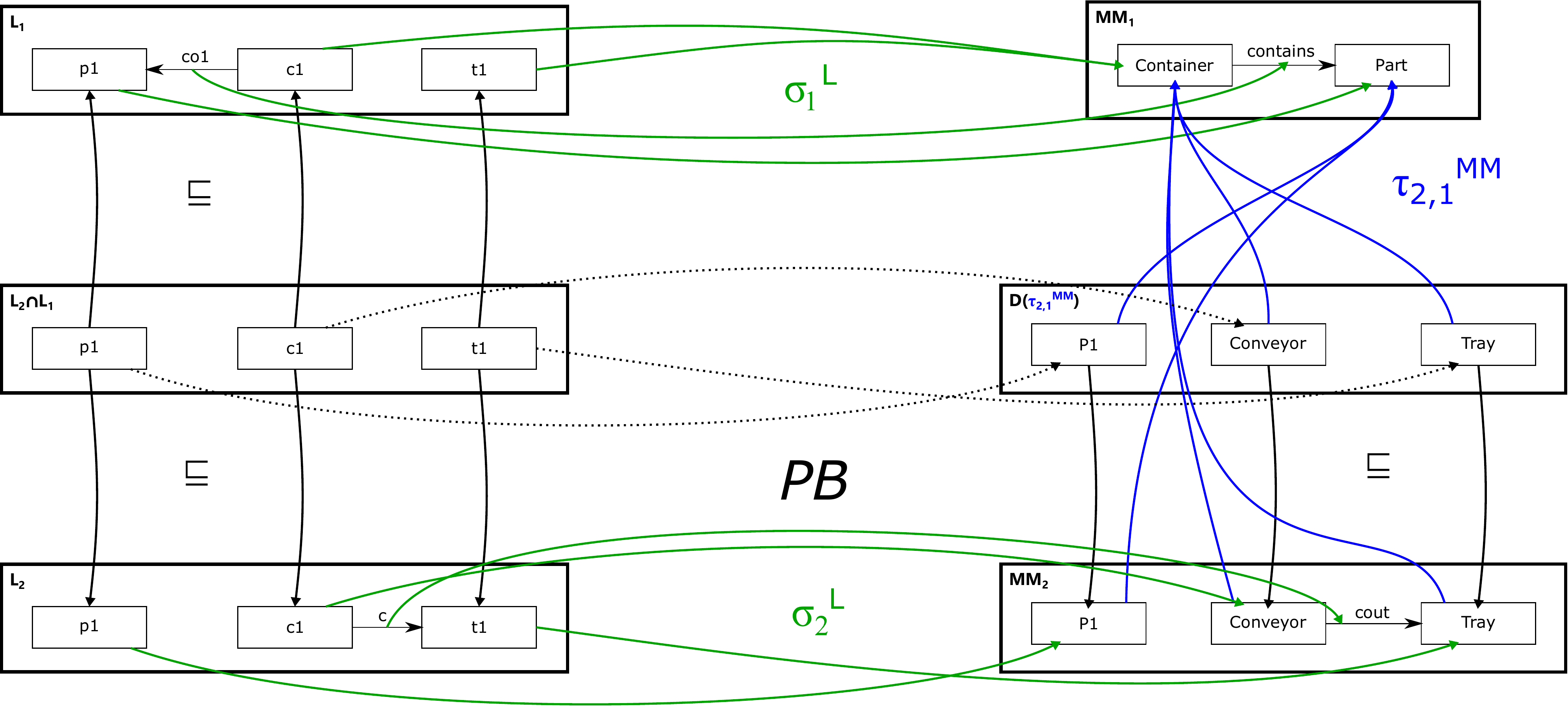}
	\caption{Detailed pullback condition in \((\graphname{\rulegraphleft}{1}, \graphname{\rulegraphleft}{2})\)}
	\label{fig:mlm-refactoring-diagram-pls-mcmt-tranfer-part-mm3-1-2}
\end{figure}

And finally, for the pair \((\graphname{\rulegraphleft}{1}, \graphname{\rulegraphleft}{2})\) we get the diagrams from Figures~\ref{fig:mlm-refactoring-diagram-formal-pls-mcmt-tranfer-part-mm3-1-2} and~\ref{fig:mlm-refactoring-diagram-pls-mcmt-tranfer-part-mm3-1-2}.

\subsection{Examples of bindings for rule matching}
\label{subsec:constructions-bindings}

We check here that the diagram from Figure~\ref{fig:mlm-binding-diagram-formal} holds for matching our example MCMT rule (\emph{Transfer Part}) into the \elementname{hammer\_config} model consistently.
That is, we use the typing chains \(\chainname{\rulegraph'} = \chain{\rulegraph'}{2}\), \(\chainname{\hierarchygraph'} = \chain{\hierarchygraph'}{2}\) and the inclusion chains \(\chainname{\rulegraphleft} = \chain{\rulegraphleft}{2}\), \(\chainname{\hierarchygraphleft} = \chain{\hierarchygraphleft}{2}\) together with the corresponding typing chain morphisms \((\chainmorph{\rulegraphleft} , id_{[2]}) : \chainname{\rulegraphleft} \to \chainname{\rulegraph'}\), \((\chainmorph{\hierarchygraphleft} , id_{[2]}) : \chainname{\hierarchygraphleft} \to \chainname{\hierarchygraph'}\), as previously constructed.

First, we describe a match of the rule typing chain \(\chainname{\rulegraph'}\) into the target typing chain \(\chainname{\hierarchygraph'}\), that is, a typing chain morphism \((\mapruletohierarchybinding , id_{[2]}) : \chainname{\rulegraph'} \to \chainname{\hierarchygraph'}\) with \(\mapruletohierarchybinding = (\mapruletohierarchybinding_{\indexone} : \graphname{\rulegraph}{\indexone} \to \graphname{\hierarchygraph}{\indexone} \mid \indexone \in [2])\).
Hence, we check the diagram first for each of the pairs \((2,0) , (2,1) , (1,0)\).
For \((2,0)\) and \((1,0)\) the condition is satisfied straightforwardly since \(\mapruletohierarchybinding_{0}\) is the identity on \(\graphname{\rulegraph}{0} = \graphname{\hierarchygraph}{0} = \elementname{Ecore}\), i.e.\ \(\mapruletohierarchybinding_{0} = id_\elementname{Ecore}\), and therefore all the typing morphisms \(\typemorph[\rulegraph]{2}{0}\), \(\typemorph[\rulegraph]{1}{0}\), \(\typemorph[\hierarchygraph]{2}{0}\) and \(\typemorph[\hierarchygraph]{1}{0}\) are total.

\begin{figure}[H]
	\centering
	\includegraphics{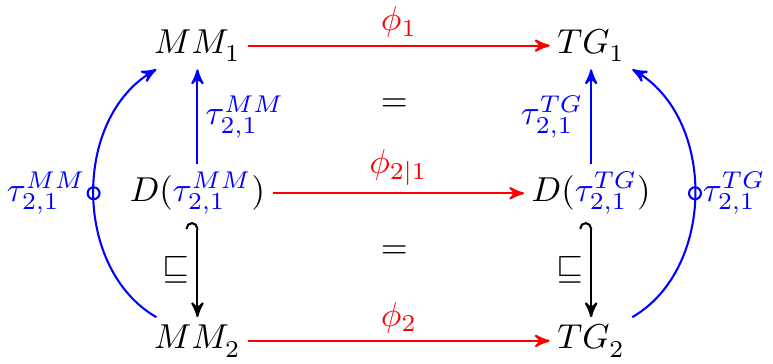}
	\caption{Commutativity condition for binding in levels \((1,2)\)}
	\label{fig:mlm-binding-diagram-formal-pls-hammer-transfer-part-1-2}
\end{figure}

\begin{figure}[H]
	\centering
	\includegraphics[width=.7\textheight,height=\linewidth,keepaspectratio,angle=270,origin=c]{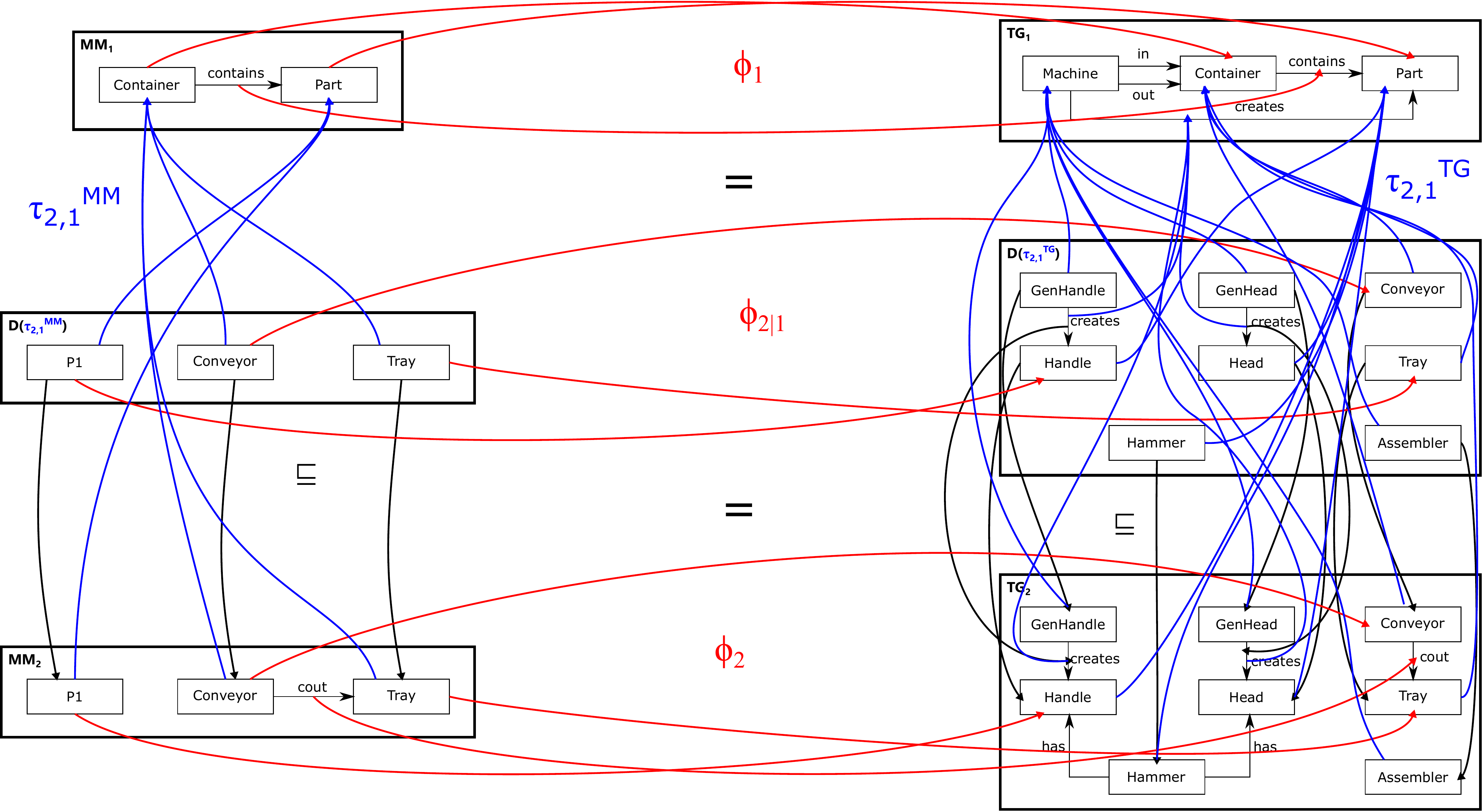}
	\caption{Detailed binding condition for levels \((1,2)\)}
	\label{fig:mlm-binding-diagram-pls-hammer-transfer-part-1-2}
\end{figure}

For the pair \((1, 2)\) we have the diagram in Figure~\ref{fig:mlm-binding-diagram-formal-pls-hammer-transfer-part-1-2} and its expanded version in Figure~\ref{fig:mlm-binding-diagram-pls-hammer-transfer-part-1-2}.

Second, we have the match \(\mapruletohierarchyleft = \mapruletohierarchybinding_3 : \graphname{\rulegraphleft}{} \to \graphname{\hierarchygraphleft}{}\) of the FROM block \(\graphname{\rulegraphleft}{} = \graphname{\rulegraph}{3}\) of the rule to the actual graph \elementname{hammer\_config} \(\graphname{\hierarchygraphleft}{} = \graphname{\hierarchygraph}{3}\).
We can also consider in this case that some of the some of the \(\tau\)'s should be labelled as \(\sigma\)'s, but we keep the same notation as before for the sake of consistency.

\begin{figure}[H]
	\centering
	\includegraphics{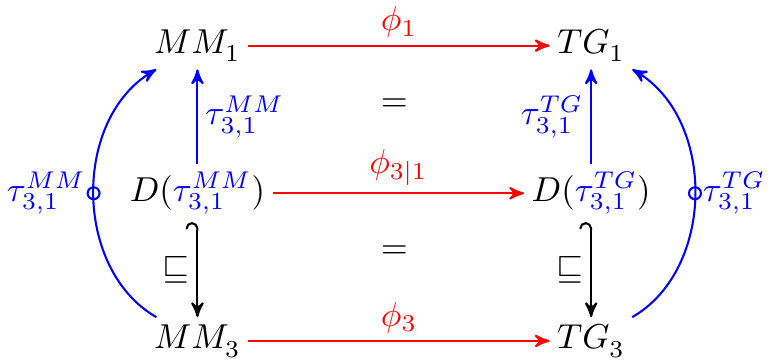}
	\caption{Commutativity condition for binding in levels \((1,3)\)}
	\label{fig:mlm-binding-diagram-formal-pls-hammer-transfer-part-1-3}
\end{figure}

\begin{figure}[H]
	\centering
	\includegraphics[width=.7\textheight,height=\linewidth,keepaspectratio,angle=270,origin=c]{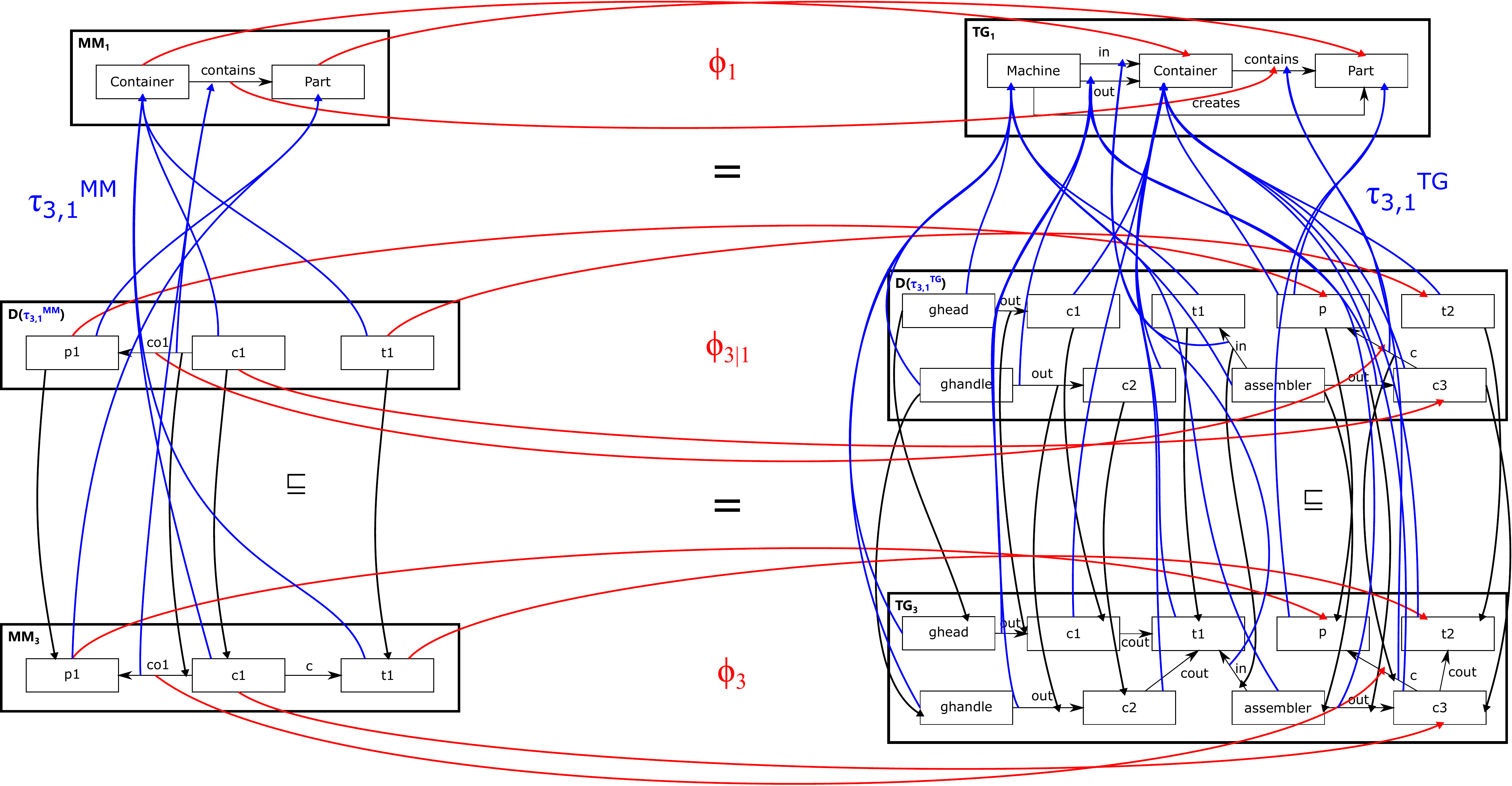}
	\caption{Detailed binding condition for levels \((1,3)\)}
	\label{fig:mlm-binding-diagram-pls-hammer-transfer-part-1-3}
\end{figure}

For one of the remaining pairs \((1, 3)\) we have the diagrams in Figures~\ref{fig:mlm-binding-diagram-formal-pls-hammer-transfer-part-1-3} and~\ref{fig:mlm-binding-diagram-pls-hammer-transfer-part-1-3}.

\begin{figure}[H]
	\centering
	\includegraphics[width=.7\textheight,height=\linewidth,keepaspectratio,angle=270,origin=c]{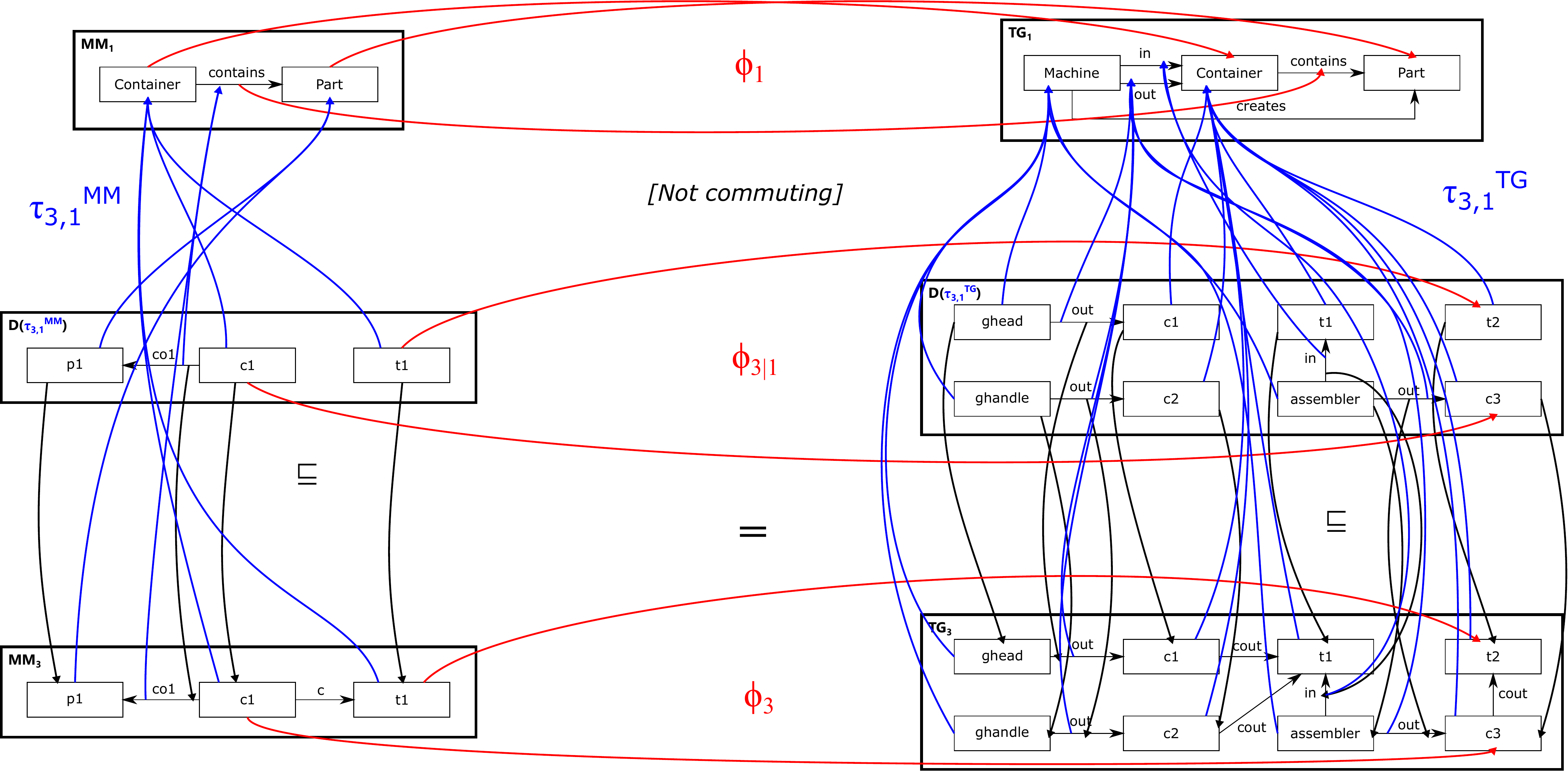}
	\caption{Detailed binding condition for levels \((1,3)\), unfulfilled}
	\label{fig:mlm-binding-diagram-pls-hammer-transfer-part-1-3-no-pb}
\end{figure}

The fact that both commutative squares are obtained reflects that the rule can actually be applied in the hierarchy.
Note that this is possible since an element \element{p}{Part} appears now in \(\graphname{\rulegraph}{3}\).
If we use the previous version of \(\graphname{\rulegraph}{3}\), the rule should not be applicable, which is ensured in these constructions by the fact that we cannot construct both commutative squares with \(\domain{{\color{cdblue}\typemorph[\rulegraph]{3}{1}}}\), as depicted in the Figure~\ref{fig:mlm-binding-diagram-pls-hammer-transfer-part-1-3-no-pb}.

\begin{figure}[H]
	\centering
	\includegraphics{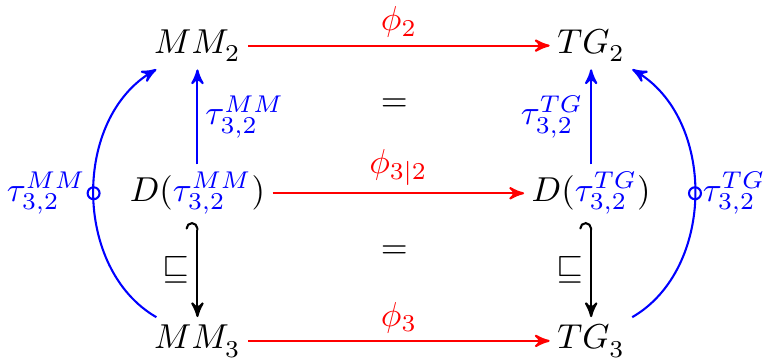}
	\caption{Commutativity condition for binding in levels \((2,3)\)}
	\label{fig:mlm-binding-diagram-formal-pls-hammer-transfer-part-2-3}
\end{figure}

\begin{figure}[H]
	\centering
	\includegraphics[width=.7\textheight,height=\linewidth,keepaspectratio,angle=270,origin=c]{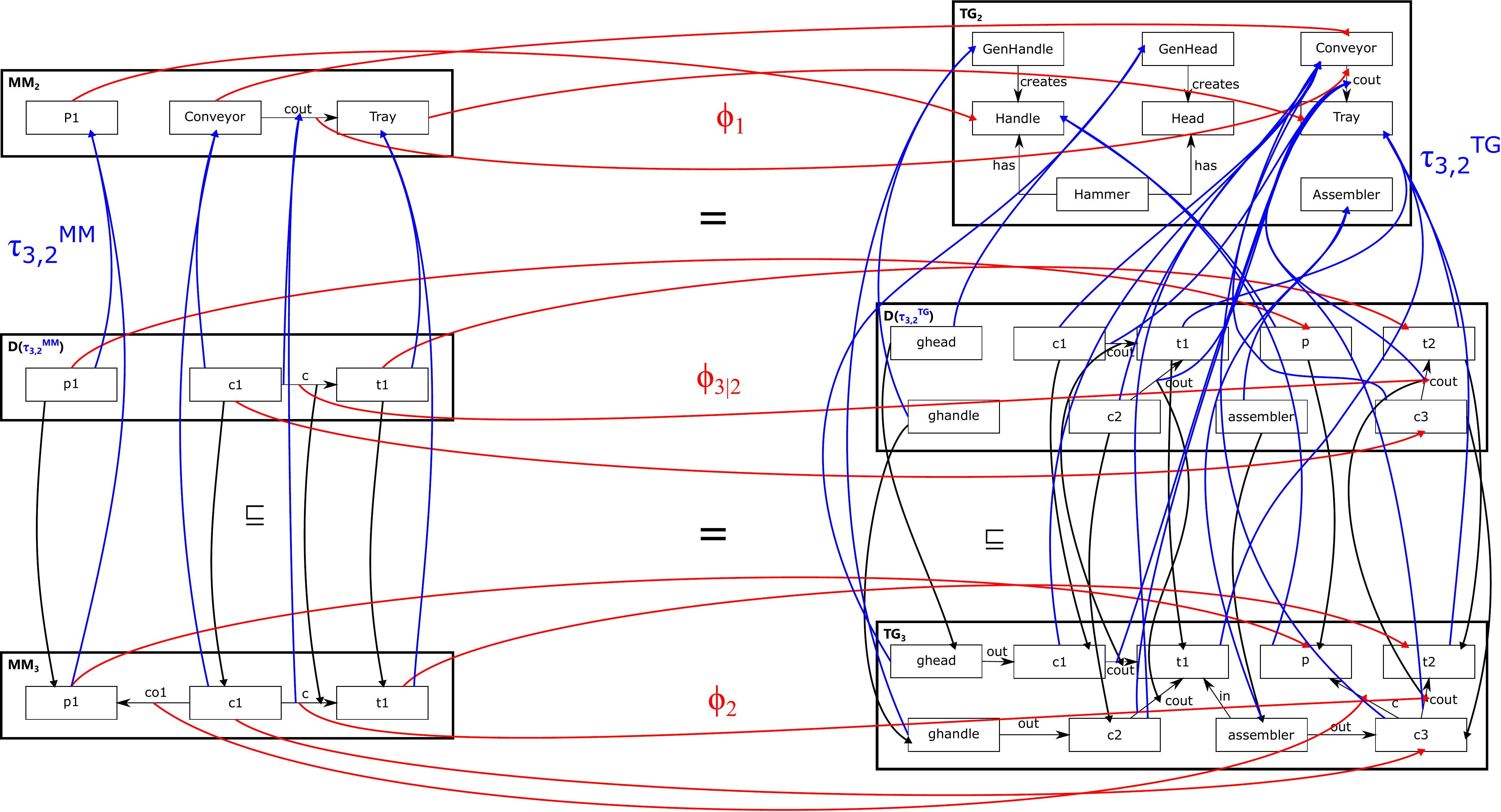}
	\caption{Detailed binding condition for levels \((2,3)\)}
	\label{fig:mlm-binding-diagram-pls-hammer-transfer-part-2-3}
\end{figure}

Finally, the same happens for the pair \((2, 3)\).
We get the diagrams in Figures~\ref{fig:mlm-binding-diagram-formal-pls-hammer-transfer-part-2-3} and~\ref{fig:mlm-binding-diagram-pls-hammer-transfer-part-2-3}, which can be constructed only if the rule is applicable.

\begin{figure}[H]
	\centering
	\includegraphics[width=.7\textheight,height=\linewidth,keepaspectratio,angle=270,origin=c]{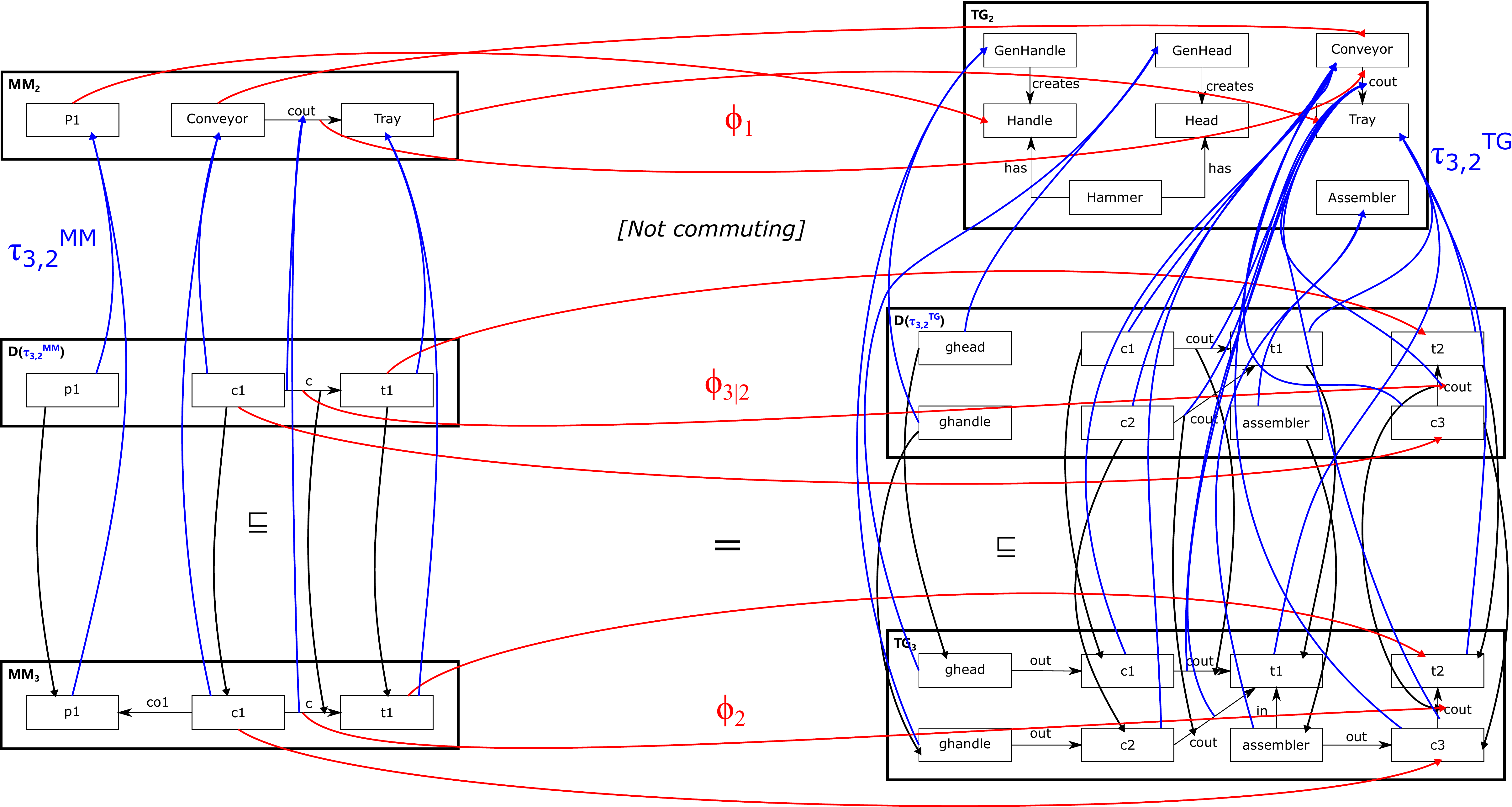}
	\caption{Detailed binding condition for levels \((2,3)\), unfulfilled}
	\label{fig:mlm-binding-diagram-pls-hammer-transfer-part-2-3-no-pb}
\end{figure}

With the initial version of \(\graphname{\rulegraph}{3}\) it is, as desired, neither possible to create both commutative squares nor to apply the rule, as Figure~\ref{fig:mlm-binding-diagram-pls-hammer-transfer-part-2-3-no-pb} depicts.

\subsection{Examples of inequality condition in rule matching}
\label{subsec:inequality-condition-rule-matching}

In condition~\ref{eq:preserve-reflect} we required an inequality for the composition of graph morphisms or, equivalently, that their domains of definition are graph and subgraph, respectively.
In order to illustrate why the requirement for this condition is inequality instead of equality, we take the following two graph chains and their morphisms, according to Figure~\ref{fig:mlm-binding-diagram-formal}, and assume that \(\maplevelone = id\), \(\graphname{\rulegraphleft}{} = \graphname{\rulegraph}{3}\) and \(\graphname{\hierarchygraphleft}{} = \graphname{\hierarchygraph}{3}\).

\begin{center}
\includegraphics[scale=.5]{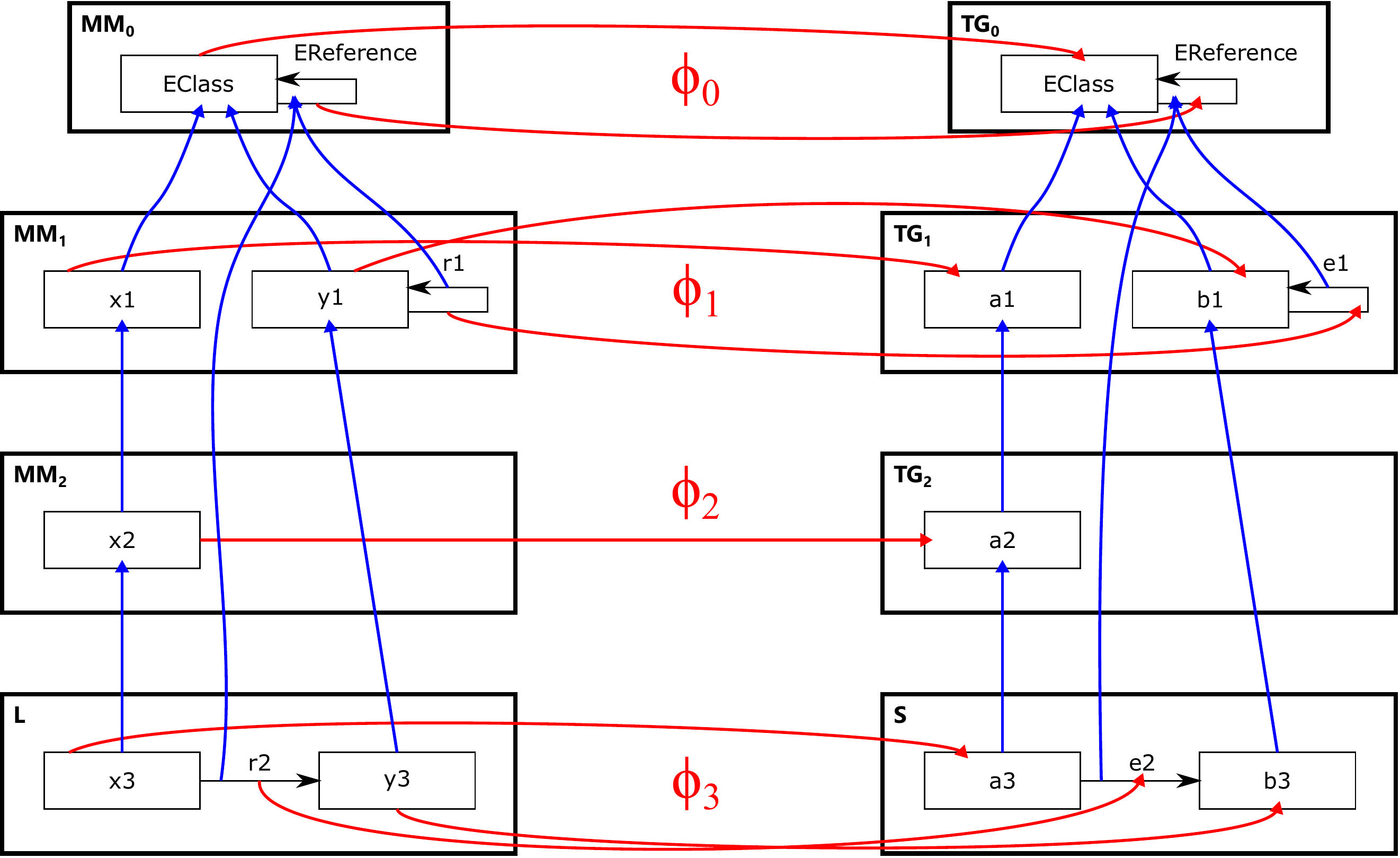}
\end{center}

In this example, the diagram in Figure~\ref{fig:mlm-binding-diagram-formal} for levels \(2\) and \(3\) looks as follows.

\begin{center}
\includegraphics[scale=.5]{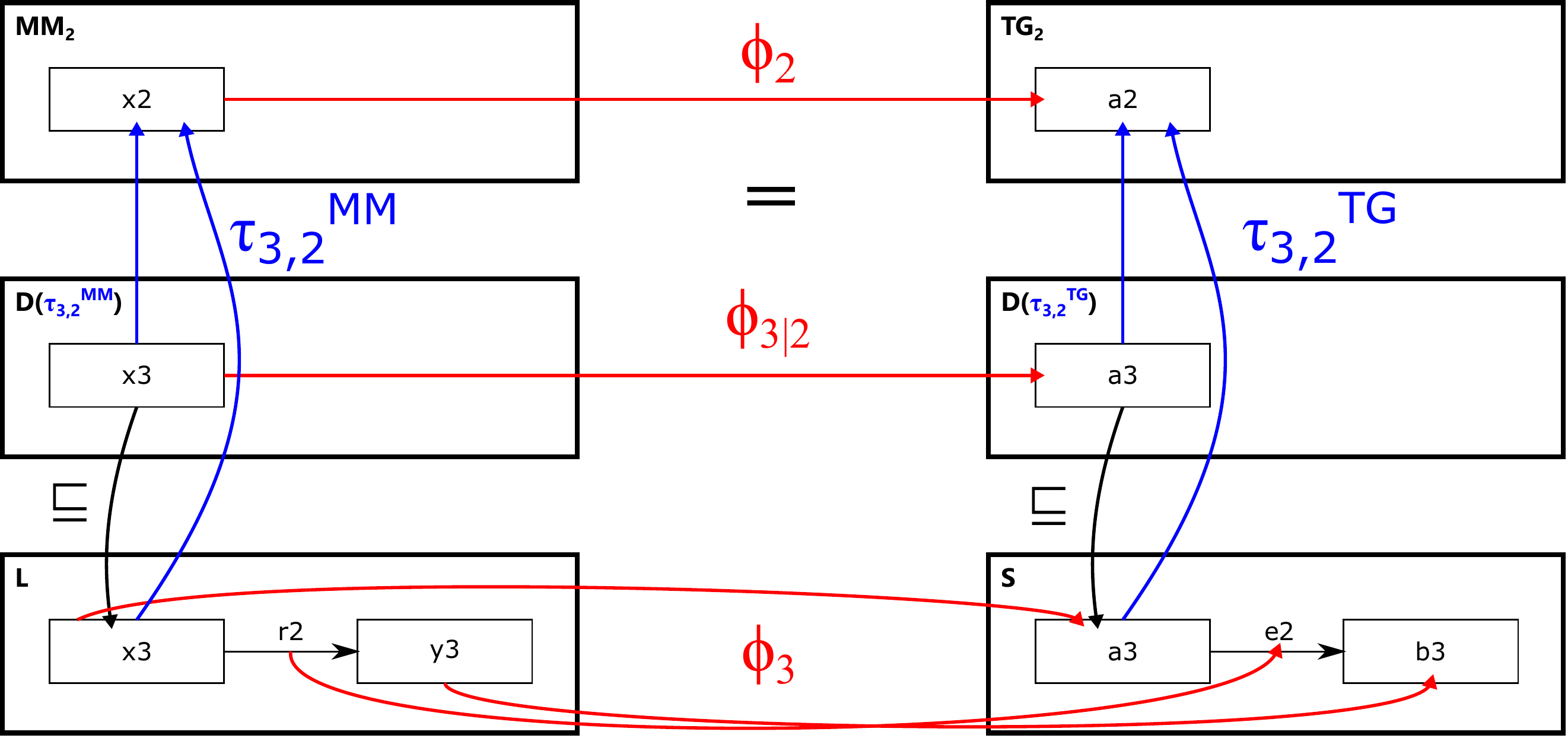}
\end{center}

The stronger condition which requires equality can be fulfilled in this case.
Recall the notation for graphs as a set of nodes plus a set of arrows plus two maps for source and target, and note that \(\varepsilon\) denotes an empty map:
\[\domain{\typemorph[\rulegraph]{3}{2} ; \morphone_2} = (\{x3\},\emptyset,\varepsilon,\varepsilon) = \domain{\morphone_3 ; \typemorph[\hierarchygraph]{3}{2}}\]
Therefore:
\[\domain{\typemorph[\rulegraph]{3}{2} ; \morphone_2} = \domain{\morphone_3 ; \typemorph[\hierarchygraph]{3}{2}} \Longleftrightarrow \typemorph[\rulegraph]{3}{2} ; \morphone_2 = \morphone_3 ; \typemorph[\hierarchygraph]{3}{2}\]

We can slightly modify the example in the following way, where a valid match could still be established as depicted.

\begin{center}
\includegraphics[scale=.5]{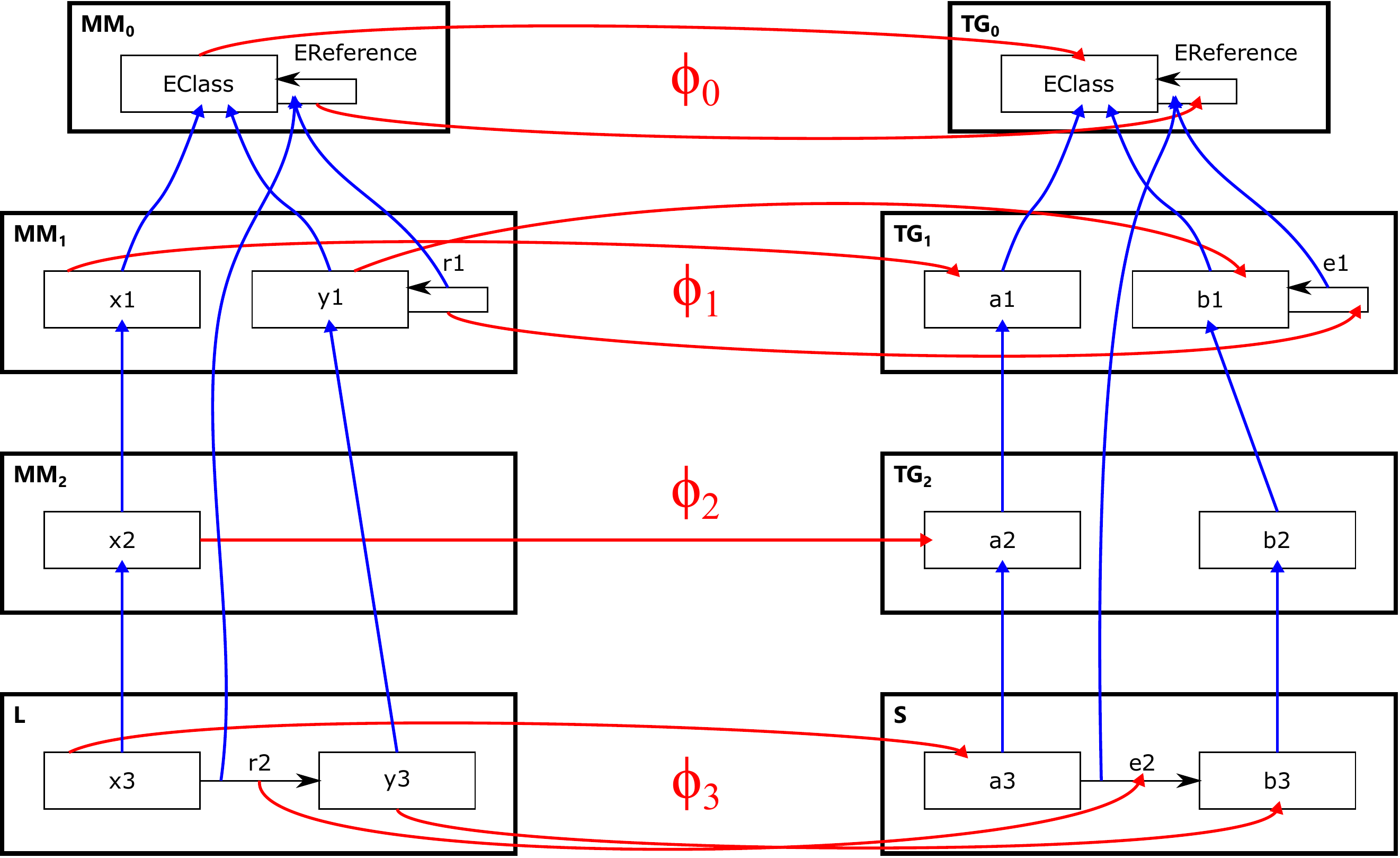}
\end{center}

Now, the diagram in Figure~\ref{fig:mlm-binding-diagram-formal} for levels \(2\) and \(3\) looks as follows.

\begin{center}
\includegraphics[scale=.5]{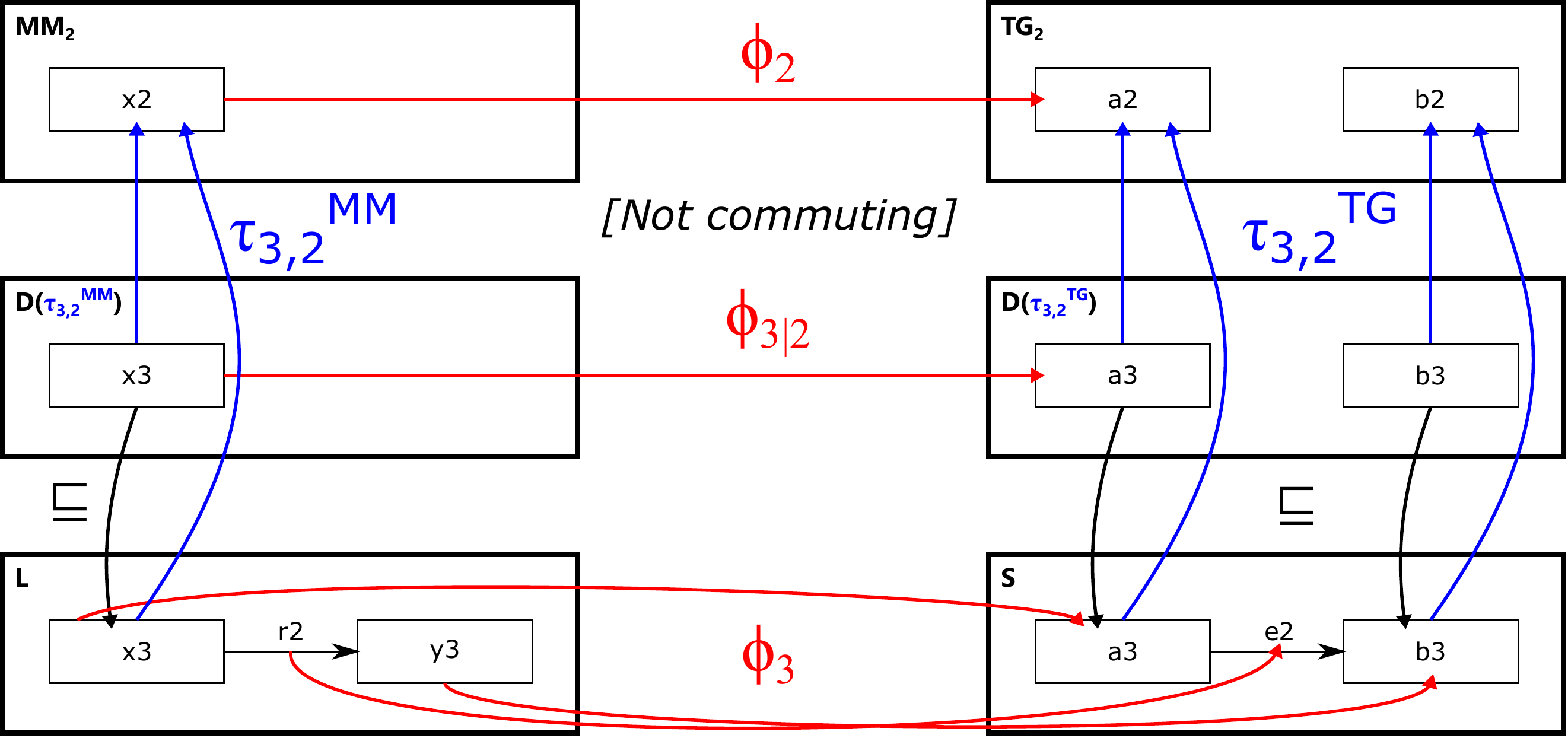}
\end{center}

And the equality condition does not fit since it does not hold for this second version:
\[\domain{\typemorph[\rulegraph]{3}{2} ; \morphone_2} = (\{x3\},\emptyset,\varepsilon,\varepsilon) \sqsubseteq (\{x3 , y3\},\emptyset,\varepsilon,\varepsilon) = \domain{\morphone_3 ; \typemorph[\hierarchygraph]{3}{2}}\]
Therefore:
\[\domain{\typemorph[\rulegraph]{3}{2} ; \morphone_2} \sqsubseteq \domain{\morphone_3 ; \typemorph[\hierarchygraph]{3}{2}} \Longleftrightarrow \typemorph[\rulegraph]{3}{2} ; \morphone_2 \preceq \morphone_3 ; \typemorph[\hierarchygraph]{3}{2}\]

%% file: thesis/06-discussion.tex
\chapter{Discussion}
\label{chap:discussion}

In this chapter we position our work in the context of the state of the art of Multilevel Modelling and Domain Specific Modelling Languages, evaluate our contributions and highlight the strengths of our approach.
In order to do so, in Section~\ref{sec:related-work} we comment on other works related to ours.
In Section~\ref{sec:tool-requirements} we use requirements for MLM approaches, which were defined in two recent publications by members of the MLM community external to our work, to evaluate the suitability of our framework.
Then, we outline future lines of research related to the work presented in this thesis in Section~\ref{sec:future-work}.
Finally, in Section~\ref{sec:conclusions} we conclude this thesis by summarizing its contributions.

\section{Related Work}
\label{sec:related-work}

This section is divided into four parts, each one discussing related works for different highlights of our approach, namely multilevel modelling (Section~\ref{subsec:mlm}), reusability of models and model transformations (Section~\ref{subsec:reusability}), multilevel model transformations (Section~\ref{subsec:multilevel-model-transformations}) and model transformations for the definition of behaviour (Section~\ref{subsec:mts-for-behaviour}).

\subsection{Multilevel modelling}
\label{subsec:mlm}

We present here other approaches that have tackled multilevel modelling by creating conceptual frameworks and tools.
Some of the aspects discussed in the text are also synthesized in Table~\ref{tab:mlm-comparison}.

Multilevel modelling started to gain momentum in the 2000s ~\cite{atkinson2001essence,atkinson2001processes,atkinson2002rearchitecting,atkinson2008reducing,delara2010automating,rossini2014formalisation}.
Even before that time, using several abstraction layers was a well-established technique for the specification of information systems~\cite{mylopoulos1980relationships,borgida1984generalization,mylopoulos1990telos,odell1994power,atkinson1997distributed,bezivin1997ontology,odell1998advanced}.
Since then, several researchers have pointed out the benefits of using multilevel modelling languages, e.g.,~\cite{delara2015dsmml,mohagheghi2010evaluating,tolvanen2016challenges,kelly2008enabling}.
Indeed, different authors have approached the formalization of multilevel modelling languages and of some other aspects related to multilevel structures (see, e.g.,~\cite{delara2010deep,rossini2014formalisation}).
Approaches such as MetaDepth~\cite{delara2010deep}, Melanee~\cite{atkinson2016melanee}, AtomPM~\cite{syriani2013atompm} or Modelverse~\cite{van2014modelverse} 
propose conceptual frameworks and tools for multilevel modelling.

The most widely used approach to the specification of multilevel frameworks is the usage of Orthogonal Classification Architecture (OCA), as defined in~\cite{atkinson2005concepts}.
This architecture implies the definition of a linguistic metamodel that captures all features of the multilevel hierarchy.
That is, the whole hierarchy becomes an instance of the linguistic metamodel.
These approaches also exploit the concept of \emph{clabject}~\cite{atkinson1997distributed}.
Since clabjects stem from the traditional object-oriented programming, their realization into a metamodelling framework requires the linguistic metamodel, that all the levels must share, and which contains it together with other elements such as \emph{field} and \emph{constraint}.
According to~\cite{delara2010mixin}, the use of a fixed topmost metamodel, as we do in our approach as an alternative to OCA, can pose a threat to the flexibility of the conceptual framework.
The main criticism is that such a framework may be intrusive, since the user is forced to use the topmost metamodel.
The authors claim that such a framework could not be used for scenarios in which the same model can have more than one ``semantics'', i.e.\ those scenarios where an element may have more than one type simultaneously.
However, our conceptual framework can be used to overcome these issues, since the user can use supplementary typing for these matters.
Furthermore, the possibility of adding new intermediate levels in a pre-existing stack favours the flexibility of switching or reusing metamodels.
This technique can also be a solution for the issues presented in~\cite{delara2015aposteriori}.

The authors of~\cite{atkinson2005concepts}, who coined the term OCA, are also the creators of a tool for MLM called Melanee~\cite{atkinson2015multi}.
This tool has been developed with a stronger focus on editing capabilities~\cite{atkinson2016melanee}, as well as possible applications into the domains of executable models~\cite{atkinson2015execution}.
This tool allows for the specification of multilevel hierarchies, with potency features like durability and mutability for attributes and relations, which offer control to the level in which attributes are instantiated.
Such fine tuning cannot be applied to classes, however, where only traditional potency is allowed.
Multiple typing is also not supported in Melanee.

MetaDepth~\cite{delara2010deep} is a deep modelling tool built on the notion of orthogonal classification and deep instantiation.
It requires the aforementioned linguistic metamodel, and support the concepts of potency formally defined in~\cite{rossini2014formalisation}.
The tool also supports several interesting features such as model transformation reuse~\cite{delara2015dsmml} and generic metamodelling~\cite{delara2010mixin}.
The authors of MetaDepth are currently developing a new framework called Kite~\cite{guerra2018quest} which may in the future inherit MetaDepth's MLM features.

Some of the formal aspects of our approach have been influenced by the DPF workbench~\cite{lamo2013dpf}, where the authors tackle multilevel modelling by means of graph morphisms and formal predicates.
The use of potency is not supported by the DPF workbench, but the approach includes standard and leap potency as part of their specification.

AToMPM~\cite{syriani2013atompm} is a modelling framework highly focused on offering cloud and web tools.
Modelverse~\cite{van2014modelverse}, which is based on AToMPM, offers multilevel modelling functionalities by implementing the concept of clabject and building a linguistic metamodel that includes a synthetic typing relation.

OMLM~\cite{igamberdiev2016integrated} uses an OCA, and extends it with a realisation dimension, which maps the modelling elements to their implementation counterparts.
They use the standard concept of potency as depth, but do not apply it to attributes, which always have potency 1, so it sticks to the standard features of MLM which are common from most proposals.

DeepTelos~\cite{jeusfeld2016deeptelos}, is an extension of Telos which allows to implement OCA by using \textit{most general concepts}, but does not natively require it.
Their proposal comes closer to ours in both the formal background and the independence from a linguistic metamodel, although the approach does require it in the case of potencies, which need a custom set of elements specified that represent the allowed instantiations.

Dual Deep Modelling~\cite{neumayr2016dual} is based on the definition of parallel hierarchies, with different depths, where relations from one to the other can be established.
They are still based on OCA and the use of potency as depth, similar to most of the approaches mentioned here.

In a similar manner, SLICER~\cite{selway2017conceptual} defines a complex and powerful set of tools to represent different relations among clabjects and the way these are instantiated, but it can still be classified as OCA since they require a linguistic metamodel that defines these concepts and relations.

The authors of the NMeta metamodel~\cite{hinkel2016nmf} developed a framework around it where this metamodel is at the top of their orthogonal stack of models.
But we believe that the fact that the metamodel defines concepts like \emph{Model} and \emph{ModelElement} which are then instantiated to create the actual stack implies that the approach is also based on the OCA principles.
The approach does not have explicit levels or potencies and uses inheritance to separate concepts from different levels of abstraction, although the authors claim that they allow for unbounded levels.

Finally, in~\cite{mallet2009automated}, the authors assume that the only way to do multilevel modelling is by using the clabject approach.
Therefore, they do not consider other alternatives to this approach for their implementation of multilevel modelling.

All these previous approaches implement their multilevel modelling solutions by using a linguistic metamodel including the clabject element, and \emph{flattening} the ontological hierarchy as an instance of this linguistic metamodel.
That is, the whole ontological stack becomes an instance of the clabject-based modelling language.
Moreover, they require the creation of supporting tools from scratch.
With our approach, multilevel modelling is realised in a different way, which improves its flexibility and removes the need for custom-made environments and tools.
Our framework does not require the definition of a specific linguistic metamodel, nor a flattening of the ontological levels, which eases its integration in two-level frameworks, thanks to the sliding window technique (see Section~\ref{sec:modelling-tooling}).
Furthermore, making other modelling techniques such as constraint definition and model transformations multilevel-aware would require to either flatten the whole ontological and linguistic hierarchies or reimplement them with multilevel capabilities for OCA-based frameworks.
And example of such constraint languages is~\cite{kantner2014specification}, where a MLM-aware version of OCL is presented as an extension to the tool Melanee.
This kind of language should provide the same features as the two-level ones, plus functionalities related to MLM concepts like potency and levels.
In our case, MultEcore could be extended with such a language, but we do not consider this kind of extension at the moment.
Instead, existing tools such as OCL (or even ATL~\cite{jouault2008atl} for MT) can be used seamlessly on the Ecore and XMI representations that MultEcore handles if the multilevel features of those models are not needed.
Alternatively, as hinted in Section~\ref{subsec:bicycle-challenge}, we could apply MCMTs for constraint definition and validation, but an in-depth research of this possibility still remains to be done.

FMMLx~\cite{frank2014multilevel} is one of the few approaches not using OCA.
Its proponents apply the concept of \emph{golden braid} with a similar realisation to ours: a topmost metamodel that defines itself, that can be transitively instantiated as many times as required to create a multilevel hierarchy.
For potency, they allow marking features as \emph{intrinsic}, which requires the specification of the level where that feature can be instantiated.

The way we specify potency can also be compared with previous formalisations of the concept, like the one proposed in~\cite{rossini2011graph} and later improved in~\cite{rossini2014formalisation}.
Table~\ref{tab:mlm-comparison} contemplates three possible types of potency: the original idea of potency as depth (\emph{traditional}), the \emph{leap} addition, which allows the instantiation to jump across levels, but only once, and the \emph{range} potency presented in this thesis, which is combined with the first one in our tool and encompasses the second.
Additionally, for the sake of completeness, we also included in Table~\ref{tab:mlm-comparison} whether the approaches support multiple typing or not.
Note that, for the sake of simplicity, the table does not distinguish if the other approaches also provide potencies for arrows and attributes, as in our approach, or just for nodes.

{
\newcommand{\myheader}[3]{\multicolumn{1}{#1 m{#2}|}{\bfseries\centering #3}}
\begin{table}[ht]
\centering
\caption {A comparison of multilevel features from different approaches}
\begin{tabular}{|m{20mm}|m{23mm}|m{21mm}|m{15mm}|m{15mm}|m{16mm}|}
    \hline
	\myheader{|}{20mm}{Approach}	& \myheader{}{23mm}{Canonical Architecture}	& \myheader{}{21mm}{Traditional potency}	& \myheader{}{15mm}{Leap potency}	& \myheader{}{15mm}{Range potency}	& \myheader{}{16mm}{Multiple typing}	\\
	\hline
	Melanee 			& OCA					& \checkmark	& ---			& ---			& ---			\\
	\hline
	Modelverse (AToMPM)	& OCA					& \checkmark	& ---			& ---			& ---			\\
	\hline
	MetaDepth			& OCA					& \checkmark	& ---			& \checkmark	& \checkmark	\\
	\hline
	DPF					& OCA					& \checkmark	& \checkmark	& ---			& ---			\\
	\hline 
	FMMLx				& Self-defining top MM	& \checkmark	& ---			& ---			& ---			\\
	\hline
	OMLM				& OCA					& \checkmark	& ---			& ---			& ---			\\
	\hline	
	DeepTelos			& OCA					& ---			& ---			& ---			& \checkmark	\\
	\hline
	Dual Deep Modelling	& OCA					& \checkmark	& ---			& ---			& ---			\\
	\hline
	SLICER				& OCA					& \checkmark	& ---			& ---			& \checkmark	\\
	\hline
	NMeta				& OCA					& ---			& ---			& ---			& ---			\\
	\hline
	This work			& Self-defining top MM	& \checkmark	& \checkmark	& \checkmark	& \checkmark	\\
	\hline
\end{tabular}
\label{tab:mlm-comparison}
\end{table}}

Notice that, unlike other approaches, our realization makes the declared potency of an element independent of the potency of its type (except for depth values different from \emph{unbounded}, where such dependency is inherent to the concept of depth).
For example, referring back to the Robolang example presented in Section~\ref{subsec:application-dimension},
the default value of the potency of an arrow \elementname{out} in \elementname{robot\_1} could be changed to some other one without being affected by the potency \typename{}{1--2--*} of its type arrow (\elementname{out} in \elementname{robolang}).
Also, this kind of potency could be used for the realisation of abstract classes by declaring values \elementname{0--0--0}, similar to~\cite{atkinson2012ontology}, although our approach natively supports abstract classes with non-zero potency, so that classes inheriting from them can be instantiated.
This allows for a clearer definition of abstract classes, in a similar manner to other tools like MetaDepth.
This equivalence between three-valued potency and other multilevel concepts (e.g.\ leap potency, mutability and durability) is outlined our work in~\cite{macias2017rearchitecting}, together with its equivalence to the potency specified by some multilevel tools.

This conceptualisation of potency binds together several ways of realising the concept contributed by different approaches.
Similarly, other works investigate the definition of precise syntax and semantics for references.
In~\cite{atkinson2015unifying}, the authors formalise uniformly all the relationships in UML class diagrams, which opens the door to a consistent and standardized way of defining types of edges for models.
In the work presented in this thesis, we formalise the most widely used relations in modelling, namely reference, inheritance and containment.
Further kinds of connections like the ones defined in~\cite{atkinson2015unifying} could be added to our formal framework by including related graph-based constructions and conditions, and possibly sanity checks to ensure their correct usage, semantically speaking.
We also refer the reader to Section~\ref{sec:tool-requirements}, for a related discussion about the definition of domain-specific associations.

Additionally, it is worth mentioning that our approach to MLM takes into account the fundamental distinction between classification and specialisation~\cite{kuhne2009contrast}.
Although it can be argued that some typing relations can be replaced by inheritance, we choose the former in several examples in this thesis for the sake of flexibility and for illustration purposes.
For example, one could consider replacing the typing relations in the Robolang example by inheritance for \elementname{Task} and its instances \elementname{GoForward}, \elementname{GoBack}, etc.  
It could be argued that this kind of replacement can reduce a whole hierarchy into two levels.
While it is outside the scope of this thesis to take a stand on the classification vs generalization question (i.e.\ instantiation vs inheritance), we refer the reader back to our arguments to exploit MLM in Section~\ref{sec:why-mlm}.
In this regard, one can consider other techniques as an alternative to MLM, such as variability modelling in Software Product Lines~\cite{gomaa2005spl}, component-based models, and in general the usage of UML profiles and stereotypes~\cite{odell1998advanced}.
But we do not believe that these techniques are a competing paradigm against MLM, but actually could be incorporated into it to achieve a higher degree of flexibility in MDSE.
Therefore, we do not restrict how a model designer would organise the concepts in her hierarchy and we leave room for the definition of good and bad practises in a similar fashion to object-oriented patterns and anti-patterns (see~\cite{kuehne2018story}).

Finally, other authors like~\cite{gerbig2017deep}, tackle the definition of DSMLs by focusing on practical aspects like their concrete syntaxes, views and notations.
In this thesis, we use concrete syntax just as a tool to ease the presentation and discussion of our case studies in Sections~\ref{subsec:application-dimension} and~\ref{subsec:extended-robolang}.
Our current and future work towards definition of concrete syntax in MultEcore by exploiting its EMF compatibility is discussed in Section~\ref{sec:future-work}.

\subsection{Reusability}
\label{subsec:reusability}

There are several alternative proposals to improve reusability of models and model transformations. 
In~\cite{chechik2016perspectives}, two perspectives on model transformation reuse are presented: programming language- and MDSE-based. 
For each perspective, the authors discuss two approaches: subtyping and mapping, and lifting and aggregating. 
In~\cite{struber2015variability} a variability-based graph transformation approach is introduced to tackle the performance problems that are introduced by systems in which a substantial number of the rules are similar to each other, a phenomenon called \emph{proliferation} in this thesis. 
In~\cite{sen2012reusable} an approach to reusable model transformations is presented, which is based on sub-typing an effective part of the existing source metamodel. 
That is, the metamodel of the models to be transformed is made a subtype of a pruned metamodel.
In this way, the models can be transformed by the same transformation rules which were written for the source metamodel.

To achieve genericness, in~\cite{sanchez2011generic,delara2014flexiblereuse} the transformation rules are typed over a generic metamodel which is called \emph{concept}.
Then, any metamodel to which there exists an embedding from the concept-metamodel (called binding) can be used with these rules.
Thus, any model in a hierarchy can be bound to the concept metamodel and get the transformation rules for free.
However, it is not always straight forward to define the embedding morphism from the concept metamodel to the metamodel.
This might be because the metamodel has several structures which have the same behaviour leading to several bindings.
This is solved by introducing syntax for the definition of cardinality in which the concept metamodel can be written in a generic way.
However, in the realisation of the concept, this is just syntactic sugar for the definition of multiple concept metamodels.
Moreover, finding reasonable embeddings due to structure mismatches (or heterogeneity) might be a challenge.
Adapters and concept inheritance could be seen as solution of this problem, as seen in~\cite{sanchez2011generic}.

The notion of \emph{concept} from~\cite{delara2014flexiblereuse} is extended in~\cite{duran2015amalgamation} to parametric models, where the parameters have both structure and behaviour.
In this case we not only have a mechanism for the reutilisation of transformations, but mechanisms for the reutilisation and composition of models with behaviour. 
The challenges in finding embeddings are however more difficult, since rules in parameter models must also be mapped to the corresponding target models.

\subsection{Multilevel model transformation}
\label{subsec:multilevel-model-transformations}

Although some operations over multilevel hierarchies have been formalized (see, e.g.,~\cite{rossini2014formalisation,delara2015dsmml}), the definition of transformations on multilevel hierarchies is not yet well handled.
To the best of our knowledge, the only proposals supporting a form of multilevel transformations are Melanee~\cite{atkinson2009infrastructure,atkinson2012towards}, MetaDepth~\cite{delara2010deep} and DeepJava~\cite{kuehne2007deepjava}.
They follow very different approaches, but none of them provides a formalization of multilevel transformations.
Here, we focus on their possibilities for multilevel hierarchy transformations.

De Lara et al.\ present in~\cite{delara2015dsmml} their extension of the Epsilon family of languages~\cite{kolovos2008epsilon} to support multilevel transformations for MetaDepth.
They define model manipulation operations, code generators working at several metalevels, and model-to-model transformations.
But all these transformations are defined at a particular level of the hierarchy with references to indirect types.
For transformations affecting multiple levels, they propose ``co-transformations''.
Their name comes from the fact that they are defined as transformations where models and their corresponding metamodels are transformed simultaneously.
Similar approaches are discussed in~\cite{herrmannsdorfer2014coupled} in the context of co-evolution. 
They also provide support for refinement of transformations, where some adaptation is required before they can be applied.

Atkinson et al.\ present in~\cite{atkinson2012towards,atkinson2015enhancing} their ideas for extending the transformation language ATL to work on multilevel models for Melanee.
Although they discuss the two-level to multilevel and multilevel to two-level transformation cases, the paper focuses on the multilevel to two-level case.
Moreover, in the current implementation of this approach, only the instances which are at the lowest level in the metamodelling hierarchy will be transformed.
For example if a modelling hierarchy contains four levels and a transformation is defined on the highest level, only the model elements on the lowest level are transformed into the target model. 
This hinders the definition of metamodels representing language behaviour at any level in the stack, that is, these metamodels need to be at the next-to-bottom level.

Lastly, the tool DeepJava~\cite{kuehne2007deepjava} extends the Java programming language with the ability to perform multiple instantiations.
Since DeepJava programs are compiled into Java, their objects can be manipulated in what could be considered a sort of imperative model transformation language which is not tailored to model-to-model transformations.

Conversely to the aforementioned approaches, other contributions like~\cite{schulz2011categorical} use Graph Theory and Category Theory as a formal background to define MT over models which define and use a DSML.
In that article, the authors use transformation rules to refactor object-oriented diagrams while co-evolving their instances semi-automatically.
They do not, however, investigate their application into multilevel hierarchies of models, neither apply them for the definition of behaviour, as opposed to what we do in our approach.

Finally, in~\cite{machado2015rule} the authors apply the double-pushout approach for higher-order transformations (HOT).
This idea seems compatible with MCMTs, and opens a possible line of research where we apply MCMT for HOT in order to perform MCMT proliferation or translation into other representations, by exploiting the formalisation that we present in this thesis.

\subsection{Model transformations for the definition of behaviour}
\label{subsec:mts-for-behaviour}

Two-level modelling tools such as Groove~\cite{rensink2003groove}, AGG~\cite{taentzer2003agg}, ATOM$^3$~\cite{delara2002multiformalism,delara2006visual}, e-Motions~\cite{rivera2009graphical}, Tiger EMF~\cite{biermann2006graphical} and MOMENT2~\cite{boronat2009moment} use in-place model transformation to deal with the behaviour specification of DSMLs.
The Groove and AGG tools do not support the definition of a specific graphical notation associated to the DSML at hand, since rule patterns are defined with graphs.
AToM$^3$ and e-Motions allow the user to specify a graphical concrete syntax for the DSMLs.
From these tools, only e-Motions support the interoperation of the models defined using it with other tools.
Moreover, e-Motions, Tiger EMF and MOMENT2 were developed for the Eclipse platform, and enable their integration with other modelling tools already defined for Eclipse because they all share the same representation for models and metamodels.
This allows users to benefit from all the available tools in EMF to achieve a complete MDSE process.
Tools like Groove, AGG and ATOM$^3$ are developed as stand-alone tools so their connection with other modelling tools requires high effort due to the lack of a common model and metamodel representation.

Systems like e-Motions and Groove focus on the definition of timed systems, and on the verification of the systems specified, providing facilities for model-checking or statistical model-checking inside their environments~\cite{ghamarian2012groove,duran2016statistical}. 
See~\cite{rensink2008explicit,ghamarian2012groove} for a discussion on Groove, or~\cite{rivera2009formal} for the use of different formal verification techniques on systems specified using in-place model transformations.
Moreover, in~\cite{hausmann2004dynamic} and~\cite{wang2012diagrammatic} two formal approaches are presented where behavioural languages with explicit time semantics are formally defined using graph theory.
Both use graph transformations as a formalisation of the MTs with which they express the behaviour of their instances.
Some of the examples in this thesis are inspired by such ideas, but we generalise to any kind of behavioural DSML, and do not consider time an exceptional feature that must be managed differently.

\section{Requirements for our tool and approach}
\label{sec:tool-requirements}

In this section we evaluate our MultEcore tool, and our approach in general, against two different sets of requirements defined in two contributions in the context of the MODELS 2018 conference~\cite{models2018}.
In both cases, we list each requirement (or a summary of it) in italics and discuss below it whether we address it and how.

The first set of requirements was published in~\cite{frank2018requirements}, and is reproduced in the following list.
All requirements are marked to have high priority except \textit{R5} (unspecified) and \textit{R6} (medium).

\begin{itemize}
	\item \textit{R1: It should be possible to define classes with a contingent level.}
	
	We understand this requirement as the possibility of having two classes on different classification levels classified by the same meta class.
	This kind of construction is possible in MultEcore by default, thanks to having range-like potency and allowing for gaps in the hierarchy.
	Moreover, we argue that our three-valued potency is capable of representing a richer variety of scenarios that the one indicated in this requirement.
	
	\item \textit{R2: It should be possible to define preliminary classes without a meta-class.}

	Defining a class without specifying its type is allowed in MultEcore, although that does not mean that the class does not have a type, but rather that the node is typed by \elementname{EClass} by default.
	Even if this is not required, we do have the same for edges, which are typed by \elementname{EReference} by default.
	
	\item \textit{R3: Intrinsic associations should allow the specification of different instantiation levels for both participating classes.}
	
	To our understanding, this requirement entails the possibility of a relation between two classes that goes across two different levels.
	An important part of the MLM community does not agree with the convenience of cross-level relations~\cite{kuehne2018story} and we are cautious about including them.
	Currently we do not allow edges between two nodes from models in different levels.
	In the future we will consider including cross-level relations, in accordance to the general agreement of the community on this matter.
	
	If this requirement is interpreted as the possibility of having a reference between two classes where the direct types of each of the three elements is located in a different level inside the same graph chain, this is, in fact, one of the key points of our formalisation of multilevel modelling (see condition~\ref{cond:non-dangling} at the end of Section~\ref{subsec:individual-typing}).
	
	\item \textit{R4: It should be possible to indicate that certain properties of a class are to be inherited to its instances.}
		
	The use of range potency for attribute declarations (that is, the first two values of our three-valued potency, since depth always equals \elementname{1}) allows us to define this kind of scenario.
	Several examples of using potency in the attribute declaration so that it can be instantiated in indirect instances of its containing node have been presented in the Bicycle case study in Section~\ref{subsec:bicycle-challenge}.
	
	\item \textit{R5: The language should provide generic association types.}
		
	Due to its focus on providing full EMF compatibility, MultEcore already offers the inheritance and containment association types by default.
	We plan to investigate the addition of further types into our framework in the future.
	
	\item \textit{R6: It should be possible to define domain-specific association types.}
	
	To our understanding, this requirement implies that the framework must have a way of providing additional constraints to associations.
	In such a case, we have provided several examples throughout this thesis where we use a supplementary hierarchy with boolean (and temporal) logics which allows us to define well-formedness constraints for a model and its instances.
	Hence we believe that this requirement, which has a lower priority in the original publication, can be also fulfilled in our approach.
	
	\item \textit{R7: A language should allow for clearly distinguishing between attributes that are regularly instantiated into individual values of instances [...] and those that serve the specification of constraints on attribute values.}
	
	We believe that this requirement is related to the possibility of having a node which both instantiates attributes defined in its transitive types and declares its own attributes to be instantiated in the levels below.
	If that is the case, our approach already contemplates such possibility, although we have not used it in our case studies.
	Moreover, fulfilling \textit{R4} allows us to also fulfil this requirement and provide fine-tuning of the level of instantiation of the attributes.
	
	\item \textit{R8: It should be possible to assign categories to operations.}
	
	We understand categories in this requirement as a kind of label or group to which operations can be assigned.
	Our approach does not consider operations as part of the model, but rather provides model transformations as way of manipulating our models (and possibly even querying them).
	However, we do agree that whichever means for manipulating the models are provided by an approach, it is useful to be able to categorise them.
	We are taking steps in that direction with the introduction of priorities and coordination for our MCMT rules, as discussed in Section~\ref{sec:future-work}.
\end{itemize}

This second list contains the requirements specified in~\cite{guerra2018quest}.
These requirements are related to flexible modelling techniques in general, not only multilevel modelling.
However, it is a good exercise for us to see how our approach can tackle them and we can gain insight to guide the future development of MultEcore.
Here, we annotate each requirement name with an apostrophe in order to maintain the original numbering in the paper while differentiating it from the ones in the previous list.

\begin{itemize}
	\item \textit{R1': Configurable inconsistency tolerance.}
	
	Our consistency rules are defined as properties through supplementary typing.
	Thanks to our flexible typing mechanism (see Section~\ref{subsec:flexibility}) these rules can be made more generic or precise as required.
	This same mechanism allows us to define models where some of the types are missing (and therefore are replaced by the default ones) or where an element can declare its type in a higher level of abstraction, favouring genericness.

	\item \textit{R2': Information extension: untyped objects, instantiating undeclared attributes, etc.}
	
	The part of this requirement related to classes and references is equivalent to the requirement \textit{R2} in the previous list, and the reader can refer to the discussion there.
	When it comes to unspecified attributes, we can address this requirement at least partially by using supplementary hierarchies.
	In general, this concept seems to be an evolution of the idea of linguistic extension, and we already developed the idea of supplementary typing as an alternative to linguistic extension.
	Further consequences of this requirement, like dynamic computation of features, remain to be investigated.
	
	\item \textit{R3': Configurable classification relation: dynamic typing, multiple typing, a-posteriori typing and multilevel modelling.}
	
	All these concepts were taken into account since the early development of our approach, and so far we do not identify a case where MultEcore cannot provide the required functionality, either in the requested manner or in an equivalent one.
	
	\item \textit{R4': Configurable generalisation relation.}
	
	In its current status, our tool supports both generalisation and classification, including multiple typing and multiple inheritance.
	Making these features configurable would not pose a challenge, as far as we can foresee.
	
	\item \textit{R5': Explicit and configurable modelling process} and \textit{R6': Process-aware extensible assistance.}
	
	These are the two only requirements in this list that MultEcore does not support either fully or partially, but the efforts to address them seem more practical than conceptual, and we believe that including process-aware features into MultEcore would not cause any conflicts with the current state of our approach.
\end{itemize}
	
In conclusion, our approach and our tool already address most of the requirements in both lists, including the ones considered as high priority in the first list and those related to conceptual matters.
Addressing the remaining ones does not seem problematic in terms of conflicts with our formal and conceptual frameworks, and we are positive that we could tackle them in the future if the modelling community agrees in their relevance.

\section{Future Work}
\label{sec:future-work}

This thesis covers topics which range from practical tool implementation to formal matters using Category Theory.
As a consequence, the scope of future work for the research presented in this thesis can go in several complementary directions.
One of these directions is the improvement of the tools developed in the context of this thesis.

First, we plan to improve the tool MultEcore, including general fixes to stability and usability as well as the addition of new functionality, like a Sirius-based concrete syntax editor that can easily exploit the multilevel features of our framework.
For example, the user could define the same syntax for direct and indirect instances of certain elements, as we do in Section~\ref{subsec:extended-robolang} in this thesis.
Some work in this direction has already started for the definition of domain-specific Coloured Petri Nets, as seen in~\cite{rodriguez2018mlmcpn}.

Second, the algorithm for proliferating MCMT rules presented in Section~\ref{subsec:proliferation} is currently being implemented as part of MultEcore's MCMT textual editor.
At the same time, we are researching the viability of using Maude~\cite{rivera2009formal} as a transformation engine for MCMTs, which would not require the proliferation and two-level MT generation steps.
Once these two lines of work are finished, we plan to develop an explicit coordination mechanism to easily create layers for MCMT rules (defined for the same or different hierarchies, but applied in the same model) in order to control the execution of behaviour.
The need for such coordination has already been identified in this thesis, for both the Robolang language (see Section~\ref{subsec:mcmts-robolang}) and its extended version (Section~\ref{subsubsec:extended-robolang-coordination}).
Another promising line of work is the possibility of amalgamation of MCMT rules defined for different hierarchies but which must be executed simultaneously and in the same model due to the double typing (application and supplementary) that our framework allows.

And third, we intend to perform further experiments related to the work presented in Section~\ref{sec:empirical-evaluation}, both for MLM rearchitecting and tool comparison and exchange.
In order to do that, we must extend first the Rearchitecter tool, by providing new heuristics, transformations and MLM concepts; as well as increasing the number of supported tools for both recommendation, importing and exporting.
We aim for this work to be a joint effort of the MLM community, which can in the future promote the convergence of MLM tools, as we presented in~\cite{macias2018convergence}.

Aside from tool-related improvements, we would like to explore the idea of reusability in more depth.
One interesting line of work related to this is the creation of \emph{families} of languages which share similar structure and behaviour and organising them on multilevel hierarchies, with MCMT rules defining the different levels of abstract behaviour.
This idea was explored at the beginning of our work, as illustrated by our Petri Nets example (see Section~\ref{subsec:petri-nets}), but was postponed in order to focus on the  the convergence of the state of the art of multilevel modelling and contribute to it with multilevel model transformations.

Additionally, we plan to study the application of MCMTs in the field of co-transformations, where several levels are transformed at once.
That is, the FROM and TO blocks in our rules could be composed of more than one level.
We believe that our formalisation is compatible with this kind of rules, and that way of exploiting a multilevel setting for model co-transformation seems to be a promising idea.

Finally, related to the more formal aspects of this work, we intend to further explore the \cat{Chain} category presented in Chapter~\ref{chap:mcmt}, and to present our results to the Graph Theory and Category Theory communities.
We are currently writing conference papers with this aim, although most of our results can already be consulted in a technical report~\cite{macias2017chains}.

\section{Conclusions}
\label{sec:conclusions}

In Chapter~\ref{chap:introduction} of this thesis we outline the research questions that guided our work and enumerate the contributions to the field of MLM that this research provides.
In this section, we revisit them and briefly comment on how the previous chapters of this thesis refer to each of these points.
We begin with Section~\ref{subsec:flexibility}, where we comment on the main driving force of our approach, namely flexibility.
Then, in Section~\ref{subsec:research-questions-revisited} we comment on our research questions and how we address them in this work.
In Section~\ref{subsec:contributions} we comment on the contributions of this work, with references to the main points of this thesis.
Finally, in Section~\ref{subsec:summary} we conclude this thesis with a summary of the work we present here.

\subsection{Flexibility and reusability}
\label{subsec:flexibility}

One of the main driving forces of our approach to multilevel modelling is flexibility and how it contributes to reusability.
Our methods and tools to build multilevel hierarchies and multilevel model transformations have a strong focus on allowing the user to define structure and behaviour in a way which is highly reusable and adaptable.
In the future, we may need to introduce recommendations, modelling guides and sanity checks as part of the approach so that our tools are not misused, following principles like the ones described in~\cite{kuehne2018story}.
But we favour flexibility at the current state of our approach, so that we can explore its reusability potential and expressive power in a multilevel setting without any strong boundaries.
Therefore, we identify six different kinds of flexibility provided by our approach:

\begin{itemize}
	\item \textbf{Level-based deep instantiation}
	Thanks to our three-valued potency, we allow individual direct types to ``jump'' over abstraction levels.
	This may cause in some extreme cases that some levels in a multilevel hierarchy are empty, i.e.\ represented by an empty graph.
	Such cases can be handled by our approach without needing any special treatment, in case we consider them useful in future scenarios.
	Otherwise, the modelling guides and sanity checks aforementioned can be directed at preventing such cases.
	\item \textbf{Evolving hierarchies}
	Thanks to the previous point, our hierarchies can be easily modified after their initial implementation.
	This not only concerns the trivial addition of new models in levels which already contain some models, but also the introduction of new intermediate abstraction levels which can be populated with models.
	\item \textbf{Multiple typing}
	While this kind of flexibility is not exclusive from our approach, we allow and thoroughly exploit the introduction of supplementary types for our models and elements, and data types for the definition of attributes.
	Supplementary types can be added or removed dynamically, without compromising the main types (and associated semantics) of the elements.
	\item \textbf{Orthogonal extension}
	The possibility of an element with multiple types being affected by different transformations for each of those types is also considered in our formalisation.
	Since our MCMTs perform graph rewriting on our objects in \cat{Chain} in the traditional way and typings are orthogonal to each other, we get the typing of the resulting objects after applying an MCMT rule without requiring any further calculations.
	\item \textbf{Horizontal flexibility}
	Our MCMT rules can, and are meant to, be applied in different branches of a multilevel hierarchy, even for branches not existing at the moment of specifying the rules, as long as the required typing relations are preserved.
	\item \textbf{Vertical flexibility}
	When applying an MCMT rule in a model, our approach allows to stretch the abstraction range as much as needed.
	That is, one typing relation in the rule can match against a transitive sequence of typing relations arbitrarily large in the target hierarchy.
	This approach gives us the desired flexibility and does not compromise the specificness of the rules for all the case studies included in this thesis.
	We do not include here some existing work regarding different notations that offer a finer control of the matching of typing relations, which can enhance even further the expressive capabilities of MCMTs.
	\item \textbf{Lazy matching} 
	As a consequence of vertical flexibility, one can see that our definition of MCMTs does not require a typing relation in an MCMT rule to reflect all the typing details of the target hierarchy.
	That is, we require rule applications to preserve typing in a confluent way but not that transitive types are reflected.
\end{itemize}

\subsection{Research questions revisited}
\label{subsec:research-questions-revisited}

In the following list we include our research questions, and provide a summarised response to each of them.

\begin{itemize}
	\item \textbf{RQ1:}
	Is there a common framework which generalises different state-of-the-art techniques and approaches to MLM in a consistent manner?
	
	\item \textit{Response:}
	We unify under our conceptual framework the MLM concepts which are considered most relevant by the community, like multiple typing, different kinds of potency, and flexible instantiation mechanisms.
	Our conceptual framework also has a strong focus on flexibility, which allows us to define highly reusable modelling hierarchies with multiple levels of abstraction and supplementary aspects.
	These features are showcased in several case studies.
	\item \textbf{RQ2:}
	Is there a formalisation of MLM which accounts for different state-of-the-art techniques and approaches to MLM?
	
	\item \textit{Response:}
	We use Category Theory and formal proofs to formalise our conceptual framework.
	This formalisation is elegant, sound and directly applicable to model transformations.
	\item \textbf{RQ3:}
	Is there an MLM transformation approach which integrates different state-of-the-art techniques and approaches to MLM?
	
	\item \textit{Response:}
	We illustrate how the notions defined in our conceptual framework for MLM can also be applied to Model Transformations, again with a strong focus on their flexibility, especially applied to the specification of DSML behaviour.
	These points are once more illustrated through motivating examples from our case studies and a formalisation using Category Theory.
\end{itemize}

\subsection{Contributions}
\label{subsec:contributions}

We list here the main contributions of our work, and refer to the places where those contributions are discussed.

\begin{itemize}
	\item An approach to Multilevel Modelling is presented in Chapter~\ref{chap:mlm}.
	The application of similar principles for Model Transformations in a multilevel setting, called Multilevel Coupled Model Transformations, is presented in Chapter~\ref{chap:mcmt}.
	The way in which this framework integrates state-of-the-art techniques, especially regarding flexibility and reusability, is discussed in Section~\ref{subsec:flexibility}.
	\item A novel formalisation of Multilevel Modelling using Graph Theory is included in Section~\ref{sec:hierarchies}.
	For the proposed Multilevel Coupled Model Transformations, we extend this formalisation using Category Theory in Section~\ref{sec:chain-category}.
	\item The implementation of our EMF-based framework is discussed in Chapter~\ref{chap:tooling}.
	It includes the aspects related to Multilevel Modelling in Section~\ref{sec:modelling-tooling} and those related to Multilevel Coupled Model Transformations in Section~\ref{sec:mcmt-tooling}.
	\item The case studies we used to analyse the strengths and weaknesses of our framework are presented in Section~\ref{sec:case-studies}.
	\item The experimental validation of MLM in general, and our approach to it in particular, is presented in Section~\ref{sec:empirical-evaluation}.
\end{itemize}

\subsection{Summary}
\label{subsec:summary}

The main goal of this thesis is to contribute to the fields of multilevel modelling and multilevel model transformations with a focus on integration rather than competition.
We also try to ``cover all bases'', from formal aspects to tooling.
We believe this work succeeds at doing so, and we support this belief with detailed explanations, formal constructions and proofs, case studies, experiments and a review of related works.